\pdfoutput=1

\documentclass[reprint,prb,nofootinbib,superscriptaddress]{revtex4-2}

\usepackage{tikz}
\usetikzlibrary{decorations,decorations.pathmorphing,decorations.markings,calc,snakes,positioning,fixedpointarithmetic,shapes,shapes.geometric,patterns}

\usepackage{etoolbox}
\AtBeginEnvironment{tikzpicture}{\catcode`$=3 } % $ % was important for auctex preview??

\tikzset{alignmid/.style={baseline={([yshift=-.5ex]current bounding box.center)}}} % adjust pictures vertically
\tikzset{every picture/.append style=alignmid}

\tikzset{
bottomzigzag/.style={postaction={draw,decorate, decoration={zigzag,amplitude=1pt,segment length=3pt,raise=1pt}}},
zigzag/.style={draw,decorate, decoration={zigzag,amplitude=1pt,segment length=3pt}},
rc/.style=rounded corners,
}

\tikzset{
    -|/.style={to path={-| (\tikztotarget)}},
    |-/.style={to path={|- (\tikztotarget)}},
}

\usepackage{ifthen}
\usepackage{xparse,pgffor}

\tikzset{
mark/.code={
\tikzset{postaction={/network/mark/.cd,#1,/tikz/.cd,decorate},decoration={name=markings,mark=at position \netmarkpos with{%+\netmarkposoff} with{
\begin{scope}[netmarktrafo]
\netmarkcode
\end{scope}
}}}
\def\netmarkpos{0.5}%\pgfdecoratedpathlength}
},
}

\def\netmarkpos{0.5}%\pgfdecoratedpathlength}
\def\netmarkposoff{0cm}
\def\netmarkcode{}

\tikzset{
netmarktrafo/.style={},
netmarkstyle/.style={solid,semithick,sharp corners},
}

\pgfkeys{
/network/mark/.cd,
sty/.code={
\tikzset{netmarkstyle/.style={#1}}
},
asty/.code={
\tikzset{netmarkstyle/.append style={#1}}
},
f/.style={asty=fill},
p/.code={ % relative positioning on decorated line
\def\netmarkpos{#1}%\pgfdecoratedpathlength}
},
poff/.code={ % absolute shift on decorated line in cm
\def\netmarkposoff{#1cm}
},
e/.code={ % mark at the End of line
\def\netmarkpos{\pgfdecoratedpathlength-0.005cm-\netmarkposoff}

\tikzset{netmarktrafo/.append style={shift={(-\netmarkwidth,0)}}}
},
s/.code={ % mark at the Start of line
\def\netmarkpos{0.005cm+\netmarkposoff}

\tikzset{netmarktrafo/.append style={shift={(\netmarkwidth,0)},xscale=-1,yscale=-1}}
},
a/.code={ % mark After end of line
\def\netmarkpos{\pgfdecoratedpathlength-0.005cm}

\tikzset{netmarktrafo/.append style={xscale=-1,shift={(-\netmarkwidth,0)}}}
},
b/.code={ % mark Before start of line
\def\netmarkpos{0.005cm}

\tikzset{netmarktrafo/.append style={xscale=-1,shift={(\netmarkwidth,0),yscale=-1}}}
},
-/.code={ % flip horizontally (mark backwards)
\tikzset{netmarktrafo/.append style={xscale=-1}}
},
r/.code={ % flip vertically (mark on the Right instead of left)
\tikzset{netmarktrafo/.append style={yscale=-1}}
},
sideoff/.code={ % shift sideways
\tikzset{netmarktrafo/.append style={shift={(0,#1)}}}
},
slab/.code={ % label next to the line
\def\netmarkwidth{0}
\def\netmarkcode{
\node[inner sep=0.04cm,netmarkstyle,draw=none] (mylabelwidthtest) at (0,0){\phantom{#1}};
\path let \p1=(mylabelwidthtest.north east), \p2=(mylabelwidthtest.south east), \n1 = {max(abs(\y1),abs(\y2))} in node[inner sep=0.04cm,netmarkstyle] at (0,\n1) {#1};
}
},
lab/.code={ % label on the line
\def\netmarkwidth{0}
\def\netmarkcode{
\node[inner sep=0.04cm,anchor=\netmarkanchor] (mylabelwidthtest) at (0,0) {\phantom{#1}};
\draw[white] (mylabelwidthtest.\pgfdecoratedangle)--(mylabelwidthtest.\pgfdecoratedangle+180);
\node[inner sep=0.04cm,anchor=\netmarkanchor,netmarkstyle] at (0,0) {#1};
}
},
arr/.code={ % arrow
\def\netmarkwidth{0.04}
\def\netmarkcode{\draw[netmarkstyle] (-0.04,0.08)--(0.04,0)--(-0.04,-0.08);}
},
rarr/.code={ % round arrow
\def\netmarkwidth{0.04}
\def\netmarkcode{\draw[netmarkstyle] (-0.04,-0.08)arc(90-180:90:0.08);}
},
dot/.code={ % dot
\def\netmarkwidth{0.08}
\def\netmarkcode{\draw[netmarkstyle] (0,0)circle(0.08);}
},
flag/.code={ % flag to the left
\def\netmarkwidth{0.06}
\def\netmarkcode{\draw[netmarkstyle] (-0.06,0)--(0,0.09)--(0.06,0)--cycle;}
},
darr/.code={ % "double arrow": half-arrow on the left
\def\netmarkwidth{0.08}
\def\netmarkcode{\draw[netmarkstyle] (-0.04,0)--(0.04,0)--(-0.04,0.08)--cycle;}
},
hcirc/.code={ % half-circle on the left
\def\netmarkwidth{0.1}
\def\netmarkcode{\draw[netmarkstyle] (-0.1,0) arc (180:0:0.1);}
},
bar/.code={ % tick, short line perpendicular to line
\def\netmarkwidth{0.05}
\def\netmarkcode{
\coordinate (a) at (0,-0.08cm-0.5*\pgflinewidth);
\coordinate (b) at (0,0.08cm+0.5*\pgflinewidth);
\draw[netmarkstyle] (a)--(b);
}
},
hbar/.code={ % tick, short perpendicular line towards the left
\def\netmarkwidth{0.05}
\def\netmarkcode{
\draw[netmarkstyle] (0, 0.5*\pgflinewidth)--++(0,0.12);
}
},
mill/.code={
\def\netmarkwidth{0.16}
\def\netmarkcode{
\draw[netmarkstyle] (0,-0.5*\pgflinewidth)--++(-0.08,-0.08)--++(0,0.08);
\draw[netmarkstyle] (0,0.5*\pgflinewidth)--++(0.08,0.08)--++(0,-0.08);
}
},
three dots/.code={
\def\netmarkwidth{0.2}
\def\netmarkcode{
\fill (-0.12,0) circle (0.5*0.05) (0,0) circle (0.5*0.05) (0.12,0) circle (0.5*0.05);
}
},
}

\tikzset{wid/.style={minimum width=#1cm}}
\tikzset{hei/.style={minimum height=#1cm}}

\tikzset{sx/.style={xshift=#1cm}}
\tikzset{sy/.style={yshift=#1cm}}

\tikzset{box/.style={draw,rectangle}}
\tikzset{fbox/.style={draw,rectangle, line width=1.1}}
\tikzset{roundbox/.style={draw,rectangle,rounded corners}}
\tikzset{froundbox/.style={draw,rectangle, rounded corners, line width=1.1}}
\tikzset{rounddiamond/.style={draw,diamond,rounded corners}}
\tikzset{dot/.style={draw, shape=circle, fill=black, scale=0.5}}

\tikzset{
netbox/.code={
\node[draw,netbdstyle] (\atomname) at (0,0) {#1};
\coordinate (\atomname-r) at (\atomname.east);
\coordinate (\atomname-l) at (\atomname.west);
\coordinate (\atomname-t) at (\atomname.north);
\coordinate (\atomname-b) at (\atomname.south);
\coordinate (\atomname-tr) at (\atomname.north east);
\coordinate (\atomname-br) at (\atomname.south east);
\coordinate (\atomname-tl) at (\atomname.north west);
\coordinate (\atomname-bl) at (\atomname.south west);
},
}

\tikzset{bdlw/.code={\tikzset{mybdstyle/.style={draw, line width=#1}}}}
\tikzset{bdcol/.code={\tikzset{mybdstyle/.append style={#1}}}}

\newcommand\setelements[1]{
\pgfkeys{/network/atom/.cd,#1}
}

\newcommand\setmarks[1]{
\pgfkeys{/network/mark/.cd,#1}
}

\newcommand\atoms[2]{
\foreach \name/\keys in {#2}{
\expandafter\atom\expandafter{\keys,#1}{\name}
}
}

\newcommand\atom[2]{
\def\atomname{#2}
\tikzset{
nettrafo/.style={},
netatompos/.style={},
netdeco/.style={},
netpostdeco/.style={},
}

\pgfkeys{/network/atom/.cd,#1}

\begin{scope}[netatompos] % shift to atom position
\begin{scope}[nettrafo] % rotate, flip and scale
\netshapecoords % set the anchor coordinates
\fill[netbackstyle] \netshapepath;
\clip \netshapepath;
\tikzset{netdeco}
\draw[netbdstyle] \netshapepath;
\end{scope}
\tikzset{netpostdeco} % draw post-decorations, not rotated, flipped, or scaled
\end{scope}

}

\pgfkeys{
/network/atom/.cd,
square/.code={

\def\netshapepath{(-\tempsize,-\tempsize)rectangle (\tempsize,\tempsize)}
\def\netshapecoords{
\node[rectangle,wid=2*\tempsize,hei=2*\tempsize,inner sep=0,transform shape](\atomname)at(0,0){};
\coordinate(\atomname-c) at (0,0);
\coordinate(\atomname-r) at (\tempsize,0);
\coordinate(\atomname-l) at (-\tempsize,0);
\coordinate(\atomname-t) at (0,\tempsize);
\coordinate(\atomname-b) at (0,-\tempsize);
\coordinate(\atomname-br) at (\tempsize,-\tempsize);
\coordinate(\atomname-tr) at (\tempsize,\tempsize);
\coordinate(\atomname-bl) at (-\tempsize,-\tempsize);
\coordinate(\atomname-tl) at (-\tempsize,\tempsize);
}},
circ/.code={

\def\netshapepath{(0,0)circle(\tempsize)}
\def\netshapecoords{
\node[circle,wid=2*\tempsize,hei=2*\tempsize,inner sep=0,transform shape](\atomname)at(0,0){};
\coordinate(\atomname-c) at (0,0);
\coordinate(\atomname-r) at (\tempsize,0);
\coordinate(\atomname-l) at (-\tempsize,0);
\coordinate(\atomname-t) at (0,\tempsize);
\coordinate(\atomname-b) at (0,-\tempsize);
}},
triang/.code={

\def\netshapepath{(-30:\tempsize)--(90:\tempsize)--(-150:\tempsize)--cycle}
\def\netshapecoords{
\node[regular polygon,regular polygon sides=3,wid=2*\tempsize,inner sep=0,transform shape](\atomname)at(0,0){};
\coordinate(\atomname-c) at (0,0);
\coordinate(\atomname-cr) at (-30:\tempsize);
\coordinate(\atomname-cl) at (-150:\tempsize);
\coordinate(\atomname-ct) at (90:\tempsize);
\coordinate(\atomname-mb) at (-90:0.5*\tempsize);
\coordinate(\atomname-mr) at (30:0.5*\tempsize);
\coordinate(\atomname-ml) at (150:0.5*\tempsize);
}},
diamond/.code={

\def\netshapepath{(0,-\tempsize)--(\tempsize,0)--(0,\tempsize)--(-\tempsize,0)--cycle}
\def\netshapecoords{
\node[rotate=45,rectangle,wid=sqrt(2)*\tempsize,hei=sqrt(2)*\tempsize,inner sep=0,transform shape](\atomname)at(0,0){};
\coordinate(\atomname-c) at (0,0);
\coordinate(\atomname-r) at (\tempsize,0);
\coordinate(\atomname-l) at (-\tempsize,0);
\coordinate(\atomname-t) at (0,\tempsize);
\coordinate(\atomname-b) at (0,-\tempsize);
}},
pentagon/.code={

\def\netshapepath{(-126:\tempsize)--(-54:\tempsize)--(18:\tempsize)--(90:\tempsize)--(162:\tempsize)--cycle}
\def\netshapecoords{
\node[regular polygon,regular polygon sides=5,wid=2*\tempsize,inner sep=0,transform shape](\atomname)at(0,0){};
\coordinate(\atomname-c) at (0,0);
\coordinate (\atomname-mb)at(-90:{\tempsize*cos(36)});
\coordinate (\atomname-mbr)at(-18:{\tempsize*cos(36)});
\coordinate (\atomname-mtr)at(54:{\tempsize*cos(36)});
\coordinate (\atomname-mtl)at(126:{\tempsize*cos(36)});
\coordinate (\atomname-mbl)at(-162:{\tempsize*cos(36)});
\coordinate (\atomname-cbr)at(-54:\tempsize);
\coordinate (\atomname-cr)at(18:\tempsize);
\coordinate (\atomname-ct)at(90:\tempsize);
\coordinate (\atomname-cl)at(162:\tempsize);
\coordinate (\atomname-cbl)at(-126:\tempsize);
}},
halfcirc/.code={

\def\netshapepath{(\tempsize,0)arc(0:180:\tempsize)--++(0,-0.04)-|cycle}
\def\netshapecoords{
\node[circle,wid=2*\tempsize,hei=2*\tempsize,inner sep=0,transform shape](\atomname)at(0,0){};
\coordinate(\atomname-c) at (0,0);
\coordinate(\atomname-r) at (\tempsize,0);
\coordinate(\atomname-l) at (-\tempsize,0);
\coordinate(\atomname-t) at (0,\tempsize);
\coordinate(\atomname-b) at (0,0);
}},
void/.code={
\def\netshapepath{}
\def\netshapecoords{
\coordinate(\atomname) at (0,0);
\coordinate(\atomname-c) at (0,0);
}},
labbox/.code={
\def\netshapepath{(0,0)}
\def\netshapecoords{}
\tikzset{netpostdeco/.append style={netbox=#1}}
},
}

\pgfkeys{
/network/atom/.cd,
p/.code={\tikzset{netatompos/.append style={shift={(#1)}}}},
xscale/.code={\tikzset{nettrafo/.append style={xscale=#1}}},
yscale/.code={\tikzset{nettrafo/.append style={yscale=#1}}},
scale/.code={\tikzset{nettrafo/.append style={scale=#1}}},
rot/.code={\tikzset{nettrafo/.append style={rotate=#1}}},
hflip/.style={xscale=-1},
vflip/.style={yscale=-1},
big/.style={scale=3/2},
small/.style={scale=2/3},
huge/.style={scale=9/4},
tiny/.style={scale=4/9},
flat/.style={yscale=2/3},
fflat/.style={yscale=4/9},
}

\tikzset{
netbdstyle/.style={line width=0.15em}, % changed from pt(default)
netdecstyle/.style={},
netpostdecstyle/.style={},
netbackstyle/.style={white},
}
\pgfkeys{
/network/atom/.cd,
bdstyle/.code={\tikzset{netbdstyle/.style={#1}}},
bdastyle/.code={\tikzset{netbdstyle/.append style={#1}}},
decstyle/.code={\tikzset{netdecstyle/.style={#1}}},
postdecstyle/.code={\tikzset{netpostdecstyle/.style={#1}}},
backstyle/.code={\tikzset{netbackstyle/.style={#1}}},
whiteblack/.style={decstyle=white,backstyle=black},
nobd/.code={bdstyle={line width=0}},
style/.style={bdastyle=#1, decstyle=#1, postdecstyle=#1},
}

\tikzset{
netbscope/.code={\begin{scope}[#1]},
netescope/.code={\end{scope}},
}

\pgfkeys{
/network/atom/.cd,
bscope/.code={\tikzset{netdeco/.append style={netbscope={#1}}}},
escope/.code={\tikzset{netdeco/.append style={netescope}}},
dec/.style 2 args={bscope={#1},#2,escope}, % for applying transformation keys to certain decorations only
vbar/.style={dec={rotate=90}{hbar}},
dcross/.style={dec={rotate=45}{hbar},dec={rotate=-45}{hbar}},
cross/.style={hbar,vbar},
sect6/.style={sect=60},
sect8/.style={sect=45},
sect10/.style={sect=36},
}

\def\regdec#1{\pgfkeys{/network/atom/.cd,#1/.code={\tikzset{netdeco/.append style={net#1}}}}}
\regdec{all}\regdec{rhalf}\regdec{rquart}\regdec{brquart}\regdec{dot}\regdec{spiral}\regdec{swirl}\regdec{hstripe}\regdec{hbar}\regdec{rrey}
\pgfkeys{/network/atom/.cd,sect/.code={\tikzset{netdeco/.append style={netsect=#1}}}}

\tikzset{
netall/.code={\fill[netdecstyle] (-0.3,-0.3)rectangle (0.3,0.3);}, % fill all
netrhalf/.code={\fill[netdecstyle] (0,-0.3)rectangle (0.3,0.3);}, % right half
netrquart/.code={\fill[netdecstyle] (0.075,-0.3)rectangle (0.3,0.3);}, % right quarter
netbrquart/.code={\fill[netdecstyle] (0,0)rectangle (0.3,-0.3);}, % bottom right quarter
netsect/.code={\fill[netdecstyle] (0,0)--(0,-0.3)arc(-90:-90+#1:0.3)--cycle;}, % section of angle #1 starting from -90
netdot/.code={\fill[netdecstyle] (0,0)circle(0.07);}, % dot in the middle
netspiral/.code={\draw[netdecstyle] plot [variable=\t,domain=0:4] ({0.075*\t*cos(pi*(\t-0.5) r)},{0.075*\t*sin(pi*(\t-0.5) r)});}, % spiral
netswirl/.code={\fill[netdecstyle] plot [variable=\t,domain=0:2] ({0.15*\t*cos(pi*(\t-0.5) r)},{0.15*\t*sin(pi*(\t-0.5) r)}) arc(-90:-450:0.3)--cycle;}, % filled swirl
nethstripe/.code={\fill[netdecstyle] (-0.3,-0.05)rectangle(0.3,0.05);}, % horizontal stripe
nethbar/.code={\draw[netdecstyle] (-0.3,0)--(0.3,0);}, % horizontal line
netrrey/.code={\draw[netdecstyle] (0,0)--(0.3,0);} % line from the middle to the right
}

\pgfkeys{
/network/atom/.cd,
postbscope/.code={\tikzset{netpostdeco/.append style={netbscope={#1}}}},
postescope/.code={\tikzset{netpostdeco/.append style={netescope}}},
postdec/.style 2 args={postbscope={#1},#2,postescope}, % for applying transformation keys to certain decorations only
arc/.code={\tikzset{netpostdeco/.append style={netarc=#1}}},
lab/.code={\tikzset{netpostdeco/.append style={netlab={#1}}}},
shadecirc/.code={\tikzset{netpostdeco/.append style=netshadecirc}},
shaderect/.code={\tikzset{netpostdeco/.append style={netshaderect={#1}}}},
debug/.code={\tikzset{netpostdeco/.append style=netdebug}},
markline/.code 2 args={\tikzset{netpostdeco/.append style={netmarkline={#1}{#2}}}},
postcirc/.code={\tikzset{netpostdeco/.append style=netpostcirc}},
}

\tikzset{
netlab/.code={
\pgfkeys{/network/atom/lab/.cd,#1}
\node[netpostdecstyle] at (\ifdefined\netlabpos\netlabpos\else\netlabang:\netlabdist\fi) {\netlabwrap{\netlabtext}};
},
netarc/.code args={#1:#2:#3}{
\draw[netpostdecstyle] (#1:#3) arc (#1:#2:#3);
},
netshadecirc/.code= {
\fill[opacity=0.4,netpostdecstyle] (0,0)circle(0.4);
},
netpostcirc/.code= {
\draw[netpostdecstyle] (0,0)circle(0.15);
},
netshaderect/.code= {
\fill[rc,opacity=0.4,netpostdecstyle] ($-1*(#1)$) rectangle (#1);
},
netdebug/.code= {
\node[red] at (0,0){\atomname};
},
netmarkline/.code 2 args= {
\draw (\atomname)edge[mark={#2}]++(#1);
},
}

\def\netlabwrap#1{#1}
\pgfkeys{
/network/atom/lab/.cd,
t/.code={\def\netlabtext{#1}}, % label text
p/.code={\def\netlabpos{#1}}, % position of the label
ang/.code={\def\netlabang{#1}},
dist/.code={\def\netlabdist{#1}},
wrap/.code={\def\netlabwrap##1{#1}}, % store in doesn't work
}

\usetikzlibrary{fadings}

\usepackage{amsmath}
\usepackage{amssymb}
\usepackage{bbold}
\usepackage{braket}
\usepackage{mathtools}
\usepackage{hyperref}

\usepackage{xstring} % for \ifeqcase in \kitepatchbase

\vfuzz=\maxdimen
\hfuzz=\maxdimen
\def\calb{\mathcal{B}}

\def\star{\operatorname{Star}}
\def\zz{\mathbb{Z}}
\def\linking{\between}

\newcommand{\breakcell}[2][t]{\begin{tabular}[#1]{@{}l@{}}#2\end{tabular}}

\tikzset{
ind/.style={mark={lab=$#1$,a}}, % normal open index label
startind/.style={mark={lab=$#1$,b}}, % normal open index label
enddots/.style={mark={three dots, a}},
bdbind/.style={line width=1.2},
sdind/.style={mark={lab=${\scriptstyle #1}$,a}}, % small open index label
redlab/.style={mark={slab=$\textcolor{red}{\scriptscriptstyle #1}$}},
redlabr/.style={mark={slab=$\textcolor{red}{\scriptscriptstyle #1}$,r}},
}

\setmarks{
i/.style={lat=$#1$,a},
ar/.style={arr,asty=fill},
vor/.style={hbar,s,poff=0.07}, % orientation of corners of plaquette in vertex liquid
}

\setelements{
dlab/.style={lab/wrap=$##1$}, % regular label for different shapes
vlab/.style={lab/wrap={$\textcolor{red}{\scriptscriptstyle ##1}$}}, % small red label indicating relation to cell complex drawing
trotstep/.style={small,circ,dot},
trottens/.style={square,dlab},
toyplaquette/.style={small,circ,all},
toyface/.style={small,circ,dlab},
toyalg/.style={circ,dcross},
comproj/.style={square,sect8},
hadamard/.style={small,square,lab/t=H},
toyedge/.style={small,circ,all},
1dedge/.style={small, diamond, rhalf},
face/.style={circ,vlab},
oface/.style={circ,swirl},
extendedtrianglab/.style args={#1:#2 and #3:#4 and #5:#6}{lab={t=$\scriptstyle{#1}$, p=#2:0.3}, lab={t=$\scriptstyle{#3}$, p=#4:0.3}, lab={t=$\scriptstyle{#5}$, p=#6:0.3}},
vertexweight/.style={small,square},
2dedge/.style = {square, sect8, dec={rotate=180}{sect8},vlab},
vertexatom/.style = {circ,swirl},
bdvertexatom/.style = {halfcirc,rhalf},
loopinsertion/.style args={#1:#2}{postdecstyle={line width=1.2},markline={#1:0.4}{flag,e,f,#2}},
indexgrow/.style={triang,flat,rhalf},
21operator/.style={triang,flat},
12operator/.style={triang,flat,all},
orientflip/.style={square,fflat,big},
barflip/.style={square,flat},
invtwoform/.style={square,small,all},
frobunit/.style={circ,small,all},
cocycle/.style={labbox=$\omega$, bdstyle={circle,inner sep=0.02cm}},
groupsym/.style={small,circ,dot,lab/t=$G$,lab/dist=0.3},
representation/.style={labbox=$R$,bdstyle={rc,inner sep=0.1cm}},
projrep/.style={labbox=$P$,bdstyle={rc,inner sep=0.07cm}},
bdedge/.style={square,rhalf},
bdtriangle/.style={triang, sect6},
fhat12/.style={small,square,brquart,vlab},
fhat01/.style={small,square,rquart,vlab},
tetrahedron/.style={square,sect8,vlab},
weight12/.style={small,flat,triang,all,vlab},
weight02/.style={small,flat,triang,rhalf,vlab},
weight01/.style={small,flat,triang,vlab},
2gonweight/.style={small,small,triang,vlab},
2gonflip/.style={small,diamond,rhalf,vlab},
delta/.style={tiny,circ,all},
doface/.style={whiteblack,circ,swirl},
cornerweight/.style={small,square,dot},
spintri/.style={triang,sect6,vlab},
spin2gon/.style={small,triang,dec={rotate=90}{rhalf},vlab},
4simplex/.style={pentagon,sect10,vlab},
4d01fhat/.style={small,square,rquart,vlab},
voledgeweight01/.style={square,small,dcross},
voledgeweight02/.style={square,small,dcross,dec={rotate=-45}{sect=90}},
voledgeweight12/.style={whiteblack,square,small,dcross},
}

\tikzset{
mapping/.style=bottomzigzag,
irrep/.style={line width=1.7},
multip/.style={zigzag},
}

\def\manifoldcol{black}
\def\manifoldbdcol{cyan}
\def\manifolddefectcol{brown}
\tikzset{
manifold/.style={fill=\manifoldcol,fill opacity=.4},
1dmanifold/.style={\manifoldcol,line width=1.5},
doublemanifold/.style={manifold,postaction=manifold},
manifoldboundary/.style={\manifoldbdcol,line width=1.5},
manifoldbdfull/.style={fill=\manifoldbdcol,fill opacity=.4},
defectline/.style={\manifolddefectcol,line width=1.5},
}

\setelements{
redlab/.style={lab/wrap=$\textcolor{red}{##1}$},
vertexlab/.style={lab/wrap=$##1$,lab/dist=0.25}, % note: double ## needed!
vertex/.style={tiny,circ,all,vertexlab},
backvertex/.style={tiny,circ,vertexlab},
midvertex/.style={bdstyle=gray,backstyle=gray,tiny,circ,vertexlab},
corner/.style={postdecstyle={line width=1.2},arc=#1:0.2},
omega/.style={postdecstyle=red,postdec={scale=0.5}{shadecirc}},
dorientin/.style={vertexlab,small,circ,dcross},
dorientout/.style={vertexlab,small,circ,dot},
defectvertex/.style={vertex,bdastyle=\manifolddefectcol,decstyle=\manifolddefectcol,all},
boundaryvertex/.style={vertex,bdastyle=\manifoldbdcol,decstyle=\manifoldbdcol,all},
colblue/.style={bdastyle=blue, decstyle=blue},
colred/.style={bdastyle=red, decstyle=red},
}

\tikzset{
backedge/.style={dashed},
midedge/.style={dotted},
or/.style={mark={p=#1,arr}},
or/.default=0.5,
ior/.style={mark={p=#1,arr,-}},
ior/.default=0.5,
weight/.style={mark={p=#1,bar}},
weight/.default=0.5,
fav/.style={mark={p=#1,hcirc}},
fav/.default=0.5,
rfav/.style={mark={p=#1,hcirc,r}},
rfav/.default=0.5,
spinedge/.style={line width=0.2cm,opacity=0.4,blue},
actualedge/.style={line width=2},
}

\newcommand{\trianglepatch}[4]{
\fill[manifold] (#1)--(#2)--(#3)--cycle;
\draw[->] ($($(#1)!0.2!(#2)$)!0.2!(#3)$)--($($(#2)!0.2!(#1)$)!0.2!(#3)$);
\node at ($($(#2)!0.5!(#1)$)!0.6!(#3)$){$#4$};
}
\newcommand{\mappingtriangle}[3]{
\atoms{void}{m12/p={$(#1)!0.5!(#2)$}, m23/p={$(#2)!0.5!(#3)$}, m13/p={$(#1)!0.5!(#3)$}, m123/p={intersection of #3--m12 and #1--m23}, y12/p={$(#1)!0.07!(#2)$}, y21/p={$(#2)!0.07!(#1)$}, y13/p={$(#1)!0.07!(#3)$}, y31/p={$(#3)!0.07!(#1)$}, y23/p={$(#2)!0.07!(#3)$}, y32/p={$(#3)!0.07!(#2)$}, x12/p={$(m123)!0.15!(m12)$}, x23/p={$(m123)!0.15!(m23)$}, x13/p={$(m123)!0.15!(m13)$}}
\trianglepatch{y12}{y21}{x12}{*}
\trianglepatch{y23}{y32}{x23}{*}
\trianglepatch{y13}{y31}{x13}{}
\fill[manifold] (y12)--(#1)--(y13)--(x13)--(y31)--(#3)--(y32)--(x23)--(y23)--(#2)--(y21)--(x12)--cycle;
\draw[dashed] (y13)--(x13)--(y31) (y32)--(x23)--(y23) (y21)--(x12)--(y12);
}
\newcommand{\trianglebdpatch}[4]{
\atoms{void}{xxxx12/p={$(#1)!0.3!(#2)$}, xxxx21/p={$(#2)!0.3!(#1)$}, xxxx13/p={$(#1)!0.4!(#3)$}, xxxx23/p={$(#2)!0.4!(#3)$}}
\fill[manifold] (xxxx12)--(xxxx21)--(xxxx23)--(#3)--(xxxx13)--cycle;
\draw[dashed] (xxxx12)--(xxxx21);
\draw[->] ($($(#1)!0.3!(#2)$)!0.2!(#3)$)--($($(#2)!0.3!(#1)$)!0.2!(#3)$);
\draw[manifoldboundary] (xxxx12)--(xxxx13) (xxxx21)--(xxxx23);
\node at ($($(#2)!0.5!(#1)$)!0.6!(#3)$){$#4$};
}

\newcommand{\boundarydefectpatch}[4]{
\atoms{void}{xxxx12/p={$(#1)!0.3!(#2)$}, xxxx21/p={$(#2)!0.3!(#1)$}, xxxx34/p={$(#3)!0.3!(#4)$}, xxxx43/p={$(#4)!0.3!(#3)$}}
\fill[manifold] (xxxx12)--(xxxx21)--(xxxx43)--(xxxx34)--cycle;
\draw[manifoldboundary] (xxxx12)--(xxxx34) (xxxx21)--(xxxx43);
\draw[dashed] (xxxx12)--(xxxx21) (xxxx43)--(xxxx34);
\atoms{defectvertex}{z/p={$(xxxx12)!0.5!(xxxx34)$}}
}
\newcommand{\boundarymediatorpatch}[4]{
\atoms{void}{xxxx12/p={$(#1)!0.3!(#2)$}, xxxx21/p={$(#2)!0.3!(#1)$}, xxxx34/p={$(#3)!0.3!(#4)$}, xxxx43/p={$(#4)!0.3!(#3)$}}
\fill[manifold] (xxxx12)--(xxxx21)--(xxxx43)--(xxxx34)--cycle;
\draw[manifoldboundary] (xxxx12)--(xxxx34) (xxxx21)--(xxxx43);
\draw[dashed] (xxxx12)--(xxxx21) (xxxx43)--(xxxx34);
}

\newcommand{\mappingtrianglebd}[3]{
\atoms{void}{m12/p={$(#1)!0.5!(#2)$}, m23/p={$(#2)!0.5!(#3)$}, m13/p={$(#1)!0.5!(#3)$}, m123/p={intersection of #3--m12 and #1--m23}, y12/p={$(#1)!0.07!(#2)$}, y21/p={$(#2)!0.07!(#1)$}, y13/p={$(#1)!0.07!(#3)$}, y31/p={$(#3)!0.07!(#1)$}, y23/p={$(#2)!0.07!(#3)$}, y32/p={$(#3)!0.07!(#2)$}, x12/p={$(m123)!0.15!(m12)$}, x23/p={$(m123)!0.15!(m23)$}, x13/p={$(m123)!0.15!(m13)$}}
\atoms{void}{z12/p={$(y12)!0.4!(x12)$}, z21/p={$(y21)!0.4!(x12)$}, z13/p={$(y13)!0.4!(x13)$}, z31/p={$(y31)!0.4!(x13)$}, z23/p={$(y23)!0.4!(x23)$}, z32/p={$(y32)!0.4!(x23)$}}
\trianglebdpatch{y12}{y21}{x12}{*}
\trianglebdpatch{y23}{y32}{x23}{*}
\trianglebdpatch{y13}{y31}{x13}{}
\fill[manifold] (z12)--(z13)--(x13)--(z31)--(z32)--(x23)--(z23)--(z21)--(x12)--cycle;
\draw[manifoldboundary] (z13)--(z12) (z23)--(z21) (z31)--(z32);
\draw[dashed] (z13)--(x13)--(z31) (z32)--(x23)--(z23) (z21)--(x12)--(z12);
}

\newcommand{\kitepatchbase}[5]{
\fill[draw, dashed, manifold] (#2)--(#1)--(#3)--(#4)--cycle;
\IfEqCase{#5}{
{0+}{\draw[->] ($($(#2)!0.2!(#1)$)!0.2!(#3)$)--($(#1)!0.2!(#4)$);}
{1+}{\draw[->] ($($(#3)!0.2!(#1)$)!0.2!(#2)$)--($(#1)!0.2!(#4)$);}
{2+}{\draw[->] ($(#3)!0.2!(#2)$)--($(#3)!0.8!(#2)$);}
{0-}{\draw[<-] ($($(#2)!0.2!(#1)$)!0.2!(#3)$)--($(#1)!0.2!(#4)$);}
{1-}{\draw[<-] ($($(#3)!0.2!(#1)$)!0.2!(#2)$)--($(#1)!0.2!(#4)$);}
{2-}{\draw[<-] ($(#3)!0.2!(#2)$)--($(#3)!0.8!(#2)$);}
}
}

\newcommand{\kitepatch}[4]{
\atoms{void}{m12/p={$(#1)!0.5!(#2)$}, m13/p={$(#1)!0.5!(#3)$}, m123/p={intersection of #3--m12 and #2--m13}}
\kitepatchbase{#1}{m12}{m13}{m123}{#4}
}

\newcommand{\kitepatchext}[4]{
\atoms{void}{xm12/p={$(#1)!0.5!(#2)$}, xm13/p={$(#1)!0.5!(#3)$}, m123/p={intersection of #3--xm12 and #2--xm13}, ym12/p={$(xm12)!0.1!(m123)$}, ym13/p={$(xm13)!0.1!(m123)$}, y1/p={$(#1)!0.1!(m123)$}}
\kitepatchbase{y1}{ym12}{ym13}{m123}{#4}
\fill[manifold] (#1)--(xm12)--(ym12)--(y1)--(ym13)--(xm13)--cycle;
}

\newcommand{\pgfextractangle}[3]{%
    \pgfmathanglebetweenpoints{\pgfpointanchor{#2}{center}}
                              {\pgfpointanchor{#3}{center}}
    \global\let#1\pgfmathresult  
}
\newcommand{\fadecorner}[3]{
\pgfextractangle{\ang}{#2}{#3}
\begin{pgfinterruptboundingbox}
\fill[\manifoldcol!40,path fading=south, transform canvas={rotate=\ang}] ([rotate=-\ang]#1)--([rotate=-\ang]#2)--([rotate=-\ang]#3)--cycle;
\end{pgfinterruptboundingbox}
\draw[dashed] (#3)--(#1)--(#2);
\node at ($(#2)!0.5!(#3)$){$\ldots$};
}

\newcommand{\fadefgon}[4]{
\pgfextractangle{\ang}{#2}{#3}
\begin{pgfinterruptboundingbox}
\fill[\manifoldcol!40,path fading=south, transform canvas={rotate=\ang}] ([rotate=-\ang]#1)--([rotate=-\ang]#2)--([rotate=-\ang]#3)--([rotate=-\ang]#4)--cycle;
\end{pgfinterruptboundingbox}
\draw[manifoldboundary] (#1)--(#4);
\draw[dashed] (#1)--(#2) (#3)--(#4);
\node at ($(#2)!0.5!(#3)$){$\ldots$};
}

\begin{document}

\title{Towards topological fixed-point models beyond gappable boundaries}

\author{Andreas Bauer}
\affiliation{Dahlem centre for Complex Quantum Systems, Freie Universit{\"a}t Berlin, Arnimallee 14, 14195 Berlin, Germany}
\author{Jens Eisert}
\affiliation{Dahlem centre for Complex Quantum Systems, Freie Universit{\"a}t Berlin, Arnimallee 14, 14195 Berlin, Germany}
\affiliation{Helmholtz-Zentrum Berlin f{\"u}r Materialien und Energie, 14109 Berlin, Germany}
\author{Carolin Wille}
\affiliation{Institut f{\"u}r Theoretische Physik, University of Cologne, 50937 Cologne, Germany}
\affiliation{Rudolf Peierls Centre for Theoretical Physics, Clarendon Laboratory, Parks Road, Oxford, OX1 3PU, UK}

\begin{abstract}
We consider fixed-point models for topological phases of matter formulated as discrete path integrals  
in the language of tensor networks. Such zero-correlation length models with an exact notion of topological invariance are known in the mathematical community as state-sum constructions or lattice topological quantum field theories. All of the established ansatzes for fixed-point models imply the existence of a gapped boundary as well as a commuting-projector Hamiltonian. Thus, they fail to capture topological phases without a gapped boundary or commuting-projector Hamiltonian, most notably chiral topological phases in $2+1$ dimensions. In this work, we present a more general fixed-point ansatz not affected by the aforementioned restrictions. Thus, our formalism opens up a possible way forward towards a microscopic fixed-point description of chiral phases and we present several strategies that may lead to concrete examples. Furthermore, we argue that our more general ansatz constitutes a universal form of topological fixed-point models, whereas established ansatzes are universal only for fixed-points of phases which admit topological boundaries.
\end{abstract}

\maketitle

\section{Introduction}
Topological phases of matter have received a lot of attention in the past decades. They are interesting for fundamental physical reasons, but are also seen as being instrumental for devising schemes for topological quantum computing. The classification of phases of matter 
is a challenging task due to a very rich underlying mathematical structure, requiring 
tools from higher algebra and category theory.

By definition, (topological) phases of matter are equivalence classes of microscopic models under locality-preserving equivalences, such as local unitary circuits. However, when people talk about classifying phases, they often describe and classify models abstractly via their lower-dimensional defects, or ``excitations''. The most prominent example of this is the classification of $2+1$-dimensional intrinsic bosonic topological phases via \emph{unitary modular tensor categories (UMTCs)}, which describe the co-dimension $2$ defects known as \emph{anyons} as well as the co-dimension $3$ defects corresponding to fusion events of the latter. Mathematically, this data of defects up to co-dimension $2$ is described by a \emph{almost-fully extended topological quantum field theory} (TQFT). The reason behind the success of those classifications is that they appear to be in one-to-one correspondence with microscopic phases
\footnote{To make this work for fermionic phases, we need to include spin structure defects into the TQFT, e.g., we need the odd-parity ground state on the non-bounding spin circle to distinguish the Kitaev chain from the trivial phase as an axiomatic $1+1$-dimensional TQFT, or vortices in a fermionic UMTC to describe the $p+ip$ superconductor in $2+1$ dimensions as 3-2-1-extended TQFT. Similarily, we need symmetry defects to distinguish SPT phases. On the other hand, the invertible bosonic $E_8$ quantum Hall state (or at least three copies of it) does not seem to be distinguished from the trivial phase by any sort of 3-2-1-extended TQFT and thus already provides a counter-example to the statement.
%Defect data of 2+1-dimensional symmetry-enriched topological phases are given by \emph{$G$-crossed braided fusion categories}, which cannot always be realized microscopically without a $3+1$-dimensional bulk, due to anomalies \cite{Bulmash2020}.
}
, even though a given TQFT could in principle be realized by more than one microscopic phase, or by none at all. Whether, why, or to which extent this \emph{unique-extension hypothesis}
\footnote{We have chosen this name since extending almost-fully extended TQFTs once more yields \emph{fully} extended TQFTs which are supposed to describe microscopic lattice models.}
holds is one of the most important fundamental questions in the field.

In order to have a systematic and reliable classification independent of the unique-extension hypothesis, or to assert the correctness of the latter, we should describe topological phases in terms of concrete microscopic realisations. Such realizations are also vital if we want to use the phase for any practical purpose, such as quantum error correction. For this approach to classification, it is favourable to work with particularly well-behaved representatives of phases, which are exactly solvable by algebraic means. A great success has been the study of so-called \emph{fixed-point models}, which are
representatives of the phase which
exhibit a \emph{zero correlation length}. The name 
stems from the idea that those models are fixed-points under so-called \emph{renormalization-group transformations} where in each step large blocks of the original model are considered as individual constituents of the resulting model. By repeated application of this `zooming out' procedure the correlation length will decrease until we hopefully converge to a model with zero correlation length.

Fixed-point models for topological phases are usually formulated as \emph{commuting-projector Hamiltonians}. All such known commuting-projector Hamiltonians are equivalent to the more mathematical \emph{state-sum} constructions, or equivalently \emph{lattice TQFTs}, which represent a discrete path integral on a triangulated Euclidean spacetime manifold. A wealth of different families of commuting-projector Hamiltonians and the corresponding state-sum constructions can be found in the literature, with the probably most known example being the \emph{Levin-Wen string-net models} \cite{Levin2004} which are equivalent to the \emph{Turaev-Viro-Barrett-Westbury state-sum} \cite{Turaev1992,Barrett1993}.

Unfortunately, all fixed-point ansatzes we are aware of suffer from a severe restriction: They can only describe phases which possess gapped/topological boundaries. Accordingly, phases for which the boundary must necessarily be gapless do not have any fixed-point descriptions to date. Those include important examples, most notably \emph{chiral} intrinsic topological phases in $2+1$ dimensions, including one of the few families of phases which have actually been realised experimentally, namely fractional quantum Hall systems. Furthermore, there exists a whole field concerned with the classification of topological phases of quadratic fermionic Hamiltonians where the absence of a gapped boundary is often almost regarded as a mere definition for a non-trivial phase. The restriction of known fixed-point ansatzes to gappable boundaries is easily explained by the fact that all established state-sum models possess a \emph{cone topological boundary}, as explained in the main text. Furthermore, it has been argued that there cannot be any commuting-projector Hamiltonians for topological phases with non-vanishing electric \cite{Kapustin2018} or thermal \cite{Kapustin2019} quantum Hall conductance. Such a non-vanishing Hall conductance usually goes hand in hand with the absence of a gapped boundary, and again makes those models incompatible with established fixed-point ansatzes.

In this work, we make a significant step towards fixed-point descriptions of phases without gapped boundaries: We provide a fixed-point ansatz which is compatible with the absence of such boundaries, and which at the same time does not give rise to commuting-projector models. We do this by formalising and generalising state-sum constructions via the language of \emph{tensor-network path integrals}, which are tensor networks describing the imaginary-time evolution in Euclidean spacetime. This is to be contrasted with the use of ground state tensor networks in the framework of \emph{MPO-injective PEPS} \cite{Bultinck2015}, which do imply (and in fact are) topological boundaries. Roughly, the difference between our ansatz and established state-sum constructions is that the latter associate tensors (such as the $F$-symbols of a fusion category) to the simplices in a triangulation. Our ansatz associates tensors to the vertices, which are allowed to depend on the \emph{star} of that vertex, that is, the combinatorics of the surrounding triangulation.

In addition to providing a more general fixed-point ansatz, we also show that this ansatz is in fact the \emph{most} general ansatz: It is \emph{universal}, in the sense that it captures any other fixed-point ansatz, no matter how complicated. We also explore the notion of universality under the assumption that a gapped boundary exists, and find that in this case ordinary triangulation-based state-sums are indeed universal. This provides us with a clean explanation for the structure of known fixed-point models, and with a clear route for finding their most general form.

To this date, we have not been able to find a concrete example for a model of our fixed-point ansatz which does not possess a gappable boundary. However, there are some strong indications that those models might exist. Most notably, quantum cellular automata exactly disentangling certain Abelian modular Walker-Wang models have been found recently \cite{Haah2018, Haah2019, Shirley2022}. Such a quantum cellular automaton yields an invertible domain wall to vacuum, which, if topologically extendable to arbitrary triangulations, would immediately yield a fixed-point model for the chiral phase. All in all, we believe that the presented ansatz has great potential to lead to a unification of conventional fixed-point models for gappable-boundary topological phases, and chiral intrinsic phases as well as free-fermionic phases.

Even though we build upon the language of tensor-network path integrals, the main ideas of this work are also understandable for readers only familiar with the notion of a state-sum constructions. Roughly, one can replace the terms ``liquid'' with ``generalized state-sum ansatz'', ``tensor'' with ``weight depending on variables within a constant-size neighbourhood'', ``index'' by ``dependence of a weight on a state-sum variable'', ``open indices'' with ``boundary configuration of the state-sum variables within a patch of triangulation''.

The structure of the remainder of this work is as follows. In Section~\ref{sec:tn_path_integral} we quickly recap tensor-network path integrals, their phases, and how tensor-network path integral fixed-point ansatzes can be formalised by a finite set of tensor-network equations for a finite set of tensor variables, a so-called \emph{liquid}. In Sections~\ref{sec:11d_simple},
\ref{sec:11d_boundary}, and \ref{sec:universal_2d}, 
we describe all the relevant ideas for the case of $1+1$ spacetime dimensions. Note that in this dimension, the classification of phases is considered mostly complete, and all phases appear to have gapped boundaries and are thus captured by existing fixed-point ansatzes. So we cannot expect our generalised fixed-point ansatz to yield any models for new phases. However, as most of our argumentation takes place on a purely diagrammatic level, the $1+1$-dimensional case is perfectly suited to introduce the general principles. 

Specifically, in Section~\ref{sec:11d_simple} and Section~\ref{sec:11d_boundary}, we look at conventional lattice TQFT which corresponds to a liquid which we call the \emph{triangle liquid} and its boundary. We argue on a purely diagrammatic level that models of this liquid possess a topological boundary and commuting-projector Hamiltonian. We also show that this liquid is a universal fixed-point ansatz for phases allowing for a topological boundary, but not for general topological phases due to a \emph{corner problem}.
In Section~\ref{sec:vertex_finegrain_failure}, we then define the \emph{vertex liquid}, which provides a more general fixed-point ansatz for topological phases in $1+1$ dimensions. In particular, in Sections~\ref{sec:vertex_finegrain_failure}, \ref{sec:cone_mapping_failure}, \ref{sec:projector_mapping_failure} we demonstrate that any attempt to emulate the vertex liquid using the triangle liquid, to construct a cone boundary for the vertex liquid, and to construct a commuting-projector Hamiltonian for vertex liquid models, must fail, at least if the constructions are supposed to work on a diagrammatic level. In Section~\ref{sec:2d_universal_finegrain}, we show that the vertex liquid is universal, in the sense that it can emulate any other topological liquid in $1+1$ dimensions on a diagrammatic level.

In Section~\ref{sec:higher_dimensions}, we describe how all the previously presented concepts carry over to higher spacetime dimensions, which happens in a mostly straight-forward manner. In Section~\ref{sec:new_models} we discuss where and how we would expect our more general liquids to yield fixed-point models beyond commuting-projector Hamiltonians and gapped boundaries. In particular, we describe how the phenomenology of chiral phases of matter fits into the picture.

\section{Phases in tensor-network path integrals}
\label{sec:tn_path_integral}
In this work we describe many-body systems in terms of tensor-network path integrals. These are \emph{not} tensor networks in the sense they are usually used for the description of quantum systems, namely so-called \emph{projective entangled pair states} (PEPS) or \emph{matrix product states} (MPS) in $1+1$ dimensions. The latter are representations of (ground) states by tensor-networks living physical space with \emph{virtual} indices contracted between the tensors and open \emph{physical} indices corresponding to the degrees of freedom. In contrast, tensor-network path integrals describe the (in our case, imaginary) time evolution in (in our case, Euclidean) spacetime, have only one type of indices, and have open indices only at a \emph{space boundary} where we cut off the tensor network. Accordingly, the main use of tensor networks is to formalise a generalised notion of topological \emph{state-sum constructions}. Note that using tensor-network path integrals instead of tensor-network states is absolutely crucial to obtain the results in this paper, since fixed-point ansatzes in terms of tensor-network states are formally the same as gapped boundaries and thus do not allow to go beyond the latter.

Phases of matter are defined as connected regions in the space of all local models, whose Hamiltonian has a \emph{spectral gap}. If we use tensor-network path integrals instead of Hamiltonians, a condition analogous to being gapped can be defined by evaluating the path integral on a spacetime annulus and consider it as an operator from the open indices at its inside to those at its outside space boundary. We say that the tensor network path integral is gapped, if this operator approaches a rank-1 operator exponentially quickly in the width of the annulus. Note that this definition also seems to automatically include the \emph{local topological order} conditions. By increasing the unit cell and blocking tensors, one can argue that every gapped model eventually converges to a \emph{fixed-point model} in the same phase, for which the annulus path integral is exactly a rank-1 operator, at least if it has some minimum width. For a detailed discussion of phases of matter in tensor-network path integrals, as well as their fixed-point models, we refer the reader to Section 2 of Ref.~\cite{liquid_intro}.

\emph{Topological} phases allow for fixed-point models which are \emph{topologically extendible}, i.e., they can be defined on arbitrary triangulations of arbitrary topological manifolds (of the same spacetime dimension) and still obey the rank-1 condition for annuli there. This implies that those fixed-point models have a notion of exact topological invariance. Namely, their tensors obey a finite set of tensor-network equations, and plugging those equations into the path integral locally changes the underlying triangulation. Those equations are powerful enough to arbitrarily change the triangulation, without being able to change the underlying topology. This will be the one and only property that characterises our topological fixed-point path integrals. In general, we will refer to a finite set of tensor-network equations for a finite set of tensor variables as a \emph{liquid}, and to the solutions of the equations as a \emph{model} of the liquid. Then, our tensor-network path integral fixed-point ansatzes are nothing but models of a \emph{topological} liquid, which is a combinatorial description of continuum manifolds. Furthermore, we will refer to the diagrams of tensor networks simply as \emph{networks}, and to the two diagrams defining an equation as a \emph{move}.

The most straight-forward way of representing a (piece-wise linear) topological manifold combinatorially is via a triangulation. For any liquid describing topological $n$-manifolds, we can construct triangulations from the networks of that liquid and vice versa. So another way to think about a liquid is as a local prescription to associate a tensor-network to triangulations. Local means that the structure of the network and the tensors at a place in the triangulation are allowed to depend on the combinatorics of the triangulation inside a constant-size neighbourhood. For established fixed-point ansatzes, this dependence is particularly trivial, namely, we place one copy of the same tensor on each $n$-simplex. For more general ansatzes, the tensor network could depend on the triangulation in an environment of combinatorial size $b$ measured in the combinatorial distance. By `fine-graining' and `blocking', we can make $b$ smaller and smaller. This is the basic idea of the \emph{universality mapping}, which will play a crucial role in later sections. However, using fine-graining, we do \emph{not} necessarily arrive at an established fixed-point ansatz. Instead, it is possible to arrive at a more complicated liquid, namely the vertex liquid which will be introduced later.

\section{The triangle liquid in \texorpdfstring{$1+1$}{1+1} dimensions}
\label{sec:11d_simple}
In this section we quickly recap a simple liquid in $1+1$ dimensions which corresponds to the conventional state-sum ansatz or lattice TQFT, and review how it gives rise to a commuting-projector Hamiltonian on a purely diagrammatic level. For a more detailed introduction we refer to Ref.~\cite{liquid_intro}. We also discuss an attempt to show that the simple liquid can capture any other fixed-point ansatz, and point out the problems and potential failure of the latter. The specific way how this attempt fails is of conceptual importance and leads us to consider a more general fixed-point ansatze in Section~\ref{sec:universal_2d}.

Note that in $1+1$ dimensions there are no intrinsic topological phases, the Kitaev chain as a fermionic intrinsic topological phase, symmetry-breaking phases, (fermionic) symmetry-protected phases based on group (super) cohomology. We do not expect topological phases beyond those, and all of those have fixed-point models of the conventional form. Even though we do not expect to gain any new models in $1+1$ dimensions, we will focus on this case for this and the following two sections as this still allows us to demonstrate all the principles of our formalism on a diagrammatic level.

\subsection{The triangle liquid}
As the name indicates, the (tensor) networks of the \emph{triangle liquid} represent triangulations of 2-manifolds. More precisely, each triangle is represented by a copy of the same 3-index tensor, and the tensors at neighbouring triangles share a \emph{bond}, i.e., a contracted index pair. Also, the triangulation is equipped with a \emph{branching structure} (or simply \emph{branching}), that is, every edge has a direction, such that the directions are non-cyclic around every triangle. Furthermore, the triangulation has an orientation and the tensors at triangles with two clockwise-directed edges different from those at triangles with two counter-clockwise-directed edges,
\begin{equation}
\label{eq:triangles}
\begin{gathered}
\begin{tikzpicture}
\atoms{vertex,redlab}{{0/lab={t=0,ang=-150}}, {1/p=60:0.8,lab={t=1,ang=90}}, {2/p=0:0.8,lab={t=2,ang=-30}}}
\draw (0)edge[or](1) (0)edge[or](2) (1)edge[or](2);
\end{tikzpicture}
\quad\rightarrow\quad
\begin{tikzpicture}
\atoms{oface}{{0/}}
\draw (0)edge[sdind=02]++(-90:0.6) (0)edge[mark={ar,s},sdind=01]++(150:0.6) (0)edge[mark={ar,s},sdind=12]++(30:0.6);
\end{tikzpicture}\;,\\
\begin{tikzpicture}
\atoms{vertex,redlab}{{2/lab={t=2,ang=-150}}, {1/p=60:0.8,lab={t=1,ang=90}}, {0/p=0:0.8,lab={t=0,ang=-30}}}
\draw (0)edge[or](1) (0)edge[or](2) (1)edge[or](2);
\end{tikzpicture}
\quad\rightarrow\quad
\begin{tikzpicture}
\atoms{oface}{{0/}}
\draw (0)edge[mark={ar,s},sdind=02]++(-90:0.6) (0)edge[sdind=01]++(150:0.6) (0)edge[sdind=12]++(30:0.6);
\end{tikzpicture}\;.
\end{gathered}
\end{equation}
For Hermitian models of the liquid, those two tensors are complex conjugates of each other.

There are a number of moves, the main one is given by
\begin{equation}
\label{eq:22pachner}
\begin{tikzpicture}
\atoms{oface}{0/, 1/p={0,-0.8}}
\draw (0)edge[mark={ar,e}](1) (0)edge[mark={ar,s},ind=a]++(135:0.5) (0)edge[mark={ar,s},ind=b]++(45:0.5) (1)edge[mark={ar,s},ind=c]++(-45:0.5) (1)edge[ind=d]++(-135:0.5);
\end{tikzpicture}=
\begin{tikzpicture}
\atoms{oface}{0/, 1/p={0.8,0}}
\draw (0)edge[mark={ar,s}](1) (0)edge[mark={ar,s},ind=a]++(135:0.5) (1)edge[mark={ar,s},ind=b]++(45:0.5) (1)edge[mark={ar,s},ind=c]++(-45:0.5) (0)edge[ind=d]++(-135:0.5);
\end{tikzpicture} \;.
\end{equation}
The move corresponds to a local change of the triangulation which itself is known as a \emph{2-2 Pachner move},
\begin{equation}
\label{eq:22pachner_move}
\begin{tikzpicture}
\atoms{vertex}{{0/lab={t=0,p=180:0.25}}}
\atoms{vertex}{{1/p={0.5,-0.5},lab={t=3,p=-90:0.25}}}
\atoms{vertex}{{2/p={1,0},lab={t=2,p=0:0.25}}}
\atoms{vertex}{{3/p={0.5,0.5},lab={t=1,p=90:0.25}}}
\draw (0)edge[or](3) (3)edge[or](2) (2)edge[or](1) (0)edge[or](1) (0)edge[or](2);
\end{tikzpicture}
\quad\leftrightarrow\quad
\begin{tikzpicture}
\atoms{vertex}{{0/lab={t=0,p=180:0.25}}}
\atoms{vertex}{{1/p={0.5,-0.5},lab={t=3,p=-90:0.25}}}
\atoms{vertex}{{2/p={1,0},lab={t=2,p=0:0.25}}}
\atoms{vertex}{{3/p={0.5,0.5},lab={t=1,p=90:0.25}}}
\draw (0)edge[or](3) (3)edge[or](2) (2)edge[or](1) (0)edge[or](1) (3)edge[or](1);
\end{tikzpicture} \;.
\end{equation}
It is known that the 2-2 Pachner move together with an additional \emph{1-3 Pachner move} form a discrete analogue of continuum homeomorphisms \cite{Pachner1991}. So, if we add versions of the move above with different edge orientations as well as the 1-3 Pachner move, we have a topological liquid. For a more detailed description of the liquid, we refer the reader to Ref.~\cite{liquid_intro}.

Models of the liquid can be constructed from finite-dimensional *-algebras. The discrete path-integrals are equivalent to what is known as \emph{lattice TQFT} \cite{Fukuma1992}.

\subsection{Commuting-projector mapping}
\label{sec:hamiltonian}
Tensor-network path integrals exhibiting topological order are a discrete-time version of the imaginary-time evolution of quantum spin systems with local Hamiltonians. If the Hamiltonian terms commute, Trotterization is not necessary, and an exact representation by a tensor-network path integral can be obtained by simply taking the product of time evolutions under the individual Hamiltonian terms.

A translation-invariant, local \emph{commuting-projector model} is given by a \emph{local ground-state projector} $P$ acting on the degrees of freedom on a lattice inside a block of some fixed size. The Hamiltonian of the model is given by
\begin{equation}
H=\sum_i (1-P_i)\;,
\end{equation}
where $i$ runs over all lattice sites, and $P_i$ denotes $P$ acting on a block centred  at site $i$, tensored with the identity everywhere else. $P$ needs to fulfil
\begin{equation}
P_iP_j=P_jP_i
\end{equation}
for all sites $i$ and $j$, which is automatic if the blocks centred  at $i$ and $j$ do not overlap. So commuting-projector models are models of a liquid, as they are given by a finite set of tensors fulfilling a finite set of equations. In $1+1$ dimensions, after sufficient fine-graining, $P$ is an operator acting on two consecutive degrees of freedom in a spin chain, i.e., a 4-index tensor,
\begin{equation}
P^{ab}_{cd}=
\begin{tikzpicture}
\atoms{comproj}{0/}
\draw (0-tr)edge[ind=b]++(45:0.3) (0-tl)edge[ind=a]++(135:0.3) (0-br)edge[ind=d]++(-45:0.3) (0-bl)edge[ind=c]++(-135:0.3);
\end{tikzpicture}\;.
\end{equation}
$P$ being a projector corresponds to a move,
\begin{equation}
\label{eq:projector_move}
\begin{tikzpicture}
\atoms{comproj}{0/, 1/p={0,0.8}}
\draw (1-tr)edge[ind=b]++(45:0.3) (1-tl)edge[ind=a]++(135:0.3) (0-br)edge[ind=d]++(-45:0.3) (0-bl)edge[ind=c]++(-135:0.3);
\draw (0-tr)to[bend right=40](1-br) (0-tl)to[bend left=40](1-bl);
\end{tikzpicture}
=
\begin{tikzpicture}
\atoms{comproj}{0/}
\draw (0-tr)edge[ind=b]++(45:0.3) (0-tl)edge[ind=a]++(135:0.3) (0-br)edge[ind=d]++(-45:0.3) (0-bl)edge[ind=c]++(-135:0.3);
\end{tikzpicture}\;,
\end{equation}
and the commutativity is a move,
\begin{equation}
\label{eq:projector_commutativity_move}
\begin{tikzpicture}
\atoms{comproj}{0/, 1/p={0.6,0.6}}
\draw (1-tr)edge[ind=b]++(45:0.3) (1-tl)edge[ind=a]++(135:0.3) (0-br)edge[ind=d]++(-45:0.3) (0-bl)edge[ind=c]++(-135:0.3) (0-tl)edge[ind=e]++(135:0.3) (1-br)edge[ind=f]++(-45:0.3);
\draw (0-tr)--(1-bl);
\end{tikzpicture}
=
\begin{tikzpicture}
\atoms{comproj}{0/, 1/p={0.6,-0.6}}
\draw (1-tr)edge[ind=b]++(45:0.3) (0-tr)edge[ind=a]++(45:0.3) (1-bl)edge[ind=d]++(-135:0.3) (0-bl)edge[ind=c]++(-135:0.3) (0-tl)edge[ind=e]++(135:0.3) (1-br)edge[ind=f]++(-45:0.3);
\draw (0-br)--(1-tl);
\end{tikzpicture}\;.
\end{equation}

Consider a regular triangulation of the plane like the following,
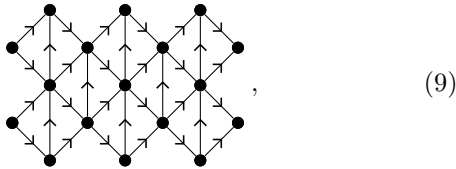
\begin{equation}
\begin{tikzpicture}
\atoms{vertex}{00x/p={0.5,-0.5}, 01x/p={1.5,-0.5}, 02x/p={2.5,-0.5},10/, 10/p={0,0}, 11/p={1,0}, 12/p={2,0}, 13/p={3,0}, 10x/p={0.5,0.5}, 11x/p={1.5,0.5}, 12x/p={2.5,0.5}, 20/p={0,1}, 21/p={1,1}, 22/p={2,1}, 23/p={3,1}, 20x/p={0.5,1.5}, 21x/p={1.5,1.5}, 22x/p={2.5,1.5}}
\foreach \x/\xp/\y/\ym in {0/1/1/0, 0/1/2/1, 1/2/1/0, 1/2/2/1, 2/3/1/0, 2/3/2/1}{
\draw (\ym\x x)edge[or](\y\x x) (\y\x)edge[or](\y\x x) (\y\x)edge[or](\ym\x x) (\ym\x x)edge[or](\y\xp) (\y\x x)edge[or](\y\xp);
};
\draw (11)edge[or](21) (12)edge[or](22);
\end{tikzpicture}\;,
\end{equation}
and the associated triangle liquid tensor network. We notice that this network has the form of a circuit, i.e., a product of local operators $P$, associated to patches
\begin{equation}
P\quad\rightarrow\quad
\begin{tikzpicture}
\atoms{vertex}{3/, 4/p={0.5,-0.5}, 5/p={1,0}, 6/p={0.5,0.5}};
\draw (3)edge[or](4) (4)edge[or](5) (3)edge[or](6) (6)edge[or](5) (4)edge[or](6);
\end{tikzpicture}\;.
\end{equation}
This suggests that we have a liquid mapping
\begin{equation}
\label{eq:comm_proj_mapping}
\begin{tikzpicture}
\atoms{comproj}{0/}
\draw (0-tr)edge[ind=b]++(45:0.3) (0-tl)edge[ind=a]++(135:0.3) (0-br)edge[ind=d]++(-45:0.3) (0-bl)edge[ind=c]++(-135:0.3);
\end{tikzpicture}
\coloneqq
\begin{tikzpicture}
\atoms{oface}{x0/p={0,0.6}, x1/p={0.8,0.6}}
\draw (x0)edge[ind=a,mark={ar,s}]++(0,0.5) (x1)edge[ind=b,mark={ar,s}]++(0,0.5) (x0)edge[ind=c]++(0,-0.5) (x1)edge[ind=d]++(0,-0.5) (x0)edge[mark={ar,e}](x1);
\end{tikzpicture}
\end{equation}
from the commuting-projector liquid to the triangle liquid. Indeed, if we plug Eq.~\eqref{eq:comm_proj_mapping} into Eq.~\eqref{eq:projector_move} or Eq.~\eqref{eq:projector_commutativity_move}, we see that those can be derived from the moves of the triangle liquid. So every model of the triangle liquid yields a commuting-projector model.

\subsection{Universality mapping and the corner problem}
\label{sec:corner_problem}
The triangle liquid is not the only liquid describing topological 2-manifolds. For example, in Ref.~\cite{liquid_intro}, we define an \emph{edge liquid} which associates a 4-index tensor to each edge. Different liquids have the advantage that they might provide simpler models for the same phases. However, we often find that liquids are equivalent on a purely diagrammatic level, such as it is the case for the triangle and the edge liquid.

The diagrammatic equivalence between two liquids can be shown via a so-called \emph{liquid mapping}, which associates to every bond dimension variable of the source liquid a collection of bond dimension variables of the target liquid, and to every tensor of a source liquid a network of the target liquid. More precisely, the source tensor indices are mapped to collections of open indices of the associated target network. The mapping has to be such that if we replace tensors by networks in the source move, we get an equation which can be derived via the target moves. If we plug the concrete tensors of a target liquid model into the networks of the mapping, we obtain a model of the source liquid.

If we want to show that two liquids are equivalent, we need a mapping both from one to the other and vice versa. The mappings need to be \emph{weak inverses} of each other meaning that applying both mappings after another needs to be equivalent to an \emph{invertible domain wall} between a liquid and itself. This ensures that the phases described by the two fixed-point ansatzes have to be in one-to-one correspondence. For a more detailed definition see Ref.~\cite{liquid_intro}. For topological liquids, it is very simple to see when two mappings are weak inverses of another and give rise to an equivalence of liquids. Namely, for this to be the case they have to be topology preserving, i.e., mapping a $n$-ball (here, $2$-ball) of network to another ball, such that mapping from the source to the target liquid and back corresponds to refining the network.

One might now be tempted to think that not only the edge liquid and triangle liquid, but \emph{all} liquids whose networks describe topological 2-manifolds are equivalent to each other, and therefore describes the same phases. To further investigate this thought we will attempt
to show that any two-dimensional topological liquid $\calb$ is equivalent to the triangle liquid via weakly invertible liquid mappings and in fact fail to do so. Instead, we will encounter fundamental difficulties which strongly indicate that not every topological liquid is equivalent to the triangle liquid and hint at the possibility of non-equivalent liquids. Based on the insights gained in this discussion, we are able to construct a more general and indeed universal liquid in Section~\ref{sec:universal_2d}.

We consider an arbitrary two-dimensional topological liquid $\calb$, which we assume to be at least as powerful as the triangle liquid. Thus, there exists a topology-preserving mapping from $\calb$ to the triangle liquid. In the other direction, a weakly invertible mapping from the triangle liquid to an arbitrary liquid $\calb$, will be called a \emph{universality mapping} since it shows that the triangle liquid is universal as a fixed-point ansatz. However, our attempt to construct such a universality mapping fails due to a reason we call the \emph{corner problem}.

There are two main ways in which an invertible mapping between two liquids $\mathcal{A}$ and $\mathcal{B}$ can fail to be a liquid mapping. First, the mapping might produce \emph{invalid} $\calb$-networks when applied to an $\mathcal{A}$-network, which do not represent a patch of a 2-manifold. For the triangle liquid introduced above, any network is valid. However, for some more general liquids $\calb$, in particular for the vertex liquid we introduce in Section~\ref{sec:universal_2d}, $\calb$-networks representing a patch of 2-manifold are subject to some constraints.
E.g., the vertex-liquid networks representing 2-manifolds are required to only have triangle plaquettes, whereas, e.g., tensor-network diagrams with 4-gon plaquettes are considered invalid.
Second, the mapped moves of $\mathcal{A}$ must be derivable from the moves of $\calb$. If the networks on both sides of the $\calb$-moves are very large compared to those of $\mathcal{A}$, it might not be possible to use them to derive the mapped moves of $\mathcal{A}$. This problem can be tackled by using large $\calb$-networks in the mapping itself. That is, the mapping will be similar to a renormalization-group transformation, i.e., a mapping with fine-graining scale $\lambda$ as introduced in Section 2 of Ref.~\cite{liquid_intro}.

Let us now attempt to construct a weakly invertible mapping from the triangle liquid to $\calb$, i.e., to associate a $\calb$-network to each triangle tensor. To construct a suitable $\calb$-network, we use the following construction. We fill a rhombus with some $\calb$-network, such that the corners have a combinatorial distance (i.e., minimum number of bonds in a connecting path) of at least $\lambda$. Then, we split the rhombus and the $\calb$-network on it into two triangles along a horizontal line,
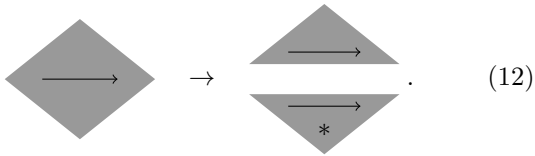
\begin{equation}
\label{eq:triangle_fine_mapping_rhombus}
\begin{tikzpicture}
\fill[manifold] (0,0)--(1,-0.8)--(2,0)--(1,0.8)--cycle;
\draw[->] (0.5,0)--(1.5,0);
\end{tikzpicture}
\quad\rightarrow\quad
\begin{tikzpicture}
\atoms{void}{x0/p={0,0.2}, x1/p={2,0.2}, x2/p={1,1}, y0/p={0,-0.2}, y1/p={2,-0.2}, y2/p={1,-1}}
\trianglepatch{x0}{x1}{x2}{}
\trianglepatch{y0}{y1}{y2}{*}
\end{tikzpicture}\;.
\end{equation}
Note that the arrow directions together with the underlying orientation are necessary to make an unambiguous choice of the cut.

Next, we combine three of the triangle-shaped network patches into a single triangle-shaped network by placing them next to each other and filling the gaps between them with some $\calb$-network as follows,
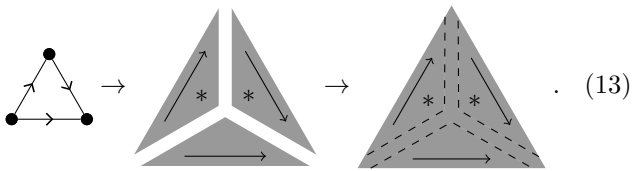
\begin{equation}
\begin{tikzpicture}
\atoms{vertex}{0/, 1/p={1,0}, 2/p=60:1}
\draw (0)edge[mark=arr](1) (1)edge[mark={arr,-}](2) (2)edge[mark={arr,-}](0);
\end{tikzpicture}
\rightarrow
\begin{tikzpicture}
\begin{scope}[shift={(-90:0.1)}]
\atoms{void}{0/, 1/p=-30:1.3, 2/p=-150:1.3}
\trianglepatch{2}{1}{0}{}
%\fill[blue!50] (0,0)--(-30:1.3)--(-150:1.3)--cycle;
\end{scope}
\begin{scope}[rotate=120,shift={(-90:0.1)}]
\atoms{void}{0/, 1/p=-30:1.3, 2/p=-150:1.3}
\trianglepatch{1}{2}{0}{*}
\end{scope}
\begin{scope}[rotate=-120,shift={(-90:0.1)}]
\atoms{void}{0/, 1/p=-30:1.3, 2/p=-150:1.3}
\trianglepatch{1}{2}{0}{*}
\end{scope}
\end{tikzpicture}
\rightarrow
% \begin{tikzpicture}
% \begin{scope}[shift={(-90:0.1)}]
% \fill[blue!50] (0,0)--(-30:1.3)--(-150:1.3)--cycle;
% \end{scope}
% \begin{scope}[rotate=120,shift={(-90:0.1)}]
% \fill[blue!50] (0,0)--(-30:1.3)--(-150:1.3)--cycle;
% \end{scope}
% \begin{scope}[rotate=-120,shift={(-90:0.1)}]
% \fill[blue!50] (0,0)--(-30:1.3)--(-150:1.3)--cycle;
% \end{scope}
% \fill[blue] (-90:0.1)--++(-30:1.3)--($(30:0.1)+(-30:1.3)$)--(30:0.1)--++(90:1.3)--($(150:0.1)+(90:1.3)$)--(150:0.1)--++(-150:1.3)--($(-90:0.1)+(-150:1.3)$)--cycle;
% \draw (-120:0.7)edge[->](-60:0.7) (60:0.7)edge[->](0:0.7) (180:0.7)edge[->](120:0.7);
% \node at (150:0.4){$*$};
% \node at (30:0.4){$*$};
% \end{tikzpicture}
\begin{tikzpicture}
\atoms{void}{0/, 1/p={2.5,0}, 2/p={60:2.5}}
\mappingtriangle{0}{2}{1}
\end{tikzpicture}
\;.
\end{equation}
The branching structure of the reference triangle on the left is important to make an unambiguous choice of such a filling.
The resulting $\calb$-network is associated to the triangle tensor for the attempted mapping,
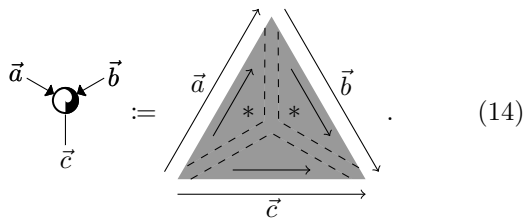
\begin{equation}
\label{eq:mapping_triangle}
\begin{tikzpicture}
\atoms{oface}{0/}
\draw (0)edge[ind=\vec{c}]++(-90:0.6) (0)edge[mark={ar,s},ind=\vec{a}]++(150:0.6) (0)edge[mark={ar,s},ind=\vec{b}]++(30:0.6);
\end{tikzpicture}
\coloneqq
\begin{tikzpicture}
\atoms{void}{0/, 1/p={2.5,0}, 2/p={60:2.5}}
\mappingtriangle{0}{2}{1}
\draw ($(0)+(150:0.2)$) edge[->,mark={slab=$\vec{a}$}] ($(2)+(150:0.2)$) ($(2)+(30:0.2)$) edge[->,mark={slab=$\vec{b}$}] ($(1)+(30:0.2)$) ($(0)+(-90:0.2)$) edge[->,mark={slab=$\vec{c}$,r}] ($(1)+(-90:0.2)$);
\end{tikzpicture}
% \begin{tikzpicture}
% \begin{scope}[shift={(-90:0.1)}]
% \fill[blue!50] (0,0)--(-30:1.3)--(-150:1.3)--cycle;
% \end{scope}
% \begin{scope}[rotate=120,shift={(-90:0.1)}]
% \fill[blue!50] (0,0)--(-30:1.3)--(-150:1.3)--cycle;
% \end{scope}
% \begin{scope}[rotate=-120,shift={(-90:0.1)}]
% \fill[blue!50] (0,0)--(-30:1.3)--(-150:1.3)--cycle;
% \end{scope}
% \fill[blue] (-90:0.1)--++(-30:1.3)--($(30:0.1)+(-30:1.3)$)--(30:0.1)--++(90:1.3)--($(150:0.1)+(90:1.3)$)--(150:0.1)--++(-150:1.3)--($(-90:0.1)+(-150:1.3)$)--cycle;
% \draw (-120:0.7)edge[->](-60:0.7) (60:0.7)edge[->](0:0.7) (180:0.7)edge[->](120:0.7);
% \node at (150:0.4){$*$};
% \node at (30:0.4){$*$};
% \atoms{void}{0/p={-150:1.3}, 1/p={-30:1.3}, 2/p=90:1.3}
% \draw ($(0)+(150:0.3)$) edge[->,mark={slab=$\vec{a}$}] ($(2)+(150:0.3)$) ($(2)+(30:0.3)$) edge[->,mark={slab=$\vec{b}$}] ($(1)+(30:0.3)$) ($(0)+(-90:0.3)$) edge[->,mark={slab=$\vec{c}$,r}] ($(1)+(-90:0.3)$);
% \end{tikzpicture}
\;.
\end{equation}
The $\calb$-network on the right has many open indices along its open boundary, their number scaling linearly with $\lambda$. The open indices along each of the three edges are ordered according to the branching and blocked together to yield one of the indices $\vec{a}$, $\vec{b}$ and $\vec{c}$, respectively. The bond dimension of the triangle-liquid model is the product of all the bond dimensions of the indices along one edge, and therefore grows exponentially in $\lambda$. Note, that in the bottom left corner (as well as in the other corners), the filling between the triangles can give rise to additional open indices. For each such index, we have to make a choice of whether to assign it to $\vec a$ or $\vec c$, and analogously for the other corners.
The construction for the mapping of the counter-clockwise triangle is analogous, gluing together three triangles with the opposite orientations.

In order to test whether this mapping works, we need to first check whether every valid triangle-liquid network patch is mapped to a valid $\calb$-network patch. If we apply the mapping to two neighbouring triangles,
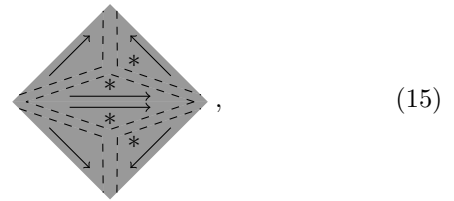
\begin{equation}
\begin{tikzpicture}
\atoms{void}{0/, 1/p={2.6,0}, 2/p={1.3,1.3}, 3/p={1.3,-1.3}}
\mappingtriangle{0}{1}{2}
\mappingtriangle{0}{1}{3}
\end{tikzpicture}\;,
\end{equation}
this is the case per construction as we just re-glue what we have cut apart in Eq.~\eqref{eq:triangle_fine_mapping_rhombus}. However, the situation is different when we consider a patch of triangles around a vertex, e.g.,
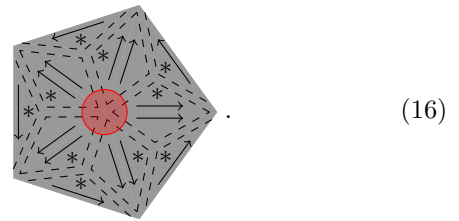
\begin{equation}
\label{eq:corner_problem_invalid_vertex}
\begin{tikzpicture}
\atoms{void}{m/}
\foreach \nr in {0,...,4}{
\atoms{void}{\nr/p=\nr*72:1.5}
};
\mappingtriangle{m}{0}{1}
\mappingtriangle{m}{1}{2}
\mappingtriangle{m}{2}{3}
\mappingtriangle{m}{3}{4}
\mappingtriangle{m}{4}{0}
\draw[red, fill=red, fill opacity=0.3] (0,0)circle(0.3);
\end{tikzpicture}
\;.
\end{equation}
The network in the red marked region around the central vertex 
does not originate from a cut which is re-glued.
We thus have no guarantee that the network in the red marked region is valid and represents a disk-like patch of 2-manifold. Note that this network also depends on the arbitrary choice of whether the open indices in the corners of Eq.~\eqref{eq:mapping_triangle} are associated to $\vec a$, $\vec b$ or $\vec c$.

%The fact that we cannot guarantee that the $\calb$-network is valid around any kind already implies that we cannot derive the Pachner moves from the $\calb$-moves. It is instructive to direcly look at how this fails. 

We also need to test whether the mapped Pachner moves can be derived from the $\calb$-moves. E.g., the mapping of the 2-2 Pachner move in Eq.~\eqref{eq:22pachner} yields
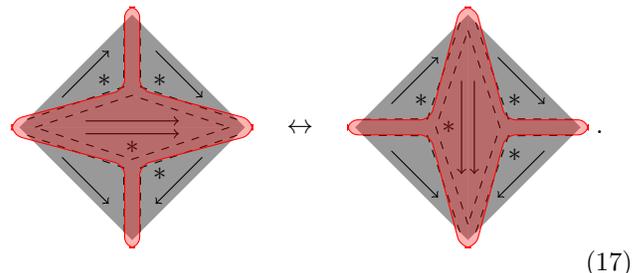
\begin{equation}
\label{eq:mapping_retriangulation}
\begin{tikzpicture}
\atoms{void}{0/p={-1.5,0}, 1/p={0,1.5}, 2/p={1.5,0}, 3/p={0,-1.5}}
\mappingtriangle{0}{1}{2}
\mappingtriangle{0}{2}{3}
\draw[red, fill=red, fill opacity=0.3, rc] (1.6,-0.1)--(1.6,0.1)--(0.1,0.5)--(0.1,1.6)--(-0.1,1.6)--(-0.1,0.5)--(-1.6,0.1)--(-1.6,-0.1)--(-0.1,-0.5)--(-0.1,-1.6)--(0.1,-1.6)--(0.1,-0.5)--cycle;
\end{tikzpicture}
\quad\leftrightarrow\quad
\begin{tikzpicture}
\atoms{void}{0/p={-1.5,0}, 1/p={0,1.5}, 2/p={1.5,0}, 3/p={0,-1.5}}
\mappingtriangle{0}{1}{3}
\mappingtriangle{1}{2}{3}
\draw[red, fill=red, fill opacity=0.3, rc] (1.6,-0.1)--(1.6,0.1)--(0.5,0.1)--(0.1,1.6)--(-0.1,1.6)--(-0.5,0.1)--(-1.6,0.1)--(-1.6,-0.1)--(-0.5,-0.1)--(-0.1,-1.6)--(0.1,-1.6)--(0.5,-0.1)--cycle;
\end{tikzpicture}\;.
\end{equation}
We notice that the $\calb$-network only changes within the red shaded area. If we want to change the $\calb$-network within some region by $\calb$-moves, the latter need to act within this region enlarged by a margin of constant size measured in the combinatorial distance. E.g., in order to perform changes within one of the red circles in the following picture, we need to apply moves within the respective larger green regions,
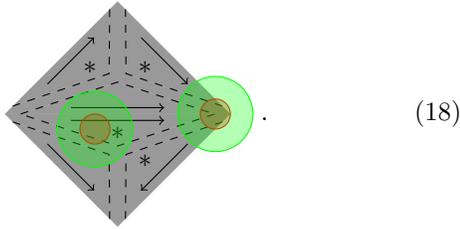
\begin{equation}
\begin{tikzpicture}
\atoms{void}{0/p={-1.5,0}, 1/p={0,1.5}, 2/p={1.5,0}, 3/p={0,-1.5}}
\mappingtriangle{0}{1}{2}
\mappingtriangle{0}{2}{3}
\draw[red, fill=red, fill opacity=0.3] (-0.3,-0.2)circle(0.2);
\draw[green, fill=green, fill opacity=0.3] (-0.3,-0.2)circle(0.5);
\draw[red, fill=red, fill opacity=0.3] (1.3,0)circle(0.2);
\draw[green, fill=green, fill opacity=0.3] (1.3,0)circle(0.5);
\end{tikzpicture}\;.
\end{equation}
For any red region far from the boundary, the green region is fully contained inside the overall $\calb$-network, if we choose the fine-graining scale $\lambda$ large enough. Consequentially, changes in these regions can be performed using the $\calb$-moves, no matter how complicated the latter are. However, this does not hold for changes right at the boundary. Thus, a mapping based on fine-graining only provides a valid mapping for an arbitrary liquid $\calb$ to the triangle liquid, if the two $\calb$-networks of every mapped move differ only in regions distant from the boundary. As we can see in Eq.~\eqref{eq:mapping_retriangulation}, the present construction almost succeeds in doing so. Most of the regions on which the two networks differ are located in the bulk of the network. The problem only arises at the corners of Eq.~\eqref{eq:mapping_retriangulation}, hence we refer to the encountered obstruction as the `corner problem'.

In addition to the corner problem, the attempted universality mapping is also incompatible with Hermiticity. We will describe in more detail how this problem arises and how to resolve it towards the end of Section~\ref{sec:finegrain_boundary}, where we circumvent the corner problem using the existence of a topological boundary.

\section{The boundary triangle liquid in \texorpdfstring{$1+1$}{1+1} dimensions}
\label{sec:11d_boundary}
In this section we describe how liquids do not only yield ansatzes for fixed-point models of topological phases, but also models for the boundaries thereof. The boundaries themselves have a topological deformability, which implies that the model has a gap as an open chain. We show that every triangle-liquid model has a standard topological boundary, using the so-called boundary cone mapping. Conversely, we find find that the existence of a boundary can be used to circumvent the corner problem in the universality mapping. This implies that all liquids with a topological boundary can be mapped (in a weakly invertible fashion) to the triangle liquid. We also show how to reduce any other sort of defect to a topological boundary using the so-called compactification mapping. Any boundary-liquid model gives rise to a matrix product state (MPS) ground state for the commuting-projector Hamiltonian derived in Section~\ref{sec:hamiltonian}. Furthermore, we introduce invertible domain walls as a kind of topological defect, which provide a way to assert whether two triangle-liquid models are in the same phase.

\subsection{The boundary triangle liquid}
\label{sec:boundary}
In this section we extend the triangle liquid to a liquid on topological $2$-manifolds with boundary, which we call \emph{boundary triangle liquid}. It is based on triangulations of 2-manifolds with boundary, such as
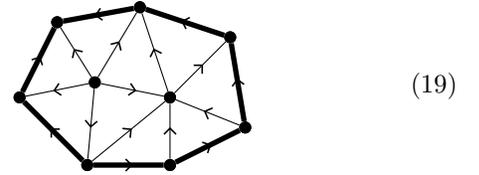
\begin{equation}
\begin{tikzpicture}
\atoms{vertex}{0/, 1/p={0.9,-0.9}, 2/p={2,-0.9}, 3/p={3,-0.4}, 4/p={2.8,0.8}, 5/p={1.6,1.2}, 6/p={0.5,1}, 7/p={1,0.2}, 8/p={2,0}}
\draw (7)edge[mark=arr](0) (7)edge[mark=arr](1) (7)edge[mark=arr](6) (7)edge[mark=arr](5) (7)edge[mark=arr](8) (8)edge[mark={arr,-}](1) (8)edge[mark={arr,-}](2) (8)edge[mark={arr,-}](3) (8)edge[mark=arr](4) (8)edge[mark=arr](5);
\draw[line width=2] (0)edge[mark={arr,-}](1)(1)edge[mark=arr](2) (6)edge[mark={arr,-}](0)  (2)edge[mark=arr](3)(3)edge[mark=arr](4)(4)edge[mark=arr](5)(5)edge[mark=arr](6);
\end{tikzpicture}
\end{equation}
for a triangulation of a disk. For better visibility, we draw the boundary edges as thicker lines. A triangle-liquid tensor is associated to each triangle in the interior. Additionally, there is one tensor associated to each boundary edge. More precisely, the tensor depends on whether the edge is oriented clockwise or counter-clockwise,
\begin{equation}
\begin{tikzpicture}
\atoms{vertex}{0/, 1/p={1,0}}
\draw[line width=2] (0)edge[ior](1) (0)edge[enddots]++(150:0.3) (1)edge[enddots]++(30:0.3);
\draw (0)edge[enddots]++(70:0.3) (1)edge[enddots]++(110:0.3);
\end{tikzpicture}
\quad\rightarrow\quad
\begin{tikzpicture}
\atoms{bdedge}{0/hflip}
\draw (0-t)edge[]++(90:0.4) (0-r)edge[line width=1.2]++(180:0.4) (0-l)edge[line width=1.2]++(0:0.4);
\end{tikzpicture}\;,
\end{equation}
and
\begin{equation}
\begin{tikzpicture}
\atoms{vertex}{0/, 1/p={1,0}}
\draw[line width=2] (0)edge[mark=arr](1) (0)edge[enddots]++(150:0.3) (1)edge[enddots]++(30:0.3);
\draw (0)edge[enddots]++(70:0.3) (1)edge[enddots]++(110:0.3);
\end{tikzpicture}
\quad\rightarrow\quad
\begin{tikzpicture}
\atoms{bdedge}{0/}
\draw (0-t)edge[mark={ar,s}]++(90:0.4) (0-l)edge[line width=1.2]++(180:0.4) (0-r)edge[line width=1.2]++(0:0.4);
\end{tikzpicture}\;,
\end{equation}
which for Hermitian models are complex conjugates. The two indices which run along the boundary are drawn with thick lines and potentially have a different bond dimension than the bulk indices, and we must not contract the two different types of indices. Note, that here the boundary is a \emph{physical boundary} where the network terminates without any open indices, in contrast to the open boundary which corresponds to cutting off a network such that open indices emerge.

In order to model homeomorphisms, we also need to be able to change the triangulation of the boundary itself. This is achieved by adding \emph{boundary Pachner moves} which attach (remove) triangles to (from) the boundary. The 2-2 boundary Pachner move,
\begin{equation}
\begin{tikzpicture}
\atoms{vertex}{0/, 1/p={0.5,0}, 2/p={1,0}};
\draw (0)edge[or](1) (1)edge[or](2) (0)edge[or,bend right=40,looseness=1.5,actualedge](2);
\draw[actualedge] (0)edge[enddots]++(135:0.3) (2)edge[enddots]++(45:0.3);
\draw[] (0)edge[enddots]++(90:0.2) (2)edge[enddots]++(90:0.2) (1)edge[enddots]++(90:0.2);
\end{tikzpicture}
\quad \leftrightarrow \quad
\begin{tikzpicture}
\atoms{vertex}{0/, 1/p={0.5,0}, 2/p={1,0}};
\draw[actualedge] (0)edge[or](1) (1)edge[or](2);
\draw[actualedge] (0)edge[enddots]++(135:0.3) (2)edge[enddots]++(45:0.3);
\draw[] (0)edge[enddots]++(90:0.2) (2)edge[enddots]++(90:0.2) (1)edge[enddots]++(90:0.2);
\end{tikzpicture}\;,
\end{equation}
\begin{equation}
\label{eq:boundary_pachner}
\begin{tikzpicture}
\atoms{bdedge}{r1/p={0,0}}
\atoms{oface}{s/p={0,0.8}}
\draw[bdbind] (r1-l)edge[ind=a]++(-0.4,0) (r1-r)edge[ind=b]++(0.4,0);
\draw (s)edge[mark={ar,s},ind=x]++(135:0.5) (s)edge[mark={ar,s},ind=y]++(45:0.5) (r1-t)edge[mark={ar,s}](s);
\end{tikzpicture}
=
\begin{tikzpicture}
\atoms{bdedge}{r1/p={0,0}}
\atoms{bdedge}{r2/p={0.7,0}}
\draw[bdbind] (r1-r)--(r2-l) (r1-l)edge[ind=a]++(-0.4,0) (r2-r)edge[ind=b]++(0.4,0);
\draw (r1-t)edge[mark={ar,s},ind=x]++(0,0.4) (r2-t)edge[mark={ar,s},ind=y]++(0,0.4); 
\end{tikzpicture}
\end{equation}
is the most important move of the liquid.
For full topological invariance, we also need a 1-3 move and different versions of the moves for different choices of branching structure. One move which is useful to consider is
\begin{equation}
\begin{tikzpicture}
\atoms{bdedge}{r1/p={0,0}}
\atoms{oface}{s/p={0,0.8}}
\draw[bdbind] (r1-l)edge[ind=a]++(-0.4,0) (r1-r)edge[ind=b]++(0.4,0);
\draw[mark={ar,s},looseness=1.5] (s)to[out=30,in=0]++(0,0.7)to[out=180,in=150](s);
\draw (r1-t)edge[mark={ar,s}](s);
\end{tikzpicture}=
\begin{tikzpicture}
\draw[bdbind] (0,0)edge[startind=a, ind=b]++(1,0);
\end{tikzpicture}
\;,
\end{equation}
which can be interpreted as removing a triangle only attached to a single vertex,
\begin{equation}
\begin{tikzpicture}
\atoms{vertex}{0/, 1/p={0,-0.5}}
\draw (0)edge[or](1) (0)edge[enddots]++(90:0.3);
\draw[actualedge] (0)edge[enddots]++(150:0.4) (0)edge[enddots]++(30:0.4);
\draw[actualedge,or,looseness=1.4] (0)to[out=-30,in=0]++(0,-0.8)to[out=180,in=-150](0);
\end{tikzpicture}
\quad \leftrightarrow \quad
\begin{tikzpicture}
\atoms{vertex}{0/}
\draw (0)edge[enddots]++(90:0.3);
\draw[actualedge] (0)edge[enddots]++(150:0.4) (0)edge[enddots]++(30:0.4);
\end{tikzpicture}\;.
\end{equation}
It is possible to find a small set of generating moves, using the principles from Ref.~\cite{liquid_intro}, but this is not the focus this work.

Similar to how the models of the triangle liquid are *-algebras, the models of the boundary triangle liquid are \emph{unitary representations} of those *-algebras. Up to technical details, the boundary triangle liquid is equivalent to the state-sum construction for topological manifolds with boundary in Ref.~\cite{Lauda2006}.

\subsection{The boundary cone mapping}
\label{sec:boundary_cone}
In this section we show that any triangle-liquid model can be extended to a boundary-triangle-liquid model on a purely diagrammatic level. In more physical terms, any fixed-point model of triangle type automatically admits a standard topological boundary.

The statement can be formalised as a liquid mapping, which we call \emph{boundary cone mapping}. The mapping maps a triangulation of a manifold with boundary to a triangulation of a closed manifold by filling each boundary circle with a `cone'. That is, for every boundary circle, we add one single new central vertex, and for each boundary edge we add a new triangle spanned by that edge and the central vertex,
\begin{equation}
\label{eq:cone_mapping_geo}
\begin{tikzpicture}
\atoms{vertex}{0/, 1/p={1,0}, 2/p={2,0}};
\draw[actualedge] (0)--(1)--(2) (0)edge[enddots]++(180:0.4) (2)edge[enddots]++(0:0.4);
\draw (0)edge[enddots]++(120:0.3) (1)edge[enddots]++(135:0.3) (1)edge[enddots]++(45:0.3) (2)edge[enddots]++(60:0.3);
\end{tikzpicture}
\quad\rightarrow\quad
\begin{tikzpicture}
\atoms{vertex}{0/, 1/p={1,0}, 2/p={2,0}, m/p={1,-1}};
\draw (0)--(1)--(2) (0)--(m) (1)--(m) (2)--(m) (0)edge[enddots]++(120:0.3) (1)edge[enddots]++(135:0.3) (1)edge[enddots]++(45:0.3) (2)edge[enddots]++(60:0.3) (0)edge[enddots]++(180:0.3) (2)edge[enddots]++(0:0.3) (m)edge[enddots]++(20:0.3) (m)edge[enddots]++(160:0.3);
\end{tikzpicture}\;.
\end{equation}
%If the boundary consists of multiple disjoint circles, we can take a separate central vertex for each circle. The representations in terms of triangle-liquid networks won't know the difference. On a global level, this amounts to gluing a disk to every boundary circle. E.g., a disk becomes a sphere,
On the level of manifolds, this amounts to removing the boundary by gluing a disk to every boundary circle. This way, a disk becomes a sphere,
\begin{equation}
\begin{tikzpicture}
\atoms{vertex}{0/p=45:1, 1/p=90:1, 2/p=135:1, 3/p=180:1, 4/p=-135:1, 5/p=-90:1, 6/p=-45:1, 7/p=0:1};
\draw[actualedge] (0)--(1)--(2)--(3)--(4)--(5)--(6)--(7)--(0);
\draw[] (0)--(2)--(7)--(3)--(6)--(4);
\end{tikzpicture}
\quad\rightarrow\quad
\begin{tikzpicture}
\atoms{vertex}{0/p=45:1, 1/p=90:1, 2/p=135:1, 3/p=180:1, 4/p=-135:1, 5/p=-90:1, 6/p=-45:1, 7/p=0:1, m/p={-0.2,0.2}};
\draw (0)--(1)--(2)--(3)--(4)--(5)--(6)--(7)--(0);
\draw (0)--(2)--(7)--(3)--(6)--(4);
\draw[backedge] (m)--(0) (m)--(1) (m)--(2) (m)--(3) (m)--(4) (m)--(5) (m)--(6) (m)--(7);
\end{tikzpicture}\;,
\end{equation}
where the drawing on the right consists of a front, and a back layer. Likewise, when the mapping is applied to an annulus, we again obtain a sphere,
\begin{equation}
\begin{tikzpicture}
\atoms{vertex}{0/p={0,0}, 1/p={2,0}, 2/p={0,2}, 3/p={2,2}, 4/p={0.5,0.7}, 5/p={1.5,0.7}, 6/p={0.5,1.3}, 7/p={1.5,1.3}};
\draw[actualedge] (0)--(4)--(6)--(2)--(0) (1)--(3)--(7)--(5)--(1);
\draw[] (0)--(1) (2)--(3) (4)--(5) (6)--(7) (0)--(5) (4)--(7) (6)--(3);
\draw[backedge] (2)to[out=-30,in=150,looseness=1.5](1);
\end{tikzpicture}
\quad\rightarrow\quad
\begin{tikzpicture}
\atoms{vertex}{0/p={0,0}, 1/p={2,0}, 2/p={0,2}, 3/p={2,2}, 4/p={0.5,0.7}, 5/p={1.5,0.7}, 6/p={0.5,1.3}, 7/p={1.5,1.3}, m1/p={-0.5,1}, m2/p={2.5,1}};
\draw[] (0)--(4)--(6)--(2) (3)--(7)--(5)--(1);
\draw[] (0)--(1) (2)--(3) (4)--(5) (6)--(7) (0)--(5) (4)--(7) (6)--(3) (m1)--(0) (m1)--(2) (m1)--(4) (m1)--(6) (m2)--(1) (m2)--(3) (m2)--(5) (m2)--(7);
\draw[backedge] (2)to[out=-30,in=150,looseness=1.5](1) (1)--(3) (0)--(2);
\end{tikzpicture}\;.
\end{equation}
Formally, the mapping is given by
\begin{equation}
\label{eq:cone_mapping}
\begin{tikzpicture}
\atoms{bdedge}{0/}
\draw (0-t)edge[mark={ar,s},ind=a]++(90:0.4) (0-l)edge[line width=1.2,ind=x]++(180:0.4) (0-r)edge[line width=1.2,ind=y]++(0:0.4);
\end{tikzpicture}
\coloneqq
\begin{tikzpicture}
\atoms{oface}{0/}
\draw (0-t)edge[mark={ar,s},ind=a]++(90:0.4) (0-l)edge[mark={ar,s},ind=x]++(180:0.4) (0-r)edge[ind=y]++(0:0.4);
\end{tikzpicture}\;,
\end{equation}
and analogously for the opposite orientation. It is easy to see that the mapped boundary Pachner moves can be derived from the bulk Pachner moves, e.g., the mapping of Eq.~\eqref{eq:boundary_pachner} directly yields a 2-2 Pachner move.
As mentioned earlier, models of the triangle liquid are *-algebras, and models of the boundary triangle liquid are unitary representations of the latter. The cone mapping in Eq.~\eqref{eq:cone_mapping} simply uses the *-algebra as a representation of itself, which is known as the \emph{regular representation}.

\subsection{Universality mapping with boundary}
\label{sec:finegrain_boundary}
In the previous section, we have seen that any model of the triangle liquid admits a cone topological boundary. Thus, this fixed-point ansatz is only suitable for topological phases which admit a gapped/topological boundary. In this section, we will conversely argue that \emph{any} liquid $\calb$ for topological 2-manifolds with boundary is equivalent to the triangle liquid. That is, the universality mapping attempted in Section~\ref{sec:corner_problem} can be made to work, if we have a topological boundary available.

In order to precisely define the universality mapping we require that the generic topological boundary liquid $\calb$ also includes a boundary defect as in Eq.~\eqref{eq:boundary_point_defect} such that the move in Eq.~\eqref{eq:2d_puncture_healing_move} holds. While the topology-changing effect of the move is essential for the mapping, the inclusion of the boundary defect on the left-hand side of the move is needed to make it work for generic models. When we map the triangle liquid to $\calb$, we need to include a tensor corresponding to the boundary defect for the triangle liquid, i.e., we consider a slight generalisation of the triangle liquid which includes further 2-index tensors called \emph{vertex weights}.

A liquid with vertex weights is defined by the additional condition that we need to insert exactly one vertex weight at one of the bonds of each plaquette that corresponds to the network around a vertex of the triangulation. E.g., a valid network representing a triangulation around a $3$-valent vertex could look like
\begin{equation}
\label{eq:vertex_weight}
\begin{tikzpicture}
\atoms{oface}{0/, 1/p={1,0}, 2/p={60:1}}
\atoms{small,square}{x/p={0.5,0}}
\draw (0)edge[mark={ar,e}](x) (x)edge[mark={ar,e}](1) (0)edge[mark={ar,e}](2) (2)edge[mark={ar,e}](1) (0)edge[enddots]++(-150:0.3) (1)edge[enddots]++(-30:0.3) (2)edge[enddots,mark={ar,s}]++(90:0.3);
\end{tikzpicture}\;.
\end{equation}

Note, that the generalisations to triangle liquids with vertex weights or boundary liquids with boundary defects are very mild. I.e., any model of a plain boundary liquid can be extended to a model of the full $\calb$ under mild conditions, namely that $\omega$ is in the image of $M$ in Eq.~\eqref{eq:interval_healing_equation} written as $Mx=\omega$. For robust bosonic intrinsic, SPT or SET topological order, per definition, the model evaluated on a tube is a rank-1 operator. Thus, the one-dimensional model arising from the circle-compactification in Eq.~\eqref{eq:tube_compactification} is trivial and corresponds to a 1-dimensional vector space. For symmetry-breaking or fermionic topological order, the corresponding graded- or super-vectorspace is still one-dimensional when we restrict to the symmetric or even-fermion-parity sector. So in those cases, we can still trivially find a solution for $x$. The only examples where we cannot find a solution for $x$ are \emph{non-robust} symmetry-breaking models where we do \emph{not} impose the symmetry, with a non-symmetric boundary projecting on one of the symmetry-broken sectors. But even with such a symmetry-breaking bulk, we \emph{can} choose the symmetric boundary, and obtain a model for $x$ fulfilling Eq.~\eqref{eq:interval_healing_equation}.

Equipped with these prerequisites we are now able to explicitly construct the mapping. To this end we take the rhombus in Eq.~\eqref{eq:triangle_fine_mapping_rhombus} and cut off the left and right corner. At the cut we terminate the $\calb$-network using the physical boundary. We then cut it into two corner-truncated triangles,
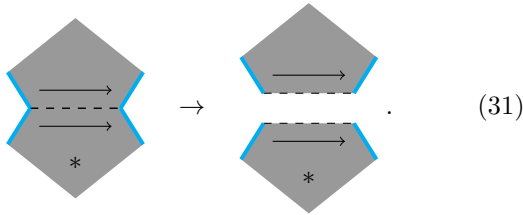
\begin{equation}
\label{eq:trianglebd_fine_mapping_rhombus}
\begin{tikzpicture}
\atoms{void}{x0/, x1/p={3,0}, x2/p={1.5,1.2}, x3/p={1.5,-1.2}}
\trianglebdpatch{x0}{x1}{x2}{}
\trianglebdpatch{x0}{x1}{x3}{*}
\end{tikzpicture}
\quad\rightarrow\quad
\begin{tikzpicture}
\atoms{void}{x0/p={0,0.2}, x1/p={3,0.2}, x2/p={1.5,1.4}, y0/p={0,-0.2}, y1/p={3,-0.2}, y2/p={1.5,-1.4}}
\trianglebdpatch{x0}{x1}{x2}{}
\trianglebdpatch{y0}{y1}{y2}{*}
\end{tikzpicture}\;.
\end{equation}

Recall that the turquoise line denotes the physical boundary, which does not have any open indices in contrast to the open boundary. Next, we take three corner-truncated triangles and fill the gaps between them,
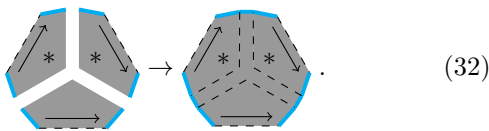
\begin{equation}
\begin{tikzpicture}
\begin{scope}[shift={(-90:0.1)}]
\atoms{void}{0/, 1/p=-30:1.3, 2/p=-150:1.3}
\trianglebdpatch{2}{1}{0}{}
\end{scope}
\begin{scope}[rotate=120,shift={(-90:0.1)}]
\atoms{void}{0/, 1/p=-30:1.3, 2/p=-150:1.3}
\trianglebdpatch{1}{2}{0}{*}
\end{scope}
\begin{scope}[rotate=-120,shift={(-90:0.1)}]
\atoms{void}{0/, 1/p=-30:1.3, 2/p=-150:1.3}
\trianglebdpatch{1}{2}{0}{*}
\end{scope}
\end{tikzpicture}
\rightarrow
\begin{tikzpicture}
\atoms{void}{0/, 1/p={2.5,0}, 2/p={60:2.5}}
\mappingtrianglebd{0}{2}{1}
\end{tikzpicture}
\;.
\end{equation}
The network obtained in this way is what we associate to the triangle tensor,
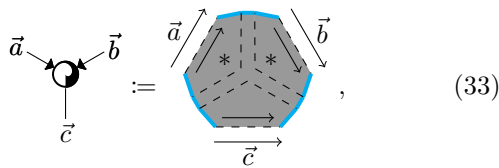
\begin{equation}
\label{eq:mapping_triangle_boundary}
\begin{tikzpicture}
\atoms{oface}{0/}
\draw (0)edge[ind=\vec{c}]++(-90:0.6) (0)edge[mark={ar,s},ind=\vec{a}]++(150:0.6) (0)edge[mark={ar,s},ind=\vec{b}]++(30:0.6);
\end{tikzpicture}
\coloneqq
\begin{tikzpicture}
\atoms{void}{0/p=180:2.5, 2/p=120:2.5, 1/p={0,0}}
\mappingtrianglebd{0}{2}{1}
\draw ($(0)+(60:0.8)+(150:0.2)$) edge[->,mark={slab=$\vec{a}$}] ($(2)-(60:0.8)+(150:0.2)$) ($(2)+(-60:0.8)+(30:0.2)$) edge[->,mark={slab=$\vec{b}$}] ($(1)-(-60:0.8)+(30:0.2)$) ($(0)+(0:0.8)+(-90:0.2)$) edge[->,mark={slab=$\vec{c}$,r}] ($(1)-(0:0.8)+(-90:0.2)$);
\end{tikzpicture}
\;,
\end{equation}
and analogous for the opposite orientation. Now, consider filling the space between two corner-truncated triangles such that there is a boundary defect on one side,
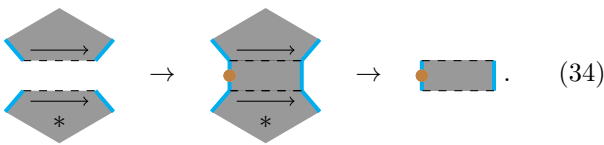
\begin{equation}
\begin{tikzpicture}
\atoms{void}{x0/p={0,0.2}, x1/p={2.4,0.2}, x2/p={1.2,0.9}, y0/p={0,-0.2}, y1/p={2.4,-0.2}, y2/p={1.2,-0.9}}
\trianglebdpatch{x0}{x1}{x2}{}
\trianglebdpatch{y0}{y1}{y2}{*}
\end{tikzpicture}
\quad\rightarrow\quad
\begin{tikzpicture}
\atoms{void}{x0/p={0,0.2}, x1/p={2.4,0.2}, x2/p={1.2,0.9}, y0/p={0,-0.2}, y1/p={2.4,-0.2}, y2/p={1.2,-0.9}}
\trianglebdpatch{x0}{x1}{x2}{}
\trianglebdpatch{y0}{y1}{y2}{*}
\boundarydefectpatch{x0}{x1}{y0}{y1}
\end{tikzpicture}
\quad\rightarrow\quad
\begin{tikzpicture}
\atoms{void}{x0/p={0,0.2}, x1/p={2.4,0.2}, y0/p={0,-0.2}, y1/p={2.4,-0.2}}
\boundarydefectpatch{x0}{x1}{y0}{y1}
\end{tikzpicture}\;.
\end{equation}
The piece in the middle is what we associate to the vertex weight,
\begin{equation}
\begin{tikzpicture}
\atoms{small,square}{x/p={0.5,0}}
\draw (x)edge[ind=\vec a]++(180:0.4) (x)edge[ind=\vec b]++(0:0.4);
\end{tikzpicture}
\coloneqq
\begin{tikzpicture}
\atoms{void}{x0/p={0,0.2}, x1/p={2.4,0.2}, y0/p={0,-0.2}, y1/p={2.4,-0.2}}
\boundarydefectpatch{x0}{x1}{y0}{y1}
\draw (0.7,0.35)edge[->,mark={slab=$\vec a$}](1.7,0.35);
\draw (0.7,-0.35)edge[->,mark={slab=$\vec b$,r}](1.7,-0.35);
\end{tikzpicture}\;.
\end{equation}

With this choice, the network around any vertex is always valid, irrespective of the configuration of surrounding triangles, e.g.,
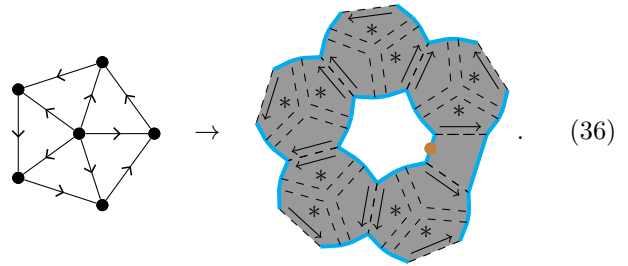
\begin{equation}
\begin{tikzpicture}
\atoms{vertex}{x/, a/p=0:1, b/p=72:1, c/p=144:1, d/p=216:1, e/p=-72:1}
\draw (a)edge[or](b) (b)edge[or](c) (c)edge[or](d) (d)edge[or](e) (e)edge[or](a) (x)edge[or](a) (x)edge[or](b) (x)edge[or](c) (x)edge[or](d) (x)edge[or](e);
\end{tikzpicture}
\quad\rightarrow\quad
\begin{tikzpicture}
\atoms{void}{m/}
\foreach \nr in {0,...,5}{
\atoms{void}{\nr/p=\nr*65:2}
};
\mappingtrianglebd{m}{0}{1}
\mappingtrianglebd{m}{1}{2}
\mappingtrianglebd{m}{2}{3}
\mappingtrianglebd{m}{3}{4}
\mappingtrianglebd{m}{4}{5}
\boundarydefectpatch{$(m)!0.07!(5)$}{$(5)!0.07!(m)$}{$(m)!0.07!(0)$}{$(0)!0.07!(m)$}
\end{tikzpicture}
\;.
\end{equation}
This is because we are just re-gluing patches of $\calb$-networks in the same way they have been cut before. If we apply the mapping to a triangulation, we obtain a manifold with a puncture at each vertex of the triangulation. Every such puncture carries exactly one boundary defect and can thus be removed using the move in Eq.~\eqref{eq:2d_puncture_healing_move}. Thus, the mapping is topology-preserving.

The next thing that we have to check is whether the mapped Pachner moves can be derived from the $\calb$-moves. For the 2-2 Pachner move we have
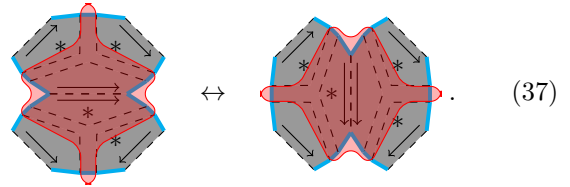
\begin{equation}
\label{eq:boundary_coarse_graining_move}
\begin{tikzpicture}
\atoms{void}{0/p={-1.5,0}, 1/p={0,1.5}, 2/p={1.5,0}, 3/p={0,-1.5}}
\mappingtrianglebd{0}{1}{2}
\mappingtrianglebd{0}{2}{3}
\draw[red,fill=red,fill opacity=0.3,rc] (0.7,0)--(0.95,0.25)--(0.1,0.7)--(0.1,1.2)--(-0.1,1.2)--(-0.1,0.7)--(-0.95,0.25)--(-0.7,0)--(-0.95,-0.25)--(-0.1,-0.7)--(-0.1,-1.2)--(0.1,-1.2)--(0.1,-0.7)--(0.95,-0.25)--cycle;
\end{tikzpicture}
\quad\leftrightarrow\quad
\begin{tikzpicture}
\atoms{void}{0/p={-1.5,0}, 1/p={0,1.5}, 2/p={1.5,0}, 3/p={0,-1.5}}
\mappingtrianglebd{0}{1}{3}
\mappingtrianglebd{1}{2}{3}
\draw[red,fill=red,fill opacity=0.3,rc] (0,0.7)--(0.25,0.95)--(0.7,0.1)--(1.2,0.1)--(1.2,-0.1)--(0.7,-0.1)--(0.25,-0.95)--(0,-0.7)--(-0.25,-0.95)--(-0.7,-0.1)--(-1.2,-0.1)--(-1.2,0.1)--(-0.7,0.1)--(-0.25,0.95)--cycle;
\end{tikzpicture}\;.
\end{equation}
The $\calb$-networks on both sides differ only within the red shaded area. Even though this area involves parts of the physical boundary, it is well isolated from the open boundary. More precisely, the combinatorial distance to the open boundary can be made arbitrarily large by choosing a larger and larger fine-graining scale $\lambda$. Thus, no matter how complicated the moves of $\calb$ are, we can use them to transform the two sides into each other.

For the 1-3 Pachner move we have
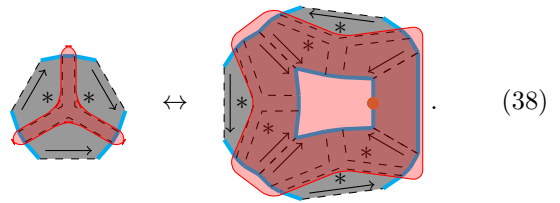
\begin{equation}
\begin{tikzpicture}
\atoms{void}{0/p=-150:1.3, 2/p=-30:1.3, 1/p=90:1.3}
\mappingtrianglebd{0}{1}{2}
\draw[red,fill=red,fill opacity=0.3,rc] ($(90:0.15)+(0:0.1)$)--++(90:0.7)--++(0:-0.2)--($(90:0.15)+(0:-0.1)$)--($(-150:0.15)+(120:0.1)$)--++(-150:0.7)--++(-60:0.2)--($(-150:0.15)+(120:-0.1)$)--($(-30:0.15)+(60:-0.1)$)--++(-30:0.7)--++(60:0.2)--($(-30:0.15)+(60:0.1)$)--cycle;
\end{tikzpicture}
\quad\leftrightarrow\quad
\begin{tikzpicture}
\atoms{void}{0/, 1/p=30:2, 2/p=135:2, 3/p=-135:2, 4/p=-30:2}
\mappingtrianglebd{1}{2}{0}
\mappingtrianglebd{2}{3}{0}
\mappingtrianglebd{3}{4}{0}
\boundarydefectpatch{$(0)!0.07!(4)$}{$(4)!0.07!(0)$}{$(0)!0.07!(1)$}{$(1)!0.07!(0)$}
%\draw[red,fill=red,fill opacity=0.3,rc] (90:1.5)--++(-150:0.3)--(150:0.6)--($(-150:1.5)+(90:0.3)$)--(-150:1.5)--++(-30:0.3)--(-90:0.6)--($(-30:1.5)+(-150:0.3)$)--(-30:1.5)--++(90:0.3)--(30:0.6)--($(90:1.5)+(-30:0.3)$)--cycle;
\draw[red,fill=red,fill opacity=0.3,rc] (122:1.5)--(148:1.6)--(180:1)--(-148:1.6)--(-122:1.5)--(-80:0.9)--(-40:1.6)--(40:1.6)--(80:0.9)--cycle;
\end{tikzpicture}\;.
\end{equation}
Again, the $\calb$-network only changes in the red shaded region which is separated from the open indices. To transform both sides into each other we also need the move in Eq.~\eqref{eq:2d_puncture_healing_move}, in order to remove the puncture on the right-hand side.

We have seen that the triangle liquid is a universal fixed-point form for two-dimensional topological liquid models for which a boundary exists. We will now argue that in addition to that, the boundary triangle liquid is also a universal fixed-point form for the boundary itself. The corresponding universality mapping,
\begin{equation}
\label{eq:mapping_triangle_boundary_boundary}
\begin{tikzpicture}
\atoms{bdedge}{0/}
\draw (0-t)edge[ind=\vec{c}]++(90:0.4) (0-l)edge[ind=\vec{a}]++(180:0.4) (0-r)edge[ind=\vec{b}]++(0:0.4);
\end{tikzpicture}
\coloneqq
\begin{tikzpicture}
\atoms{void}{0/, 1/p=-60:2.5, 2/p=0:2.5}
\mappingtrianglebd{0}{1}{2}
\draw ($(0)+(-60:0.8)+(-150:0.2)$) edge[->,mark={slab=$\vec{a}$,r}] ($(1)-(-60:0.8)+(-150:0.2)$) ($(2)+(-120:0.8)+(-30:0.2)$) edge[->,mark={slab=$\vec{b}$}] ($(1)-(-120:0.8)+(-30:0.2)$) ($(0)+(0:0.8)+(90:0.2)$) edge[->,mark={slab=$\vec{c}$}] ($(2)-(0:0.8)+(90:0.2)$);
\end{tikzpicture}
\;,
\end{equation}
is nothing but the universality mapping in Eq.~\eqref{eq:mapping_triangle_boundary} combined with the cone mapping in Eq.~\eqref{eq:cone_mapping}. As a consequence, all moves of the boundary triangle liquid are derivable from the moves of the triangle liquid which in turn are derivable from the moves of $\calb$. Furthermore, it is easy to see that the universality mapping applied to a triangulation of a manifold with boundary yields the same manifold with boundary again,
\begin{equation}
\begin{multlined}
\begin{tikzpicture}
\atoms{vertex}{0/, 1/p={0.8,0.1}, 2/p={1.6,0.1}, 3/p={2.4,0}};
\draw[actualedge] (0)edge[or](1) (1)edge[or](2) (2)edge[or](3) (0)edge[enddots]++(-0.3,-0.1) (3)edge[enddots]++(0.3,-0.1);
\draw (0)edge[enddots]++(90:0.4) (1)edge[enddots]++(130:0.4) (1)edge[enddots]++(80:0.4) (2)edge[enddots]++(135:0.4) (2)edge[enddots]++(90:0.4) (2)edge[enddots]++(45:0.4) (3)edge[enddots]++(80:0.4);
\end{tikzpicture}\\
\quad\rightarrow\quad
\begin{tikzpicture}
\clip (123:3.5)arc(123:57:3.5)--(57:2)arc(57:123:2)--cycle;
\atoms{void}{x/, 0/p=120:8, 1/p=100:8, 2/p=80:8, 3/p=60:8, y/p=140:8, z/p=40:8};
\mappingtrianglebd{0}{x}{1}
\mappingtrianglebd{1}{x}{2}
\mappingtrianglebd{2}{x}{3}
\mappingtrianglebd{3}{x}{z}
\mappingtrianglebd{y}{x}{0}
\end{tikzpicture}\;.
\end{multlined}
\end{equation}

There is one problem with the universality mapping described above though, namely that it is not compatible with Hermiticity. The latter is a property in liquids with an orientation, i.e., liquids where 1) each index of each tensor is either input or output, 2) in the networks inputs can only be connected to outputs and vice versa, and 3) for each tensor, there is an orientation-reversed dual with inputs and outputs exchanged. A Hermitian model is one where each tensor and its dual are complex conjugates of another. For the triangle liquid, the clockwise and counter-clockwise triangles are duals, and the input indices are marked with ingoing arrows.

For the Hermiticity of the triangle liquid to follow from Hermiticity of $\calb$, we would need the two truncated triangles in Eq.~\eqref{eq:trianglebd_fine_mapping_rhombus} to be the same apart from orientation reversal. However, there is no guarantee that it is possible to cut the truncated rhombus such that this is the case. Instead, if we want to derive Hermiticity, we have to consider a slight generalisation of the triangle liquid. Namely, we have to insert an additional 2-index \emph{edge weight} at each bond (i.e., at each edge of the triangulation), e.g.,
\begin{equation}
\begin{tikzpicture}
\atoms{vertex}{0/, 1/p={1,0}, 2/p={0.5,-0.5}, 3/p={0.5,0.5}}
\draw (0)edge[or](3) (3)edge[or](1) (1)edge[or](2) (0)edge[or](2) (0)edge[or](1);
\draw (0)edge[enddots]++(135:0.4) (0)edge[enddots]++(-135:0.4) (1)edge[enddots]++(45:0.4) (1)edge[enddots]++(-45:0.4) (2)edge[enddots]++(-150:0.4) (2)edge[enddots]++(-90:0.4) (2)edge[enddots]++(-30:0.4) (3)edge[enddots]++(45:0.4) (3)edge[enddots]++(135:0.4);
\end{tikzpicture}
\quad\rightarrow\quad
\begin{tikzpicture}
\atoms{oface}{0/p={0,0.5}, 1/p={0,-0.5}}
\atoms{square,small,flat}{x/}
\draw (0)edge[mark={ar,e}](x) (x)edge[mark={ar,e}](1) (0)edge[mark={ar,s},enddots]++(135:0.5) (0)edge[mark={ar,s},enddots]++(45:0.5) (1)edge[mark={ar,s},enddots]++(-45:0.5) (1)edge[enddots]++(-135:0.5);
\end{tikzpicture}
\;.
\end{equation}
Obviously, we have to also modify the moves by inserting edge weights, e.g., one on each side of the 2-2 Pachner move. For Hermitian models, the clockwise and counter-clockwise triangle tensors are complex conjugates of another as before, and additionally the edge weight is a Hermitian matrix.

In Eq.~\eqref{eq:trianglebd_fine_mapping_rhombus}, we cut a truncated rhombus into two halves which couldn't be guaranteed to be equal up to orientation reversal. With the modified liquid, we can instead start with two equal halves (up to orientation reversal), fill the gap between them,
\begin{equation}
\label{eq:edgeweight_universal_prep}
\begin{tikzpicture}
\atoms{void}{x0/p={0,0.2}, x1/p={2.4,0.2}, x2/p={1.2,0.9}, y0/p={0,-0.2}, y1/p={2.4,-0.2}, y2/p={1.2,-0.9}}
\trianglebdpatch{x0}{x1}{x2}{}
\trianglebdpatch{y0}{y1}{y2}{}
\end{tikzpicture}
\quad\rightarrow\quad
\begin{tikzpicture}
\atoms{void}{x0/p={0,0.2}, x1/p={2.4,0.2}, x2/p={1.2,0.9}, y0/p={0,-0.2}, y1/p={2.4,-0.2}, y2/p={1.2,-0.9}}
\trianglebdpatch{x0}{x1}{x2}{}
\trianglebdpatch{y0}{y1}{y2}{}
\boundarymediatorpatch{x0}{x1}{y0}{y1}
\end{tikzpicture}
\quad\rightarrow\quad
\begin{tikzpicture}
\atoms{void}{x0/p={0,0.2}, x1/p={2.4,0.2}, y0/p={0,-0.2}, y1/p={2.4,-0.2}}
\boundarymediatorpatch{x0}{x1}{y0}{y1}
\end{tikzpicture}\;,
\end{equation}
and then use this gap as the edge weight,
\begin{equation}
\label{eq:edgeweight_universal_mapping}
\begin{tikzpicture}
\atoms{square,small,flat}{x/}
\draw (x-t)edge[mark={ar,s},ind=\vec a]++(90:0.3) (x-b)edge[ind=\vec b]++(-90:0.3);
\end{tikzpicture}
\coloneqq
\begin{tikzpicture}
\atoms{void}{x0/p={0,0.2}, x1/p={2.4,0.2}, y0/p={0,-0.2}, y1/p={2.4,-0.2}}
\boundarymediatorpatch{x0}{x1}{y0}{y1}
\draw (0.8,0.3)edge[mark={slab=$\vec a$},->] (1.6,0.3) (0.8,-0.3)edge[mark={slab=$\vec b$,r},->] (1.6,-0.3);
\end{tikzpicture}\;.
\end{equation}

Since the two halves in Eq.~\eqref{eq:edgeweight_universal_prep} are equal up to orientation reversal, the $\calb$-networks constructed for the clockwise and counter-clockwise triangle tensor in Eq.~\eqref{eq:mapping_triangle_boundary} will also be equal up to orientation reversal. Furthermore, even if the $\calb$-network in Eq.~\eqref{eq:edgeweight_universal_mapping} is not reflection symmetric, it can be chosen so within an arbitrarily large neighbourhood of the open boundary, and thus the original and reflected network are related by $\calb$-moves for a sufficient fine-graining scale. Thus, the Hermiticity of the resulting edge-weight-triangle-liquid model can be derived from the Hermiticity of the $\calb$-model.

If an edge weight $X$ is not only Hermitian but also positive semi-definite, then we can find a square root $X=AA^\dagger$, and include $A$ and $A^\dagger$ to the adjacent triangle tensors. This way, we obtain a Hermitian model of the ordinary triangle liquid without edge weight. However, if $X$ has negative eigenvalues, then we cannot do so. There are, in fact, models where this is the case. E.g., let the triangle tensor be the $\zz_2$ group algebra where the output is multiplied with a Pauli $Z$ matrix, let the vertex weight be the scalar $\frac12$, and the edge weight be the Pauli $Z$ matrix. In the network representing a triangulation, we get $\zz_2$ algebra tensors on all triangles and Pauli $Z$ matrices on all edges between a clockwise and a counter-clockwise triangle. So this is a discrete gauge theory summing over a $1$-cocycle $A$ with action $(-1)^{(A, \omega_1)}$. $\omega_1$ is a 1-cycle representing the first Stiefel-Whitney class as defined in Appendix~\ref{sec:classical_appendix}, and in this case consists of the edges between a clockwise and a counter-clockwise triangle.

\subsection{Ground state tensor networks}
\label{sec:tensor_network_gs}
Consider a model of the boundary triangle liquid, and a network consisting only of the tensors associated to the boundary edges,
\begin{equation}
\label{eq:ground_state_mps}
\begin{tikzpicture}
\atoms{vertex}{0/, 1/p={0.8,0.1}, 2/p={1.6,0.1}, 3/p={2.4,0}};
\draw[actualedge] (0)edge[or](1) (1)edge[or](2) (2)edge[or](3) (0)edge[enddots]++(-0.3,-0.1) (3)edge[enddots]++(0.3,-0.1);
\end{tikzpicture}
\quad\rightarrow\quad
\begin{tikzpicture}
\atoms{bdedge}{0/, 1/p={0.8,0}, 2/p={1.6,0}};
\draw[bdbind] (0-r)--(1-l) (1-r)--(2-l) (0-l)edge[enddots]++(-0.3,0) (2-r)edge[enddots]++(0.3,0);
\draw (0-t)edge[mark={ar,s}]++(0,0.3) (1-t)edge[mark={ar,s}]++(0,0.3) (2-t)edge[mark={ar,s}]++(0,0.3);
\end{tikzpicture}
\end{equation}
of some arbitrarily large boundary circle. The following move,
\begin{equation}
\begin{tikzpicture}
\atoms{bdedge}{0/, 1/p={0.8,0}}
\atoms{comproj}{p/p={0.4,0.8}}
\draw[bdbind] (0-r)--(1-l) (0-l)edge[ind=x]++(-0.3,0) (1-r)edge[ind=y]++(0.3,0);
\draw (p-tl)edge[ind=a]++(120:0.3) (p-tr)edge[ind=b]++(50:0.3);
\draw[mark={ar,s},rc] (0-t)--++(90:0.2)--(p-bl);
\draw[mark={ar,s},rc] (1-t)--++(90:0.2)--(p-br);
\end{tikzpicture}
=
\begin{tikzpicture}
\atoms{bdedge}{0/, 1/p={0.8,0}}
\atoms{oface}{x0/p={0,0.6}, x1/p={0.8,0.6}}
\draw[bdbind] (0-r)--(1-l) (0-l)edge[ind=x]++(-0.3,0) (1-r)edge[ind=y]++(0.3,0);
\draw (x0)edge[ind=a,mark={ar,s}]++(0,0.5) (x1)edge[ind=b,mark={ar,s}]++(0,0.5) (0-t)edge[mark={ar,s}](x0) (1-t)edge[mark={ar,s}](x1) (x0)edge[mark={ar,e}](x1);
\end{tikzpicture}
=
\begin{tikzpicture}
\atoms{bdedge}{0/, 1/p={0.8,0}}
\draw[bdbind] (0-r)--(1-l) (0-l)edge[ind=x]++(-0.3,0) (1-r)edge[ind=y]++(0.3,0);
\draw (0-t)edge[ind=a,mark={ar,s}]++(0,0.3) (1-t)edge[ind=b,mark={ar,s}]++(0,0.3);
\end{tikzpicture}\;,
\end{equation}
defined via the commuting-projector mapping from Eq.~\eqref{eq:comm_proj_mapping}, can be derived from the moves in Eq.~\eqref{eq:boundary_pachner} and a 3-1 Pachner move of the triangle liquid. That is, the tensor in Eq.~\eqref{eq:ground_state_mps} is invariant under applying the local ground state projectors of the corresponding commuting-projector model. Thus, the tensor is a ground state of the model on a triangulated circle. Tensor networks of the form in Eq.~\eqref{eq:ground_state_mps} are known as \emph{matrix product states (MPS)} and are commonly used to represent groundstates of gapped local one-dimensional Hamiltonians. Note, that different boundaries for the triangle liquid yield different MPS representations for different families of ground states.% For the symmetry-breaking models, the irreducible representation boundaries correspond to ground states with completely broken symmetry.

\section{The vertex liquid in \texorpdfstring{$1+1$}{1+1} dimensions}
\label{sec:universal_2d}
In the previous sections, we have discussed liquids with topological boundaries and found that the triangle liquid constitutes a universal ansatz for the latter. In this section, we will consider a more general liquid, the \emph{vertex liquid}. Accordingly, we show that we cannot map the triangle liquid to the latter by means of a universality mapping, i.e., the discussed corner problem indeed occurs. Similar mechanisms also lead to the failure of the boundary cone mapping and the commuting projector mapping. Finally, we show that the vertex liquid is a universal for two-dimensional topological liquid, irrespective of the existence of topological boundaries. To do so, we construct a universality mapping from the vertex liquid to an arbitrary topological liquid which does not suffer from a corner problem. As the vertex liquid has more complicated moves which makes it difficult to find concrete non-trivial models, we provide a strategy called the \emph{operator ansatz} to simplify the description of a liquid at the end of this section.

\subsection{The vertex liquid}
\label{sec:vertex_liquid}
As the name indicates, the vertex liquid associates one tensor to each vertex of a triangulation. The edges of the triangulation correspond to bonds of the tensor network, such that the diagram of the network looks like the drawing of the triangulation itself, e.g.,
\begin{equation}
\begin{tikzpicture}
\atoms{vertex}{0/, 1/p=60:1, 2/p=0:1}
\draw (0)--(1)--(2)--(0) (0)edge[enddots]++(-80:0.7) (0)edge[enddots]++(-150:0.7) (0)edge[enddots]++(140:0.7) (1)edge[enddots]++(40:0.7) (1)edge[enddots]++(140:0.7) (2)edge[enddots]++(-120:0.7) (2)edge[enddots]++(-60:0.7) (2)edge[enddots]++(0:0.7) (2)edge[enddots]++(60:0.7);
\end{tikzpicture}
\quad\rightarrow\quad
\begin{tikzpicture}
\atoms{vertexatom}{0/, {1/p=60:1}, {2/p=0:1}}
\draw (0)--(1)--(2)--(0) (0)edge[enddots]++(-80:0.7) (0)edge[enddots]++(-150:0.7) (0)edge[enddots]++(140:0.7) (1)edge[enddots]++(40:0.7) (1)edge[enddots]++(140:0.7) (2)edge[enddots]++(-120:0.7) (2)edge[enddots]++(-60:0.7) (2)edge[enddots]++(0:0.7) (2)edge[enddots]++(60:0.7);
\end{tikzpicture}
\;.
\end{equation}
As the vertices in a triangulation have different numbers of adjacent edges, they are represented by different tensors with different numbers of indices. As the number of indices is apparent from the diagrammatic calculus, we use the same shape for these different tensors. E.g., in the network above, all $3$ tensors are different, as they have $4$-, $5$-, and $6$ indices, respectively. Naively, we would need an infinite amount of tensors to represent triangulations with arbitrary valencies, but we can restrict to a finite set of adjacencies as shown in Appendix~\ref{sec:appendix_links}.

We again consider a triangulation equipped with a branching structure. The tensor associated to a vertex depends on its \emph{star}, that is, the configuration of triangles containing the vertex, including their edge directions. This dependency is indicated by marking the bonds of the tensor with an ingoing arrow, if the corresponding triangle edge is pointing inwards. Also, for triangles with two inward-pointing or two outward-pointing edges, the direction of the third edge opposite to the vertex will be signified by adding a tick to one of the two bonds. The following example illustrates the marking,
\begin{equation}
\begin{tikzpicture}
\atoms{vertex}{0/, 1/p=0:1, 2/p=72:1, 3/p=144:1, 4/p=216:1, 5/p=-72:1}
\draw (0)edge[ior](1) (0)edge[ior](2) (0)edge[or](3) (0)edge[or](4) (0)edge[ior](5) (1)edge[or](2) (2)edge[or](3) (3)edge[ior](4) (4)edge[ior](5) (5)edge[or](1);
\end{tikzpicture}
\quad\rightarrow\quad
\begin{tikzpicture}
\atoms{vertexatom}{0/}
\draw (0)edge[mark=vor,mark={ar,s}]++(0:0.7) (0)edge[mark={ar,s},mark=vor]++(72:0.7) (0)edge[mark=vor,mark={vor,r}]++(144:0.7) (0)edge[mark={vor,r}]++(-144:0.7) (0)edge[mark={ar,s}]++(-72:0.7);
\end{tikzpicture}\;.
\end{equation}
For Hermitian models, tensors whose stars are related by a reflection are complex conjugates of each other.
The bond dimension of an index at an edge is allowed to depend on the \emph{star} of that edge, i.e., the configuration of the two adjacent triangles including the edge directions. As a consequence, we are only allowed to contract index pairs corresponding to the opposite corners of the same star. More generally, vertex-liquid networks which do not represent branching-structure triangulations, since they contain non-triangular plaquettes non-matching decorations/markings, are considered \emph{invalid}.

We now describe the moves of the liquid. Instead of giving a finite set of generating moves, it is more convenient to describe a larger infinite set of moves which can be derived. The arguments in Appendix~\ref{sec:appendix_links} then make it clear that it is always possible to find a finite generating subset. While for the triangle liquid, any equation between two planar, disk-topology networks defines a derived move, this is not possible for the vertex liquid, since valid vertex liquid networks are restricted to triangular plaquettes. E.g., consider the following equation 
\begin{equation}
\label{eq:vertex_not_move}
\begin{tikzpicture}
\atoms{vertexatom}{0/p={-0.5,0}, 1/p={0.5,0}}
\draw (0)edge[mark={ar,s}](1) (0)edge[ind=a,mark={ar,s}]++(-120:0.6) (0)edge[ind=b]++(180:0.6) (0)edge[ind=c,mark={ar,s}]++(120:0.6) (1)edge[ind=d,mark=vor]++(45:0.6) (1)edge[ind=e]++(-45:0.6);
\end{tikzpicture}
=
\begin{tikzpicture}
\atoms{vertexatom}{0/p={0,-0.5}, 1/p={0,0.5}}
\draw (0)edge[mark={ar,e}](1) (0)edge[ind=a,mark={ar,s}]++(-120:0.6) (0)edge[ind=b]++(180:0.6) (1)edge[ind=c,mark={ar,s}]++(120:0.6) (1)edge[ind=d]++(45:0.6) (0)edge[ind=e]++(-45:0.6);
\end{tikzpicture}\;
\end{equation}
as a move that replaces the network on the left, embedded in some larger vertex-liquid network, with the network on the right. On the left-hand side of the equation, the right-most tensor corresponds to the corner of a triangle which has edges $d$ and $e$. Applying the move inserts an additional edge between $d$ and $e$, so the triangle becomes a $4$-gon which yields an invalid network. Thus, the equation above cannot define a vertex-liquid move.

Instead, all moves of the vertex liquid must satisfy the following condition: For each pair of consecutive open indices, the numbers of bonds separating them along the boundary on the left and on the right side have to be equal. In Eq.~\eqref{eq:vertex_not_move}, $d$ and $e$ are separated by zero bonds on the left, but by one bond on the right. A straight-forward way to obtain moves which do obey the condition is to take Pachner moves and represent each involved vertex by a tensor. E.g., a 2-2 Pachner move yields
\begin{equation}
\label{eq:vertex_pachner_move}
\begin{tikzpicture}
\atoms{vertexatom}{0/, 1/p={1,1}, 2/p={-1,1}, 3/p={0,2}};
\draw (0)edge[mark={ar,s}](1) (1)edge[mark={ar,s}](3) (3)edge[mark={ar,s}](2) (2)edge[mark={ar,e}](0) (0)edge[mark={ar,s}](3) (0)edge[ind=a,mark={ar,s}]++(-30:0.5) (0)edge[ind=b,mark={ar,s},mark={vor,r}]++(-150:0.5) (1)edge[ind=c,mark={ar,s}]++(0:0.5) (2)edge[ind=d,mark={ar,s}]++(-120:0.5) (2)edge[ind=e]++(120:0.5) (3)edge[ind=f,mark={ar,s}]++(10:0.5) (3)edge[ind=g]++(60:0.5) (3)edge[ind=h,mark=vor]++(90:0.5) (3)edge[ind=i,mark={ar,s}]++(120:0.5) (3)edge[ind=j]++(170:0.5);
\end{tikzpicture}
=
\begin{tikzpicture}
\atoms{vertexatom}{0/, 1/p={1,1}, 2/p={-1,1}, 3/p={0,2}};
\draw (0)edge[mark={ar,s}](1) (1)edge[mark={ar,s}](3) (3)edge[mark={ar,s}](2) (2)edge[mark={ar,e}](0) (1)edge[mark={ar,s}](2) (0)edge[ind=a,mark={ar,s}]++(-30:0.5) (0)edge[ind=b,mark={ar,s},mark={vor,r}]++(-150:0.5) (1)edge[ind=c,mark={ar,s}]++(0:0.5) (2)edge[ind=d,mark={ar,s}]++(-120:0.5) (2)edge[ind=e]++(120:0.5) (3)edge[ind=f,mark={ar,s}]++(10:0.5) (3)edge[ind=g]++(60:0.5) (3)edge[ind=h,mark=vor]++(90:0.5) (3)edge[ind=i,mark={ar,s}]++(120:0.5) (3)edge[ind=j]++(170:0.5);
\end{tikzpicture}\;.
\end{equation}
We can write down one such move for each matching quadruple of stars for the vertices at the corners. But there are also many other kinds of moves allowed, and we will study those in more detail in Appendix~\ref{sec:vertex_moves}. Along the lines of Appendix~\ref{sec:appendix_links}, there exists a finite set of generating moves, but this is not the focus of this work.

Note that the vertex liquid is not the only way to construct a liquid more general than the triangle liquid. In Appendix~\ref{sec:extended_triangle_liquid}, we give another example for such a liquid, which might be instructive to have a look at. However, the triangle liquid seems like the simplest and most natural ansatz for a universal topological liquid, despite its algebraic structure still being pretty complicated.

\subsection{Failure of mapping from the triangle liquid}
\label{sec:vertex_finegrain_failure}
In Section~\ref{sec:corner_problem}, we have seen that when we attempt to apply a universality mapping to map to an arbitrary liquid $\calb$ from the triangle liquid, we potentially encounter the so-called corner problem and can neither guarantee that the networks obtained via the mapping are valid nor derive the moves via the mapping. In this section we demonstrate that this problem actually occurs for the vertex liquid and thus show that there cannot be any topology-preserving mapping from the triangle liquid to the vertex liquid. More precisely, we show that any attempted mapping does neither yield valid networks, nor is compatible with the moves of the vertex liquid.

The attempted mapping associates to the triangle a connected, planar vertex-liquid network with disk topology, represented by a triangular area
\begin{equation}
\label{eq:mapping_triangle_fail}
\begin{tikzpicture}
\atoms{oface}{0/}
\draw (0)edge[ind=\vec{c}]++(-90:0.6) (0)edge[mark={ar,s},ind=\vec{a}]++(150:0.6) (0)edge[mark={ar,s},ind=\vec{b}]++(30:0.6);
\end{tikzpicture}
\coloneqq
\begin{tikzpicture}
\atoms{void}{0/, 1/p={1.5,0}, 2/p={60:1.5}}
\fill[manifold,draw, dashed] (0)--(1)--(2)--cycle;
\draw ($(0)+(150:0.2)$) edge[->,mark={slab=$\vec{a}$}] ($(2)+(150:0.2)$) ($(2)+(30:0.2)$) edge[->,mark={slab=$\vec{b}$}] ($(1)+(30:0.2)$) ($(0)+(-90:0.2)$) edge[->,mark={slab=$\vec{c}$,r}] ($(1)+(-90:0.2)$);
\end{tikzpicture}\;.
\end{equation}
Here, $\vec a$ corresponds to a sequence $a_0,\ldots,a_x$ of open indices on the boundary of the vertex-liquid network, ordered according to the arrow direction, and $\vec b$ and $\vec c$ are given analogously. Consider the bottom right corner and the vertex-liquid network connecting the index components $b_x$ and $c_x$. In this corner, this network could look like
\begin{equation}
\label{eq:finegrain_mapping_vertex_corner}
\begin{tikzpicture}
\atoms{void}{0/, 1/p=120:1.5, 2/p=180:1.5}
\fadecorner{0}{1}{2}
\atoms{vertexatom}{x/p={150:0.7}}
\draw (x)edge[ind=c_x,mark={ar,s}]++(-90:0.7) (x)edge[ind=b_x]++(30:0.7);
\end{tikzpicture}\;,\text{or}\quad
\begin{tikzpicture}
\atoms{void}{0/, 1/p=120:1.8, 2/p=180:1.8}
\fadecorner{0}{1}{2}
\atoms{vertexatom}{x/p={138:1.2}, y/p={162:1.2}}
\draw (y)edge[ind=c_x,mark={ar,s}]++(-90:0.5) (x)edge[ind=b_x]++(30:0.5) (x)edge[mark={ar,s}](y);
\end{tikzpicture}\;.
\end{equation}
There could also be more bonds separating $b_x$ and $c_x$ or different ingoing/outgoing arrows. However, no matter what specific network we choose for the mapping, the resulting networks will be invalid at $l$-valent networks for some $l$. E.g., for a 6-valent vertex and the first of the mappings in Eq.~\eqref{eq:finegrain_mapping_vertex_corner}, we obtain
\begin{equation}
\begin{tikzpicture}
\atoms{void}{m/}
\foreach \ang in {0,60,...,300}{
\atoms{void}{x\ang/p={\ang:1.5}}
};
\fadecorner{m}{x0}{x60}
\fadecorner{m}{x60}{x120}
\fadecorner{m}{x120}{x180}
\fadecorner{m}{x180}{x240}
\fadecorner{m}{x240}{x300}
\fadecorner{m}{x300}{x0}
\foreach \ang in {0,60,...,300}{
\atoms{vertexatom}{\ang/p={\ang+30:0.7}}
};
\draw (0)edge[mark={ar,s}](300) (300)edge[mark={ar,s}](240) (240)edge[mark={ar,s}](180) (180)edge[mark={ar,s}](120) (120)edge[mark={ar,s}](60) (60)edge[mark={ar,s}](0);
\end{tikzpicture}\;.
\end{equation}
This network is invalid as it contains a plaquette which is a $6$-gon rather than a triangle.

Also, again the Pachner move can not be derived. If we consider the top corner of the move in Eq.~\eqref{eq:failed_mapped_move} for the first mapping candidate in Eq.~\eqref{eq:finegrain_mapping_vertex_corner} we obtain
\begin{equation}
\label{eq:vertex_finegrain_failure}
\begin{tikzpicture}
\atoms{void}{0/, 1/p=-135:1.5, 2/p=-90:1.5, 3/p=-45:1.5}
\fadecorner{0}{1}{2}
\fadecorner{0}{2}{3}
\atoms{vertexatom}{x/p={-112.5:0.9}, y/p={-67.5:0.9}}
\draw (x)edge[ind=a_x,mark={ar,s}]++(135:0.6) (y)edge[ind=b_0]++(45:0.6) (x)edge[mark={ar,e}](y);
\end{tikzpicture}
=
\begin{tikzpicture}
\atoms{void}{0/, 1/p=-135:1.5, 2/p=-45:1.5}
\fadecorner{0}{1}{2}
\atoms{vertexatom}{x/p={-90:0.6}}
\draw (x)edge[ind=a_x,mark={ar,s}]++(135:0.7) (x)edge[ind=b_0]++(45:0.7);
\end{tikzpicture}
\;.
\end{equation}
This move cannot be derived from vertex-liquid moves, as $a_x$ and $b_0$ are two consecutive open indices which are separated by one bond on the left-hand side, but by no bond on the right-hand side.

\subsection{Failure of boundary cone mapping}
\label{sec:cone_mapping_failure}
In this section we discuss why there is no analogue to the boundary cone mapping from Section~\ref{sec:boundary_cone} for the vertex liquid. In order to be able to talk about a hypothetical boundary cone mapping, we need to extend the vertex liquid with a boundary first, and the simplest way to do so is the following \emph{boundary vertex liquid}. This liquid associates tensors to the boundary vertices of a triangulation of a 2-manifold with boundary, e.g.,
\begin{equation}
\begin{tikzpicture}
\atoms{vertex}{0/}
\draw[actualedge] (0)edge[or, enddots]++(180:0.7) (0)edge[ior,enddots]++(0:0.7);
\draw (0)edge[or, enddots]++(135:0.6) (0)edge[ior,enddots]++(90:0.6) (0)edge[or,enddots]++(45:0.6);
\end{tikzpicture}
\quad\rightarrow\quad
\begin{tikzpicture}
\atoms{bdvertexatom}{0/}
\draw[bdbind] (0)--++(180:0.5) (0)edge[mark={ar,s}]++(0:0.5);
\draw (0)--++(135:0.5) (0)edge[mark={ar,s}]++(90:0.5) (0)--++(45:0.5);
\end{tikzpicture}\;.
\end{equation}
The boundary Pachner moves are implemented by moves such as
\begin{equation}
\label{eq:boundary_vertex_move}
\begin{tikzpicture}
\atoms{bdvertexatom}{0/, 1/p={0.8,0}, 2/p={1.6,0}}
\draw[bdbind] (0)edge[mark={ar,e}](1) (1)edge[mark={ar,e}](2) (0)edge[ind=x,mark={ar,s}]++(180:0.5) (2)edge[ind=y]++(0:0.5);
\draw (0)edge[ind=a]++(135:0.5) (0)edge[ind=b,mark={ar,s}]++(90:0.5) (1)edge[ind=c]++(120:0.5) (1)edge[ind=d,mark={vor,r}]++(60:0.5) (2)edge[ind=e]++(100:0.5) (2)edge[ind=f,mark={ar,s}]++(65:0.5) (2)edge[ind=g]++(30:0.5);
\end{tikzpicture}
=
\begin{tikzpicture}
\atoms{bdvertexatom}{0/, 2/p={1,0}}
\atoms{vertexatom}{1/p={0.5,0.5}}
\draw[bdbind] (0)edge[mark={ar,e}](2) (0)edge[ind=x,mark={ar,s}]++(180:0.5) (2)edge[ind=y]++(0:0.5);
\draw (0)edge[mark={ar,e}](1) (1)edge[mark={ar,e}](2);
\draw (0)edge[ind=a]++(135:0.5) (0)edge[ind=b,mark={ar,s}]++(90:0.5) (1)edge[ind=c]++(120:0.5) (1)edge[ind=d,mark={vor,r}]++(60:0.5) (2)edge[ind=e]++(100:0.5) (2)edge[ind=f,mark={ar,s}]++(65:0.5) (2)edge[ind=g]++(30:0.5);
\end{tikzpicture}\;.
\end{equation}
%Analogous to how the vertex liquid is universal for all topological liquids in $1+1$ dimensions as we will see in Section~\ref{sec:2d_universal_finegrain}, one can show that any $1+1$-dimensional topological liquid with boundary can be mapped from the boundary vertex liquid. 
Note that under the mild assumption of an extra boundary point defect fulfilling Eq.~\eqref{eq:2d_puncture_healing_move}, we can apply the universality mapping in Section~\ref{sec:finegrain_boundary} to obtain a weakly invertible mapping from the simpler boundary triangle liquid. Such a point defect exists in all physically relevant models, so any vertex-liquid model with boundary-vertex-liquid boundary can also be written as a triangle-liquid model with boundary-triangle-liquid boundary. So the boundary vertex liquid is useless for the task of capturing new phases.

Assume there was a boundary cone mapping from the boundary vertex liquid (with boundary point defect) to the vertex liquid. Then we could combine it with the universality mapping in Section~\ref{sec:finegrain_boundary} to obtain a weakly invertible mapping from the triangle liquid to the vertex liquid. As such a mapping does not exist according to Section~\ref{sec:vertex_finegrain_failure}, also the boundary cone mapping cannot exist. We will nonetheless give a direct argument for why there is no boundary cone mapping, as it is instructive and does not depend on the existence of a boundary point defect.

Assume there was a mapping from the boundary vertex liquid to the vertex liquid. Consider the network that this mapping associates to a boundary vertex tensor. This is a connected planar network with the topology of a disk whose boundary is half-open, half-physical,
\begin{equation}
\begin{tikzpicture}
\atoms{bdvertexatom}{0/}
\draw[bdbind] (0)edge[ind=\vec a]++(180:0.5) (0)edge[mark={ar,s}, ind=\vec b]++(0:0.5);
\draw (0)edge[mark={ar,s},ind=\vec x]++(90:0.5);
\end{tikzpicture}
\coloneqq
\begin{tikzpicture}
\fill[manifold] (0,0)rectangle(1,1);
\draw[manifoldboundary] (0,0)--(1,0);
\draw (-0.2,0)edge[->,mark={slab=$\vec a$}](-0.2,1);
\draw (1.2,0)edge[->,mark={slab=$\vec b$,r}](1.2,1);
\draw (0,1.2)edge[->,mark={slab=$\vec x$}](1,1.2);
\end{tikzpicture} \;.
\end{equation}
Here, $\vec a$ and $\vec b$ correspond to the sequences of open indices $a_0,\ldots,a_x$ and $b_0,\ldots,b_x$, respectively, of the network on the right which are ordered according to the arrow directions. We again consider different possible realizations of the mapping. E.g., along the physical boundary, from the index components $a_0$ to $b_0$, the vertex-liquid network on the right could look like
\begin{equation}
\label{eq:attempted_cone_mapping}
\begin{tikzpicture}
\atoms{void}{0/p={-0.4,-0.3}, 1/p={-0.4,0.7}, 2/p={0.4,0.7}, 3/p={0.4,-0.3}}
\fadefgon{3}{2}{1}{0}
\atoms{vertexatom}{0/}
\draw (0)edge[ind=a_0]++(180:0.6) (0)edge[mark={ar,s},ind=b_0]++(0:0.6);
\end{tikzpicture}
\;,\quad\text{or}\quad
\begin{tikzpicture}
\atoms{void}{0/p={-0.7,-0.3}, 1/p={-0.7,0.7}, 2/p={0.7,0.7}, 3/p={0.7,-0.3}}
\fadefgon{3}{2}{1}{0}
\atoms{vertexatom}{0/p={-0.3,0}, 1/p={0.3,0}}
\draw (0)edge[mark={ar,s}](1) (0)edge[ind=a_0]++(180:0.6) (1)edge[mark={ar,s},ind=b_0]++(0:0.6);
\end{tikzpicture}\;,
\end{equation}
or like a network with more bonds separating $a_0$ and $b_0$ and with another choice of arrow directions.

We again show, that for every possible choice of the mapping, there exist boundary vertex liquid networks which result in invalid vertex liquid networks under the mapping.  To this end, we consider a network with a boundary circle consisting of, e.g., four boundary vertices. For, e.g., the first case in Eq.~\eqref{eq:attempted_cone_mapping} we obtain
\begin{equation}
\begin{tikzpicture}
\atoms{bdvertexatom}{{0/p={-90:0.5},rot=180}, {1/p={0:0.5},rot=-90}, {2/p={90:0.5},rot=0}, {3/p={180:0.5},rot=90}}
\draw (0-l)edge[mark={ar,e}](1-r) (1-l)edge[mark={ar,e}](2-r) (2-l)edge[mark={ar,e}](3-r) (3-l)edge[mark={ar,e}](0-r);
\draw (0)edge[mark={ar,s}]++(-90:0.5) (1)edge[mark={ar,s}]++(0:0.5) (2)edge[mark={ar,s}]++(90:0.5) (3)edge[mark={ar,s}]++(180:0.5);
\end{tikzpicture}
\coloneqq
\begin{tikzpicture}
\foreach \nr in {0,1,2,3}{
\atoms{void}{\nr/p={45+90*\nr:0.4}, x\nr/p={45+90*\nr:1.3}}
};
\fadefgon{0}{x0}{x1}{1}
\fadefgon{1}{x1}{x2}{2}
\fadefgon{2}{x2}{x3}{3}
\fadefgon{3}{x3}{x0}{0}
\atoms{vertexatom}{{0/p={-90:0.7}}, {1/p={0:0.7}}, {2/p={90:0.7}}, {3/p={180:0.7}}}
\draw (0)edge[mark={ar,e}](1) (1)edge[mark={ar,e}](2) (2)edge[mark={ar,e}](3) (3)edge[mark={ar,e}](0);
\end{tikzpicture}\;.
\end{equation}
The resulting network has a non-triangular 4-gon plaquette and is hence invalid. For the second case in Eq.~\eqref{eq:attempted_cone_mapping} we would even obtain an 8-gon.

Moreover, there is a problem when we apply the mapping to boundary vertex liquid moves such as Eq.~\eqref{eq:boundary_vertex_move}. E.g., for the first case in Eq.~\eqref{eq:attempted_cone_mapping}, we obtain
\begin{equation}
\begin{tikzpicture}
\atoms{void}{0/p={-0.4,-0.3}, 1/p={-0.4,0.7}, 2/p={0.4,0.7}, 3/p={0.4,-0.3}, 4/p={1.2,0.7}, 5/p={1.2,-0.3}, 6/p={2,0.7}, 7/p={2,-0.3}}
\fadefgon{3}{2}{1}{0}
\fadefgon{5}{4}{2}{3}
\fadefgon{7}{6}{4}{5}
\atoms{vertexatom}{0/, 1/p={0.8,0}, 2/p={1.6,0}}
\draw (0)edge[mark={ar,e}](1) (1)edge[mark={ar,e}](2) (0)edge[ind=x_0,mark={ar,s}]++(180:0.5) (2)edge[ind=y_0]++(0:0.5);
\end{tikzpicture}
=
\begin{tikzpicture}
\atoms{void}{0/p={-0.4,-0.3}, 1/p={-0.4,0.7}, 2/p={0.4,0.7}, 3/p={0.4,-0.3}, 4/p={1.2,0.7}, 5/p={1.2,-0.3}, 6/p={2,0.7}, 7/p={2,-0.3}}
\fadefgon{3}{2}{1}{0}
\fadefgon{5}{4}{2}{3}
\atoms{vertexatom}{0/, 2/p={0.8,0}}
\draw (0)edge[mark={ar,e}](2) (0)edge[ind=x_0,mark={ar,s}]++(180:0.5) (2)edge[ind=y_0]++(0:0.5);
\end{tikzpicture}\;.
\end{equation}
This equation cannot be derived from the moves of the vertex liquid as $x_0$ and $y_0$ are consecutive open indices which are separated by two bonds on the left-hand side but by only one bond on the right-hand side.

Let us provide some more intuition for why the boundary cone mapping works for the triangle liquid, but does not for the vertex liquid. The networks of both the triangle and the vertex liquid represent triangulations, however, for the triangle liquid we can have $x$-valent vertices for arbitrarily high $x$, whereas for the vertex liquid we have $x<l$ for some constant $l$. In terms of triangulations, the boundary cone mapping fills a boundary circle consisting of $b$ edges by adding the corresponding $b$-gon and then dividing it into triangles in a pizza-like manner, e.g. for $b=21$, we have
\begin{equation}
\begin{tikzpicture}
\foreach[remember= \x as \eval] \x in {0,1,...,20}{
\atoms{vertex}{\x/p={\x*360/21:1.5}}
\ifnum\x=0\relax
\else
\draw[actualedge] (\number\numexpr\x-1\relax)--(\x);
\fi
}
\draw[actualedge] (20)--(0);
\end{tikzpicture}
\quad\rightarrow\quad
\begin{tikzpicture}
\atoms{vertex}{c/}
\foreach[remember= \x as \eval] \x in {0,1,...,20}{
\atoms{vertex}{\x/p={\x*360/21:1.5}}
\draw (c)--(\x);
\ifnum\x=0\relax
\else
\draw (\number\numexpr\x-1\relax)--(\x);
\fi
}
\draw (20)--(0);
\end{tikzpicture}\;.
\end{equation}
On the left is a circle of boundary edges surrounded by 2-manifold on the outside, and a hole in the middle. The right side denotes filling that hole with a triangulation of a disk. The essence of why the boundary cone mapping works is that the network associated to the right side has a one-dimensional structure, in particular, the distance between any of the triangles to the boundary is a constant independent of $b$. This construction does not work for the vertex liquid, as there is a $b$-valent vertex on the right, so the corresponding vertex-liquid network is invalid for $b>l$.

If we want to close the hole generated by the boundary by a vertex liquid network, we need to fill the interior of the $b$-edge circle by a triangulation with some maximum adjacency $l$, such as for $l=7$,
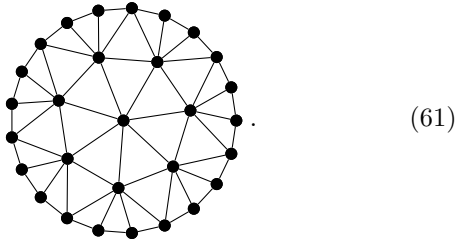
\begin{equation}
\label{eq:hyperbolic_triangulation}
\begin{tikzpicture}
\foreach[remember= \x as \eval] \x in {0,1,...,20}{
\atoms{vertex}{\x/p={\x*360/21:1.5}}
\ifnum\x=0\relax
\else
\draw (\number\numexpr\x-1\relax)--(\x);
\fi
}
\draw (20)--(0);
\atoms{vertex}{m/}
\foreach[remember= \x as \eval] \x in {0,1,...,6}{
\atoms{vertex}{c\x/p={\x*360/7+360/42:0.9}}
\draw (m)--(c\x);
\ifnum\x=0\relax
\else
\draw (c\number\numexpr\x-1\relax)--(c\x);
\fi
}
\draw (c6)--(c0);
\draw (c0)--(20) (c0)--(0) (c0)--(1) (c0)--(2) (c1)--(2) (c1)--(3) (c1)--(4) (c1)--(5) (c2)--(5) (c2)--(6) (c2)--(7) (c2)--(8) (c3)--(8) (c3)--(9) (c3)--(10) (c3)--(11) (c4)--(11) (c4)--(12) (c4)--(13) (c4)--(14) (c5)--(14) (c5)--(15) (c5)--(16) (c5)--(17) (c6)--(17) (c6)--(18) (c6)--(19) (c6)--(20);
\end{tikzpicture}\;.
\end{equation}
Closing off the boundary with such a triangulation cannot be formalized as a liquid mapping though, as the combinatorial distance between some vertex in the middle and the boundary becomes arbitrarily large for large $b$.

To see the last point, define \emph{removing a layer} of a triangulation of a disk as removing all the triangles which contain a boundary edge or a boundary vertex. Removing a layer cannot decrease the size $b$ of the boundary more than by a constant factor, depending on the maximum allowed adjacency $l$. Thus, we need to remove at least $\sim \log(b)$ layers until nothing is left. On the other hand, the number of layers we have to remove is proportional to the combinatorial distance of some vertices in the middle to the boundary. So this distance grows unboundedly with increasing $b$.

Note that the tensor network in Eq.~\eqref{eq:hyperbolic_triangulation} has a hyperbolic geometry with a constant negative curvature, similar to a \emph{MERA} state which is usually used for states with conformal instead of topological symmetry.
%\caro{The last argument (formally) relies on the finite adjacency of the vertex liquid which we put in by hand and one may argue that we can simply allow for arbitrary adjacencies. Should we comment on sth like curvature here?} added a comment

\subsection{Failure of the commuting-projector mapping}
\label{sec:projector_mapping_failure}
In this section, we argue that there also is no analogue of the commuting-projector mapping from Section~\ref{sec:hamiltonian} for the vertex liquid. We assume that such a mapping exists and then show that it does neither yield valid vertex-liquid networks, nor is it compatible with the vertex-liquid moves. The hypothetical mapping associates to the projector a connected planar vertex-liquid network with disk topology,
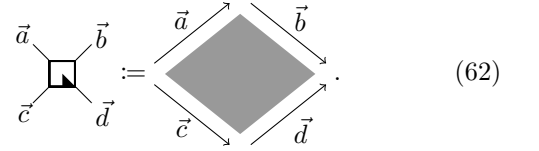
\begin{equation}
\label{eq:vertex_comproj_mapping_ansatz}
\begin{tikzpicture}
\atoms{comproj}{0/}
\draw (0-tr)edge[ind=\vec b]++(45:0.3) (0-tl)edge[ind=\vec a]++(135:0.3) (0-br)edge[ind=\vec d]++(-45:0.3) (0-bl)edge[ind=\vec c]++(-135:0.3);
\end{tikzpicture}
\coloneqq
\begin{tikzpicture}
\atoms{void}{0/, 1/p={1,-0.8}, 2/p={2,0}, 3/p={1,0.8}}
\fill[manifold] (0)--(1)--(2)--(3)--cycle;
\draw ($(0)+(135:0.2)$)edge[->, mark={slab=$\vec a$}] ($(3)+(135:0.2)$);
\draw ($(0)+(-135:0.2)$)edge[->, mark={slab=$\vec c$,r}] ($(1)+(-135:0.2)$);
\draw ($(3)+(45:0.2)$)edge[->, mark={slab=$\vec b$}] ($(2)+(45:0.2)$);
\draw ($(1)+(-45:0.2)$)edge[->, mark={slab=$\vec d$,r}] ($(2)+(-45:0.2)$);
\end{tikzpicture}\;.
\end{equation}
Here, $\vec a$ corresponds to a sequence $a_0,\ldots,a_x$ of open indices ordered according to the direction of the arrow on the right, and the same holds for $\vec b$, $\vec c$, and $\vec d$. For some choices of the hypothetical mapping, the left corner of the network connecting the index components $a_0$ and $c_0$ looks like
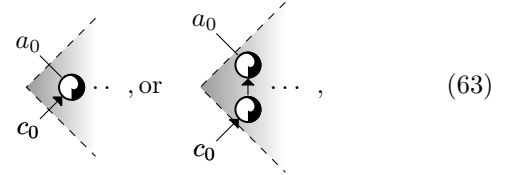
\begin{equation}
\label{eq:comproj_mapping_fail}
\begin{tikzpicture}
\atoms{void}{m/, 1/p=-45:1.3, 2/p=45:1.3}
\fadecorner{m}{1}{2}
\atoms{vertexatom}{0/p=0:0.6}
\draw (0)edge[ind=a_0]++(135:0.6) (0)edge[mark={ar,s},ind=c_0]++(-135:0.6);
\end{tikzpicture}\;,\text{or}\quad
\begin{tikzpicture}
\atoms{void}{m/, 1/p=-45:1.6, 2/p=45:1.6}
\fadecorner{m}{1}{2}
\atoms{vertexatom}{0/p=-25:0.7, 1/p=25:0.7}
\draw (1)edge[ind=a_0]++(135:0.6) (0)edge[mark={ar,s},ind=c_0]++(-135:0.6) (0)edge[mark={ar,e}](1);
\end{tikzpicture}\;,
\end{equation}
while other choices of mapping can have more bonds separating $a_0$ and $c_0$. The same holds for the other corners with the index pairs $(a_x,b_0)$, $(c_x,d_0)$, and $(b_x,d_x)$.

The plaquettes in commuting-projector-liquid networks can be arbitrary large $l$-gons, whereas in the vertex-liquid networks only triangle plaquettes are allowed. However, applying any hypothetical mapping to $l$-gon plaquettes of a commuting-projector-liquid network, yields vertex-liquid networks with $m$-gon plaquettes, such that $m$ gets arbitrarily large when $l$ does. E.g., a $4$-gon plaquette of the commuting-projector liquid,
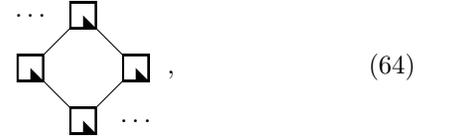
\begin{equation}
\begin{tikzpicture}
\atoms{comproj}{0/, 1/p={0.7,0.7}, 2/p={-0.7,0.7}, 3/p={0,1.4}}
\draw (0-tr)--(1-bl) (1-tl)--(3-br) (0-tl)--(2-br) (2-tr)--(3-bl);
\node at (0.7,0){$\ldots$};
\node at (-0.7,1.4){$\ldots$};
\end{tikzpicture}\;,
\end{equation}
yields a $4$-gon plaquette of the vertex liquid for the mapping given by the first diagram in Eq.~\eqref{eq:comproj_mapping_fail}, and to a plaquette with more than four edges in the other cases. Thus, no matter what vertex-liquid network we take on the right hand side in Eq.~\eqref{eq:vertex_comproj_mapping_ansatz}, the mapping does not yield valid networks.

Moreover, the moves of the commuting-projector liquid can map between $l$-gon plaquettes for different $l$, whereas the moves of the vertex liquid map triangle plaquettes to triangle plaquettes. Therefore, the mapped commuting-projector-liquid moves cannot be derived from the vertex-liquid moves. E.g., the move in Eq.~\eqref{eq:projector_move} for the first case in Eq.~\eqref{eq:comproj_mapping_fail} in the left corner yields
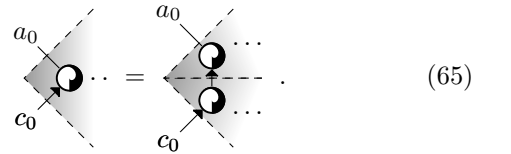
\begin{equation}
\label{eq:comproj_mapping_move_fail}
\begin{tikzpicture}
\atoms{void}{m/, 1/p=-45:1.3, 2/p=45:1.3}
\fadecorner{m}{1}{2}
\atoms{vertexatom}{0/p=0:0.6}
\draw (0)edge[ind=a_0]++(135:0.6) (0)edge[mark={ar,s},ind=c_0]++(-135:0.6);
\end{tikzpicture}
=
\begin{tikzpicture}
\atoms{void}{m/, 1/p=-45:1.3, 2/p=45:1.3, 3/p=0:1.3}
\fadecorner{m}{1}{3}
\fadecorner{m}{3}{2}
\atoms{vertexatom}{0/p=-25:0.7, 1/p=25:0.7}
\draw (1)edge[ind=a_0]++(135:0.6) (0)edge[mark={ar,s},ind=c_0]++(-135:0.6) (0)edge[mark={ar,e}](1);
\end{tikzpicture}\;.
\end{equation}
The resulting move cannot be derived by the vertex-liquid moves, as the consecutive open indices $a_0$ and $c_0$ are separated by $0$ bonds on the left, but by $1$ bond on the right.
\subsection{Universality mapping for the vertex liquid}
\label{sec:2d_universal_finegrain}
In Section~\ref{sec:vertex_finegrain_failure}, we have seen
that the corner problem described in Section~\ref{sec:corner_problem} actually appears when trying to map the vertex liquid from the triangle liquid. In this section, we look at the converse direction and argue that \emph{any} liquid $\calb$ describing topological 2-manifolds can be mapped from the vertex liquid using a universality mapping. That is, we show that the vertex liquid is a universal fixed-point ansatz for $2$-dimensional topological order, which can emulate any other ansatz. This can be done without running into a corner problem and without any extra conditions on $\calb$.

In order to construct the mapping we again resort to a cutting and gluing procedure. We start with a $\calb$-network on a triangle, such that the corners have combinatorial distance $\lambda$ and divide this network into $3$ kite-shaped parts,
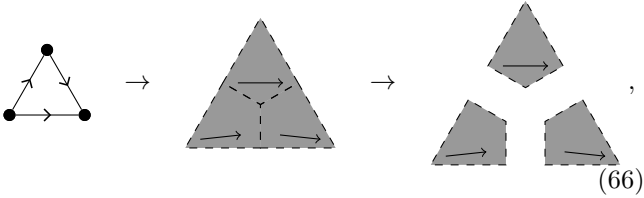
\begin{equation}
\begin{tikzpicture}
\atoms{vertex}{0/, 1/p=60:1, 2/p=0:1}
\draw (0)edge[or](1) (1)edge[or](2) (0)edge[or](2);
\end{tikzpicture}
\quad\rightarrow\quad
\begin{tikzpicture}
\atoms{void}{0/, 1/p=60:2, 2/p=0:2}
\kitepatch{0}{1}{2}{1-}
\kitepatch{1}{2}{0}{2+}
\kitepatch{2}{1}{0}{1+}
\end{tikzpicture}
\quad\rightarrow\quad
\begin{tikzpicture}
\begin{scope}[shift={(-150:0.3)}]
\atoms{void}{0/, 1/p=60:2, 2/p=0:2}
\kitepatch{0}{1}{2}{1-}
\end{scope}
\begin{scope}[shift={(90:0.3)}]
\atoms{void}{0/, 1/p=60:2, 2/p=0:2}
\kitepatch{1}{2}{0}{2+}
\end{scope}
\begin{scope}[shift={(-30:0.3)}]
\atoms{void}{0/, 1/p=60:2, 2/p=0:2}
\kitepatch{2}{1}{0}{1+}
\end{scope}
\end{tikzpicture}\;,
\end{equation}
which form the building blocks of our construction. Note, that all three kite-networks are different as the arrows indicate. The cuts of the triangle which define the detailed structure of the building blocks can be chosen unambiguously using the branching structure of the reference triangle on the left. We can associate the orientation-reversed network and decomposition to the counter-clockwise triangle.

To construct the $\calb$-network associated to the $l$-index vertex tensor, we eventually want to glue together $l$ kite-networks around a vertex. Before that, we need to choose a consistent way to glue the kite-networks along the edges. We do this for each star of the edges, i.e., every configuration of two triangles adjacent to the edge. We decompose both triangles into kite-networks, pick the two pairs of kites adjacent to the edge, and place them next to the edge such that they are separated by a gap along the edge. We then fill this gap with some $\calb$-network, and then cut the network along a line perpendicular to the edge, e.g.,
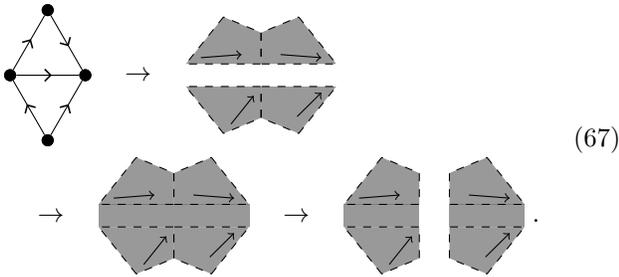
\begin{equation}
\label{eq:kite_gap_filling}
\begin{multlined}
\begin{tikzpicture}
\atoms{vertex}{0/, 1/p=60:1, 2/p=0:1, 3/p=-60:1}
\draw (0)edge[or](1) (1)edge[or](2) (0)edge[or](2) (3)edge[or](0) (3)edge[or](2);
\end{tikzpicture}
\quad\rightarrow\quad
\begin{tikzpicture}
\atoms{void}{0/p={0,0.15}, 1/p={1,1.4}, 2/p={2,0.15}, 3/p={0,-0.15}, 4/p={1,-1.4}, 5/p={2,-0.15}}
\kitepatch{0}{1}{2}{1-}
\kitepatch{2}{1}{0}{1+}
\kitepatch{3}{4}{5}{2-}
\kitepatch{5}{4}{3}{0+}
\end{tikzpicture}
\\\rightarrow\quad
\begin{tikzpicture}
\atoms{void}{0/p={0,0.15}, 1/p={1,1.4}, 2/p={2,0.15}, 3/p={0,-0.15}, 4/p={1,-1.4}, 5/p={2,-0.15}}
\kitepatch{0}{1}{2}{1-}
\kitepatch{2}{1}{0}{1+}
\kitepatch{3}{4}{5}{2-}
\kitepatch{5}{4}{3}{0+}
\fill[manifold] (0)--(2)--(5)--(3)--cycle;
\end{tikzpicture}
\quad\rightarrow\quad
\begin{tikzpicture}
\atoms{void}{0/p={0,0.15}, 1/p={1,1.4}, 2/p={2,0.15}, 3/p={0,-0.15}, 4/p={1,-1.4}, 5/p={2,-0.15}}
\kitepatch{0}{1}{2}{1-}
\kitepatch{3}{4}{5}{2-}
\fill[manifold] (0)--($(0)!0.5!(2)$)--($(3)!0.5!(5)$)--(3)--cycle;
\begin{scope}[xshift=0.4cm]
\atoms{void}{0/p={0,0.15}, 1/p={1,1.4}, 2/p={2,0.15}, 3/p={0,-0.15}, 4/p={1,-1.4}, 5/p={2,-0.15}}
\kitepatch{2}{1}{0}{1+}
\kitepatch{5}{4}{3}{0+}
\fill[manifold] (2)--($(0)!0.5!(2)$)--($(3)!0.5!(5)$)--(5)--cycle;
\end{scope}
\end{tikzpicture}
\;.
\end{multlined}
\end{equation}
Now, for each star of a vertex, we decompose the surrounding triangles into kite-networks, and keep only the kite-networks adjacent to the vertex. We fill the gaps between the kite networks according to Eq.~\eqref{eq:kite_gap_filling}, such that only a small gap around the central vertex remains. At last, we fill this remaining gap with $\calb$-network, e.g.,
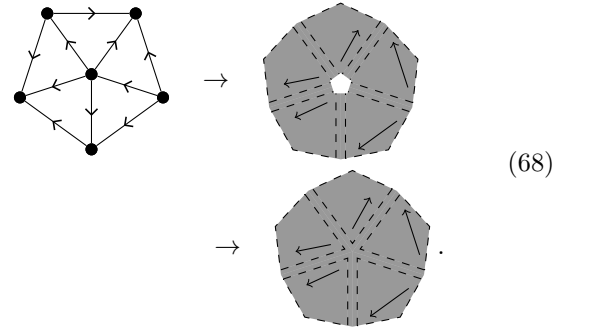
\begin{equation}
\label{eq:vertex_mapping_preparation}
\begin{multlined}
\begin{tikzpicture}
\atoms{vertex}{x/, a/p=-90:1, b/p=-162:1, c/p=126:1, d/p=54:1, e/p=-18:1}
\draw (a)edge[or](b) (b)edge[ior](c) (c)edge[or](d) (d)edge[ior](e) (e)edge[or](a) (x)edge[or](a) (x)edge[or](b) (x)edge[or](c) (x)edge[or](d) (x)edge[ior](e);
\end{tikzpicture}
\quad\rightarrow\quad
\begin{tikzpicture}
\atoms{void}{c/}
\foreach \nr in {0,...,4}{
\atoms{void}{\nr/p={(-90+\nr*72)}:2}
\atoms{void}{x\nr/p={(-90+\nr*72+36)}:0.15}
};
\kitepatchext{c}{0}{1}{2+}
\kitepatchext{c}{1}{2}{2-}
\kitepatchext{c}{2}{3}{0-}
\kitepatchext{c}{3}{4}{1-}
\kitepatchext{c}{4}{0}{0-}
\fill[white] (x0)--(x1)--(x2)--(x3)--(x4)--cycle;
\end{tikzpicture}
\\
\rightarrow\quad
\begin{tikzpicture}
\atoms{void}{c/}
\foreach \nr in {0,...,4}{
\atoms{void}{\nr/p={(-90+\nr*72)}:2}
};
\kitepatchext{c}{0}{1}{2+}
\kitepatchext{c}{1}{2}{2-}
\kitepatchext{c}{2}{3}{0-}
\kitepatchext{c}{3}{4}{1-}
\kitepatchext{c}{4}{0}{0-}
\end{tikzpicture}\;.
\end{multlined}
\end{equation}
The $\calb$-network constructed in this way is used for the mapping, i.e., 
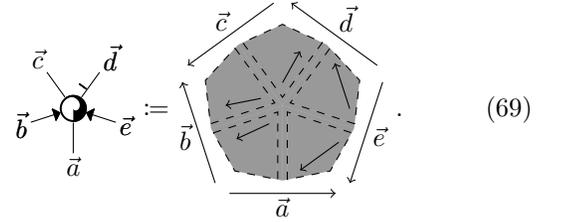
\begin{equation}
\label{eq:vertex_universality_mapping}
\begin{tikzpicture}
\atoms{vertexatom}{0/}
\draw (0)edge[ind=\vec a]++(-90:0.6) (0)edge[ind=\vec b,mark={ar,s}]++(-162:0.6) (0)edge[ind=\vec c]++(126:0.6) (0)edge[ind=\vec d,mark={vor,r}]++(54:0.6) (0)edge[ind=\vec e,mark={ar,s}]++(-18:0.6);
\end{tikzpicture}
\coloneqq
\begin{tikzpicture}
\atoms{void}{c/}
\foreach \nr/\arrow/\lab in {0/->/a,1/<-/e,2/->/d,3/->/c,4/<-/b}{
\atoms{void}{\nr/p={(-90+\nr*72)}:2}
\draw[\arrow,mark={slab=$\vec \lab$,r}] ($({-90+\nr*72-36}:1.2)+({-90+\nr*72}:0.2)$)--($({-90+\nr*72+36}:1.2)+({-90+\nr*72}:0.2)$);
};
\kitepatchext{c}{0}{1}{2+}
\kitepatchext{c}{1}{2}{2-}
\kitepatchext{c}{2}{3}{0-}
\kitepatchext{c}{3}{4}{1-}
\kitepatchext{c}{4}{0}{0-}
%\draw ($(0)+(90+36:0.2)$)edge[->,mark={slab=$\vec a$,r}]($(1)+(90+36:0.2)$) ($(1)+(162+36:0.2)$)edge[<-,mark={slab=$\vec b$,r}]($(2)+(162+36:0.2)$) ($(2)+(-126+36:0.2)$)edge[->,mark={slab=$\vec c$,r}]($(3)+(-126+36:0.2)$) ($(3)+(-54+36:0.2)$)edge[<-,mark={slab=$\vec d$,r}]($(4)+(-54+36:0.2)$) ($(4)+(18+36:0.2)$)edge[->,mark={slab=$\vec e$,r}]($(0)+(18+36:0.2)$);
\end{tikzpicture}\;.
\end{equation}

We now verify our claim, that this prescription defines a valid mapping. To this end, we consider any patch of vertex-liquid network and the associated $\calb$-network, e.g.,
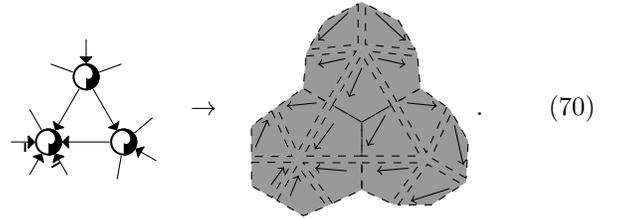
\begin{equation}
\begin{tikzpicture}
\atoms{vertexatom}{0/, 1/p={1,0}, 2/p={60:1}}
\draw (0)edge[mark={ar,s}](1) (1)edge[mark={ar,s}](2) (0)edge[mark={ar,s}](2);
\draw (0)edge[mark={ar,s},mark=vor]++(-60:0.5) (0)edge[mark={ar,s}]++(-120:0.5) (0)edge[]++(120:0.5) (0)edge[mark={ar,s},mark={vor,r}]++(180:0.5) (1)edge[]++(-100:0.5) (1)edge[mark={ar,s}]++(-30:0.5) (1)edge[]++(40:0.5) (2)edge[]++(160:0.5) (2)edge[mark={ar,s}]++(90:0.5) (2)edge[]++(20:0.5);
\end{tikzpicture}
\quad\rightarrow\quad
\begin{tikzpicture}
\atoms{void}{0/p=-150:1, 1/p=-30:1, 2/p=90:1, 3/p=-90:1.7, 4/p=30:1.7, 5/p=150:1.7, 6/p={$(0)+(180:1.2)$}, 7/p={$(0)+(-120:1.2)$}, 8/p={$(1)+(-30:1.2)$}, 9/p={$(2)+(90:1.2)$}}
\kitepatchext{0}{1}{2}{1+}
\kitepatchext{1}{2}{0}{2-}
\kitepatchext{2}{0}{1}{0-}

\kitepatchext{0}{2}{5}{2-}
\kitepatchext{0}{5}{6}{2+}
\kitepatchext{0}{6}{7}{1+}
\kitepatchext{0}{7}{3}{0+}
\kitepatchext{0}{3}{1}{1+}

\kitepatchext{1}{0}{3}{0-}
\kitepatchext{1}{3}{8}{2+}
\kitepatchext{1}{8}{4}{2+}
\kitepatchext{1}{4}{2}{2+}

\kitepatchext{2}{1}{4}{1-}
\kitepatchext{2}{4}{9}{2+}
\kitepatchext{2}{9}{5}{2-}
\kitepatchext{2}{5}{0}{0-}
\end{tikzpicture}
\;.
\end{equation}
At all the points and lines where the different $\calb$-network patches meet, we just re-glue parts which have been cut before. Thus, the resulting $\calb$-network is valid everywhere.

Next we check that the mapped moves of the vertex liquid can be derived from the $\calb$-moves. We consider a move of the vertex liquid, such as
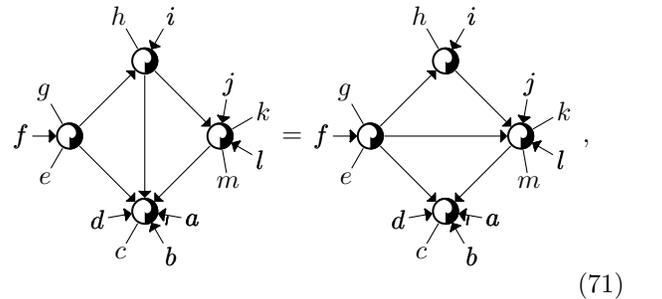
\begin{equation}
\label{eq:vertex_pachner_move_exmp}
\begin{tikzpicture}
\atoms{vertexatom}{0/, 1/p={1,1}, 2/p={-1,1}, 3/p={0,2}};
\draw (0)edge[mark={ar,s}](1) (1)edge[mark={ar,s}](3) (3)edge[mark={ar,s}](2) (2)edge[mark={ar,e}](0) (0)edge[mark={ar,s}](3);
\draw (0)edge[ind=a,mark={ar,s},mark=vor]++(-10:0.5) (0)edge[ind=b,mark={ar,s}]++(-60:0.5) (0)edge[ind=c]++(-120:0.5) (0)edge[ind=d,mark={ar,s}]++(-170:0.5) (1)edge[ind=m]++(-80:0.5) (1)edge[ind=l,mark={ar,s}]++(-30:0.5) (1)edge[ind=k]++(30:0.5) (1)edge[ind=j,mark={ar,s}]++(80:0.5) (3)edge[ind=i,mark={ar,s}]++(60:0.5) (3)edge[ind=h]++(120:0.5) (2)edge[ind=g]++(120:0.5) (2)edge[mark={ar,s},ind=f]++(180:0.5) (2)edge[ind=e]++(-120:0.5);
\end{tikzpicture}
=
\begin{tikzpicture}
\atoms{vertexatom}{0/, 1/p={1,1}, 2/p={-1,1}, 3/p={0,2}};
\draw (0)edge[mark={ar,s}](1) (1)edge[mark={ar,s}](3) (3)edge[mark={ar,s}](2) (2)edge[mark={ar,e}](0) (1)edge[mark={ar,s}](2);
\draw (0)edge[ind=a,mark={ar,s},mark=vor]++(-10:0.5) (0)edge[ind=b,mark={ar,s}]++(-60:0.5) (0)edge[ind=c]++(-120:0.5) (0)edge[ind=d,mark={ar,s}]++(-170:0.5) (1)edge[ind=m]++(-80:0.5) (1)edge[ind=l,mark={ar,s}]++(-30:0.5) (1)edge[ind=k]++(30:0.5) (1)edge[ind=j,mark={ar,s}]++(80:0.5) (3)edge[ind=i,mark={ar,s}]++(60:0.5) (3)edge[ind=h]++(120:0.5) (2)edge[ind=g]++(120:0.5) (2)edge[mark={ar,s},ind=f]++(180:0.5) (2)edge[ind=e]++(-120:0.5);
\end{tikzpicture}\;,
\end{equation}
which, after applying the mapping, corresponds to  the following equation between $\calb$-networks,
\begin{equation}
\begin{tikzpicture}
\atoms{void}{0/, 1/p={1,1}, 2/p={-1,1}, 3/p={0,2}, 4/p={-1.2,-0.2}, 5/p={1.2,-0.2}, 6/p={1.2,2.5}, 7/p={-1.2,2.5}, 8/p={$(0)+(-120:1.4)$}, 9/p={$(0)+(-60:1.4)$}, 10/p={$(1)+(-30:1.4)$}, 11/p={$(1)+(30:1.4)$}, 12/p={$(2)+(180:1.2)$}}
\kitepatchext{0}{1}{3}{1+}
\kitepatchext{1}{3}{0}{2-}
\kitepatchext{3}{0}{1}{0-}

\kitepatchext{0}{2}{3}{0+}
\kitepatchext{3}{0}{2}{2+}
\kitepatchext{2}{3}{0}{1-}

\kitepatchext{0}{2}{4}{0+}
\kitepatchext{0}{4}{8}{2-}
\kitepatchext{0}{8}{9}{2+}
\kitepatchext{0}{9}{5}{0+}
\kitepatchext{0}{5}{1}{1+}

\kitepatchext{1}{0}{5}{0-}
\kitepatchext{1}{5}{10}{2+}
\kitepatchext{1}{10}{11}{2-}
\kitepatchext{1}{11}{6}{2-}
\kitepatchext{1}{6}{3}{0+}

\kitepatchext{3}{1}{6}{2+}
\kitepatchext{3}{6}{7}{2-}
\kitepatchext{3}{7}{2}{2+}

\kitepatchext{2}{3}{7}{1-}
\kitepatchext{2}{7}{12}{2+}
\kitepatchext{2}{12}{4}{2-}
\kitepatchext{2}{4}{0}{1-}
\draw[red,fill=red,fill opacity=0.3,rc] (0,-0.2)--(1.2,1)--(0,2.2)--(-1.2,1)--cycle;
\end{tikzpicture}
=
\begin{tikzpicture}
\atoms{void}{0/, 1/p={1,1}, 2/p={-1,1}, 3/p={0,2}, 4/p={-1.2,-0.2}, 5/p={1.2,-0.2}, 6/p={1.2,2.5}, 7/p={-1.2,2.5}, 8/p={$(0)+(-120:1.4)$}, 9/p={$(0)+(-60:1.4)$}, 10/p={$(1)+(-30:1.4)$}, 11/p={$(1)+(30:1.4)$}, 12/p={$(2)+(180:1.2)$}}
\kitepatchext{0}{1}{2}{1+}
\kitepatchext{1}{2}{0}{2-}
\kitepatchext{2}{0}{1}{0-}

\kitepatchext{1}{2}{3}{0+}
\kitepatchext{3}{1}{2}{2+}
\kitepatchext{2}{3}{1}{1-}

\kitepatchext{0}{2}{4}{0+}
\kitepatchext{0}{4}{8}{2-}
\kitepatchext{0}{8}{9}{2+}
\kitepatchext{0}{9}{5}{0+}
\kitepatchext{0}{5}{1}{1+}

\kitepatchext{1}{0}{5}{0-}
\kitepatchext{1}{5}{10}{2+}
\kitepatchext{1}{10}{11}{2-}
\kitepatchext{1}{11}{6}{2-}
\kitepatchext{1}{6}{3}{0+}

\kitepatchext{3}{1}{6}{2+}
\kitepatchext{3}{6}{7}{2-}
\kitepatchext{3}{7}{2}{2+}

\kitepatchext{2}{3}{7}{1-}
\kitepatchext{2}{7}{12}{2+}
\kitepatchext{2}{12}{4}{2-}
\kitepatchext{2}{4}{0}{1-}
\draw[red,fill=red,fill opacity=0.3,rc] (0,-0.2)--(1.2,1)--(0,2.2)--(-1.2,1)--cycle;
\end{tikzpicture}\;.
\end{equation}
We observe that the two $\calb$-networks only differ inside the red shaded region. This region is separated from the open boundary by a combinatorial distance which scales linearly with the fine-graining scale $\lambda$. Thus, no matter how complicated the moves of $\calb$ are, we can use them to transform the networks on the left-hand side to the network on the right-hand side (or vice versa), if we choose a sufficiently large fine-graining scale $\lambda$.

Having provided an argument that \emph{any} liquid representing topological 2-manifolds is equivalent to the vertex liquid via a universality mapping, let us now consider two concrete examples. First, we consider the case of $\calb$ being the triangle liquid. We can choose a mapping where we replace the $n$-index vertex tensor with a cycle of $n$ triangle tensors, such as
\begin{equation}
\label{eq:vertex_triangle_mapping}
\begin{tikzpicture}
\atoms{vertexatom}{0/}
\draw (0)edge[ind=a]++(-90:0.5) (0)edge[ind=b,mark={ar,s}]++(-162:0.5) (0)edge[ind=c]++(126:0.5) (0)edge[ind=d,mark={vor,r}]++(54:0.5) (0)edge[ind=e,mark={ar,s}]++(-18:0.5);
\end{tikzpicture}
\coloneqq
\begin{tikzpicture}
\atoms{oface}{a/p={-90:0.5}, b/p={-162:0.5}, c/p={126:0.5}, d/p={54:0.5}, e/p={-18:0.5}}
\draw (a)edge[mark={ar,s}](b) (b)edge[mark={ar,s}](c) (c)edge[mark={ar,s}](d) (d)edge[mark={ar,s}](e) (e)edge[mark={ar,s}](a) (e)edge[ind=e,mark={ar,s}]++(-18:0.5) (a)edge[ind=a]++(-90:0.5) (b)edge[ind=b,mark={ar,s}]++(-162:0.5) (c)edge[ind=c]++(126:0.5) (d)edge[ind=d]++(54:0.5);
\end{tikzpicture}\;.
\end{equation}
Be aware that the tensor on the left-hand side is a vertex liquid tensor, whereas the tensors on the right-hand side are triangle liquid tensors, even though we use the same symbols. If we use the mapping on, e.g., the vertex-liquid move in Eq.~\eqref{eq:vertex_pachner_move_exmp}, we obtain
\begin{equation}
\begin{gathered}
\begin{tikzpicture}
\atoms{void}{0/, 1/p={1.5,1.5}, 2/p={-1.5,1.5}, 3/p={0,3}};
\atoms{vertexatom}{4/p={$(0)+(140:0.6)$}, 5/p={$(0)+(-170:0.6)$}, 6/p={$(0)+(-120:0.6)$}, 7/p={$(0)+(-60:0.6)$}, 8/p={$(0)+(-10:0.6)$}, 9/p={$(0)+(40:0.6)$}, 10/p={$(0)+(90:0.6)$}, 11/p={$(1)+(-130:0.6)$}, 12/p={$(1)+(-80:0.6)$}, 13/p={$(1)+(-30:0.6)$}, 14/p={$(1)+(30:0.6)$}, 15/p={$(1)+(80:0.6)$}, 16/p={$(1)+(130:0.6)$}, 17/p={$(3)+(-35:0.6)$}, 18/p={$(3)+(60:0.6)$}, 19/p={$(3)+(120:0.6)$}, 20/p={$(3)+(-145:0.6)$}, 21/p={$(2)+(55:0.6)$}, 22/p={$(2)+(120:0.6)$}, 23/p={$(2)+(180:0.6)$}, 24/p={$(2)+(-120:0.6)$}, 25/p={$(2)+(-55:0.6)$}, 26/p={$(3)+(-90:0.6)$}}
\draw (26)edge[mark={ar,e}](10) (25)edge[mark={ar,e}](4) (21)edge[mark={ar,e}](20) (17)edge[mark={ar,e}](16) (11)edge[mark={ar,e}](9);
\draw (4)edge[mark={ar,s}](5) (5)edge[mark={ar,s}](6) (6)edge[mark={ar,s}](7) (7)edge[mark={ar,s}](8) (8)edge[mark={ar,s}](9) (9)edge[mark={ar,s}](10) (10)edge[mark={ar,s}](4);
\draw (11)edge[mark={ar,s}](12) (12)edge[mark={ar,s}](13) (13)edge[mark={ar,s}](14) (14)edge[mark={ar,s}](15) (15)edge[mark={ar,s}](16) (16)edge[mark={ar,s}](11);
\draw (17)edge[mark={ar,s}](18) (18)edge[mark={ar,s}](19) (19)edge[mark={ar,s}](20) (20)edge[mark={ar,s}](26) (26)edge[mark={ar,s}](17);
\draw (21)edge[mark={ar,s}](22) (22)edge[mark={ar,s}](23) (23)edge[mark={ar,s}](24) (24)edge[mark={ar,s}](25) (25)edge[mark={ar,s}](21);
\draw (8)edge[ind=a,mark={ar,s}]++(-10:0.5) (7)edge[ind=b,mark={ar,s}]++(-60:0.5) (6)edge[ind=c]++(-120:0.5) (5)edge[ind=d,mark={ar,s}]++(-170:0.5) (12)edge[ind=m]++(-80:0.5) (13)edge[ind=l,mark={ar,s}]++(-30:0.5) (14)edge[ind=k]++(30:0.5) (15)edge[ind=j,mark={ar,s}]++(80:0.5) (18)edge[ind=i,mark={ar,s}]++(60:0.5) (19)edge[ind=h]++(120:0.5) (22)edge[ind=g]++(120:0.5) (23)edge[mark={ar,s},ind=f]++(180:0.5) (24)edge[ind=e]++(-120:0.5);
\end{tikzpicture}
\\=
\begin{tikzpicture}
\atoms{void}{0/, 1/p={1.5,1.5}, 2/p={-1.5,1.5}, 3/p={0,3}};
\atoms{vertexatom}{4/p={$(0)+(140:0.6)$}, 5/p={$(0)+(-170:0.6)$}, 6/p={$(0)+(-120:0.6)$}, 7/p={$(0)+(-60:0.6)$}, 8/p={$(0)+(-10:0.6)$}, 9/p={$(0)+(40:0.6)$}, 10/p={$(2)+(0:0.6)$}, 11/p={$(1)+(-130:0.6)$}, 12/p={$(1)+(-80:0.6)$}, 13/p={$(1)+(-30:0.6)$}, 14/p={$(1)+(30:0.6)$}, 15/p={$(1)+(80:0.6)$}, 16/p={$(1)+(130:0.6)$}, 17/p={$(3)+(-35:0.6)$}, 18/p={$(3)+(60:0.6)$}, 19/p={$(3)+(120:0.6)$}, 20/p={$(3)+(-145:0.6)$}, 21/p={$(2)+(55:0.6)$}, 22/p={$(2)+(120:0.6)$}, 23/p={$(2)+(180:0.6)$}, 24/p={$(2)+(-120:0.6)$}, 25/p={$(2)+(-55:0.6)$}, 26/p={$(1)+(180:0.6)$}}
\draw (26)edge[mark={ar,e}](10) (25)edge[mark={ar,e}](4) (21)edge[mark={ar,e}](20) (17)edge[mark={ar,e}](16) (11)edge[mark={ar,e}](9);
\draw (4)edge[mark={ar,s}](5) (5)edge[mark={ar,s}](6) (6)edge[mark={ar,s}](7) (7)edge[mark={ar,s}](8) (8)edge[mark={ar,s}](9) (9)edge[mark={ar,s}](4);
\draw (11)edge[mark={ar,s}](12) (12)edge[mark={ar,s}](13) (13)edge[mark={ar,s}](14) (14)edge[mark={ar,s}](15) (15)edge[mark={ar,s}](16) (16)edge[mark={ar,s}](26) (26)edge[mark={ar,s}](11);
\draw (17)edge[mark={ar,s}](18) (18)edge[mark={ar,s}](19) (19)edge[mark={ar,s}](20) (20)edge[mark={ar,s}](17);
\draw (21)edge[mark={ar,s}](22) (22)edge[mark={ar,s}](23) (23)edge[mark={ar,s}](24) (24)edge[mark={ar,s}](25) (25)edge[mark={ar,s}](10) (10)edge[mark={ar,s}](21);
\draw (8)edge[ind=a,mark={ar,s}]++(-10:0.5) (7)edge[ind=b,mark={ar,s}]++(-60:0.5) (6)edge[ind=c]++(-120:0.5) (5)edge[ind=d,mark={ar,s}]++(-170:0.5) (12)edge[ind=m]++(-80:0.5) (13)edge[ind=l,mark={ar,s}]++(-30:0.5) (14)edge[ind=k]++(30:0.5) (15)edge[ind=j,mark={ar,s}]++(80:0.5) (18)edge[ind=i,mark={ar,s}]++(60:0.5) (19)edge[ind=h]++(120:0.5) (22)edge[ind=g]++(120:0.5) (23)edge[mark={ar,s},ind=f]++(180:0.5) (24)edge[ind=e]++(-120:0.5);
\end{tikzpicture}
\;.
\end{gathered}
\end{equation}
This equation corresponds to a retriangularisation of a disk, which can be performed by a sequence of Pachner moves. As a consequence the mapped vertex liquid move can be derived from the triangle liquid moves. The same holds true for all other vertex liquid moves.

As a second example, we take for $\calb$ the extended triangle liquid presented in Appendix~\ref{sec:extended_triangle_liquid}. In this case, we need a slightly larger network,
\begin{equation}
\label{eq:vertex_extended_triangle_mapping}
\begin{tikzpicture}
\atoms{vertexatom}{0/}
\draw (0)edge[ind=aa']++(-90:0.5) (0)edge[ind=b'b,mark={ar,s}]++(-162:0.5) (0)edge[ind=cc']++(126:0.5) (0)edge[ind=dd',mark={vor,r}]++(54:0.5) (0)edge[ind=e'e,mark={ar,s}]++(-18:0.5);
\end{tikzpicture}
\coloneqq
\begin{tikzpicture}
\foreach \nr/\lab in {0/a, 1/b, 2/c, 3/d, 4/e}{
\atoms{oface}{
{\lab/p={-90-72*\nr:0.5},extendedtrianglab=5:{90-72*\nr} and 6:{-30-72*\nr} and 6:{-150-72*\nr}},
{x\lab/p={-90-72*\nr:1},extendedtrianglab=6:{-90-72*\nr} and 6:{30-72*\nr} and 6:{150-72*\nr}},
{y\lab/p={-110-72*\nr:1.5},extendedtrianglab=6:{90-72*\nr} and 6:{-30-72*\nr} and 6:{-150-72*\nr}},
{z\lab/p={-70-72*\nr:1.5},extendedtrianglab=6:{90-72*\nr} and 6:{-30-72*\nr} and 6:{-150-72*\nr}}}
\draw (\lab)--(x\lab) (x\lab)--(y\lab) (x\lab)--(z\lab) (y\lab)edge[ind=\lab]++(-90-72*\nr:0.5) (z\lab)edge[ind=\lab ']++(-90-72*\nr:0.5);
};
\draw (a)edge[](b) (b)edge[](c) (c)edge[](d) (d)edge[](e) (e)edge[](a) (ya)--(zb) (yb)--(zc) (yc)--(zd) (yd)--(ze) (ye)--(za);
\end{tikzpicture}\;.
\end{equation}
We cannot take a mapping of the same form as Eq.~\eqref{eq:vertex_triangle_mapping}, because for this mapping, the retriangularisation resulting from, e.g., Eq.~\eqref{eq:vertex_pachner_move_exmp} requires Pachner moves involving triangles which contain boundary vertices or edges. If we use the mapping in Eq.~\eqref{eq:vertex_extended_triangle_mapping} instead, the corresponding retriangularisation can be performed with Pachner moves acting only on triangles which are distant from the boundary.

\subsection{Operator ansatz}
The vertex liquid is quite complicated, as its models consist of a large set of tensors fulfilling a large set of tensor-network equations. More precisely, there is one tensor for every vertex star, and one move for every combination of vertex star for the four corners of a 2-2 Pacher move. This makes it hard to find models in practice, even though we would like to point out that fundamentally, at each finite bond dimension, we are looking for a finite set of complex numbers fulfilling a finite set of polynomial equations, just as for any linear-algebraic structure.

It does not seem possible to simplify the liquid much on a purely combinatorial level. However, using very weak assumption on the tensor type, i.e., the interpretation of the diagrams in terms of computations, we can obtain a significant simplification. The assumptions needed for this hold at least for ordinary array tensors, but also for symmetric tensors or fermionic tensors. We will refer to this simplification as the \emph{operator ansatz}.

While a more detailed explanation of this concept is given in Appendix~\ref{sec:vertex_operator_ansatz}, the basic idea is the following. We find that if we go from a star $L_1$ to another star $L_2$ which only differs locally at a few edges, the tensors associated to $L_1$ and $L_2$ are related by applying an operator to indices nearby of where the change happens. This is a consequence of the vertex-liquid moves, but also of the fact that the product $AB$ of two matrices does not change if we restrict the support of $B$ to the image of $A$.

So instead of providing a tensor for each star, we can equivalently specify the tensor for one chosen simple star and all the operators changing the star, which are tensors themselves. By applying the operators corresponding to a sequence of deformations, we can construct all vertex tensors from the one for the simple star. Most interestingly, the operator implementing the change from $L_1$ to $L_2$ only depends on the local change itself, but not on the full data of $L_1$ and $L_2$. So, instead of specifying one tensor for every star, it suffices to provide one tensor for each move locally changing the stars. Different sequences of moves might yield the same change between two stars, which yields coherence tensor-network equations for the operators implementing the moves. In Appendix~\ref{sec:vertex_operator_ansatz}, we use this operator ansatz to gain insight into the array-tensor vertex liquid models. While we do not manage to arrive at a full classification yet, we find that a large part of the vertex-liquid models are specified by a symmetric Frobenius algebra.

\section{Liquids in higher dimensions}
\label{sec:higher_dimensions}
In this section we describe how the arguments in the previous sections generalise to higher spacetime dimensions.

\subsection{The triangle liquid in higher dimensions}
\label{sec:universal_higherd}
The generalisation of the discussed liquids to higher dimensions is straight-forward in large parts. We start with discussing the higher-dimensional generalisation of the triangle liquid which we refer to as the \emph{simplex liquid}. The $n$-simplex liquid associates one $n+1$-index tensor to every $n$-simplex of a triangulation of a $n$-manifold. E.g., for $n=3$, we have a $4$-index tensor associated to a tetrahedron,
\begin{equation}
\begin{tikzpicture}
\atoms{vertex,redlab}{{0/lab={t=0,ang=-135}}, {1/p={1,0},lab={t=1,ang=-45}}, {2/p={0,1},lab={t=2,ang=135}}, {3/p={1,1},lab={t=3,ang=45}}}
\draw (1)edge[or=0.3](2) (1)edge[or](3) (2)edge[or](3) (0)edge[or](1) (0)edge[or](2) (0)edge[backedge,or=0.3](3);
\end{tikzpicture}
\quad\rightarrow\quad
\begin{tikzpicture}
\atoms{tetrahedron}{{0/lab={t=0123,p=135:0.4}}}
\draw (0-b)edge[redlab=123, ind=T]++(0,-0.6) (0-r)edge[redlabr=023,ind=T]++(0.6,0) (0-t)edge[redlabr=013,ind=T]++(0,0.6) (0-l)edge[redlab=012,ind=T]++(-0.6,0);
\end{tikzpicture}\;.
\end{equation}
The triangulation carries a branching structure, that is, every edge is directed such that the directions are never cyclic around any triangle. The branching structure allows to distinguish between the different indices of the tensor, and label the $n$-simplices as `left-handed' or `right-handed', relative to an underlying orientation of the manifold. Left-handed and right-handed $n$-simplices are represented by different tensors, which for a Hermitian model are complex conjugates of each other. Tensors at $n$-simplices sharing a common $n-1$-simplex have a contracted index pair.

The Pachner moves in $n$ dimensions can be obtained by considering the boundary of the $n+1$-simplex, and dividing it into two parts consisting of $x$ and $y$ $n$-simplices, respectively, such that $x+y=n+2$. E.g., for $n=3$, one Pachner move is given by dividing a the boundary of a 4-simplex into two parts containing three and two simplices each,
\begin{equation}
\label{eq:23pachner_complex}
\begin{tikzpicture}
\atoms{vertex,redlab}{{0/lab={t=0,ang=180}}, {1/p={1,-1.2},lab={t=1,ang=-90}}, {2/p={1.3,-0.4},lab={t=2,ang=-160}}, {3/p={1,1.2},lab={t=3,ang=90}}, {4/p={2,0},lab={t=4,ang=0}}}
\draw (0)edge[or](1) (0)edge[or](2) (0)edge[or](3) (0)edge[backedge,or](4) (1)edge[or](2) (1)edge[or](4) (2)edge[or](3) (2)edge[or](4) (3)edge[or](4);
\end{tikzpicture}
\quad\leftrightarrow\quad
\begin{tikzpicture}
\atoms{vertex,redlab}{{0/lab={t=0,ang=180}}, {1/p={1,-1.2},lab={t=1,ang=-90}}, {2/p={1.3,-0.4},lab={t=2,ang=-160}}, {3/p={1,1.2},lab={t=3,ang=90}}, {4/p={2,0},lab={t=4,ang=0}}}
\draw (0)edge[or](1) (0)edge[or](2) (0)edge[or](3) (0)edge[backedge,or](4) (1)edge[or](2) (1)edge[or](4) (2)edge[or](3) (2)edge[or](4) (3)edge[or](4) (1)edge[or=0.6,midedge,bend left=10](3);
\end{tikzpicture}\;.
\end{equation}
Here, the left-hand side represents two tetrahedra glued at the $024$ face and the right-hand side represents three tetrahedra sharing the $13$ edge. As an equation between networks the move corresponds to
\begin{equation}
\label{eq:23pachner}
\begin{tikzpicture}
\atoms{tetrahedron}{{0/rot=90,lab={t=0124,p=-45:0.4}}, {1/p={0,0.8},rot=180,lab={t=0124,p=45:0.4}}}
\draw (0-r)edge[redlab=024](1-t) (0-b)edge[sdind=124]++(0:0.3) (0-t)edge[sdind=014]++(180:0.3) (0-l)edge[sdind=012]++(-90:0.3) (1-l)edge[sdind=023]++(0:0.3) (1-b)edge[sdind=234]++(90:0.3) (1-r)edge[sdind=034]++(180:0.3);
\end{tikzpicture}
=
\begin{tikzpicture}
\atoms{tetrahedron}{{0/lab={t=0123,p=-135:0.4}}, {1/p={2,0},rot=90,lab={t=1234,p=-45:0.4}}, {2/p={1,1},rot=45,lab={t=0134,p=90:0.4}}}
\draw (0-b)edge[looseness=0.6,out=-90,in=-90,redlabr=123](1-l) (0-t)edge[out=90,in=-135,redlab=013](2-l) (0-l)edge[sdind=012]++(180:0.4) (0-r)edge[sdind=023]++(0:0.2) (1-b)edge[sdind=234]++(0:0.4) (1-t)edge[sdind=124]++(180:0.2) (2-r)edge[sdind=034]++(45:0.4) (2-t)edge[sdind=014]++(135:0.4) (1-r)edge[out=90,in=-45,redlabr=134](2-b);
\end{tikzpicture}\;.
\end{equation}
For a more detailed description of the simplex liquid in 2+1 dimensions, including a simplified version and the connection to established descriptions of non-chiral topological order via state sums or commuting projector Hamiltonians we refer the reader to Ref.~\cite{liquid_intro}.

\subsection{The vertex liquid in higher dimensions}
\label{sec:vertex_liquid_higher}
The vertex liquid has a generalization to $n$ dimensions as well. The generalization is straight-forward apart from one subtlety, namely that the vertex-liquid networks in higher dimensions describe triangulations with not only a branching structure, but also a \emph{dual branching}, as defined below. The latter is trivial in the case of triangulations of $1+1$-dimensional oriented manifolds.

To describe the dual branching, we first need to introduce some definitions.
The \emph{link} of an $x$-simplex $X$ in an $n$-dimensional triangulation is defined as the triangulation of an $n-x-1$-sphere formed by the $n-x-1$-simplices which together with $X$ span an $n$-simplex.
The \emph{star} $\star(X)$ is the configuration of all $n$-simplices containing $X$. We also consider all sub-simplices of those $n$-simplices as part of the star, which is sometimes called ``closed star'' in the literature. The sub-simplices containing $X$ are called the \emph{internal} simplices. We will usually think of the link and star as simplicial complexes on their own rather than as sub-complexes of the triangulation. Note that $X$ can be contained in an $n$-simplex multiple times, in which case the star contains multiple copies of this $n$-simplex, and analogous reasoning applies to the link. As such, the link and star contain the same combinatorial information, as the latter is simply spanned by the former together with an $x$-simplex. However, if the triangulation has a branching structure, then we will also equip the edges of the star with directions, including the edges which are not part of the link.

Even though we are not aware of an official explicit proof in the literature, the consensus seems to be that it suffices to restrict to triangulations with a finite but large enough set of allowed stars, also in higher dimensions. More discussion on this can be found at the end of Appendix~\ref{sec:appendix_links}. Note that this is necessary and sufficient for the higher-dimensional vertex liquids to be determined by a finite set of tensors subject to a finite set of axioms.
A \emph{dual branching} of an $x$-simplex $X$ is an identification of the star of $X$ with its canonical representative. If this star does not have any symmetries, then there is only one possible dual branching, otherwise there are as many as there are elements of the symmetry group of the star. Hence, there is a unique dual branching for the edges in a $2$-dimensional triangulation, as the two adjacent triangles cannot have any symmetries. Note, that reflection symmetries are not allowed if the triangulation is equipped with an orientation.

An \emph{$x$-dual-branched triangulation} is one where all the $y$-simplices with $y\geq x$ carry a dual branching. More precisely, we equip the star of an $z$-simplex for $z<x$ in a $x$-dual-branched triangulation with the dual branchings of its internal $z$-simplices with $z\geq x$. Consequently, also the standard representatives of stars should be equipped with such dual branchings. Thus, in order to define the dual branching we should proceed inductively, starting defining standard representatives for the $n-1$-simplex stars, then defining standard representatives for the $n-2$-simplex stars containing dual branchings of the internal $n-1$-simplices therein, and so on.

The vertex liquid in $n$ dimensions represents $1$-dual-branched, branched triangulations as a network with one tensor at every vertex, depending on the star of that vertex, including the $1$-dual-branching of the internal simplices of the star. At every edge, there is a bond, whose dimension is allowed to depend on the star of that edge.

The moves correspond to Pachner moves of the triangulations, and consist of the tensors at all the vertices involved in the Pachner move. This is a sufficient set of generating moves, however, 
we can write down a larger family of moves which can be consistently added. To this end we go from the triangulation to its dual cellulation, where vertices become $n$-cells whose boundary is dual to a $n-1$-dimensional triangulation. Then, every pair of $n$-sphere cellulations, which do not differ in the vicinity of the boundary, defines a move. Not differing in the vicinity of the boundary means that the star of all boundary vertices (restricted to the interior) stays the same.

Let us consider the case of $n=2+1$ to be more specific. Possible vertex links are two-dimensional triangulations such as the tetrahedron, the octahedron, or a cube with diagonal edges over all $6$ faces. The corresponding stars can be obtained by adding a central vertex spanning simplices with all existing simplices. These stars carry a branching, i.e., non-cyclic edge directions. They also carry a $1$-dual-branching, meaning that the configuration of tetrahedra surrounding every internal edge is identified with a canonical configuration. E.g., if there are $l$ surrounding tetrahedra with edge directions allowing for full rotation invariance, then there are $l$ possible choices for the dual branching of that edge. The dual branching of internal triangles is trivial since the two adjacent tetrahedra cannot have any symmetries due to the orientation of the 3-manifold.

The moves are Pachner moves, such as the $2-3$ Pachner move in Eq.~\eqref{eq:23pachner_complex} which has five tensors on both the left and right. There is one such move for each choice of stars of the involved vertices. More generally, consistent moves are dual to recellulations which do not change the vicinity of the boundary. An example of such a move (not drawing branching and dual branching) is
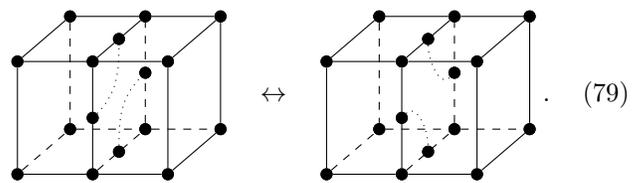
\begin{equation}
\label{eq:vertex_3d_move}
\begin{tikzpicture}
\atoms{vertex}{0/, 1/p={1,0}, 2/p={2,0}, 3/p={0,1.5}, 4/p={1,1.5}, 5/p={2,1.5}}
\begin{scope}[shift={(0.7,0.6)}]
\atoms{vertex}{x0/, x1/p={1,0}, x2/p={2,0}, x3/p={0,1.5}, x4/p={1,1.5}, x5/p={2,1.5}}
\end{scope}
\atoms{vertex}{a/p={$(1)!0.5!(4)$}, b/p={$(1)!0.5!(x1)$}, c/p={$(4)!0.5!(x4)$}, d/p={$(x1)!0.5!(x4)$}}
\draw (0)--(1)--(2)--(5)--(4)--(3)--(0) (1)--(4) (x2)--(x5)--(x4)--(x3) (2)--(x2) (3)--(x3) (4)--(x4) (5)--(x5);
\draw[backedge] (x3)--(x0)--(x1)--(x2) (x1)--(x4) (0)--(x0) (1)--(x1);
\draw[midedge] (a)to[out=45,in=-90](c) (b)to[out=90,in=-135](d);
\end{tikzpicture}
\quad\leftrightarrow\quad
\begin{tikzpicture}
\atoms{vertex}{0/, 1/p={1,0}, 2/p={2,0}, 3/p={0,1.5}, 4/p={1,1.5}, 5/p={2,1.5}}
\begin{scope}[shift={(0.7,0.6)}]
\atoms{vertex}{x0/, x1/p={1,0}, x2/p={2,0}, x3/p={0,1.5}, x4/p={1,1.5}, x5/p={2,1.5}}
\end{scope}
\atoms{vertex}{a/p={$(1)!0.5!(4)$}, b/p={$(1)!0.5!(x1)$}, c/p={$(4)!0.5!(x4)$}, d/p={$(x1)!0.5!(x4)$}}
\draw (0)--(1)--(2)--(5)--(4)--(3)--(0) (1)--(4) (x2)--(x5)--(x4)--(x3) (2)--(x2) (3)--(x3) (4)--(x4) (5)--(x5);
\draw[backedge] (x3)--(x0)--(x1)--(x2) (x1)--(x4) (0)--(x0) (1)--(x1);
\draw[midedge] (a)to[out=45,in=90](b) (c)to[out=-90,in=-135](d);
\end{tikzpicture}\;.
\end{equation}
It consists of two vertex tensors on each side, with three indices contracted between them. The dual $3$-cells correspond to vertices of the original triangulation whose link is given by
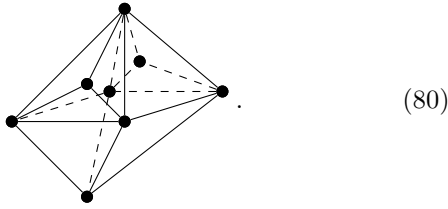
\begin{equation}
\begin{tikzpicture}
\atoms{vertex}{0/, 1/p={1.5,0}, 2/p={1.3,0.4}, 3/p={2.8,0.4}, 4/p={1,-1}, 5/p={1.5, 1.5}, a/p={1,0.5}, b/p={1.7, 0.8}}
\draw (0)--(1)--(3) (0)--(4) (1)--(4) (3)--(4) (0)--(5) (1)--(5) (3)--(5) (0)--(a) (1)--(a) (5)--(a);
\draw[backedge] (3)--(2)--(0) (2)--(4) (2)--(5) (2)--(b) (3)--(b) (5)--(b);
\end{tikzpicture}\;.
\end{equation}

An example for a recellulation which does \emph{not} give rise to a move of the $3$-dimensional vertex liquid is
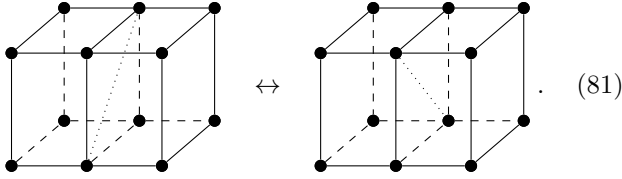
\begin{equation}
\begin{tikzpicture}
\atoms{vertex}{0/, 1/p={1,0}, 2/p={2,0}, 3/p={0,1.5}, 4/p={1,1.5}, 5/p={2,1.5}}
\begin{scope}[shift={(0.7,0.6)}]
\atoms{vertex}{x0/, x1/p={1,0}, x2/p={2,0}, x3/p={0,1.5}, x4/p={1,1.5}, x5/p={2,1.5}}
\end{scope}
\draw (0)--(1)--(2)--(5)--(4)--(3)--(0) (1)--(4) (x2)--(x5)--(x4)--(x3) (2)--(x2) (3)--(x3) (4)--(x4) (5)--(x5);
\draw[backedge] (x3)--(x0)--(x1)--(x2) (x1)--(x4) (0)--(x0) (1)--(x1);
\draw[midedge] (1)--(x4);
\end{tikzpicture}
\quad\leftrightarrow\quad
\begin{tikzpicture}
\atoms{vertex}{0/, 1/p={1,0}, 2/p={2,0}, 3/p={0,1.5}, 4/p={1,1.5}, 5/p={2,1.5}}
\begin{scope}[shift={(0.7,0.6)}]
\atoms{vertex}{x0/, x1/p={1,0}, x2/p={2,0}, x3/p={0,1.5}, x4/p={1,1.5}, x5/p={2,1.5}}
\end{scope}
\draw (0)--(1)--(2)--(5)--(4)--(3)--(0) (1)--(4) (x2)--(x5)--(x4)--(x3) (2)--(x2) (3)--(x3) (4)--(x4) (5)--(x5);
\draw[backedge] (x3)--(x0)--(x1)--(x2) (x1)--(x4) (0)--(x0) (1)--(x1);
\draw[midedge] (4)--(x1);
\end{tikzpicture}\;.
\end{equation}
Even though both cellulations have the same boundary, some of the vertices are adjacent to an interior edge on the left but not on the right (or vice versa).

\subsection{Commuting-projector mapping}
\label{sec:higher_commuting_projector}
The commuting-projector mapping from Section~\ref{sec:hamiltonian} generalizes straight-forwardly to the simplex liquids in $n$ dimensions. The commuting-projector model has degrees of freedom associated to the $n-1$-simplices of a $n-1$-dimensional triangulation. Consider a star $L$ of a vertex in this $n-1$-dimensional triangulation, i.e., a configuration of $n-1$-simplices around a vertex. E.g., for $n=3$, one such configuration is given by
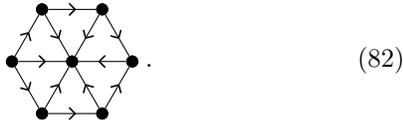
\begin{equation}
\begin{tikzpicture}
\atoms{vertex}{0/, 1/p=0:0.8, 2/p=60:0.8, 3/p=120:0.8, 4/p=180:0.8, 5/p=-120:0.8, 6/p=-60:0.8}
\draw (1)edge[ior](2) (2)edge[ior](3) (3)edge[ior](4) (4)edge[or](5) (5)edge[or](6) (6)edge[or](1) (0)edge[ior](1) (0)edge[ior](2) (0)edge[ior](3) (0)edge[ior](4) (0)edge[ior](5) (0)edge[ior](6);
\end{tikzpicture}\;.
\end{equation}
For each such $L$, there is a projector acting on the degrees of freedom around the vertex, and the Hamiltonian is given by (minus) the sum over the projectors for all vertices. The projector for a given $L$ is obtained from the $n$-simplex liquid model by evaluating the latter on the $n$-dimensional triangulation spanned by the exterior simplices of $L$ and an edge. E.g., for the example above in dimension $n=3$, we obtain a triangulation with $6$ tetrahedra,
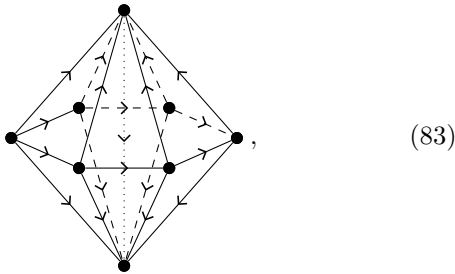
\begin{equation}
\label{eq:3d_projector_mapping}
\begin{tikzpicture}
\atoms{vertex}{{0/p={-1.5,0}}, {1/p={1.5,0}}, {2/p={-0.6,0.4}}, {3/p={0.6,0.4}}, {4/p={-0.6,-0.4}}, {5/p={0.6,-0.4}}, {6/p={0,1.7}}, {7/p={0,-1.7}}}
\draw (0)edge[or](2) (2)edge[dashed,or](3) (3)edge[dashed,or](1) (0)edge[or](4) (4)edge[or](5) (5)edge[or](1) (0)edge[or](6) (1)edge[or](6) (2)edge[dashed,or](6) (3)edge[dashed,or](6) (4)edge[or](6) (5)edge[or](6) (0)edge[or](7) (1)edge[or](7) (2)edge[dashed,or](7) (3)edge[dashed,or](7) (4)edge[or](7) (5)edge[or](7);
\draw (6)edge[dotted,or](7);
\end{tikzpicture}\;,
\end{equation}
sharing a central edge. The simplex-liquid model for this cellulation evaluates to a $12$-index tensor, which is a projector from the $6$ bottom indices to the $6$ top indices.

On the other hand, an attempted commuting-projector mapping for the vertex liquid must necessarily fail for reasons analogous to those given in Section~\ref{sec:projector_mapping_failure}. E.g., we could look at a pair of open indices right at the bottom and top of one of the edges on the equator of Eq.~\eqref{eq:3d_projector_mapping} which correspond to a plaquette of the network surrounding that edge. Since in a product of projectors, an edge can be adjacent to arbitrarily many projectors, this plaquette cannot always be triangular. Moreover, the projector move equating Eq.~\eqref{eq:3d_projector_mapping} with two stacked copies mapped to the vertex liquid would change the plaquette from the triangular to something else, and thus cannot be derived from vertex-liquid moves.
\subsection{Boundary liquid and cone mapping}
\label{sec:higher_dim_boundary}
The boundary triangle liquid can be straight-forwardly generalized to a \emph{boundary simplex liquid} in $n$ dimensions. The liquid is based on triangulations of manifolds with boundary. In addition to associating one tensor to each bulk $n$-simplex, another tensor is associated to the boundary $n-1$-simplices. The boundary tensor has one additional index which is connected to a bulk-tensor index. E.g., for $n=3$, we have an additional triangle tensor,
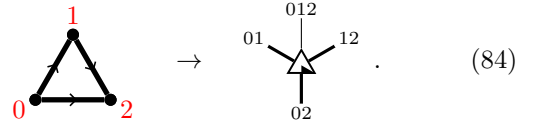
\begin{equation}
\begin{tikzpicture}
\atoms{vertex,redlab}{{0/lab={t=0,ang=-150}}, {1/p={1,0},lab={t=2,ang=-30}}, {2/p={60:1},lab={t=1,ang=90}}}
\draw[actualedge] (0)edge[or](1) (0)edge[or](2) (2)edge[or](1);
\end{tikzpicture}
\quad\rightarrow\quad
\begin{tikzpicture}
\atoms{bdtriangle}{0/}
\draw[bdbind] (0-mb)edge[sdind=02]++(-90:0.4) (0-mr)edge[sdind=12]++(30:0.4) (0-ml)edge[sdind=01]++(150:0.4);
\draw (0-ct)edge[sdind=012]++(90:0.4);
\end{tikzpicture}\;.
\end{equation}
The index labelled $012$ is contracted with the indices of the tetrahedron tensors in the bulk, whereas the indices labelled by $01$, $02$ and $12$ are contracted with indices of other boundary tensors and have a potentially different bond dimension than the $012$ index. In addition to the bulk Pachner moves, there are boundary Pachner moves that correspond to attaching (removing) a bulk $n$-simplex to (from) the boundary. E.g., in $n=3$ dimensions a boundary Pachner move is
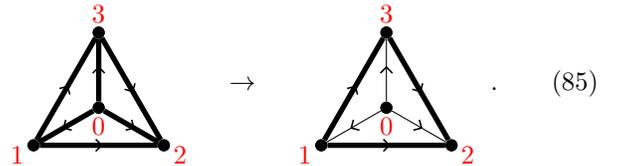
\begin{equation}
\begin{tikzpicture}
\atoms{vertex,redlab}{{0/lab={t=0,ang=-90}}, {1/p={-150:1},lab={t=1,ang=-150}}, {2/p={-30:1},lab={t=2,ang=-30}}, {3/p={90:1},lab={t=3,ang=90}}}
\draw[actualedge] (1)edge[or](3) (3)edge[or](2) (1)edge[or](2) (0)edge[or](1) (0)edge[or](2) (0)edge[or](3);
\end{tikzpicture}
\quad\rightarrow\quad
\begin{tikzpicture}
\atoms{vertex,redlab}{{0/lab={t=0,ang=-90}}, {1/p={-150:1},lab={t=1,ang=-150}}, {2/p={-30:1},lab={t=2,ang=-30}}, {3/p={90:1},lab={t=3,ang=90}}}
\draw[actualedge] (1)edge[or](3) (3)edge[or](2) (1)edge[or](2);
\draw (0)edge[or](1) (0)edge[or](2) (0)edge[or](3);
\end{tikzpicture}\;.
\end{equation}
As a tensor-network equation this corresponds to
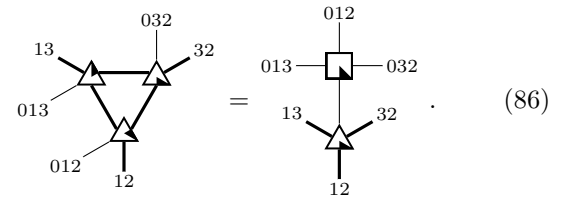
\begin{equation}
\label{eq:3d_boundary_pachner}
\begin{tikzpicture}
\atoms{bdtriangle}{{0/p={-90:0.5},rot=120,hflip}, {1/p=30:0.5}, {2/p=150:0.5, rot=120}}
\draw[bdbind] (0-mb)--(1-mb) (1-ml)--(2-mb) (2-ml)--(0-ml) (0-mr)edge[sdind=12]++(-90:0.4) (1-mr)edge[sdind=32]++(30:0.4) (2-mr)edge[sdind=13]++(150:0.4);
\draw (0-ct)edge[sdind=012]++(-150:0.4) (1-ct)edge[sdind=032]++(90:0.4) (2-ct)edge[sdind=013]++(-150:0.4);
\end{tikzpicture}
=
\begin{tikzpicture}
\atoms{bdtriangle}{0/}
\atoms{tetrahedron}{1/p={0,1}}
\draw[bdbind] (0-mb)edge[sdind=12]++(-90:0.4) (0-mr)edge[sdind=32]++(30:0.4) (0-ml)edge[sdind=13]++(150:0.4);
\draw (0-ct)--(1-b) (1-t)edge[sdind=012]++(90:0.4) (1-r)edge[sdind=032]++(0:0.4) (1-l)edge[sdind=013]++(180:0.4);
\end{tikzpicture}\;.
\end{equation}
In the same way, the boundary cone mapping from Section~\ref{sec:boundary_cone} can be generalized, mapping the boundary $n-1$-simplex to the bulk $n$-simplex. E.g., for $n=3$, we have
\begin{equation}
\begin{tikzpicture}
\atoms{bdtriangle}{0/}
\draw[bdbind] (0-mb)edge[ind=a]++(-90:0.4) (0-mr)edge[ind=b]++(30:0.4) (0-ml)edge[ind=c]++(150:0.4);
\draw (0-ct)edge[ind=x]++(90:0.4);
\end{tikzpicture}
\coloneqq
\begin{tikzpicture}
\atoms{tetrahedron}{0/}
\draw (0-b)edge[ind=x]++(-90:0.4) (0-r)edge[ind=b]++(0:0.4) (0-t)edge[ind=a]++(90:0.4) (0-l)edge[ind=c]++(180:0.4);
\end{tikzpicture}
\;.
\end{equation}
When using this assignment for the boundary Pachner move in Eq.~\eqref{eq:3d_boundary_pachner}, the move reduces to the bulk Pachner move in Eq.~\eqref{eq:23pachner_complex}. As a consequence, every simplex-liquid model has a cone topological boundary obtained from the cone mapping.

Meanwhile, a boundary cone mapping is not possible for the vertex liquid, for reasons analogous to the ones given in Section~\ref{sec:cone_mapping_failure}. We cannot fill boundary triangulations of arbitrary size with a bulk triangulation of constant thickness, due to the restricted finite set of allowed links. The best we can get is again a hyperbolic geometry with a triangulation of logarithmic depth. In higher dimensions, we have the additional problem that closing the boundary with a cone is not a topological operation, as we get a singularity at the centre of the cone if the boundary is not a $n-1$-sphere (which is always the case at least for the connected components in $1+1$ dimensions). This is not a problem for the simplex liquid, as the latter can in fact be defined on $n$-manifolds with singularities, but the vertex liquid cannot since all vertex links are supposed to be triangulations of spheres. 

\subsection{Universality mappings with and without topological boundary}
\label{sec:universality_higherdim}
The universality mapping in Section~\ref{sec:finegrain_boundary} from the triangle liquid to any topological boundary liquid $\calb$ generalizes to $n$ spacetime dimensions as well. Roughly, instead of a triangle with the three corners removed, the mapping consists in filling an $n$-simplex with the neighborhood of all the $n-2$-simplices removed. Gluing those together according to an $n$-dimensional triangulation, we obtain the original $n$-manifold with the neighborhood of the $n-2$-skeleton of the triangulation removed. To fill in the removed bits, $\calb$ must include topology-changing moves attaching $x$-handles for $x\leq n-2$. Those moves hold generically if we decorate them with a some additional defects. A detailed construction of the boundary universality mapping in $2+1$ dimensions can be found in Appendix~\ref{sec:universality_boundary_3d}.

Also the stronger universality mapping in Section~\ref{sec:2d_universal_finegrain} from the vertex liquid to any topological liquid $\calb$ works in higher dimensions. As in the $1+1$-dimensional case, the mapping proceeds by filling the $n$-cell dual to a vertex of a triangulation. More precisely, the filling is obtained by a multi-step cutting-and-gluing procedure which involves decomposing the $n$-cells into higher-dimensional kites. A detailed description of the process for $2+1$ dimensions can be found in Appendix~\ref{sec:higher_dimensions_vertex_finegrain}.
\section{Fixed-point models for new phases?}
\label{sec:new_models}
In the previous sections we have developed a framework to describe fixed-point models which do not necessarily have a topological boundary and are more general on a diagrammatic level than the simplex liquids used in conventional state-sum constructions. Thus, the crucial question which presents itself is, whether there are phases which can be captured by our new fixed-point ansatz, but not by any of the conventional ones.
Unfortunately, we cannot give a definitive affirmative answer to that question yet. However, in the following we will list several indications for why believe that this indeed is the case. Most importantly, \emph{chiral} phases can not be described by conventional state-sum constructions and are therefore a hot candidate to be represented by a vertex-liquid model.

\subsection{Unphysical topological invariants}
\label{sec:unphysical}
Let us start by noting that we can consider different \emph{tensor types} for the tensors which constitute a liquid model. Roughly speaking, a tensor type corresponds to an interpretation of the Penrose diagrams in terms of data and computations \cite{liquid_intro, tensor_type}. Physically relavant tensor types are conventional tensors (arrays of complex numbers), fermionic tensors (tensors with a $\mathbb Z_2$ grading and Grassmann variables), or tensors with symmetries. We can also consider liquid models for other tensor types corresponding to locally computable topological invariants, which do not have any immediate physical relevance. For such tensor types, it is indeed possible to show that vertex-liquid models are more powerful than simplex-liquid models.

For this purpose, consider a very ``specialized'' tensor type where the tensors themselves are diagrams of the vertex liquid, modulo the moves of the latter. The tensor product is given by the disjoint union of diagrams and contraction is given by connecting open indices in the diagrams. Then assigning to each tensor its own diagram defines a ``tautological'' model of the vertex liquid for the specialized tensor type. Asking whether which model can be rewritten as a simplex-liquid model is equivalent to asking whether the simplex liquid and vertex liquid are equivalent, which is not the case by the arguments of Section~\ref{sec:vertex_finegrain_failure}.

As another example, consider so-called all-scalar liquid models, that is, liquid models where all the tensors are scalars. These models are physically trivial without any degrees of freedom, however, they can still be in non-trivial exact phases. Note, that topological simplex-liquid models in $n$ spacetime dimensions with one scalar associated to the $n$-simplex are necessarily trivial. This follows from the fact, that some Pachner moves change the number of $n$-simplices and hence the scalar associated to the $n$-simplex needs to be $1$. However, if we allow for a more generalised variant of the simplex liquid, that associates scalars also to the lower-dimensional simplices, we find that the Pachner moves define a set of constraints for this set of scalars, which have non-trivial solutions. In Appendix~\ref{sec:classical_appendix}, we show that they all give rise to models equivalent to an exponentiated-Euler-characteristic liquid model associating a scalar $\alpha$ to odd-dimensional and $\alpha^{-1}$ to even-dimensional simplices. However, if we consider vertex-liquid models, we can show that there exist models, which are inequivalent to any exponentiated-Euler-characteristic liquid model as is shown explicitly in  Appendix~\ref{sec:classical_appendix}. Thus, in the case of all-scalar liquids, vertex-liquid models are indeed more general than simplex liquid models.

%caro: move this to a footnote?
% We also describe how the study of all-scalar liquid models coincides with the theory of \emph{characteristic classes}, which also yields important insights for gravitational anomalies and how they can be phrased using liquid models.

%
%A model of the triangle liquid is the same as a mapping from the triangle liquid to the vertex liquid. One such model is given, e.g., by associating the empty vertex-liquid network to all triangle-liquid tensors. This model is obviously not in the same phase as the vertex-liquid model mentioned above. Models in the same phase would correspond to topology-preserving, hence invertible, mappings from the triangle to the vertex liquid which cannot exist as we argued in

\subsection{Chiral phases, commuting-projector models, and topological boundaries}
Let us now discuss where we would expect vertex-liquid models to describe topological phases which are physically relevant. This does not appear to be the case in $1+1$ dimensions, since there bosonic or fermionic intrisic or symmetry-protected topological phases of matter have been classified in general (also non-fixed-point models) to a relatively satisfactory level of rigour. Since all these phases have gapped/topological boundaries they are captured by triangle-liquid models and there is no need to consider vertex-liquid models. However, in $2+1$ dimensions, there is an important class of intrinsic topological phases of matter, namely \emph{chiral} phases.

It has been conjectured that intrinsic topological phases in $2+1$ dimensions are specified by the UMTC $\mathcal{M}$ describing their anyon content together with a number $c\in \mathbb{Q}$ known as the \emph{chiral central charge} \cite{Wen2015}. $\mathcal{M}$ determines $c$ $\mod 8$, and phases for the same $\mathcal{M}$ but different $c$ are related by stacking with the so-called \emph{$E_8$ phase}, which is an invertible intrinsic topological phase with trivial anyon content and $c=8$. Chiral phases are the ones for which $c\neq 0$, or whose UMTC is not a \emph{Drinfeld center}. They have several properties which make them incompatible with established fixed-point ansatzes.

Most importantly, chiral phases do not have gapped/topological boundaries. A non-microscopic description of such boundaries is given by the \emph{Lagrangian algebra} in the UMTC describing which anyon condense at the boundary. Such Lagrangian algebras are in one-to-one correspondence with fusion categories whose Drinfel\'d centre is $\mathcal{M}$. Note that the Drinfel\'d centre of any fusion category always has a vanishing chiral central charge, so there are no topological boundaries for chiral phases. Also in the study of non-fixed-point microscopic models for chiral phases, it turns out that no matter what boundary we use, it will always be gapless. In contrast, established fixed-point ansatzes always have topological boundaries via the boundary cone mapping, as we have seen in Section~\ref{sec:boundary_cone} and \ref{sec:higher_dim_boundary}.

Furthermore, the non-zero chiral central charge of chiral phases is closely related to a non-zero \emph{thermal Hall conductance}. It has been argued in Ref.~\cite{Kapustin2019} that chiral topological phases with non-zero thermal Hall conductance cannot be realized as quantum spin systems with commuting-projector Hamiltonians. This parallels, but is different from the no-go theorem of Ref.~\cite{Kapustin2018} proving the vanishing of the \emph{electric Hall conductance} for commuting-projector models. In contrast, established fixed-point ansatzes always allow for commuting-projector Hamiltonians via a commuting-projector mapping, as we have seen in Section~\ref{sec:hamiltonian} and \ref{sec:higher_commuting_projector}.

Moreover, there seems to be an obstruction to represent ground states of chiral Hamiltonians with tensor-networks \cite{Dubail2015}, and certainly the classification of phases via \emph{MPO-injective PEPS} is restricted to non-chiral phases \cite{Bultinck2015}. In contrast, established fixed-point ansatzes always have tensor-network representations for their ground states. As argued in Section~\ref{sec:tensor_network_gs}, topological boundaries and tensor-network representations of ground states are one and the same thing in our framework. A standard tensor-network representation can therefore be obtained using the boundary cone mapping, but there are also other representations corresponding to other boundaries.

All of the above restrictions can be circumvented using our more general fixed-point ansatz. As argued in Sections~\ref{sec:cone_mapping_failure},\ref{sec:higher_dim_boundary}, \ref{sec:projector_mapping_failure}, and \ref{sec:higher_commuting_projector}, the constructions for topological boundaries as well as commuting-projector models fail for vertex-liquid models. Furthermore, we have seen in Sections~\ref{sec:2d_universal_finegrain}, and \ref{sec:higher_dimensions_vertex_finegrain} that the vertex liquid is a universal fixed-point ansatz. That is, if there are any fixed-point models for chiral phases, then there must also be vertex-liquid models of those phases.

We would like to stress once more that the failure of the commuting-projector and the boundary cone mapping is on a purely diagrammatic level, and does not concretely imply that there are models of the vertex liquid which do not possess commuting-projector Hamiltonians or topological boundaries. However, given how well the properties of chiral phases fit to the vertex liquid, it is tempting to believe that such models exist and a potential route to arrive at such a model is sketched at the end of this section.

\subsection{Chiral anomaly and projective tensors}
\label{sec:projective}
We expect the path integrals for chiral phases to also obey topological deformability due to the fact that they can be described by a \emph{$3-2-1$-extended axiomatic TQFT}, namely the \emph{Reshetikin-Turaev construction} \cite{Reshetikhin1991} based on a UMTC. In addition, several chiral phases  have microscopic field-theory models with topological invariance, namely \emph{Chern-Simons theories} \cite{Witten1989}.

However, there is one important detail in which chiral phases differ from non-chiral ones. The topological invariance for chiral phases with $c\neq 0$ holds only up to phase pre-factors. Chern-Simons theories for those phases have a so-called \emph{framing anomaly}, also known as \emph{chiral anomaly}. This means that the partition function associated to a manifold does not only depend on its topology, but changes by a phase factor when the metric is changed. As different triangulations for combinatorial manifolds play a similar role to different metrics for continuum manifolds, this suggests that a hypothetical discrete version of a chiral Chern-Simons theory should obey re-triangulization invariance only up to phase factors. 

In the field theoretic setting, there are two different approaches to describe the chiral anomaly. The first is to define the model on the boundary of a bulk in one dimension higher, the second is to introduce an additional structure (similar to an orientation or a spin structure) known as \emph{Atiyah 2-framing}. While we will discuss both of these approaches and their respective discrete analogues momentarily, we would like to emphasise, that on the level of tensor-network path integrals, there is in principle no immediate need to incorporate the chiral anomaly into the model. Instead, we could simply accept the fact, that Pachner moves only hold up to prefactors. A more formal way to say this, is to work with \emph{projective tensors}, i.e., equivalence classes of tensors modulo scalar pre-factors. Projective tensors define a consistent interpretation of the diagrammatic calculus of tensor networks which is formalised by the fact that they form a \emph{compact closed category}, or more generally, a \emph{tensor type} (cf.~Ref.~\cite{tensor_type}). Since quantum mechanics ultimately predicts probability distributions with a fixed normalisation, global scalar pre-factors are generally irrelevant in physical models.

If we do not want to work with projective tensors, we need to work out the phase factors depend on the metric in Chern-Simons theory. As mentioned above, the latter can be removed by introducing an \emph{Atiyah 2-framing}. It is known that Atiyah 2-framings are equivalent to \emph{$P_1$-structures} via \emph{obstruction theory}, similar to how orientations are equivalent to $\omega_1$-structures and spin structures are equivalent to $\omega_2$-structures, where $\omega_1$, and $\omega_2$ are \emph{characteristic classes}, namely the first and second Stiefel-Whitney class\footnote{For an introduction to (simplicial) cohomology and characteristic classes, we refer the reader to Appendix~\ref{sec:classical_appendix}.}. Concretely, the phase factor we get from a model with chiral anomaly by changing the metric between two $3$-manifolds $X_1$ and $X_2$ via a $4$-manifold $Y$ is given by
\begin{equation}
\label{eq:chiral_anomaly}
e^{2\pi i \frac{c}{24} \int_Y P_1}\;,
\end{equation}
where $c$ is the chiral central charge and $P_1$ is the first \emph{Pontryagin class}.

One way to make the model independent of the metric is to include a bounding $4$-manifold $Y$ and the above phase factor into its definition. This agrees with a common definition of an anomalous model -- a model which is defined on the boundary of a model in one higher dimension which is somewhat trivial, in this case, the integration over a characteristic class. 

Another option is to redefine the model on $3$-manifolds $X$ with a $P_1$-structure $F$ by including a phase
\begin{equation}
e^{2\pi i \frac{c}{24} \int_X F}\;.
\end{equation}
The obtained model will not depend on the metric, but only on $F$, which is related to $P_1$ as follows. Usually, an $X$-structure for a degree-$i$ characteristic class $X$ is a choice of $i-1$-cocycle $F$ such that $dF=X$. However, a $P_1$-structure on $3$-manifolds cannot be exactly defined this way since $P_1$ is of degree $4$ and does not exist on $3$-manifolds, but is represented by a $4$-cocycle on $4$-manifolds. Instead, we observe, that a continuous change of metric of a $3$-manifold can be viewed as a $4$-manifold $Y$ bounding the initial and final $3$-manifolds $X_1$ and $X_2$. Now, the change $\Delta F$ corresponding to the change of metric is given by the value of $P_1$ on this changing $4$-manifold. Thus, a $P_1$-structure is represented by an arbitrary $3$-cocycle $F$, and the trivialization of $P_1$ is implemented only by how $F$ changes when we change the metric.

The above viewpoint on chiral anomaly can be made more formal when working combinatorially with triangulations. As discussed, a hypothetical liquid model with chiral anomaly will have invariance under a Pachner move $M$ only up to a phase factor
\begin{equation}
\label{eq:pachner_anomaly}
e^{2\pi i \frac{c}{24} \Delta F(M)}\;,
\end{equation}
where $\Delta F(M)$ is an integer (or at least a rational number) depending only on what $M$ looks like in a constant-size environment. Pachner moves in $2+1$ dimensions can be viewed as gluing $4$-simplices to the $3+1$-dimensional triangulation in a $4$th dimension. A sequence of Pachner moves then yields a $4$-dimensional triangulation bounding the initial and final $2+1$-dimensional triangulations. As explained in Appendix~\ref{sec:classical_appendix}, $P_1$ comes with a local formula for a $\zz$-valued (or at least $\mathbb{Q}$-valued) $0$-cycle with its value on a vertex depending on the star of the latter, and the anomaly in Eq.~\eqref{eq:chiral_anomaly} gives rise to an all-scalar vertex liquid model on $3+1$-dimensional triangulations.
For a $3+1$-dimensional triangulation coming from a sequence of $2+1$-dimensional moves, we can redistribute the value of $P_1$ associated to the vertices such that there is one value for each move $M$. While it is a priori not obvious that this is possible, the form of the local formula in Ref.~\cite{Gaifullin2004} ensures this. $\Delta F(M)$ is then exactly given by this value, and accordingly the phase factor in Eq.~\eqref{eq:pachner_anomaly} is given by redistributing the anomaly all-scalar liquid model in Eq.~\eqref{eq:chiral_anomaly}. The topological invariance of the $3+1$-dimensional all-scalar liquid model is in accordance with the fact that the phase factors for sequences of moves connecting the same two triangulations have to be equal.

Alternatively, as in the continuum case, we could implement the anomaly via a bounding 4-manifold, i.e., we would define the model at the boundary of the $4$-dimensional all-scalar (and therefore physically trivial) anomaly liquid model from Eq.~\eqref{eq:chiral_anomaly}. The moves of the combined liquid are boundary Pachner moves, attaching or removing a $4$-simplex from the boundary. The phase factor in Eq.~\eqref{eq:pachner_anomaly} we get from the corresponding $2+1$-dimensional Pachner move on the boundary is canceled by the contribution of the $4$-simplex to the all-scalar bulk liquid model in Eq.~\eqref{eq:chiral_anomaly}.

In both cases, the evaluation of the model would depend on additional data, either on the $P_1$-structure $F$ or on the bounding $4$-manifold. In fact, $\smallint P_1$ is a cobordism invariant (such as all hermitian all-scalar liquid models), and so the evaluation only depends on the cobordism class of the bounding $4$-manifold. A popular viewpoint is that different $P_1$-structures encode different cobordism classes of bounding $4$-manifolds.

To end our discussion on chiral anomalies, we remark that for certain phases, the anomaly is less severe. To see this, note that according to Eq.~\eqref{eq:p1mod3} in the appendix following from the \emph{Hirzebruch signature theorem}, $\int_Y P_1$ is divisible by $3$ on any closed $4$-manifold $Y$. In fact, the value of $\int_y P_1/3$ precisely labels the different $4$-dimensional cobordism classes. Thus, for phases with $c=0\mod 8$, the evaluation of the model does not depend on the bounding $4$-manifold, so in a way one could say the anomaly vanishes \emph{globally}. Those phases are precisely the invertible phases consisting of copies of the $E_8$ phase with trivial anyon content. It is important to note that this does not imply that the $E_8$ phase has no anomaly in the sense that we could define the model without $P_1$-structure or bounding $4$-manifold, since $P_1\mod 3$ does not have a locally computable trivialization as explained in Section~\ref{sec:local_versus_global}. However, the phase obtained by stacking three copies of $E_8$ does not have an anomaly, and can be defined as a Chern-Simons theory without dependence on a 2-framing \cite{Randal2020}.

\subsection{Infinite-bond-dimension tensors}
\label{sec:infinite}

In Section 2 of Ref.~\cite{liquid_intro} we described a fine-graining procedure which might converge to a zero-correlation-length fixed-point path integral. If we apply this procedure to a gapped path integral which can be topologically extended, this fixed-point model is a model of a liquid describing topological $n$-manifolds. By further fine-graining we can reshape this liquid model into a vertex-liquid model as described in Section~\ref{sec:2d_universal_finegrain} and Section~\ref{sec:higher_dimensions_vertex_finegrain}. If we apply this procedure to a $2+1$-dimensional non-fixed-point tensor-network path integral in a chiral phase and the fine-graining sequence converges, we obtain a vertex-liquid fixed-point model for this phase. 

It seems hard to imagine that this fine-graining procedure has no notion of convergence at all, given the variety of different invertible domain walls $S_b$ we are allowed to apply at each fine-graining scale $\lambda$. However, it is well conceivable that the limit of the sequence does not converge to a tensor with \emph{finite} bond dimension. So one might consider fixed-point models consisting of tensors with infinite bond dimension. The latter are represented by arrays for which some indices take arbitrary natural numbers as values instead of numbers in a finite range $\{0,\ldots,d-1\}$. The contraction of two indices with infinite bond dimension involves an infinite sum which does not necessarily need to converge. If we want to ensure that the evaluation of arbitrary tensor networks is well-defined, we have to impose some \emph{decay condition} for the entries of the tensor, such as imposing that the tensor is normalizable with an appropriate norm, as described in Section 2 of Ref.~\cite{liquid_intro}, or Ref.~\cite{tensor_type}.

An interesting example for an infinite-tensor liquid model is the standard state-sum construction \cite{Chen2011} for the $2+1$-dimensional group cohomology SPT phases for the non-trivial (discontinuous) $U(1)$ 3-cocycles. This model has been studied and found to have a non-zero electric Hall conductance in Ref.~\cite{Demarco2021}. At first glance, the state sum looks like a simplex-liquid model, and evades the no-go theorem in Ref.~\cite{Kapustin2018} not by being a vertex-liquid model, but by the fact that the latter only rules out \emph{finite-dimensional} commuting-projector models. However, there seem to be problems with formally interpreting this model as an infinite-bond-dimension simplex-liquid model. After going to the Fourier basis, the homogenuous $U(1)$ $3$-cocycle becomes a tensor with $4$ $\zz$-valued indices satisfying a decay condition. However, at each vertex there will be a $\zz$-multiplication tensor which does not obey any decay condition. Potentially, this can be resolved by transferring some of the decay from the cocycle tensors to the $\zz$-multiplication tensors. If we do so, however, the state-sum has the form of a vertex liquid, and the modified $\zz$-multiplication tensor at a vertex does not define a commutative algebra anymore and prevents us from reshaping the model into a simplex-liquid model.

% Accordingly, the naive cone boundary of the simplicial state sum seems to be physically invalid, as it would be symmetry-breaking with a continuous degeneracy.

%In contrast to projective tensors, we are not aware of any precise argument of why infinite tensors would be necessary for vertex-liquid models of chiral topological phases. Indeed, the argumentation via QCAs in Section~\ref{sec:disentangling_CYWW} suggests that chiral fixed-point models are possible with finite bond dimension. 

%Still, infinite bond dimension is a possibility one can take into account, and in fact the resulting models would still be exactly solvable in practice. 

%We would like to remark that taking finite versus infinite tensors is independent of taking vertex- versus simplex-liquid models. In principle, there can be non-chiral liquid models with infinite tensors as well as chiral liquid models with finite tensors. The cone boundary for simplex-liquid models and the universality of vertex-liquid models is independent of the tensor type, and thus holds equally for finite tensors as for infinite tensors. So even if we allow infinite tensors, we cannot avoid going to a vertex liquid if we want to represent chiral phases, or other phases without topological boundary.

\subsection{Chiral liquids from disentangling Crane-Yetter-Walker-Wang models}
\label{sec:disentangling_CYWW}
A construction which might help finding a chiral vertex-liquid model realizing a specific UMTC is the so-called \emph{Crane-Yetter-Walker-Wang (CYWW) model}. This is a $3+1$-dimensional simplicial state-sum \cite{Crane1993} or equivalently a $3$-dimensional commuting-projector Hamiltonian \cite{Walker2011} which is strongly believed to be in a trivial phase, whereas its cone boundary is believed to represent a chiral topological phase. If we can disentangle the bulk, we obtain a standalone $2+1$-dimensional chiral model. More precisely, the CYWW model can be defined for an arbitrary \emph{unitary braided fusion category}, and is trivial only if this braided fusion category is \emph{modular} (i.e., an UMTC). The chiral phase at the boundary then realizes exactly this UMTC, which is potentially chiral.

A more formal way to obtain the $2+1$-dimensional model is to assume that the modular CYWW model is in a trivial exact phase, such that there exists an invertible (topological fixed-point) domain wall to the trivial model (vacuum). Such a domain wall is also a boundary, more precisely an \emph{invertible boundary}. We can now consider a thin layer of CYWW model in between the cone boundary on one side and the invertible boundary on the other side. The CYWW model restricted to such configurations can be interpreted as a $2+1$-dimensional model, namely the (potentially) chiral model. In other words, we apply a compactification mapping with an interval between cone- and invertible boundary,
\begin{equation}
\label{eq:cyww_sandwich_compactification}
\begin{tikzpicture}
\atoms{vertex,bdastyle=blue,decstyle=blue}{1/p={0,1}}
\atoms{vertex,bdastyle=red,decstyle=red}{0/}
\draw[1dmanifold] (0)--(1);
\node[blue] (s) at (0,1.5){\breakcell{Cone\\boundary}};
\node (b) at (1,0.5){CYWW bulk};
\node[red] (i) at (0,-0.5){\breakcell{Invertible\\boundary}};
\end{tikzpicture}\;.
\end{equation}
If we do not want to work with projective tensors, i.e., we do not want to ignore phase pre-factors, then the compactified $2+1$-dimensional model can not be entirely standalone, but has to be defined as the boundary of the $e^{2\pi i\frac{c}{24}P_1}$ all-scalar liquid model, representing the chiral anomaly. In this case, the invertible domain wall is not to vacuum, but to the $e^{2\pi i\frac{c}{24}P_1}$, and the compactification is is as follows,
\begin{equation}
\label{eq:cyww_compactification_anomaly}
\begin{tikzpicture}
\atoms{vertex,bdastyle=blue,decstyle=blue}{1/p={0,1}}
\atoms{vertex,bdastyle=red,decstyle=red}{0/}
\draw[1dmanifold] (0)--(1);
\draw[brown,line width=1.5] (0)--++(0,-1);
\node[blue] at (0,1.5){\breakcell{Cone\\boundary}};
\node at (1,0.5){CYWW bulk};
\node[red] at (-1,0){\breakcell{Invertible\\domain wall}};
\node[brown] at (1,-0.5){\breakcell{$e^{2\pi i \frac{c}{24}P_1}$\\anomaly}};
\end{tikzpicture}
\quad\rightarrow
\begin{tikzpicture}
\atoms{circ,small,bdastyle=blue,decstyle=blue}{1/}
\atoms{vertex,bdastyle=red,decstyle=red}{0/}
\draw[brown,line width=1.5] (0)--++(0,-1);
\node at (0,0.3){\textcolor{red}{Compactified} \textcolor{blue}{model}};
\node[brown] at (1,-0.5){\breakcell{$e^{2\pi i \frac{c}{24}P_1}$\\anomaly}};
\end{tikzpicture}
\;.
\end{equation}
We will investigate the CYWW model in more detail in Appendix~\ref{sec:CYWW_detail}.

Surely, we cannot expect the hypothetical invertible boundary to be simpler than the hypothetical chiral liquid model itself. In fact, the invertible boundary itself must be a vertex-liquid model and cannot be a simplex-liquid model, otherwise the same would hold for the compactified model. However, this approach offers significantly more guidance as opposed to free search for chiral vertex-liquid models, since we can aim to realize one specific UMTC.

One more concrete way towards invertible CYWW boundaries is to make use of the fact that many modular chiral CYWW models can be disentangled via so-called \emph{quantum cellular automata (QCAs)}. Specifically, such QCAs have been found for some abelian UMTCs using the \emph{stabilizer formalism}, starting from the three-fermion UMTC in Ref.~\cite{Haah2018}, later generalized to odd-prime stabilizers in Ref.~\cite{Haah2019}
\footnote{The disentangled models are not CYWW models but conjectured to be local unitarily equivalent to CYWW models for some abelian UMTCs.}
and most recently (shortly after a first version of this manuscript was submitted to the arXiv) in Ref.~\cite{Shirley2022} for more UMTCs including the semion and $U(1)_4$ UMTCs. A QCA is a unitary which maps local operators, acting non-trivially only in some real-space support, to local operators on the same support enlarged by a constant-size margin. The QCAs disentangle the CYWW models in the sense that each projector of the corresponding commuting-projector Hamiltonian is mapped onto an on-site projector ($(1-Z)/2$ where $Z$ is the Pauli-$Z$ operator) on a single qubit or qu-$d$-it. Even though the families of QCAs mentioned in the paragraph above are restricted to regular spacial grids, we conjecture that they can be defined for the CYWW models on arbitrary spacial triangulations/cellulations. In fact, for the case of the 3-fermion CYWW model, a constructive existence proof for the QCA on arbitrary cubulations has been presented in Ref.~\cite{Fidkowski2019}. 

%In this context, a QCA is considered trivial if it can be represented by a constant-depth local unitary circuit. The above mentioned QCAs disentangle modular CYWW models in the following way. The ground state of the CYWW model is expressed as a \emph{Pauli stabilizer state}, and the QCA is a \emph{Clifford operation} which maps each stabilizer to a Pauli $Z$ operator acting on a single qubit (or qu-$d$-it in the case of Ref.~\cite{Haah2019}). Accordingly, the 

As such, a QCA is \emph{not} a microscopic realisation of an invertible domain wall or a continuous gapped path. However, a simple and compelling argument \cite{Piroli2020} shows that any abstract QCA has a concrete representation as conjugation with a \emph{simple PEPO}. If we close all open indices on the disentangled side of this PEPO with the $\ket{0}$ vector, and contract all open indices on the entangled side with the indices coming from the bulk CYWW tensor-network path integral, we indeed obtain a boundary for the corresponding CYWW tensor-network path integrals. The topological invariance of this boundary follows from the fact that it is invariant under applying the local ground-state projector at any point. The latter corresponds to removing/attaching a diamond-like cell of the CYWW spacetime bulk from the boundary. Showing full topological invariance corresponds to removing/attaching an arbitrary bulk cell, and would require a little more work.

Furthermore, the simpleness condition of this PEPO resembles the invertibility condition for the boundary. Unfortunately though, the simpleness condition is a global condition where all open PEPO indices are physical. Thus the condition is unaffected by stacking a standalone $3$-dimensional tensor-network path integral onto the $3$-dimensional PEPO, and thus cannot guarantee invertibility of the resulting boundary. The simple QCA PEPO in Ref.~\cite{Piroli2020} is obtained by acting with a fixed rank-1 tensor-network superoperator on an arbitrary (as long as the result is non-zero) choice of $3$-dimensional PEPO. Depending on this choice, we might indeed obtain a PEPO corresponding to an invertible boundary.

\section{Conclusions and outlook}
\label{sec:outlook}
In this work, we have suggested a new ansatz for fixed point models of topological phases, in terms of tensor-network path integrals with an exact topological invariance. In contrast to other fixed-point ansatzes, it is compatible with the absence of commuting-projector Hamiltonians and gapped boundaries. The inability of fixed-point models to capture chiral topological phases has been the major caveat with this approach to classification. Our ansatz seems like a promising step towards overcoming this restriction. As such,
we provide a new instrument for the community to tackle this long-standing question in the
study of quantum phases of matter.

The suggestion of a new fixed-point ansatz is by no means the
only point of this work. In fact, the main technical result is our notion of universality for topological fixed-point ansatzes. Using a tool called a universality mapping, we show that some fixed-point ansatzes can emulate any other fixed-point ansatz for a certain type of topological order. We apply this procedure to topological order in general as well as topological order with topological boundary. In the former case, the vertex liquid is a universal fixed-point ansatz, whereas in the latter case, we arrive at a variant of the simplex liquid, which is indeed equivalent to the most general versions of established fixed-point ansatzes. This way, we are able to provide a very clear explanation why those ansatzes look the way they do. Mathematically, the universality mapping from the vertex liquid could be phrased as providing a normal form of locally computable quantum invariants of piece-wise linear manifolds.

The most urgent task remaining is to find a concrete example for a fixed-point model of our ansatz which cannot be captured by existing ansatzes, or equivalently, does not possess a topological boundary. We have given two main indications for such examples to exist. First, we have argued that a fixed-point model for intrinsic chiral phases can be obtained from an invertible boundary disentangling the modular CYWW model. One possible approach of finding such an invertible boundary would be to combine the disentangling QCAs in Refs.~\cite{Haah2018, Haah2019, Shirley2022} with the construction in Ref.~\cite{Piroli2020}.

Second, we have been looking at all-scalar \emph{classical} invariants, i.e., models consisting of scalars only. As we have seen in Appendix~\ref{sec:classical_appendix}, such all-scalar liquid models correspond to characteristic classes. Combinatorial formulas for the latter are mostly known, and it seems necessary to use vertex liquid models. It is therefore conceivable that there also exist locally computable \emph{quantum} invariants which require the vertex liquid. Moreover, we saw that the chiral anomaly of chiral phases is captured by an all-scalar vertex-liquid model. Therefore, an approach to vertex-liquid models for chiral phases could be to develop a $\zz$-valued local formula for the first Pontryagin class in $4$-manifold triangulations and look at possible (non-all-scalar) boundaries thereof. To this end, one could use constructions similar to Ref.~\cite{Gaifullin2004}, but taking into account the dependency on a branching and dual branching.

\subsection*{Acknowledgments}
AB would like to thank Nathanan Tantivasadakarn for pointing out Ref.~\cite{Fidkowski2019}.
We thank the DFG, 
CRC 183, for which it is a Berlin-Cologne internode publication within project B01, and
EI 519/15-1, for support.

\bibliography{universal_state_sums_refs}{}

\begin{thebibliography}{42}%
\makeatletter
\providecommand \@ifxundefined [1]{%
 \@ifx{#1\undefined}
}%
\providecommand \@ifnum [1]{%
 \ifnum #1\expandafter \@firstoftwo
 \else \expandafter \@secondoftwo
 \fi
}%
\providecommand \@ifx [1]{%
 \ifx #1\expandafter \@firstoftwo
 \else \expandafter \@secondoftwo
 \fi
}%
\providecommand \natexlab [1]{#1}%
\providecommand \enquote  [1]{``#1''}%
\providecommand \bibnamefont  [1]{#1}%
\providecommand \bibfnamefont [1]{#1}%
\providecommand \citenamefont [1]{#1}%
\providecommand \href@noop [0]{\@secondoftwo}%
\providecommand \href [0]{\begingroup \@sanitize@url \@href}%
\providecommand \@href[1]{\@@startlink{#1}\@@href}%
\providecommand \@@href[1]{\endgroup#1\@@endlink}%
\providecommand \@sanitize@url [0]{\catcode `\\12\catcode `\$12\catcode
  `\&12\catcode `\#12\catcode `\^12\catcode `\_12\catcode `\%12\relax}%
\providecommand \@@startlink[1]{}%
\providecommand \@@endlink[0]{}%
\providecommand \url  [0]{\begingroup\@sanitize@url \@url }%
\providecommand \@url [1]{\endgroup\@href {#1}{\urlprefix }}%
\providecommand \urlprefix  [0]{URL }%
\providecommand \Eprint [0]{\href }%
\providecommand \doibase [0]{https://doi.org/}%
\providecommand \selectlanguage [0]{\@gobble}%
\providecommand \bibinfo  [0]{\@secondoftwo}%
\providecommand \bibfield  [0]{\@secondoftwo}%
\providecommand \translation [1]{[#1]}%
\providecommand \BibitemOpen [0]{}%
\providecommand \bibitemStop [0]{}%
\providecommand \bibitemNoStop [0]{.\EOS\space}%
\providecommand \EOS [0]{\spacefactor3000\relax}%
\providecommand \BibitemShut  [1]{\csname bibitem#1\endcsname}%
\let\auto@bib@innerbib\@empty
%</preamble>
\bibitem [{\citenamefont {Levin}\ and\ \citenamefont {Wen}(2005)}]{Levin2004}%
  \BibitemOpen
  \bibfield  {author} {\bibinfo {author} {\bibfnamefont {M.~A.}\ \bibnamefont
  {Levin}}\ and\ \bibinfo {author} {\bibfnamefont {X.-G.}\ \bibnamefont
  {Wen}},\ }\bibfield  {title} {\bibinfo {title} {String-net condensation: A
  physical mechanism for topological phases},\ }\href
  {https://doi.org/10.1103/PhysRevB.71.045110} {\bibfield  {journal} {\bibinfo
  {journal} {Phys. Rev. B}\ }\textbf {\bibinfo {volume} {71}},\ \bibinfo
  {pages} {045110} (\bibinfo {year} {2005})},\ \Eprint
  {https://arxiv.org/abs/cond-mat/0404617} {arXiv:cond-mat/0404617}
  \BibitemShut {NoStop}%
\bibitem [{\citenamefont {Turaev}\ and\ \citenamefont
  {Viro}(1992)}]{Turaev1992}%
  \BibitemOpen
  \bibfield  {author} {\bibinfo {author} {\bibfnamefont {V.~G.}\ \bibnamefont
  {Turaev}}\ and\ \bibinfo {author} {\bibfnamefont {O.~Y.}\ \bibnamefont
  {Viro}},\ }\bibfield  {title} {\bibinfo {title} {State sum invariants of
  3-manifolds and quantum 6j-symbols},\ }\href
  {https://doi.org/10.1016/0040-9383(92)90015-A} {\bibfield  {journal}
  {\bibinfo  {journal} {Topology}\ }\textbf {\bibinfo {volume} {31}},\ \bibinfo
  {pages} {865} (\bibinfo {year} {1992})}\BibitemShut {NoStop}%
\bibitem [{\citenamefont {Barrett}\ and\ \citenamefont
  {Westbury}(1996)}]{Barrett1993}%
  \BibitemOpen
  \bibfield  {author} {\bibinfo {author} {\bibfnamefont {J.~W.}\ \bibnamefont
  {Barrett}}\ and\ \bibinfo {author} {\bibfnamefont {B.~W.}\ \bibnamefont
  {Westbury}},\ }\bibfield  {title} {\bibinfo {title} {Invariants of
  piecewise-linear 3-manifolds},\ }\href
  {https://doi.org/10.1090/S0002-9947-96-01660-1} {\bibfield  {journal}
  {\bibinfo  {journal} {Trans. Amer. Math. Soc.}\ }\textbf {\bibinfo {volume}
  {348}},\ \bibinfo {pages} {3997} (\bibinfo {year} {1996})},\ \Eprint
  {https://arxiv.org/abs/hep-th/9311155} {arXiv:hep-th/9311155} \BibitemShut
  {NoStop}%
\bibitem [{\citenamefont {Kapustin}\ and\ \citenamefont
  {Fidkowski}(2020)}]{Kapustin2018}%
  \BibitemOpen
  \bibfield  {author} {\bibinfo {author} {\bibfnamefont {A.}~\bibnamefont
  {Kapustin}}\ and\ \bibinfo {author} {\bibfnamefont {L.}~\bibnamefont
  {Fidkowski}},\ }\bibfield  {title} {\bibinfo {title} {{Local commuting
  projector Hamiltonians and the quantum Hall effect}},\ }\href
  {https://doi.org/10.1007/s00220-019-03444-1} {\bibfield  {journal} {\bibinfo
  {journal} {Commun. Math. Phys.}\ }\textbf {\bibinfo {volume} {373}},\
  \bibinfo {pages} {763} (\bibinfo {year} {2020})},\ \Eprint
  {https://arxiv.org/abs/1810.07756} {arXiv:1810.07756} \BibitemShut {NoStop}%
\bibitem [{\citenamefont {Kapustin}\ and\ \citenamefont
  {Spodyneiko}(2020)}]{Kapustin2019}%
  \BibitemOpen
  \bibfield  {author} {\bibinfo {author} {\bibfnamefont {A.}~\bibnamefont
  {Kapustin}}\ and\ \bibinfo {author} {\bibfnamefont {L.}~\bibnamefont
  {Spodyneiko}},\ }\bibfield  {title} {\bibinfo {title} {Thermal hall
  conductance and a relative topological invariant of gapped two-dimensional
  systems},\ }\href {https://doi.org/10.1103/PhysRevB.101.045137} {\bibfield
  {journal} {\bibinfo  {journal} {Phys. Rev. B}\ }\textbf {\bibinfo {volume}
  {101}},\ \bibinfo {pages} {045137} (\bibinfo {year} {2020})},\ \Eprint
  {https://arxiv.org/abs/1905.06488} {arXiv:1905.06488} \BibitemShut {NoStop}%
\bibitem [{\citenamefont {Bultinck}\ \emph {et~al.}(2017)\citenamefont
  {Bultinck}, \citenamefont {Marien}, \citenamefont {Williamson}, \citenamefont
  {Sahinoglu}, \citenamefont {Haegeman},\ and\ \citenamefont
  {Verstraete}}]{Bultinck2015}%
  \BibitemOpen
  \bibfield  {author} {\bibinfo {author} {\bibfnamefont {N.}~\bibnamefont
  {Bultinck}}, \bibinfo {author} {\bibfnamefont {M.}~\bibnamefont {Marien}},
  \bibinfo {author} {\bibfnamefont {D.~J.}\ \bibnamefont {Williamson}},
  \bibinfo {author} {\bibfnamefont {M.~B.}\ \bibnamefont {Sahinoglu}}, \bibinfo
  {author} {\bibfnamefont {J.}~\bibnamefont {Haegeman}},\ and\ \bibinfo
  {author} {\bibfnamefont {F.}~\bibnamefont {Verstraete}},\ }\bibfield  {title}
  {\bibinfo {title} {Anyons and matrix product operator algebras},\ }\href
  {https://doi.org/10.1016/j.aop.2017.01.004} {\bibfield  {journal} {\bibinfo
  {journal} {Ann. Phys.}\ }\textbf {\bibinfo {volume} {378}},\ \bibinfo {pages}
  {183} (\bibinfo {year} {2017})},\ \Eprint {https://arxiv.org/abs/1511.08090}
  {arXiv:1511.08090} \BibitemShut {NoStop}%
\bibitem [{\citenamefont {Haah}\ \emph {et~al.}(2018)\citenamefont {Haah},
  \citenamefont {Fidkowski},\ and\ \citenamefont {Hastings}}]{Haah2018}%
  \BibitemOpen
  \bibfield  {author} {\bibinfo {author} {\bibfnamefont {J.}~\bibnamefont
  {Haah}}, \bibinfo {author} {\bibfnamefont {L.}~\bibnamefont {Fidkowski}},\
  and\ \bibinfo {author} {\bibfnamefont {M.~B.}\ \bibnamefont {Hastings}},\
  }\href@noop {} {\bibinfo {title} {Nontrivial quantum cellular automata in
  higher dimensions}} (\bibinfo {year} {2018}),\ \Eprint
  {https://arxiv.org/abs/1812.01625} {arXiv:1812.01625} \BibitemShut {NoStop}%
\bibitem [{\citenamefont {Haah}(2021)}]{Haah2019}%
  \BibitemOpen
  \bibfield  {author} {\bibinfo {author} {\bibfnamefont {J.}~\bibnamefont
  {Haah}},\ }\bibfield  {title} {\bibinfo {title} {Clifford quantum cellular
  automata: Trivial group in 2d and witt group in 3d},\ }\href
  {https://doi.org/10.1063/5.0022185} {\bibfield  {journal} {\bibinfo
  {journal} {J. Math. Phys.}\ }\textbf {\bibinfo {volume} {62}},\ \bibinfo
  {pages} {092202} (\bibinfo {year} {2021})},\ \Eprint
  {https://arxiv.org/abs/1907.02075} {arXiv:1907.02075} \BibitemShut {NoStop}%
\bibitem [{\citenamefont {Shirley}\ \emph {et~al.}(2022)\citenamefont
  {Shirley}, \citenamefont {Chen}, \citenamefont {Dua}, \citenamefont
  {Ellison}, \citenamefont {Tantivasadakarn},\ and\ \citenamefont
  {Williamson}}]{Shirley2022}%
  \BibitemOpen
  \bibfield  {author} {\bibinfo {author} {\bibfnamefont {W.}~\bibnamefont
  {Shirley}}, \bibinfo {author} {\bibfnamefont {Y.-A.}\ \bibnamefont {Chen}},
  \bibinfo {author} {\bibfnamefont {A.}~\bibnamefont {Dua}}, \bibinfo {author}
  {\bibfnamefont {T.~D.}\ \bibnamefont {Ellison}}, \bibinfo {author}
  {\bibfnamefont {N.}~\bibnamefont {Tantivasadakarn}},\ and\ \bibinfo {author}
  {\bibfnamefont {D.~J.}\ \bibnamefont {Williamson}},\ }\href@noop {} {\bibinfo
  {title} {Three-dimensional quantum cellular automata from chiral semion
  surface topological order and beyond}} (\bibinfo {year} {2022}),\ \Eprint
  {https://arxiv.org/abs/2202.05442} {arXiv:2202.05442} \BibitemShut {NoStop}%
\bibitem [{\citenamefont {Bauer}\ \emph {et~al.}(2022)\citenamefont {Bauer},
  \citenamefont {Eisert},\ and\ \citenamefont {Wille}}]{liquid_intro}%
  \BibitemOpen
  \bibfield  {author} {\bibinfo {author} {\bibfnamefont {A.}~\bibnamefont
  {Bauer}}, \bibinfo {author} {\bibfnamefont {J.}~\bibnamefont {Eisert}},\ and\
  \bibinfo {author} {\bibfnamefont {C.}~\bibnamefont {Wille}},\ }\bibfield
  {title} {\bibinfo {title} {A unified diagrammatic approach to topological
  fixed point models},\ }\href
  {https://doi.org/10.21468/SciPostPhysCore.5.3.038} {\bibfield  {journal}
  {\bibinfo  {journal} {SciPost Phys. Core}\ }\textbf {\bibinfo {volume} {5}},\
  \bibinfo {pages} {38} (\bibinfo {year} {2022})},\ \Eprint
  {https://arxiv.org/abs/2011.12064} {arXiv:2011.12064} \BibitemShut {NoStop}%
\bibitem [{\citenamefont {Pachner}(1991)}]{Pachner1991}%
  \BibitemOpen
  \bibfield  {author} {\bibinfo {author} {\bibfnamefont {U.}~\bibnamefont
  {Pachner}},\ }\bibfield  {title} {\bibinfo {title} {P.~l. homeomorphic
  manifolds are equivalent by elementary shellings},\ }\href
  {https://doi.org/10.1016/S0195-6698(13)80080-7} {\bibfield  {journal}
  {\bibinfo  {journal} {Europ. J. Comb.}\ }\textbf {\bibinfo {volume} {12}},\
  \bibinfo {pages} {129 } (\bibinfo {year} {1991})}\BibitemShut {NoStop}%
\bibitem [{\citenamefont {Fukuma}\ \emph {et~al.}(1994)\citenamefont {Fukuma},
  \citenamefont {Hosono},\ and\ \citenamefont {Kawai}}]{Fukuma1992}%
  \BibitemOpen
  \bibfield  {author} {\bibinfo {author} {\bibfnamefont {M.}~\bibnamefont
  {Fukuma}}, \bibinfo {author} {\bibfnamefont {S.}~\bibnamefont {Hosono}},\
  and\ \bibinfo {author} {\bibfnamefont {H.}~\bibnamefont {Kawai}},\ }\bibfield
   {title} {\bibinfo {title} {Lattice topological field theory in two
  dimensions},\ }\href {https://doi.org/10.1007/BF02099416} {\bibfield
  {journal} {\bibinfo  {journal} {Commun. Math. Phys.}\ }\textbf {\bibinfo
  {volume} {161}},\ \bibinfo {pages} {157} (\bibinfo {year} {1994})},\ \Eprint
  {https://arxiv.org/abs/hep-th/9212154} {arXiv:hep-th/9212154} \BibitemShut
  {NoStop}%
\bibitem [{\citenamefont {Lauda}\ and\ \citenamefont
  {Pfeiffer}(2007)}]{Lauda2006}%
  \BibitemOpen
  \bibfield  {author} {\bibinfo {author} {\bibfnamefont {A.~D.}\ \bibnamefont
  {Lauda}}\ and\ \bibinfo {author} {\bibfnamefont {H.}~\bibnamefont
  {Pfeiffer}},\ }\bibfield  {title} {\bibinfo {title} {State sum construction
  of two-dimensional open-closed topological quantum field theories},\ }\href
  {https://doi.org/10.1142/S0218216507005725} {\bibfield  {journal} {\bibinfo
  {journal} {J. Knot. Theor. Ram.}\ }\textbf {\bibinfo {volume} {16}},\
  \bibinfo {pages} {1121} (\bibinfo {year} {2007})},\ \Eprint
  {https://arxiv.org/abs/math/0602047} {arXiv:math/0602047} \BibitemShut
  {NoStop}%
\bibitem [{\citenamefont {Bauer}\ and\ \citenamefont
  {Nietner}(2022)}]{tensor_type}%
  \BibitemOpen
  \bibfield  {author} {\bibinfo {author} {\bibfnamefont {A.}~\bibnamefont
  {Bauer}}\ and\ \bibinfo {author} {\bibfnamefont {A.}~\bibnamefont
  {Nietner}},\ }\href@noop {} {\bibinfo {title} {Tensor types and their usage
  in physics}} (\bibinfo {year} {2022}),\ \Eprint
  {https://arxiv.org/abs/2208.01135} {arXiv:2208.01135} \BibitemShut {NoStop}%
\bibitem [{\citenamefont {Wen}(2016)}]{Wen2015}%
  \BibitemOpen
  \bibfield  {author} {\bibinfo {author} {\bibfnamefont {X.-G.}\ \bibnamefont
  {Wen}},\ }\bibfield  {title} {\bibinfo {title} {A theory of 2+1d bosonic
  topological orders},\ }\href {https://doi.org/10.1093/nsr/nwv077} {\bibfield
  {journal} {\bibinfo  {journal} {Natl. Sci. Rev.}\ }\textbf {\bibinfo {volume}
  {3}},\ \bibinfo {pages} {68} (\bibinfo {year} {2016})},\ \Eprint
  {https://arxiv.org/abs/1506.05768} {arXiv:1506.05768} \BibitemShut {NoStop}%
\bibitem [{\citenamefont {Dubail}\ and\ \citenamefont
  {Read}(2015)}]{Dubail2015}%
  \BibitemOpen
  \bibfield  {author} {\bibinfo {author} {\bibfnamefont {J.}~\bibnamefont
  {Dubail}}\ and\ \bibinfo {author} {\bibfnamefont {N.}~\bibnamefont {Read}},\
  }\bibfield  {title} {\bibinfo {title} {Tensor network trial states for chiral
  topological phases in two dimensions and a no-go theorem in any dimension},\
  }\href {https://doi.org/10.1103/PhysRevB.92.205307} {\bibfield  {journal}
  {\bibinfo  {journal} {Phys. Rev. B}\ }\textbf {\bibinfo {volume} {92}},\
  \bibinfo {pages} {205307} (\bibinfo {year} {2015})},\ \Eprint
  {https://arxiv.org/abs/1307.7726} {arXiv:1307.7726} \BibitemShut {NoStop}%
\bibitem [{\citenamefont {Reshetikhin}\ and\ \citenamefont
  {Turaev}(1991)}]{Reshetikhin1991}%
  \BibitemOpen
  \bibfield  {author} {\bibinfo {author} {\bibfnamefont {N.}~\bibnamefont
  {Reshetikhin}}\ and\ \bibinfo {author} {\bibfnamefont {V.}~\bibnamefont
  {Turaev}},\ }\bibfield  {title} {\bibinfo {title} {Invariants of 3-manifolds
  via link polynomials and quantum groups},\ }\href
  {https://doi.org/10.1007/BF01239527} {\bibfield  {journal} {\bibinfo
  {journal} {Invent. Math.}\ }\textbf {\bibinfo {volume} {103}},\ \bibinfo
  {pages} {547–597} (\bibinfo {year} {1991})}\BibitemShut {NoStop}%
\bibitem [{\citenamefont {Witten}(1989)}]{Witten1989}%
  \BibitemOpen
  \bibfield  {author} {\bibinfo {author} {\bibfnamefont {E.}~\bibnamefont
  {Witten}},\ }\bibfield  {title} {\bibinfo {title} {Quantum field theory and
  the jones polynomial},\ }\href {https://doi.org/10.1007/BF01217730}
  {\bibfield  {journal} {\bibinfo  {journal} {Comm. Math. Phys.}\ }\textbf
  {\bibinfo {volume} {121}},\ \bibinfo {pages} {351} (\bibinfo {year}
  {1989})}\BibitemShut {NoStop}%
\bibitem [{\citenamefont {Gaifullin}(2004)}]{Gaifullin2004}%
  \BibitemOpen
  \bibfield  {author} {\bibinfo {author} {\bibfnamefont {A.~A.}\ \bibnamefont
  {Gaifullin}},\ }\bibfield  {title} {\bibinfo {title} {Local formulae for
  combinatorial pontrjagin classes},\ }\href {https://doi.org/10.4213/im502}
  {\bibfield  {journal} {\bibinfo  {journal} {Izv. Math.}\ }\textbf {\bibinfo
  {volume} {68}},\ \bibinfo {pages} {861–910} (\bibinfo {year} {2004})},\
  \Eprint {https://arxiv.org/abs/math/0407035} {arXiv:math/0407035}
  \BibitemShut {NoStop}%
\bibitem [{\citenamefont {Randal-Williams}\ \emph {et~al.}(2020)\citenamefont
  {Randal-Williams}, \citenamefont {Tsui},\ and\ \citenamefont
  {Wen}}]{Randal2020}%
  \BibitemOpen
  \bibfield  {author} {\bibinfo {author} {\bibfnamefont {O.}~\bibnamefont
  {Randal-Williams}}, \bibinfo {author} {\bibfnamefont {L.}~\bibnamefont
  {Tsui}},\ and\ \bibinfo {author} {\bibfnamefont {X.-G.}\ \bibnamefont
  {Wen}},\ }\href@noop {} {\bibinfo {title} {Quantization of chern-simons
  topological invariants for h-type and l-type quantum systems}} (\bibinfo
  {year} {2020}),\ \Eprint {https://arxiv.org/abs/2008.02613}
  {arXiv:2008.02613} \BibitemShut {NoStop}%
\bibitem [{\citenamefont {Chen}\ \emph {et~al.}(2013)\citenamefont {Chen},
  \citenamefont {Gu}, \citenamefont {Liu},\ and\ \citenamefont
  {Wen}}]{Chen2011}%
  \BibitemOpen
  \bibfield  {author} {\bibinfo {author} {\bibfnamefont {X.}~\bibnamefont
  {Chen}}, \bibinfo {author} {\bibfnamefont {Z.-C.}\ \bibnamefont {Gu}},
  \bibinfo {author} {\bibfnamefont {Z.-X.}\ \bibnamefont {Liu}},\ and\ \bibinfo
  {author} {\bibfnamefont {X.-G.}\ \bibnamefont {Wen}},\ }\bibfield  {title}
  {\bibinfo {title} {Symmetry protected topological orders and the group
  cohomology of their symmetry group},\ }\href
  {https://doi.org/10.1103/PhysRevB.87.155114} {\bibfield  {journal} {\bibinfo
  {journal} {Phys. Rev. B}\ }\textbf {\bibinfo {volume} {87}},\ \bibinfo
  {pages} {155114} (\bibinfo {year} {2013})},\ \Eprint
  {https://arxiv.org/abs/1106.4772} {arXiv:1106.4772} \BibitemShut {NoStop}%
\bibitem [{\citenamefont {DeMarco}\ and\ \citenamefont
  {Wen}(2021)}]{Demarco2021}%
  \BibitemOpen
  \bibfield  {author} {\bibinfo {author} {\bibfnamefont {M.}~\bibnamefont
  {DeMarco}}\ and\ \bibinfo {author} {\bibfnamefont {X.-G.}\ \bibnamefont
  {Wen}},\ }\href@noop {} {\bibinfo {title} {{A commuting projector model with
  a non-zero quantized Hall conductance}}} (\bibinfo {year} {2021}),\ \Eprint
  {https://arxiv.org/abs/2102.13057} {arXiv:2102.13057} \BibitemShut {NoStop}%
\bibitem [{\citenamefont {Crane}\ and\ \citenamefont
  {Yetter}(1993)}]{Crane1993}%
  \BibitemOpen
  \bibfield  {author} {\bibinfo {author} {\bibfnamefont {L.}~\bibnamefont
  {Crane}}\ and\ \bibinfo {author} {\bibfnamefont {D.~N.}\ \bibnamefont
  {Yetter}},\ }\href {https://doi.org/10.1142/9789812796387_0005} {\bibinfo
  {title} {{A categorical construction of 4D TQFTs}}} (\bibinfo {year}
  {1993}),\ \Eprint {https://arxiv.org/abs/hep-th/9301062}
  {arXiv:hep-th/9301062} \BibitemShut {NoStop}%
\bibitem [{\citenamefont {Walker}\ and\ \citenamefont
  {Wang}(2012)}]{Walker2011}%
  \BibitemOpen
  \bibfield  {author} {\bibinfo {author} {\bibfnamefont {K.}~\bibnamefont
  {Walker}}\ and\ \bibinfo {author} {\bibfnamefont {Z.}~\bibnamefont {Wang}},\
  }\bibfield  {title} {\bibinfo {title} {(3+1)-tqfts and topological
  insulators},\ }\href {https://doi.org/10.1007/s11467-011-0194-z} {\bibfield
  {journal} {\bibinfo  {journal} {Front. Phys.}\ }\textbf {\bibinfo {volume}
  {7}},\ \bibinfo {pages} {150} (\bibinfo {year} {2012})},\ \Eprint
  {https://arxiv.org/abs/1104.2632} {arXiv:1104.2632} \BibitemShut {NoStop}%
\bibitem [{\citenamefont {Fidkowski}\ \emph {et~al.}(2020)\citenamefont
  {Fidkowski}, \citenamefont {Haah},\ and\ \citenamefont
  {Hastings}}]{Fidkowski2019}%
  \BibitemOpen
  \bibfield  {author} {\bibinfo {author} {\bibfnamefont {L.}~\bibnamefont
  {Fidkowski}}, \bibinfo {author} {\bibfnamefont {J.}~\bibnamefont {Haah}},\
  and\ \bibinfo {author} {\bibfnamefont {M.~B.}\ \bibnamefont {Hastings}},\
  }\bibfield  {title} {\bibinfo {title} {An exactly solvable model for a $4+1d$
  beyond-cohomology symmetry protected topological phase},\ }\href
  {https://doi.org/10.1103/PhysRevB.101.155124} {\bibfield  {journal} {\bibinfo
   {journal} {Phys. Rev. B}\ }\textbf {\bibinfo {volume} {101}},\ \bibinfo
  {pages} {155124} (\bibinfo {year} {2020})},\ \Eprint
  {https://arxiv.org/abs/1912.05565} {arXiv:1912.05565} \BibitemShut {NoStop}%
\bibitem [{\citenamefont {Piroli}\ and\ \citenamefont
  {Cirac}(2020)}]{Piroli2020}%
  \BibitemOpen
  \bibfield  {author} {\bibinfo {author} {\bibfnamefont {L.}~\bibnamefont
  {Piroli}}\ and\ \bibinfo {author} {\bibfnamefont {J.~I.}\ \bibnamefont
  {Cirac}},\ }\bibfield  {title} {\bibinfo {title} {Quantum cellular automata,
  tensor networks, and area laws},\ }\href
  {https://doi.org/10.1103/PhysRevLett.125.190402} {\bibfield  {journal}
  {\bibinfo  {journal} {Phys. Rev. Lett.}\ }\textbf {\bibinfo {volume} {125}},\
  \bibinfo {pages} {190402} (\bibinfo {year} {2020})},\ \Eprint
  {https://arxiv.org/abs/2007.15371} {arXiv:2007.15371} \BibitemShut {NoStop}%
\bibitem [{\citenamefont {Cooper}(1988)}]{Cooper1988}%
  \BibitemOpen
  \bibfield  {author} {\bibinfo {author} {\bibfnamefont {W.~P.}\ \bibnamefont
  {Cooper}, \bibfnamefont {D.;~Thurston}},\ }\bibfield  {title} {\bibinfo
  {title} {Triangulating 3-manifolds using 5 vertex link types},\ }\href
  {https://doi.org/10.1016/0040-9383(88)90004-3} {\bibfield  {journal}
  {\bibinfo  {journal} {Topology}\ }\textbf {\bibinfo {volume} {27}},\ \bibinfo
  {pages} {23} (\bibinfo {year} {1988})}\BibitemShut {NoStop}%
\bibitem [{\citenamefont {Kohan}()}]{Kohan2020}%
  \BibitemOpen
  \bibfield  {author} {\bibinfo {author} {\bibfnamefont {M.}~\bibnamefont
  {Kohan}},\ }\href {https://mathoverflow.net/q/379606} {\bibinfo {title}
  {Local complexity of triangulations}},\ \bibinfo {howpublished}
  {MathOverflow}\BibitemShut {NoStop}%
\bibitem [{\citenamefont {Bauer}\ \emph {et~al.}(2019)\citenamefont {Bauer},
  \citenamefont {Eisert},\ and\ \citenamefont {Wille}}]{tensor_lattice}%
  \BibitemOpen
  \bibfield  {author} {\bibinfo {author} {\bibfnamefont {A.}~\bibnamefont
  {Bauer}}, \bibinfo {author} {\bibfnamefont {J.}~\bibnamefont {Eisert}},\ and\
  \bibinfo {author} {\bibfnamefont {C.}~\bibnamefont {Wille}},\ }\bibfield
  {title} {\bibinfo {title} {Towards a mathematical formalism for classifying
  phases of matter},\ }\href@noop {} {\  (\bibinfo {year} {2019})},\ \Eprint
  {https://arxiv.org/abs/1903.05413} {arXiv:1903.05413} \BibitemShut {NoStop}%
\bibitem [{\citenamefont {Levitt}\ and\ \citenamefont
  {Rourke}(1978)}]{Levitt1978}%
  \BibitemOpen
  \bibfield  {author} {\bibinfo {author} {\bibfnamefont {N.}~\bibnamefont
  {Levitt}}\ and\ \bibinfo {author} {\bibfnamefont {C.}~\bibnamefont
  {Rourke}},\ }\bibfield  {title} {\bibinfo {title} {The existence of
  combinatorial formulae for characteristic classes},\ }\href
  {https://doi.org/10.1090/S0002-9947-1978-0494134-6} {\bibfield  {journal}
  {\bibinfo  {journal} {Transactions of the American Mathematical Society}\
  }\textbf {\bibinfo {volume} {239}},\ \bibinfo {pages} {391} (\bibinfo {year}
  {1978})}\BibitemShut {NoStop}%
\bibitem [{\citenamefont {Goldstein}\ and\ \citenamefont
  {Turner}(1976)}]{Goldstein1976}%
  \BibitemOpen
  \bibfield  {author} {\bibinfo {author} {\bibfnamefont {R.~Z.}\ \bibnamefont
  {Goldstein}}\ and\ \bibinfo {author} {\bibfnamefont {E.~C.}\ \bibnamefont
  {Turner}},\ }\bibfield  {title} {\bibinfo {title} {A formula for
  stiefel-whitney homology classes},\ }\href
  {https://doi.org/10.1090/S0002-9939-1976-0415643-5} {\bibfield  {journal}
  {\bibinfo  {journal} {Proc. Amer. Math. Soc.}\ }\textbf {\bibinfo {volume}
  {58}},\ \bibinfo {pages} {339} (\bibinfo {year} {1976})}\BibitemShut
  {NoStop}%
\bibitem [{\citenamefont {Gaifullin}\ and\ \citenamefont
  {Gorodkov}(2019)}]{Gaifullin2018}%
  \BibitemOpen
  \bibfield  {author} {\bibinfo {author} {\bibfnamefont {A.~A.}\ \bibnamefont
  {Gaifullin}}\ and\ \bibinfo {author} {\bibfnamefont {D.}~\bibnamefont
  {Gorodkov}},\ }\bibfield  {title} {\bibinfo {title} {An explicit local
  combinatorial formula for the first pontryagin class},\ }\href
  {https://doi.org/10.1070/RM9920} {\bibfield  {journal} {\bibinfo  {journal}
  {Russ. Math. Surv.}\ }\textbf {\bibinfo {volume} {74}},\ \bibinfo {pages} {6}
  (\bibinfo {year} {2019})}\BibitemShut {NoStop}%
\bibitem [{\citenamefont {Steenrod}(1947)}]{Steenrod1947}%
  \BibitemOpen
  \bibfield  {author} {\bibinfo {author} {\bibfnamefont {N.~E.}\ \bibnamefont
  {Steenrod}},\ }\bibfield  {title} {\bibinfo {title} {Products of cocycles and
  extensions of mappings},\ }\href {https://doi.org/10.2307/1969172} {\bibfield
   {journal} {\bibinfo  {journal} {Ann. Math.}\ }\textbf {\bibinfo {volume}
  {48}},\ \bibinfo {pages} {290} (\bibinfo {year} {1947})}\BibitemShut
  {NoStop}%
\bibitem [{\citenamefont {Gaiotto}\ and\ \citenamefont
  {Kapustin}(2016)}]{Gaiotto2015}%
  \BibitemOpen
  \bibfield  {author} {\bibinfo {author} {\bibfnamefont {D.}~\bibnamefont
  {Gaiotto}}\ and\ \bibinfo {author} {\bibfnamefont {A.}~\bibnamefont
  {Kapustin}},\ }\bibfield  {title} {\bibinfo {title} {{Spin TQFTs and
  fermionic phases of matter}},\ }\href
  {https://doi.org/10.1142/S0217751X16450445} {\bibfield  {journal} {\bibinfo
  {journal} {Int. J. Mod. Phys. A}\ }\textbf {\bibinfo {volume} {31}},\
  \bibinfo {pages} {1645044} (\bibinfo {year} {2016})},\ \Eprint
  {https://arxiv.org/abs/1505.05856} {arXiv:1505.05856} \BibitemShut {NoStop}%
\bibitem [{\citenamefont {May}(1999)}]{May1999}%
  \BibitemOpen
  \bibfield  {author} {\bibinfo {author} {\bibfnamefont {J.}~\bibnamefont
  {May}},\ }\href
  {https://www.math.uchicago.edu/~may/CONCISE/ConciseRevised.pdf} {\emph
  {\bibinfo {title} {A Concise Course in Algebraic Topology}}},\ Chicago
  Lectures in Mathematics\ (\bibinfo  {publisher} {University of Chicago
  Press},\ \bibinfo {year} {1999})\BibitemShut {NoStop}%
\bibitem [{\citenamefont {Brown}(1982)}]{Brown1982}%
  \BibitemOpen
  \bibfield  {author} {\bibinfo {author} {\bibfnamefont {E.~H.}\ \bibnamefont
  {Brown}},\ }\bibfield  {title} {\bibinfo {title} {The cohomology of $bso_n$
  and $bo_n$ with integer coefficients},\ }\href
  {http://www.jstor.org/stable/2044298} {\bibfield  {journal} {\bibinfo
  {journal} {Proc. Amer. Math. Soc.}\ }\textbf {\bibinfo {volume} {85}},\
  \bibinfo {pages} {283} (\bibinfo {year} {1982})}\BibitemShut {NoStop}%
\bibitem [{\citenamefont {Milnor}\ and\ \citenamefont
  {Stasheff}(1974)}]{Milnor1974}%
  \BibitemOpen
  \bibfield  {author} {\bibinfo {author} {\bibfnamefont {J.~W.}\ \bibnamefont
  {Milnor}}\ and\ \bibinfo {author} {\bibfnamefont {J.~D.}\ \bibnamefont
  {Stasheff}},\ }\href {https://books.google.de/books?id=vJbKCwAAQBAJ} {\emph
  {\bibinfo {title} {Characteristic Classes}}},\ \bibinfo {series} {Annals of
  Mathematics Studies}, Vol.~\bibinfo {volume} {76}\ (\bibinfo  {publisher}
  {Princeton University Press},\ \bibinfo {year} {1974})\BibitemShut {NoStop}%
\bibitem [{\citenamefont {Hirzebruch}\ \emph {et~al.}(1966)\citenamefont
  {Hirzebruch}, \citenamefont {Hirzebruch}, \citenamefont {Schwarzenberger},\
  and\ \citenamefont {Borel}}]{Hirzebruch1966}%
  \BibitemOpen
  \bibfield  {author} {\bibinfo {author} {\bibfnamefont {F.}~\bibnamefont
  {Hirzebruch}}, \bibinfo {author} {\bibfnamefont {R.}~\bibnamefont
  {Hirzebruch}}, \bibinfo {author} {\bibfnamefont {R.}~\bibnamefont
  {Schwarzenberger}},\ and\ \bibinfo {author} {\bibfnamefont {A.}~\bibnamefont
  {Borel}},\ }\href
  {https://hirzebruch.mpim-bonn.mpg.de/id/eprint/114/1/M2_Topological\%20methods.pdf}
  {\emph {\bibinfo {title} {Topological Methods in Algebraic Geometry}}},\
  Classics in mathematics\ (\bibinfo  {publisher} {Springer-Verlag},\ \bibinfo
  {year} {1966})\BibitemShut {NoStop}%
\bibitem [{\citenamefont {Rowell}\ \emph {et~al.}(2009)\citenamefont {Rowell},
  \citenamefont {Stong},\ and\ \citenamefont {Wang}}]{Rowell2007}%
  \BibitemOpen
  \bibfield  {author} {\bibinfo {author} {\bibfnamefont {E.}~\bibnamefont
  {Rowell}}, \bibinfo {author} {\bibfnamefont {R.}~\bibnamefont {Stong}},\ and\
  \bibinfo {author} {\bibfnamefont {Z.}~\bibnamefont {Wang}},\ }\bibfield
  {title} {\bibinfo {title} {On classification of modular tensor categories},\
  }\href {https://doi.org/10.1007/s00220-009-0908-z} {\bibfield  {journal}
  {\bibinfo  {journal} {Comm. Math. Phys.}\ }\textbf {\bibinfo {volume}
  {292}},\ \bibinfo {pages} {343} (\bibinfo {year} {2009})},\ \Eprint
  {https://arxiv.org/abs/0712.1377} {arXiv:0712.1377} \BibitemShut {NoStop}%
\bibitem [{\citenamefont {Burnell}\ \emph {et~al.}(2014)\citenamefont
  {Burnell}, \citenamefont {Chen}, \citenamefont {Fidkowski},\ and\
  \citenamefont {Vishwanath}}]{Burnell2013}%
  \BibitemOpen
  \bibfield  {author} {\bibinfo {author} {\bibfnamefont {F.~J.}\ \bibnamefont
  {Burnell}}, \bibinfo {author} {\bibfnamefont {X.}~\bibnamefont {Chen}},
  \bibinfo {author} {\bibfnamefont {L.}~\bibnamefont {Fidkowski}},\ and\
  \bibinfo {author} {\bibfnamefont {A.}~\bibnamefont {Vishwanath}},\ }\bibfield
   {title} {\bibinfo {title} {Exactly soluble model of a 3d symmetry protected
  topological phase of bosons with surface topological order},\ }\href
  {https://doi.org/10.1103/PhysRevB.90.245122} {\bibfield  {journal} {\bibinfo
  {journal} {Phys. Rev. B}\ }\textbf {\bibinfo {volume} {90}},\ \bibinfo
  {pages} {245122} (\bibinfo {year} {2014})},\ \Eprint
  {https://arxiv.org/abs/1302.7072} {arXiv:1302.7072} \BibitemShut {NoStop}%
\bibitem [{\citenamefont {Williamson}\ and\ \citenamefont
  {Wang}(2017)}]{Williamson2016}%
  \BibitemOpen
  \bibfield  {author} {\bibinfo {author} {\bibfnamefont {D.~J.}\ \bibnamefont
  {Williamson}}\ and\ \bibinfo {author} {\bibfnamefont {Z.}~\bibnamefont
  {Wang}},\ }\bibfield  {title} {\bibinfo {title} {Hamiltonian models for
  topological phases of matter in three spatial dimensions},\ }\href
  {https://doi.org/10.1016/j.aop.2016.12.018} {\bibfield  {journal} {\bibinfo
  {journal} {Ann. Phys.}\ }\textbf {\bibinfo {volume} {377}},\ \bibinfo {pages}
  {311} (\bibinfo {year} {2017})},\ \Eprint {https://arxiv.org/abs/1606.07144}
  {arXiv:1606.07144} \BibitemShut {NoStop}%
\bibitem [{\citenamefont {Chen}\ and\ \citenamefont {Tata}(2021)}]{Chen2021}%
  \BibitemOpen
  \bibfield  {author} {\bibinfo {author} {\bibfnamefont {Y.-A.}\ \bibnamefont
  {Chen}}\ and\ \bibinfo {author} {\bibfnamefont {S.}~\bibnamefont {Tata}},\
  }\href@noop {} {\bibinfo {title} {Higher cup products on hypercubic lattices:
  application to lattice models of topological phases}} (\bibinfo {year}
  {2021}),\ \Eprint {https://arxiv.org/abs/2106.052747} {arXiv:2106.052747}
  \BibitemShut {NoStop}%
\end{thebibliography}%

\newpage
\appendix

\section{Restricting valency of vertices in triangulations}
\label{sec:appendix_links}
In this appendix, we discuss whether any manifold can be triangulated with a finite set of vertex stars, such that we can restrict to a finite set of tensors and moves in the vertex liquid. Equivalently, we discuss whether we can restrict the valency $l$, i.e., the number of $n$-simplices adjacent to a vertex, by some global upper bound. We first discuss the case of $2$ dimensions and later comment on higher dimensions. In 2 dimensions, we show that it suffices to restrict ourselves to triangulations with $l$-valent vertices for $l\leq 10$. More precisely, we show that every manifold admits an $l\leq 9$ triangulation, and that every two such triangulations of the same manifold are connected via Pachner moves such that all intermediate triangulations satisfy $l\leq 10$. Note that we did not make too much effort in optimizing the number $10$, and most likely smaller bounds like $8$ or $9$ are possible. It is clear that we need to allow $3\leq l\leq 7$ in order to have 1-3 Pachner moves and in order to represent negative curvature.

The first step consists in mapping each triangulation to an equivalent one with $l\leq 9$. To this end, we fist go to the dual lattice, which is a $3$-valent cellulation consisting of $n$-gons with arbitrarily high $n$. We then triangulate each $n$-gon in a zig-zag manner, e.g., for $n=11$,
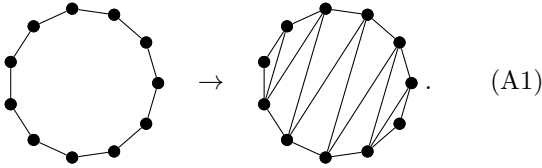
\begin{equation}
\begin{tikzpicture}
\foreach \n in {0,...,10}{
\atoms{vertex}{\n/p={\n*360/11:1}}
}
\draw (0)--(1)--(2)--(3)--(4)--(5)--(6)--(7)--(8)--(9)--(10)--(0);
\end{tikzpicture}
\quad\rightarrow\quad
\begin{tikzpicture}
\foreach \n in {0,...,10}{
\atoms{vertex}{\n/p={\n*360/11:1}}
}
\draw (0)--(1)--(2)--(3)--(4)--(5)--(6)--(7)--(8)--(9)--(10)--(0);
\draw (0)--(9)--(1)--(8)--(2)--(7)--(3)--(6)--(4);
\end{tikzpicture}\;.
\end{equation}
Every vertex on the boundary of this zig-zag triangulation is adjacent to at most $3$ triangles. As the dual cellulation is trivalent, vertices in the overall triangulation are adjacent to $l\leq 3+3+3=9$ triangles.

As a second step we realize that pairing up adjacent triangles in the zig-zag triangulation (without leaving any gaps) and then performing a 2-2 Pachner move on all pairs rotates the zig-zag pattern by half a unit, e.g.,
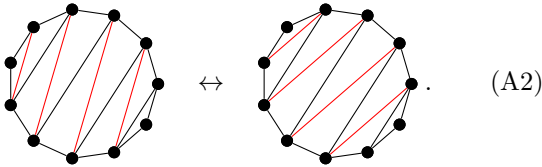
\begin{equation}
\begin{tikzpicture}
\foreach \n in {0,...,10}{
\atoms{vertex}{\n/p={\n*360/11:1}}
}
\draw (0)--(1)--(2)--(3)--(4)--(5)--(6)--(7)--(8)--(9)--(10)--(0);
\draw (0)--(9) (1)--(8) (2)--(7) (3)--(6);
\draw[red] (9)--(1) (8)--(2) (7)--(3) (6)--(4);
\end{tikzpicture}
\quad\leftrightarrow\quad
\begin{tikzpicture}
\foreach \n in {0,...,10}{
\atoms{vertex}{\n/p={\n*360/11:1}}
}
\draw (0)--(1)--(2)--(3)--(4)--(5)--(6)--(7)--(8)--(9)--(10)--(0);
\draw (9)--(0) (8)--(1) (7)--(2) (6)--(3);
\draw[red] (0)--(8) (1)--(7) (2)--(6) (3)--(5);
\end{tikzpicture}
\;.
\end{equation}
When we perform all the 2-2 Pachner moves in a sequence we obtain intermediate triangulations where a boundary vertex is adjacent to $4$ triangles, but not more. There are (at least) two \emph{tip vertices} at the boundary which are only connected to a single triangle, and we can rotate those to an arbitrary position using the moves described above. As argued before, we can do so using 2-2 Pachner moves involving only triangulations with $l\leq 3+3+4=10$.

2-2 Pachner moves of the original triangulation involve two vertices in the dual cellulation,
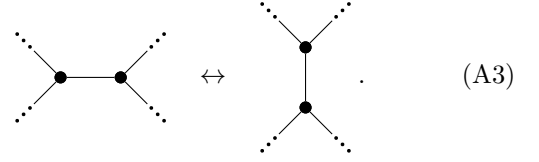
\begin{equation}
\begin{tikzpicture}
\atoms{vertex}{0/p={-0.4,0}, 1/p={0.4,0}}
\draw (0)--(1) (0)edge[mark={three dots,a}]++(135:0.5) (0)edge[mark={three dots,a}]++(-135:0.5) (1)edge[mark={three dots,a}]++(45:0.5) (1)edge[mark={three dots,a}]++(-45:0.5);
\end{tikzpicture}
\quad\leftrightarrow\quad
\begin{tikzpicture}
\atoms{vertex}{0/p={0,-0.4}, 1/p={0,0.4}}
\draw (0)--(1) (1)edge[mark={three dots,a}]++(135:0.5) (0)edge[mark={three dots,a}]++(-135:0.5) (1)edge[mark={three dots,a}]++(45:0.5) (0)edge[mark={three dots,a}]++(-45:0.5);
\end{tikzpicture}\;.
\end{equation}
Such a move can be performed using only 2-2 Pachner moves on the dual zig-zag-triangulated triangulation as follows: First, we rotate the ziz-zag triangulation of each of the $4$ involved plaquettes such that its tip vertex coincides with one of the involved vertices. We are hence left with the following,
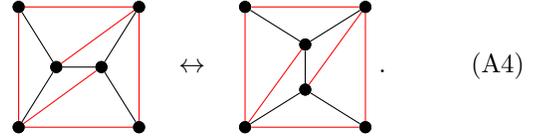
\begin{equation}
\begin{tikzpicture}
\atoms{vertex}{0/p={-0.3,0}, 1/p={0.3,0}, 2/p={-0.8,-0.8}, 3/p={0.8,-0.8}, 4/p={0.8,0.8}, 5/p={-0.8,0.8}}
\draw (2)--(0)--(1)--(3) (4)--(1) (5)--(0);
\draw[red] (2)--(3)--(4)--(5)--(2) (2)--(1) (4)--(0);
\end{tikzpicture}
\quad\leftrightarrow\quad
\begin{tikzpicture}
\atoms{vertex}{0/p={0,-0.3}, 1/p={0,0.3}, 2/p={-0.8,-0.8}, 3/p={0.8,-0.8}, 4/p={0.8,0.8}, 5/p={-0.8,0.8}}
\draw (2)--(0)--(1)--(5) (4)--(1) (3)--(0);
\draw[red] (2)--(3)--(4)--(5)--(2) (2)--(1) (4)--(0);
\end{tikzpicture}\;.
\end{equation}
The two pictures are the same combinatorially. We see that after the rotation of the tip vertices, the above move becomes trivial and merely corresponds to a regrouping of which zig-zag triangles belong to which dual cells.

The situation for the 1-3 Pachner moves is similar. In the dual cellulation, the move becomes
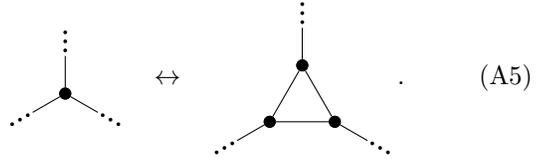
\begin{equation}
\begin{tikzpicture}
\atoms{vertex}{0/}
\draw (0)edge[mark={three dots,a}]++(-30:0.5) (0)edge[mark={three dots,a}]++(90:0.5) (0)edge[mark={three dots,a}]++(-150:0.5);
\end{tikzpicture}
\quad\leftrightarrow\quad
\begin{tikzpicture}
\atoms{vertex}{1/p=-30:0.5, 3/p=90:0.5, 2/p=-150:0.5}
\draw (1)--(2)--(3)--(1);
\draw (1)edge[mark={three dots,a}]++(-30:0.5) (3)edge[mark={three dots,a}]++(90:0.5) (2)edge[mark={three dots,a}]++(-150:0.5);
\end{tikzpicture}\;.
\end{equation}
After zig-zag triangulating all the plaquettes and rotating the tip vertices towards where the move happens, we are left with
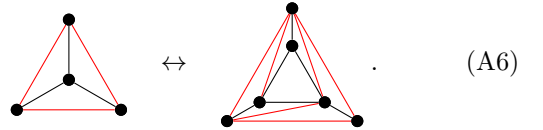
\begin{equation}
\begin{tikzpicture}
\atoms{vertex}{0/, 1/p=-30:0.8, 2/p=90:0.8, 3/p=-150:0.8}
\draw[red] (1)--(2)--(3)--(1);
\draw (0)--(1) (0)--(2) (0)--(3);
\end{tikzpicture}
\quad\leftrightarrow\quad
\begin{tikzpicture}
\atoms{vertex}{1/p=-30:0.5, 3/p=90:0.5, 2/p=-150:0.5, a/p=-30:1, b/p=-150:1, c/p=90:1}
\draw (1)--(2)--(3)--(1) (1)--(a) (2)--(b) (3)--(c);
\draw[red] (1)--(c)--(a) (1)--(b)--(a) (2)--(c)--(b);
\end{tikzpicture}\;.
\end{equation}
This is nothing but a sequence of two 1-3 Pachner moves.
We see that the Pachner moves of the original triangulation can be mimicked by Pachner moves of the dual-zig-zag-triangulated triangulation by rotating the tip vertices and applying 1-3 Pachner moves, all of which involve only $l\leq 10$ triangulations.

It is conceivable that the above argumentation can be generalized to $n$ dimensions. First, we consider the dual cellulation where each vertex is adjacent to $n+1$ $n$-cells. Then we triangulate each $n$-cell with a ``zig-zag'' triangulation with no vertices in the interior and with a global upper bound $L$ for the number of $n$-simplices adjacent to each boundary vertex. The number of $n$-simplices adjacent to a vertex in the overall resulting triangulation is thus upper bounded by $(n+1)L$. Unfortunately, proving the existence of a zig-zag triangulation in general dimensions does not appear to be an easy task, and we have not yet found the best way to do this. Nonetheless, the existence of a sufficient upper bound for the vertex valency or a sufficient finite set of vertex stars seems to be widely accepted. Unfortunately, there there is no good reference where this is proven, but discussions can be found in Refs.~\cite{Cooper1988,Kohan2020}.

So far we restricted our discussion in general dimensions to sufficient valency bounds to triangulate any manifold, but we would also like to be able to connect any two triangulations of the same manifold by Pachner moves such that all intermediate triangulations satisfy the bound. A set satisfying the latter in $n$ dimensions can be obtained from the former in $n+1$ dimensions. As $n$-dimensional stars, we can just take all the different subsets of $n$-simplices of the $n+1$-dimensional stars forming $n$-disks. For two (bounded) triangulations $T_1$ and $T_2$ of the same manifold $M$, we take a bounded triangulation of $M\times [0,1]$ restricting to $T_1$ at $M\times 0$ and to $T_2$ at $M\times 1$. Such a triangulation corresponds to a sequence of Pachner moves mapping $T_1$ to $T_2$. The stars of intermediate triangulations will indeed be the suitable subsets of $n$-simplices of the allowed $n+1$-dimensional stars.

\section{Liquids for defects}
\label{sec:defects}
In this appendix we outline how to represent arbitrary types of topological defects by liquid models. We also discuss how arbitrary defects can be reduced to boundaries via a \emph{compactification mapping}. Finally, we discuss topology-changing moves including defects, as well as invertible domain walls.

\subsection{Liquids on higher order manifolds}
\label{sec:higher_order_manifolds}
Just like topological fixed-point models for physical boundaries are models of a liquid on manifolds with boundary, topological defects are modeled by a liquid on some general type of \emph{higher order manifold} (cf.~Ref.~\cite{tensor_lattice}). Intuitively, a higher order manifold is a composite of manifolds of different dimensions, meeting, terminating at, and embedded into each other. For example, one type of higher order manifold is given by $2+1$-dimensional manifolds with an embedded $0+1$-dimensional manifold. Liquid models on this type of higher order manifold are fixed-point models for an anyon worldline/ribbon operator inside a $2+1$-dimensional topological model. Note that the notion of a defect here is much more general as what is usually referred to as a topological defect, including as examples the boundary itself, lines where three $2$-manifolds meet, points inside the boundary of a $4$-manifold, or lines inside a $3$-manifold where a $2$-manifold emerges which is not embedded in the $3$-manifold.

We will illustrate all the concepts in this and following sections focussing on one simple example for a type of higher order manifold, which also appears in Section~\ref{sec:finegrain_boundary} of the main text -- a 2-manifold terminating at a 1-manifold corresponding to a boundary which itself has an embedded 0-manifold corresponding to \emph{boundary defects}, e.g.,
\begin{equation}
\label{eq:boundary_point_defect}
\begin{tikzpicture}
\draw[manifoldboundary,manifold] (0,0)ellipse(2cm and 1cm);
\draw[manifoldboundary,fill=white] (-1,0)circle(0.4);
\draw[manifoldboundary,fill=white] (1,0)circle(0.4);
\atoms{defectvertex}{0/p={$(-1,0)+(120:0.4)$}, 1/p={100:2cm and 1cm}, 2/p={-30:2cm and 1cm}}
\end{tikzpicture}\;.
\end{equation}

In general, the type of a topological defect in a higher order manifold is specified by its dimension $d$, and its \emph{link}. The link (not to be confused with the link of a simplex in a triangulation, which is similar but different) is another higher order manifold which is obtained by considering
the normal space of the defect at an arbitrary point, and then taking the set of points of a fixed distance $\epsilon$ from the origin within that normal space. Intuitively, we take the intersection of the higher order manifold with a perpendicular $n-d$-sphere around a point of the defect, when (locally) embedded in $n$-dimensional Euclidean space for a large enough $n$.
E.g., for a $d=0+1$-dimensional anyon worldline in a $n=2+1$-dimensional spacetime, the normal space is a plane perpendicular to the anyon world line, and the distance-$\epsilon$ set, and hence the link, is simply a circle. For a boundary (in any dimension), the normal space is a half-line and the distance-$\epsilon$ set is a single point. For the brown $0+0$-dimensional boundary defect above, the normal space is a half-plane, hence the link is a half-circle, or equivalently, an interval,
\begin{equation}
\label{eq:boundary_defect}
\begin{tikzpicture}
\fill[manifold] (0,0)rectangle(1,2);
\draw[manifoldboundary] (1,0)--(1,2);
\atoms{defectvertex}{x/p={1,1}}
\draw[dashed] (x)circle(0.6);
\end{tikzpicture}
\quad\rightarrow\quad
\begin{tikzpicture}
\atoms{boundaryvertex}{0/, 1/p={0,0.8}}
\draw[1dmanifold] (0)--(1);
\end{tikzpicture}
\;.
\end{equation}

Let us give an example of a liquid describing higher order manifolds of the type above. It consists of tensors describing the bulk, the boundary and the boundary defects and a collection of moves which allow topological deformations. For the bulk and boundary part, we will take the boundary triangle liquid from Section~\ref{sec:boundary}. Then we add one 2-index tensor describing the boundary defects and one move which allows to move the defects along the boundary. The defects are represented by special vertices in the boundary to which we associate the 
2-index tensor, i.e.,
\begin{equation}
\begin{tikzpicture}
\atoms{vertex}{0/, 1/p={1.6,0}}
\atoms{defectvertex}{2/p={0.8,0}}
\draw[actualedge] (0)edge[or](2) (2)edge[or](1) (0)edge[enddots]++(180:0.3) (1)edge[enddots]++(0:0.3);
\draw (0)edge[enddots]++(120:0.4) (0)edge[enddots]++(80:0.4) (2)edge[enddots]++(110:0.4) (2)edge[enddots]++(70:0.4) (1)edge[enddots]++(100:0.4) (1)edge[enddots]++(60:0.4);
\end{tikzpicture}
\quad\rightarrow\quad
\begin{tikzpicture}
\atoms{bdedge}{0/, 1/p={1,0}}
\atoms{circ,small,all}{2/p={0.5,0}}
\draw[bdbind] (0)edge[mark={ar,e}](2) (2)--(1) (0)edge[enddots]++(180:0.5) (1)edge[enddots]++(0:0.5);
\draw (0)edge[enddots]++(90:0.4) (1)edge[enddots]++(90:0.4);
\end{tikzpicture}
\;.
\end{equation}
To guarantee topological invariance, we need a move that changes the position of the boundary defects,
\begin{equation}
\begin{tikzpicture}
\atoms{bdedge}{0/}
\atoms{circ,small,all}{2/p={0.5,0}}
\draw[bdbind] (0)edge[mark={ar,e}](2) (2)edge[ind=b]++(0:0.5) (0)edge[ind=a]++(180:0.4);
\draw (0)edge[ind=x]++(90:0.4);
\end{tikzpicture}
=
\begin{tikzpicture}
\atoms{bdedge}{0/}
\atoms{circ,small,all}{2/p={-0.5,0}}
\draw[bdbind] (0)edge[](2) (0)edge[ind=b]++(0:0.5) (2)edge[ind=a,mark={ar,s}]++(180:0.4);
\draw (0)edge[ind=x]++(90:0.4);
\end{tikzpicture}\;.
\end{equation}
Note that the equation is linear in the defect tensor, so its models (solutions) for array tensors consist of a whole sub-vector space. It is in fact a feature of any kind of $0+0$-dimensional defect that their models form a vector space.

\subsection{The compactification mapping}
\label{sec:compactification}
A crucial observation for the study of defect models is that models of a $d$-dimensional defect with link $C$ within some higher-order-manifold liquid are in one-to-one correspondence with models of ordinary $d+1$-dimensional boundary liquids via the following \emph{compactification mapping}. Loosely speaking, this is a mapping which maps an ordinary $d+1$-manifold $M$ to the higher-order manifold $M\times C$. More precisely, it is a mapping from networks of an ordinary $d+1$-dimensional liquid to networks of the original higher-order-manifold liquid which only extend in the $C$-direction by some constant combinatorial distance. In restricted settings, this is also known as ``folding trick'', or ``dimensional reduction''. As an example consider the $d=0$-dimensional defect in Eq.~\eqref{eq:boundary_defect} and its link. We map a network representing a $0+1$-dimensional manifold $M$, e.g., a line, to a network representing the $2$-dimensional manifold $M \times C$, i.e., a line times a strip of small, constant width,
\begin{equation}
\label{eq:compactification1d}
\begin{tikzpicture}
\draw[\manifoldbdcol,line width=4] (0,0)--(2,0);
\draw[\manifoldcol,line width=2] (0,0)--(2,0);
\end{tikzpicture}
\quad\rightarrow\quad
\begin{tikzpicture}
\fill[manifold] (0,0)--(2,0)--(2,0.4)--(0,0.4)--cycle;
\draw[manifoldboundary] (0,0)--(2,0) (2,0.4)--(0,0.4);
\end{tikzpicture}\;.
\end{equation}
A concrete example for a compactification mapping of this kind is the mapping from the one-dimensional liquid in Section~\ref{sec:universal_1d} to the boundary triangle liquid,
\begin{equation}
\begin{tikzpicture}
\atoms{1dedge}{0/}
\draw (0-l)edge[ind=a_0a_1]++(180:0.4) (0-r)edge[ind=b_0b_1]++(0:0.4);
\end{tikzpicture}
\coloneqq
\begin{tikzpicture}
\atoms{bdedge}{0/, {1/p={0,0.8}}}
\draw[] (0)edge[mark={ar,s}](1);
\draw[line width=2] (0-l)edge[ind=a_0]++(180:0.4) (0-r)edge[ind=b_0]++(0:0.4) (1-l)edge[ind=a_1]++(180:0.4) (1-r)edge[ind=b_1]++(0:0.4);
\end{tikzpicture}\;.
\end{equation}

We can extend the domain of the compactification mapping to $d+1$-dimensional manifold with boundary, by closing off $M\times C$ with $\partial M \times \operatorname{Cone}(C)$. Here, $\operatorname{Cone}(C)$ is $C\times [0,1]$ with $C\times 1$ contracted to a single point identified with the defect itself. For the defect link in Eq.~\eqref{eq:boundary_defect}, we close off the thin strip with a point defect,
\begin{equation}
\label{eq:compactification_boundary}
\begin{tikzpicture}
\draw[\manifoldbdcol,line width=4] (0,0)--(2,0);
\draw[\manifoldcol,line width=2] (0,0)--(2,0);
\atoms{circ,small,all,bdastyle=\manifoldbdcol,decstyle=\manifolddefectcol}{x/p={2,0}}
\end{tikzpicture}
\quad\rightarrow\quad
\begin{tikzpicture}
\fill[manifold] (0,0)--(1.5,0)to[out=0,in=-135](2,0.3)to[out=135,in=0](1.5,0.6)--(0,0.6)--cycle;
\draw[manifoldboundary] (0,0)--(1.5,0)to[out=0,in=-135](2,0.3)to[out=135,in=0](1.5,0.6)--(0,0.6);
\atoms{defectvertex}{x/p={2,0.3}}
\end{tikzpicture}\;.
\end{equation}
Further, we can consider coupling the higher-order-manifold liquid to the $d+1$-dimensional ordinary liquid by letting the latter terminate at the defect. There is a mapping from this coupled liquid to the higher-order-manifold liquid without the defect, in our case,
\begin{equation}
\label{eq:compactification_coupling}
\begin{tikzpicture}
\fill[manifold] (0,0)rectangle(0.7,1);
\draw[manifoldboundary] (0.7,0)--(0.7,1);
\draw[\manifoldbdcol,line width=4] (0.7,0.5)--++(0.7,0);
\draw[\manifoldcol,line width=2]  (0.7,0.5)--++(0.7,0);
\end{tikzpicture}
\quad\rightarrow\quad
\begin{tikzpicture}
\fill[manifold] (0,0)--(0.7,0)to[out=90,in=180](1,0.3)--++(0.3,0)--(1.3,0.7)--++(-0.3,0)to[out=180,in=-90](0.7,1)--(0,1)--cycle;
\draw[manifoldboundary] (0.7,0)to[out=90,in=180](1,0.3)--++(0.3,0) (0.7,1)to[out=-90,in=180]++(0.3,-0.3)--++(0.3,0);
\end{tikzpicture}
\;.
\end{equation}

Furthermore, there is a mapping from the higher-order-manifold liquid to the coupled liquid where the $d+1$-dimensional liquid additionally has a boundary, by attaching to the defect a thin layer of $d+1$-manifold terminated by the boundary, in our case,
\begin{equation}
\label{eq:compactification_backwards}
\begin{tikzpicture}
\fill[manifold] (0,0)rectangle(1,1.5);
\draw[manifoldboundary] (1,0)--(1,1.5);
\atoms{defectvertex}{x/p={1,0.75}}
\end{tikzpicture}
\quad\rightarrow\quad
\begin{tikzpicture}
\fill[manifold] (0,0)rectangle(1,1.5);
\draw[manifoldboundary] (1,0)--(1,1.5);
\draw[\manifoldbdcol,line width=4] (1,0.75)--(1.5,0.75);
\draw[\manifoldcol,line width=2] (1,0.75)--(1.5,0.75);
\atoms{circ,small,all,bdastyle=\manifoldbdcol,decstyle=\manifolddefectcol}{x/p={1.5,0.75}}
\end{tikzpicture}
\;.
\end{equation}

Now, using the mapping in Eq.~\eqref{eq:compactification1d}, we get an ordinary $d+1$-dimensional liquid model coupled to the original liquid model via Eq.~\eqref{eq:compactification_coupling}. We can use Eq.~\eqref{eq:compactification_boundary} to get a model for a boundary of the $d+1$-dimensional liquid from the model of the defect. Vice versa, we can use Eq.~\eqref{eq:compactification_backwards} to obtain a model of the defect from a model of the boundary. We notice that applying first Eq.~\eqref{eq:compactification_backwards} and then Eq.~\eqref{eq:compactification_boundary} together with Eq.~\eqref{eq:compactification_coupling} yields the trivial mapping, so defect models and boundary models of the $d+1$-dimensional model are in one-to-one correspondence.

\subsection{Topology-changing moves}
Another way of enhancing topological liquids except for adding defects is to add topology-changing moves. An interesting example is to add arbitrary \emph{surgery moves}, which yields liquid models for \emph{invertible} topological phases only. We can also add defects and topology-changing moves. In this section, we will focus on one specific topology-changing moves for the boundary defect liquid which will be used in Section~\ref{sec:finegrain_boundary}, namely,
\begin{equation}
\label{eq:2d_puncture_healing_move}
\begin{tikzpicture}
\fill[manifold] (0,0)circle(0.8);
\draw[manifoldboundary,fill=white] (0,0)circle(0.4);
\atoms{defectvertex}{0/p=-90:0.4}
\end{tikzpicture}
=
\begin{tikzpicture}
\fill[manifold] (0,0)circle(0.8);
\end{tikzpicture}
\;.
\end{equation}
Both sides represent a (higher order) manifold with a circle of open boundary on the outside, but the left-hand side has a puncture with physical boundary on the inside. The brown dot on the blue line represents the $0+0$-dimensional defect on the boundary. The equation symbolizes an arbitrary move between two concrete networks representing the two corresponding topologies. Which of these topology-changing moves we chose does not matter since they can all be derived from each other using the topology-preserving moves.

As a concrete example, consider the boundary-triangle liquid with boundary defects described above in this appendix. A simple version of a move as above is given by
\begin{equation}
\begin{tikzpicture}
\atoms{defectvertex}{0/}
\draw[looseness=1.5,actualedge,ior] (0)to[out=30,in=0]++(0,0.9)to[out=180,in=150](0);
\draw (0)edge[enddots]++(-90:0.4) (0)edge[enddots]++(-10:0.4) (0)edge[enddots]++(-60:0.4) (0)edge[enddots]++(-120:0.4) (0)edge[enddots]++(-170:0.4);
\end{tikzpicture}
\quad\leftrightarrow\quad
\begin{tikzpicture}
\atoms{vertex}{0/, 1/p={0,0.5}}
\draw[looseness=1.5,ior] (0)to[out=30,in=0]++(0,0.9)to[out=180,in=150](0);
\draw (0)edge[or](1);
\draw (0)edge[enddots]++(-90:0.4) (0)edge[enddots]++(-10:0.4) (0)edge[enddots]++(-60:0.4) (0)edge[enddots]++(-120:0.4) (0)edge[enddots]++(-170:0.4);
\end{tikzpicture}\;,
\end{equation}
where the left-hand side represents a `1-gon hole' inside a two-dimensional triangulation, while a self-glued triangle fills the hole on the right-hand side. In terms of networks this move reads
\begin{equation}
\begin{tikzpicture}
\atoms{bdedge}{r1/}
\atoms{circ,small,all}{x/p={0,-0.5}}
\draw[bdbind,looseness=2] (r1-r)edge[out=0,in=0,mark={ar,e}](x) (r1-l)edge[out=180,in=180](x);
\draw (r1-t)edge[mark={ar,s},ind=x]++(90:0.3);
\end{tikzpicture}
=
\begin{tikzpicture}
\atoms{oface}{s/}
\draw (s)edge[mark={ar,s},ind=x]++(90:0.5);
\draw[looseness=2,mark={ar,s}] (s)to[out=-30,in=0](0,-0.6)to[out=180,in=-150](s);
\end{tikzpicture}
\;.
\end{equation}
We can use compactification mappings to better understand the structure of the equations that correspond to topology-changing moves such as the move in Eq.~\eqref{eq:2d_puncture_healing_move}. For the right-hand side of Eq.~\eqref{eq:2d_puncture_healing_move} we consider the compactification mapping,
\begin{equation}
\label{eq:tube_compactification}
\begin{tikzpicture}
\draw[\manifoldcol,line width=3] (0,0)--(1.5,0);
\end{tikzpicture}
\quad\rightarrow\quad
\begin{tikzpicture}
\fill[manifold,looseness=0.8] (0,0)--(1.5,0)to[out=180,in=180](1.5,0.6)--(0,0.6)to[out=180,in=180]cycle;
\fill[manifold,looseness=0.8] (0,0)--(1.5,0)to[out=0,in=0](1.5,0.6)--(0,0.6)to[out=0,in=0]cycle;
\end{tikzpicture}\;,
\end{equation}
needed to calculate models for point defects in the bulk, whose link is a circle,
\begin{equation}
\begin{tikzpicture}
\fill[manifold] (0,0)rectangle(2,2);
\atoms{defectvertex}{x/p={1,1}}
\draw[dashed] (x)circle(0.6);
\end{tikzpicture}
\quad\rightarrow\quad
\begin{tikzpicture}
\draw[1dmanifold] (0,0)circle(0.6);
\end{tikzpicture}
\;.
\end{equation}
At endpoints of the line, we can close the tube by a little cup, this time without a point defect at the end,
\begin{equation}
\begin{tikzpicture}
\atoms{circ,small,all,bdastyle=\manifoldcol,decstyle=\manifoldcol}{x/p={2,0}}
\draw[\manifoldcol,line width=3] (0,0)--(2,0);
\end{tikzpicture}
\quad\rightarrow\quad
\begin{tikzpicture}
\fill[manifold,looseness=0.8] (0,0)--(2,0)to[out=0,in=-90](2.4,0.3)to[out=90,in=0](2,0.6)--(0,0.6)to[out=180,in=180]cycle;
\fill[manifold,looseness=0.8] (0,0)--(2,0)to[out=0,in=-90](2.4,0.3)to[out=90,in=0](2,0.6)--(0,0.6)to[out=0,in=0]cycle;
\end{tikzpicture}\;.
\end{equation}
Using this mapping, we can obtain a model on 1-manifolds with boundary from a model on 2-manifolds. For the left-hand side of Eq.~\eqref{eq:2d_puncture_healing_move} we consider the domain wall between the interval-compactification mapping and the circle-compactification mapping,
\begin{equation}
\begin{tikzpicture}
\draw[\manifoldbdcol,line width=4] (1,0)--(2,0);
\draw[\manifoldcol,line width=2] (1,0)--(2,0);
\draw[\manifoldcol,line width=3] (0,0)--(1,0);
\atoms{circ,small,all,bdastyle=\manifoldbdcol,decstyle=\manifoldcol}{x/p={1,0}}
\end{tikzpicture}
\quad\rightarrow\quad
\begin{tikzpicture}
\fill[manifold] (0,0)--(1,0)to[out=0,in=180](1.5,0.1)--(2,0.1)--(2,0.5)--(1.5,0.5)to[out=180,in=0](1,0.6)--(0,0.6)to[out=0,in=0]cycle;
\fill[manifold] (0,0)--(1,0)to[out=180,in=180](1,0.6)--(0,0.6)to[out=180,in=180]cycle;
\draw[manifoldboundary] (2,0.1)--(1.5,0.1)to[out=180,in=0](1,0)to[out=180,in=180](1,0.6)to[out=0,in=180](1.5,0.5)--(2,0.5);
\end{tikzpicture}
\;,
\end{equation}
where the right-hand side corresponds to a tube where we cut out a half-disk with physical boundary on one side.
The compactified version of the move in Eq.~\eqref{eq:2d_puncture_healing_move} yields
\begin{equation}
\label{eq:interval_healing_equation}
\begin{tikzpicture}
\draw[\manifoldbdcol,line width=4] (1,0)--(2,0);
\draw[\manifoldcol,line width=2] (1,0)--(2,0);
\draw[\manifoldcol,line width=3] (0,0)--(1,0);
\atoms{circ,small,all,bdastyle=\manifoldbdcol,decstyle=\manifoldcol}{x/p={1,0}}
\atoms{circ,small,all,bdastyle=\manifoldbdcol,decstyle=\manifolddefectcol}{x/p={2,0}}
\end{tikzpicture}
=
\begin{tikzpicture}
\atoms{circ,small,all,bdastyle=\manifoldcol,decstyle=\manifoldcol}{x/p={1,0}}
\draw[\manifoldcol,line width=3] (0,0)--(1,0);
\end{tikzpicture}\;.
\end{equation}
To interpret the algebraic structure of this move, we note that as we see in Section~\ref{sec:universal_1d}, every model of a one-dimensional liquid is equivalent to a projector. For quantum systems these are (super-)linear maps which can be restricted to their support, such that a model is simply given by a (super-)vector space. The $0+1$-dimensional boundaries are vectors $x$ and $w$, and the domain wall is a matrix $M$. The above equation is then of the form $Mx=w$, so a boundary defect fulfilling Eq.~\eqref{eq:2d_puncture_healing_move} exists if $w$ is in the image of $M$.

\subsection{Invertible domain walls}
\label{sec:domain_walls}
An important type of ``defect'' is given by \emph{domain walls}. For $n$-dimensional topological fixed-point models such as the triangle-liquid models in $n=2$, a domain wall is a $n-1$-dimensional topological defect separating two models. Along the lines of Section~\ref{sec:higher_order_manifolds}, the link of this defect consists of two points, each corresponding to one of the two models. The corresponding compactification mapping on the models acts as stacking the two models and identifying them with a single model. Models for domain walls are then in one-to-one correspondence with boundaries of the stacked model, which is well-known as the `folding trick'. 

Of especial importance are \emph{invertible} domain walls, which obey further, topology-changing moves as described in Section 2 of Ref.~\cite{liquid_intro}. These moves allow to generate isolated islands (bubbles) of either model inside of the other, and joining or splitting those bubbles. In $1+1$ dimensions, the moves look like
\begin{equation}
\label{eq:invertible_domain1}
\begin{tikzpicture}
\fill[blue!30] (0,0)circle(1);
\clip (0,0)circle(1);
\draw[line width=2,fill=red!30] (-1.1,0)circle(0.8);
\draw[line width=2,fill=red!30] (1.1,0)circle(0.8);
\end{tikzpicture}
=
\begin{tikzpicture}
\fill[red!30] (0,0)circle(1);
\clip (0,0)circle(1);
\draw[line width=2,fill=blue!30] (0,-1.1)circle(0.8);
\draw[line width=2,fill=blue!30] (0,1.1)circle(0.8);
\end{tikzpicture}\;,
\end{equation}
\begin{equation}
\label{eq:invertible_domain2}
\begin{tikzpicture}
\fill[red!30] (0,0)circle(0.8);
\draw[line width=2,fill=blue!30] (0,0)circle(0.4);
\end{tikzpicture}
=
\begin{tikzpicture}
\fill[red!30] (0,0)circle(0.8);
\end{tikzpicture}\;,
\end{equation}
and
\begin{equation}
\label{eq:invertible_domain3}
\begin{tikzpicture}
\fill[blue!30] (0,0)circle(0.8);
\draw[line width=2,fill=red!30] (0,0)circle(0.4);
\end{tikzpicture}
=
\begin{tikzpicture}
\fill[blue!30] (0,0)circle(0.8);
\end{tikzpicture}\;.
\end{equation}
In general dimensions, invertibility corresponds to moves
\begin{equation}
B^{br}_x\times B_{n-x} = B_x\times B^{rb}_{n-x}
\end{equation}
for all $0\leq x\leq n$. Here, $B_x$ is the $x$-ball, $B^{br}_x$ is a red $x$-ball with a smaller blue $x$-ball in the centre separated by a domain wall, and $B^{rb}_x$ is the same $x$-ball but with blue and red exchanged. Note, that the open boundaries on both sides are topologically equal, even though geometrically they are not.

Invertible domain walls are so important because they are central to the very definition of (exact) phases of fixed-point models. Two models are in the same exact phase if there exists an invertible domain wall between them, and if the models are topological liquid models, then the domain wall is itself a liquid model for a topological invertible domain wall. Just as the liquid moves give rise to constraints on the liquid models, invertible domain walls give rise to equivalence relations. We should keep in mind that as discussed in Section 2 of Ref.~\cite{liquid_intro}, there might be technical differences between exact phases of fixed-point models and phases defined via continuous gapped paths of path integrals or Hamiltonians. 

The universality arguments for bulk liquids presented in this work can be extended to define universal domain wall liquids. Recall that universality in the bulk implies that fixed-point models are classified by a single set of tensors and equations. Analogously universality of the domain wall liquid implies that the phase equivalence of two models can be determined via a single set of domain wall tensors and equations.

As a concrete example, let us discuss a domain wall liquid for the triangle liquid. As argued in Section~\ref{sec:compactification}, domain wall liquids can be obtained by considering boundary liquids for the compactified liquid, which in our case consists of stacking two triangle liquid models into a single one. The boundary triangle liquid after stacking is determined by a boundary tensor,
\begin{equation}
\label{eq:domain_wall_stack_tensor}
\begin{tikzpicture}
\atoms{bdedge}{0/}
\draw[bdbind] (0-l)--++(180:0.4) (0-r)--++(0:0.4);
\draw[red] (0-t)edge[mark={ar,s}]++(90:0.4);
\draw[blue] (0-b)--++(-90:0.4);
\end{tikzpicture}\;,
\end{equation}
whose bulk index is divided into a red and a blue part, corresponding to the two different stacked models. It has to obey the equations coming from the boundary Pachner move with the triangle tensor being a tensor product of a red and blue triangle tensor. Via weakly invertible liquid mappings, this liquid is equivalent to a slightly refined liquid consisting of one separate boundary for the blue and red liquid. The boundaries form a domain wall by `interacting' in the following way. The boundary bond dimensions are equal and the two different boundary tensors (here red and blue) commute,
\begin{equation}
\label{eq:domain_wall_commutativity}
\begin{tikzpicture}
\atoms{bdedge}{{0/bdastyle=red,decstyle=red}, {1/p={1,0},bdastyle=blue,decstyle=blue}}
\draw[bdbind] (0-r)--(1-l) (0-l)--++(180:0.4) (1-r)--++(0:0.4);
\draw[red] (0-t)edge[mark={ar,s}]++(90:0.4);
\draw[blue] (1-b)--++(-90:0.4);
\end{tikzpicture}
=
\begin{tikzpicture}
\atoms{bdedge}{{0/p={1,0},bdastyle=red,decstyle=red}, {1/bdastyle=blue,decstyle=blue}}
\draw[bdbind] (0-l)--(1-r) (1-l)--++(180:0.4) (0-r)--++(0:0.4);
\draw[red] (0-t)edge[mark={ar,s}]++(90:0.4);
\draw[blue] (1-b)--++(-90:0.4);
\end{tikzpicture}\;.
\end{equation}
The equivalence of this \emph{domain-wall triangle liquid} with the less elegant liquid determined by Eq.~\eqref{eq:domain_wall_stack_tensor} can be shown explicitly via weakly invertible liquid mappings. The mapping form the former to the latter is given by
\begin{equation}
\label{eq:domain_wall_simpl_mapping}
\begin{tikzpicture}
\atoms{bdedge, colred}{0/}
\draw[bdbind] (0-l)--++(180:0.4) (0-r)--++(0:0.4);
\draw[red] (0-t)edge[mark={ar,s}]++(90:0.4);
\end{tikzpicture}
\coloneqq
\begin{tikzpicture}
\atoms{bdedge}{0/}
\atoms{oface, colblue}{1/p={0,-0.6}}
\draw[bdbind] (0-l)--++(180:0.4) (0-r)--++(0:0.4);
\draw[red] (0-t)edge[mark={ar,s}]++(90:0.4);
\draw[blue,mark={ar,e}] (1)to[out=-45,in=0]++(-90:0.5)to[out=180,in=-135](1);
\draw[blue] (0)edge[mark={ar,e}](1);
\end{tikzpicture}\;,
\end{equation}
and analogously with red and blue exchanged. The boundary Pachner move in Eq.~\eqref{eq:boundary_pachner} is derived by
\begin{equation}
\begin{multlined}
\begin{tikzpicture}
\atoms{bdedge,colred}{0/}
\atoms{oface, colred}{2/p={0,0.6}}
\draw[bdbind] (0-l)--++(180:0.4) (0-r)--++(0:0.4);
\draw[red] (0-t)edge[mark={ar,s}]++(90:0.4);
\draw[red] (2)edge[mark={ar,s}]++(135:0.5) (2)edge[mark={ar,s}]++(45:0.5);
\end{tikzpicture}
\overset{\ref{eq:domain_wall_simpl_mapping}}{=}
\begin{tikzpicture}
\atoms{bdedge}{0/}
\atoms{oface, colblue}{1/p={0,-0.6}}
\atoms{oface, colred}{2/p={0,0.6}}
\draw[bdbind] (0-l)--++(180:0.4) (0-r)--++(0:0.4);
\draw[red] (0-t)edge[mark={ar,s}]++(90:0.4);
\draw[red] (2)edge[mark={ar,s}]++(135:0.5) (2)edge[mark={ar,s}]++(45:0.5);
\draw[blue,mark={ar,e}] (1)to[out=-45,in=0]++(-90:0.5)to[out=180,in=-135](1);
\draw[blue] (0)edge[mark={ar,e}](1);
\end{tikzpicture}
\overset{\ref{eq:boundary_pachner}}{=}
\begin{tikzpicture}
\atoms{bdedge}{0/}
\atoms{oface, colblue}{1/p={0,-0.6}, 1x/p={-0.4,-1}, 1y/p={0.4,-1}}
\atoms{oface, colred}{2/p={0,0.6}}
\draw[bdbind] (0-l)--++(180:0.4) (0-r)--++(0:0.4);
\draw[red] (0-t)edge[mark={ar,s}]++(90:0.4);
\draw[red] (2)edge[mark={ar,s}]++(135:0.5) (2)edge[mark={ar,s}]++(45:0.5);
\draw[blue,mark={ar,e}] (1x)to[out=-45,in=0]++(-90:0.5)to[out=180,in=-135](1x);
\draw[blue,mark={ar,e}] (1y)to[out=-45,in=0]++(-90:0.5)to[out=180,in=-135](1y);
\draw[blue] (0)edge[mark={ar,e}](1) (1)edge[mark={ar,e}](1x) (1)edge[mark={ar,e}](1y);
\end{tikzpicture}
\\
\overset{\ref{eq:22pachner}}{=}
\begin{tikzpicture}
\atoms{bdedge}{0/, 0x/p={1,0}}
\atoms{oface, colblue}{1/p={0,-0.6}, 1x/p={1,-0.6}}
\draw[bdbind] (0-l)--++(180:0.3) (0-r)--(0x-l) (0x-r)--++(0:0.3);
\draw[red] (0-t)edge[mark={ar,s}]++(90:0.4) (0x-t)edge[mark={ar,s}]++(90:0.4);
\draw[blue,mark={ar,e}] (1x)to[out=-45,in=0]++(-90:0.5)to[out=180,in=-135](1x);
\draw[blue,mark={ar,e}] (1)to[out=-45,in=0]++(-90:0.5)to[out=180,in=-135](1);
\draw[blue] (0)edge[mark={ar,e}](1) (0x)edge[mark={ar,e}](1x);
\end{tikzpicture}
\overset{\ref{eq:domain_wall_simpl_mapping}}{=}
\begin{tikzpicture}
\atoms{bdedge, colred}{0/, 0x/p={0.8,0}}
\draw[bdbind] (0-l)--++(180:0.3) (0-r)--(0x-l) (0x-r)--++(0:0.3);
\draw[red] (0-t)edge[mark={ar,s}]++(90:0.4) (0x-t)edge[mark={ar,s}]++(90:0.4);
\end{tikzpicture}
\;,
\end{multlined}
\end{equation}
where in the second equation we used the moves of the blue triangle liquid, and in the third equation, we used the moves for the triangle boundary liquid of the stacked model. The commutativity in Eq.~\eqref{eq:domain_wall_commutativity} can be derived by
\begin{equation}
\begin{multlined}
\begin{tikzpicture}
\atoms{bdedge,colred}{0/}
\atoms{bdedge,colblue}{1/p={0.8,0}}
\draw[bdbind] (0-l)--++(180:0.3) (0-r)--(1-l) (1-r)--++(0:0.3);
\draw[red] (0-t)edge[mark={ar,s}]++(90:0.4);
\draw[blue] (1-b)edge[]++(-90:0.4);
\end{tikzpicture}
\overset{\ref{eq:domain_wall_simpl_mapping}}{=}
\begin{tikzpicture}
\atoms{bdedge}{0/, 1/p={0.8,0}}
\atoms{oface, colblue}{0x/p={0,-0.6}}
\atoms{oface, colred}{1x/p={0.8,0.6}}
\draw[bdbind] (0-l)--++(180:0.3) (0-r)--(1-l) (1-r)--++(0:0.3);
\draw[blue,mark={ar,e}] (0x)to[out=-45,in=0]++(-90:0.5)to[out=180,in=-135](0x);
\draw[red,mark={ar,e}] (1x)to[out=45,in=0]++(90:0.5)to[out=180,in=135](1x);
\draw[blue] (0-b)edge[mark={ar,e}](0x);
\draw[red] (1-t)edge[mark={ar,s}](1x);
\draw[red] (0-t)edge[mark={ar,s}]++(90:0.4);
\draw[blue] (1-b)edge[]++(-90:0.4);
\end{tikzpicture}
\overset{\ref{eq:boundary_pachner}}{=}
\begin{tikzpicture}
\atoms{bdedge}{0/}
\atoms{oface, colblue}{1/p={0,-0.6}, 1x/p={-0.4,-1}}
\atoms{oface, colred}{2/p={0,0.6}, 0x/p={0.4,1}}
\draw[bdbind] (0-l)--++(180:0.4) (0-r)--++(0:0.4);
\draw[red] (0-t)edge[mark={ar,s}]++(90:0.4);
\draw[red] (2)edge[mark={ar,s}]++(135:0.5) (2)edge[mark={ar,s}](0x);
\draw[blue,mark={ar,e}] (1x)to[out=-45,in=0]++(-90:0.5)to[out=180,in=-135](1x);
\draw[red,mark={ar,e}] (0x)to[out=45,in=0]++(90:0.5)to[out=180,in=135](0x);
\draw[blue] (0)edge[mark={ar,e}](1) (1)edge[mark={ar,e}](1x) (1)--++(-45:0.5);
\end{tikzpicture}
\\
\overset{\ref{eq:22pachner}}{=}
\begin{tikzpicture}
\atoms{bdedge}{0/}
\draw[bdbind] (0-l)--++(180:0.4) (0-r)--++(0:0.4);
\draw[red] (0-t)edge[mark={ar,s}]++(90:0.4);
\draw[blue] (0-b)edge[]++(-90:0.4);
\end{tikzpicture}
=
\begin{tikzpicture}
\atoms{bdedge,colred}{0/}
\atoms{bdedge,colblue}{1/p={-0.8,0}}
\draw[bdbind] (1-l)--++(180:0.3) (1-r)--(0-l) (0-r)--++(0:0.3);
\draw[red] (0-t)edge[mark={ar,s}]++(90:0.4);
\draw[blue] (1-b)edge[]++(-90:0.4);
\end{tikzpicture}
\;,
\end{multlined}
\end{equation}
where the last equation holds for symmetry reasons.

The reverse mapping from the boundary triangle liquid for the stack to the domain wall triangle liquid is even more straight forward,
\begin{equation}
\begin{tikzpicture}
\atoms{bdedge}{0/}
\draw[bdbind] (0-l)--++(180:0.4) (0-r)--++(0:0.4);
\draw[red] (0-t)edge[mark={ar,s}]++(90:0.4);
\draw[blue] (0-b)edge[]++(-90:0.4);
\end{tikzpicture}
\coloneqq
\begin{tikzpicture}
\atoms{bdedge,colred}{0/}
\atoms{bdedge,colblue}{1/p={0.8,0}}
\draw[bdbind] (0-l)--++(180:0.3) (0-r)--(1-l) (1-r)--++(0:0.3);
\draw[red] (0-t)edge[mark={ar,s}]++(90:0.4);
\draw[blue] (1-b)edge[]++(-90:0.4);
\end{tikzpicture}\;.
\end{equation}
The mapped boundary Pachner move for the stacked model can be derived from the boundary Pachner moves of the individual liquids plus the commutativity move in Eq.~\eqref{eq:domain_wall_commutativity}. We have thus shown the equivalence of the two versions of the domain wall liquid.

If we want to impose invertibility of the domain wall liquid, we have to implement Eq.~\eqref{eq:invertible_domain1}, Eq.~\eqref{eq:invertible_domain2}, and Eq.~\eqref{eq:invertible_domain3} as concrete moves. The simplest moves representing the corresponding topology changes are given by
\begin{equation}
\begin{tikzpicture}
\atoms{bdedge,bdastyle=red,decstyle=red}{0/rot=90, {1/p={0.8,0},rot=90}}
\draw[red] (0)edge[mark={ar,e}](1);
\draw[line width=2] (0-l)--++(-90:0.4) (0-r)--++(90:0.4) (1-l)--++(-90:0.4) (1-r)--++(90:0.4);
\end{tikzpicture}
=
\begin{tikzpicture}
\atoms{bdedge,bdastyle=blue,decstyle=blue}{0/, {1/p={0,0.8}}}
\draw[blue] (0)edge[mark={ar,s}](1);
\draw[line width=2] (0-l)--++(180:0.4) (0-r)--++(0:0.4) (1-l)--++(180:0.4) (1-r)--++(0:0.4);
\end{tikzpicture}\;,
\end{equation}
\begin{equation}
\begin{tikzpicture}
\atoms{bdedge,bdastyle=red,decstyle=red}{0/rot=90}
\draw[line width=2,looseness=2] (0-r)to[out=90,in=90](0.5,0)to[out=-90,in=-90](0-l);
\draw[red] (0-t)edge[mark={ar,s}]++(180:0.4);
\end{tikzpicture}
=
\begin{tikzpicture}
\atoms{oface,bdastyle=red,decstyle=red}{0/}
\draw[red,mark={ar,e},looseness=2] (0)to[out=90,in=90](0.5,0)to[in=-90,out=-90](0);
\draw[red] (0)edge[mark={ar,s}]++(180:0.5);
\end{tikzpicture}\;,
\end{equation}
and
\begin{equation}
\begin{tikzpicture}
\atoms{bdedge,bdastyle=blue,decstyle=blue}{0/rot=90}
\draw[line width=2,looseness=2] (0-r)to[out=90,in=90](0.5,0)to[out=-90,in=-90](0-l);
\draw[blue] (0-t)--++(180:0.4);
\end{tikzpicture}
=
\begin{tikzpicture}
\atoms{oface,bdastyle=blue,decstyle=blue}{0/}
\draw[blue,mark={ar,e},looseness=2] (0)to[out=90,in=90](0.5,0)to[in=-90,out=-90](0);
\draw[blue] (0)edge[mark={ar,s}]++(180:0.5);
\end{tikzpicture}\;.
\end{equation}

Now, we have argued in Section~\ref{sec:finegrain_boundary} that the boundary triangle liquid is universal for all topological liquids in $1+1$ dimensions exhibiting a topological boundary. Consequently, the domain-wall triangle liquid model given above is a universal domain wall liquid between liquid models with topological boundary. We have therefore obtained a complete classification of topological phases with topological boundary in the sense that we have identified them with a fixed set of tensorial variables subject to a fixed set of tensor-network equations (the triangle liquid), modulo equivalence via another fixed set of tensor variables and equivalences (the domain-wall triangle liquid).

Note that analogously, we can define a \emph{domain-wall vertex liquid}, show that it is universal for domain walls between general liquid models, and therefore obtain a classification of general topological liquid models in $1+1$ dimensions. At this point we should remind the reader that we do not expect any physically meaningful vertex-liquid models beyond triangle-liquid models in $1+1$ dimensions, but the situation might be different in higher dimensions.

\section{The extended triangle liquid}
\label{sec:extended_triangle_liquid}
In this appendix, we will present an alternative example for a liquid which cannot be emulated by the triangle liquid. The \emph{extended triangle liquid} might appear to some readers as a more straight-forward generalization of the triangle liquid, as it still associates tensors to the triangles of a triangulation. However, the tensors are not all the same, but depend on the combinatorics of a distance-1 neighborhood of the triangle. After introducing the liquid, we will explicitly see how the corner problem obstructs a possible topology-preserving mapping from the triangle liquid.
\subsection{The liquid}
As announced, the extended triangle liquid represents a triangulation by a Poincar\'e dual network just like the triangle liquid. However, the tensor at a triangle depends on the valencies, i.e., the number of adjacent edges, of its corner vertices. Diagrammatically, we will draw the tensors in the same way as for the triangle liquid, just that we add labels indicating the valencies, e.g.,
\begin{equation}
\begin{tikzpicture}
\atoms{vertex}{0/, 1/p={1,0}, 2/p=60:1}
\draw (0)--(1) (1)--(2) (2)--(0);
\draw (0)edge[enddots]++(-100:0.5) (0)edge[enddots]++(160:0.5) (1)edge[enddots]++(-80:0.5) (1)edge[enddots]++(-30:0.5) (1)edge[enddots]++(20:0.5) (2)edge[enddots]++(0:0.5) (2)edge[enddots]++(50:0.5) (2)edge[enddots]++(90:0.5) (2)edge[enddots]++(140:0.5) (2)edge[enddots]++(180:0.5);
\end{tikzpicture}
\quad\rightarrow\quad
\begin{tikzpicture}
\atoms{oface, lab={t=$4$,p=-150:0.35},lab={t=$5$,p=-30:0.35},lab={t=$7$,p=90:0.35}}{0/}
\draw (0)--++(30:0.7) (0)--++(-90:0.7) (0)--++(150:0.7);
\end{tikzpicture}
\;.
\end{equation}
A model of this liquid consists of one tensor for every triple of numbers, such as $\{4,5,7\}$ in the example above. General triangulations can have vertices with arbitrarily high valencies, so in order to represent any triangulation we would need an infinite number of different tensors. However, we can without loss of generality restrict to a finite set of different adjacencies, as we show in Appendix~\ref{sec:appendix_links}.

The moves of the liquid are again based on the Pachner moves. However, the 2-2 Pachner move changes the valency of the four vertices at the corners. If we apply a 2-2 Pachner move to a triangulation, this changes not only the tensors at the two triangles directly involved in the move, but also those at the triangles adjacent to the four vertices at the corners. E.g., for the Pachner move below, the tensors at the triangles marked in red \emph{and} blue change
\begin{equation}
\label{eq:extended_triangle_22pachner}
\begin{gathered}
\begin{tikzpicture}
\atoms{vertex}{0/, 1/p={0,1.5}, 2/p={-0.75,0.75}, 3/p={0.75,0.75}, {4/p={$(2)+(-100:1.2)$},lab={t=6,ang=-100}}, {5/p={$(2)+(-150:1)$},lab={t=5,ang=-150}}, {6/p={$(2)+(150:1)$},lab={t=8, ang=150}}, {7/p={$(2)+(100:1)$},lab={t=9, ang=100}}, {8/p={$(1)+(60:1)$},lab={t=5, ang=60}}, {9/p={$(3)+(-90:1.2)$},lab={t=7,ang=-90}}, {10/p={$(3)+(-50:1)$},lab={t=8,ang=-50}}, {11/p={$(3)+(-10:1)$},lab={t=7,ang=-10}}, {12/p={$(3)+(30:1)$},lab={t=6,ang=30}}, {13/p={$(3)+(70:1)$},lab={t=8,ang=70}}}
\fill[blue,opacity=0.3] (4-c)--(5-c)--(6-c)--(7-c)--(8-c)--(13-c)--(12-c)--(11-c)--(10-c)--(9-c)--cycle (0-c)--(3-c)--(1-c)--(2-c)--cycle;
\fill[red,opacity=0.3] (0-c)--(2-c)--(1-c)--(3-c)--cycle;
\draw (0)--(2)--(1)--(3)--(0) (4)--(5)--(6)--(7)--(8)--(13)--(12)--(11)--(10)--(9)--(4);
\draw (2)--(4) (2)--(5) (2)--(6) (2)--(7) (1)--(7) (1)--(8) (1)--(13) (3)--(9) (3)--(10) (3)--(11) (3)--(12) (3)--(13) (0)--(9) (0)--(4);
\draw (0)--(1);
\end{tikzpicture}
\\\quad\leftrightarrow\quad
\begin{tikzpicture}
\atoms{vertex}{0/, 1/p={0,1.5}, 2/p={-0.75,0.75}, 3/p={0.75,0.75}, {4/p={$(2)+(-100:1.2)$},lab={t=6,ang=-100}}, {5/p={$(2)+(-150:1)$},lab={t=5,ang=-150}}, {6/p={$(2)+(150:1)$},lab={t=8, ang=150}}, {7/p={$(2)+(100:1)$},lab={t=9, ang=100}}, {8/p={$(1)+(60:1)$},lab={t=5, ang=60}}, {9/p={$(3)+(-90:1.2)$},lab={t=7,ang=-90}}, {10/p={$(3)+(-50:1)$},lab={t=8,ang=-50}}, {11/p={$(3)+(-10:1)$},lab={t=7,ang=-10}}, {12/p={$(3)+(30:1)$},lab={t=6,ang=30}}, {13/p={$(3)+(70:1)$},lab={t=8,ang=70}}}
\fill[blue,opacity=0.3] (4-c)--(5-c)--(6-c)--(7-c)--(8-c)--(13-c)--(12-c)--(11-c)--(10-c)--(9-c)--cycle (0-c)--(3-c)--(1-c)--(2-c)--cycle;
\fill[red,opacity=0.3] (0-c)--(2-c)--(1-c)--(3-c)--cycle;
\draw (0)--(2)--(1)--(3)--(0) (4)--(5)--(6)--(7)--(8)--(13)--(12)--(11)--(10)--(9)--(4);
\draw (2)--(4) (2)--(5) (2)--(6) (2)--(7) (1)--(7) (1)--(8) (1)--(13) (3)--(9) (3)--(10) (3)--(11) (3)--(12) (3)--(13) (0)--(9) (0)--(4);
\draw (2)--(3);
\end{tikzpicture}\;.
\end{gathered}
\end{equation}
Here, the labels at the boundary vertices indicate their valencies. The tensor network equation corresponding to the move above is given by
\begin{equation}
\begin{tikzpicture}
\atoms{void}{x0/, x1/p={0,1.5}, x2/p={-0.75,0.75}, x3/p={0.75,0.75}}
\atoms{oface}{
{0/p={$(x2)+(-80:0.8)$},extendedtrianglab=5:20 and 6:120 and 6:-100},
{1/p={$(x2)+(-125:0.8)$},extendedtrianglab=6:70 and 5:180 and 6:-70},
{2/p={$(x2)+(180:0.8)$},extendedtrianglab=6:0 and 8:120 and 5:-120},
{3/p={$(x2)+(125:0.8)$},extendedtrianglab=6:-60 and 9:70 and 8:-170},
{4/p={$(x2)+(80:0.8)$},extendedtrianglab=6:-120 and 9:130 and 6:-10},
{5/p={$(x1)+(120:0.7)$},extendedtrianglab=6:-60 and 5:60 and 9:180},
{6/p={$(x1)+(60:0.7)$},extendedtrianglab=6:-120 and 5:120 and 8:0},
{7/p={$(x3)+(110:0.8)$},extendedtrianglab=8:50 and 7:-50 and 6:180},
{8/p={$(x3)+(66:0.8)$},extendedtrianglab=7:-120 and 8:120 and 6:0},
{9/p={$(x3)+(22:0.8)$},extendedtrianglab=7:-160 and 6:90 and 7:-30},
{10/p={$(x3)+(-22:0.8)$},extendedtrianglab=7:160 and 7:30 and 8:-50},
{11/p={$(x3)+(-66:0.8)$},extendedtrianglab=7:120 and 8:0 and 7:-120},
{12/p={$(x3)+(-110:0.8)$},extendedtrianglab=7:-70 and 7:50 and 5:-160},
{13/p={$(x0)+(-90:0.5)$},extendedtrianglab=5:100 and 6:-150 and 7:-10},
{a/p={-0.4,0.75},extendedtrianglab=5:-50 and 6:180 and 6:60},
{b/p={0.4,0.75},extendedtrianglab=7:0 and 5:-135 and 6:135}}
\draw (0)--(1)--(2)--(3)--(4)--(5)--(6)--(7)--(8)--(9)--(10)--(11)--(12)--(13)--(0);
\draw (a)--(b) (a)--(0) (a)--(4) (b)--(7) (b)--(12);
\draw (1)--++(-125:0.5) (2)--++(180:0.5) (3)--++(125:0.5) (5)--++(120:0.5) (6)--++(60:0.5) (8)--++(66:0.5) (9)--++(22:0.5) (10)--++(-22:0.5) (11)--++(-66:0.5) (13)--++(-90:0.5);
\end{tikzpicture}
=
\begin{tikzpicture}
\atoms{void}{x0/, x1/p={0,1.5}, x2/p={-0.75,0.75}, x3/p={0.75,0.75}}
\atoms{oface}{{0/p={$(x2)+(-80:0.8)$},extendedtrianglab=4:10 and 7:120 and 6:-100},
{1/p={$(x2)+(-125:0.8)$},extendedtrianglab=7:70 and 5:180 and 6:-70},
{2/p={$(x2)+(180:0.8)$},extendedtrianglab=7:0 and 8:120 and 5:-120},
{3/p={$(x2)+(125:0.8)$},extendedtrianglab=7:-60 and 9:70 and 8:-170},
{4/p={$(x2)+(80:0.8)$},extendedtrianglab=7:-120 and 9:130 and 5:10},
{5/p={$(x1)+(120:0.7)$},extendedtrianglab=5:-60 and 5:60 and 9:180},
{6/p={$(x1)+(60:0.7)$},extendedtrianglab=5:-120 and 5:120 and 8:0},
{7/p={$(x3)+(110:0.8)$},extendedtrianglab=8:50 and 8:-50 and 5:170},
{8/p={$(x3)+(66:0.8)$},extendedtrianglab=8:-120 and 8:120 and 6:0},
{9/p={$(x3)+(22:0.8)$},extendedtrianglab=8:-160 and 6:90 and 7:-30},
{10/p={$(x3)+(-22:0.8)$},extendedtrianglab=8:160 and 7:30 and 8:-50},
{11/p={$(x3)+(-66:0.8)$},extendedtrianglab=8:120 and 8:0 and 7:-120},
{12/p={$(x3)+(-110:0.8)$},extendedtrianglab=7:-70 and 8:50 and 4:-160},
{13/p={$(x0)+(-90:0.5)$},extendedtrianglab=4:100 and 6:-150 and 7:-10},
{a/p={0,0.4},extendedtrianglab=4:-100 and 7:150 and 8:30},
{b/p={0,1.15},extendedtrianglab=5:100 and 7:-150 and 8:-30}}
\draw (0)--(1)--(2)--(3)--(4)--(5)--(6)--(7)--(8)--(9)--(10)--(11)--(12)--(13)--(0);
\draw (a)--(b) (a)--(0) (a)--(12) (b)--(4) (b)--(7);
\draw (1)--++(-125:0.5) (2)--++(180:0.5) (3)--++(125:0.5) (5)--++(120:0.5) (6)--++(60:0.5) (8)--++(66:0.5) (9)--++(22:0.5) (10)--++(-22:0.5) (11)--++(-66:0.5) (13)--++(-90:0.5);
\end{tikzpicture}
\;.
\end{equation}
As there are many different 'embeddings' of a Pachner move into a neighbourhood with different adjacencies at the corner vertices, there is a large number of different moves, namely one for each combination of valencies in Eq.~\eqref{eq:extended_triangle_22pachner}. This is no conceptual problem since the set of moves is still finite when we restrict the vertex valencies as hinted at above, but it makes finding concrete models very difficult.

\subsection{Failure mapping from the triangle liquid}
Let us now show that there cannot be a topology-preserving mapping from the triangle liquid to the extended triangle liquid. Such an attempted mapping associates to the triangle an extended-triangle-liquid network as depicted in Eq.~\eqref{eq:mapping_triangle_fail} for an analogous mapping to the vertex liquid. For different possible choices of the mapping, the bottom right corner could look like
\begin{equation}
\label{eq:finegrain_extended_triangle_corner}
\begin{tikzpicture}
\atoms{void}{0/, 1/p=120:1.5, 2/p=180:1.5}
\fadecorner{0}{1}{2}
\atoms{circ,lab={t=$\scriptstyle{6}$,p=-30:0.3}}{x/p={150:0.7}}
\draw (x)edge[ind=c_x]++(-90:0.7) (x)edge[ind=b_x]++(30:0.7);
\end{tikzpicture}\;,\text{or}\quad
\begin{tikzpicture}
\atoms{void}{0/, 1/p=120:1.5, 2/p=180:1.5}
\fadecorner{0}{1}{2}
\atoms{circ,lab={t=$\scriptstyle{5}$,p=-30:0.3}}{x/p={150:0.7}}
\draw (x)edge[ind=c_x]++(-90:0.7) (x)edge[ind=b_x]++(30:0.7);
\end{tikzpicture}\;,\text{or}\quad
\begin{tikzpicture}
\atoms{void}{0/, 1/p=120:2, 2/p=180:2}
\fadecorner{0}{1}{2}
\atoms{circ}{{x/p={138:1.2},lab={t=$\scriptstyle{11}$,p=138+180:0.3}}, {y/p={162:1.2},lab={t=$\scriptstyle{11}$,p=162+180:0.3}}}
\draw (y)edge[ind=c_x]++(-90:0.5) (x)edge[ind=b_x]++(30:0.5) (x)--(y);
\end{tikzpicture}\;,
\end{equation}
or $b_x$ and $c_x$ could be separated by more bonds and the valency labels in the corner could take other values. The point is, that regardless of the specific choice we make for the mapping, no choice can result in a valid extended triangle liquid networks for all initial networks of the triangle liquid. This is due to the fact that the triangle liquid allows for $l$-valent vertices for \emph{any} $l$ whereas the tensors at the corner of the mapping have fixed adjacency labels. As a result, for every choice of the mapping, there exists a small patch of the triangle liquid with a vertex of adjacency $l$, such that the adjacency $l$ does not match the adjacency labels of the network obtained from the mapping.

E.g., for the first case in Eq.~\eqref{eq:finegrain_extended_triangle_corner} and an $l=5$ vertex, we obtain a $5$-gon plaquette with tensors valency-6 tensors
\begin{equation}
\begin{tikzpicture}
\atoms{void}{m/}
\foreach \nr in {0,1,...,4}{
\atoms{void}{x\nr/p={\nr*72:1.5}}
};
\fadecorner{m}{x0}{x1}
\fadecorner{m}{x1}{x2}
\fadecorner{m}{x2}{x3}
\fadecorner{m}{x3}{x4}
\fadecorner{m}{x4}{x0}
\foreach \nr in {0,1,...,4}{
\atoms{circ,lab={t=$\scriptstyle{6}$,p={180+\nr*72+36}:0.3}}{\nr/p={\nr*72+36:0.7}}
};
\draw[] (0)--(1)--(2)--(3)--(4)--(0);
\end{tikzpicture}\;.
\end{equation}
The valencies do not match and hence the resulting network is invalid.

Another problem with the attempted mapping is that the mapped moves of the triangle liquid cannot be derived from the extended-triangle-liquid moves. This problem again originates from the fact that the tensors of the triangle liquid do not carry any information about the adjacency of vertices, whereas this is the case for the extended triangle liquid. E.g., applying the mapping to the 2-2 Pachner move yields
\begin{equation}
\label{eq:failed_mapped_move}
\begin{tikzpicture}
\atoms{void}{0/p={-1,0}, 1/p={0,1}, 2/p={1,0}, 3/p={0,-1}}
\draw[dashed, fill, manifold] (0)--(1)--(2)--cycle;
\draw[dashed, fill, manifold] (0)--(2)--(3)--cycle;
\draw ($(0)+(135:0.2)$)edge[->, mark={slab=$\vec a$}] ($(1)+(135:0.2)$);
\draw ($(0)+(-135:0.2)$)edge[->, mark={slab=$\vec c$,r}] ($(3)+(-135:0.2)$);
\draw ($(1)+(45:0.2)$)edge[->, mark={slab=$\vec b$}] ($(2)+(45:0.2)$);
\draw ($(3)+(-45:0.2)$)edge[<-, mark={slab=$\vec d$,r}] ($(2)+(-45:0.2)$);
\end{tikzpicture}
=
\begin{tikzpicture}
\atoms{void}{0/p={-1,0}, 1/p={0,1}, 2/p={1,0}, 3/p={0,-1}}
\draw[dashed, fill, manifold] (0)--(1)--(3)--cycle;
\draw[dashed, fill, manifold] (1)--(2)--(3)--cycle;
\draw ($(0)+(135:0.2)$)edge[->, mark={slab=$\vec a$}] ($(1)+(135:0.2)$);
\draw ($(0)+(-135:0.2)$)edge[->, mark={slab=$\vec c$,r}] ($(3)+(-135:0.2)$);
\draw ($(1)+(45:0.2)$)edge[->, mark={slab=$\vec b$}] ($(2)+(45:0.2)$);
\draw ($(3)+(-45:0.2)$)edge[<-, mark={slab=$\vec d$,r}] ($(2)+(-45:0.2)$);
\end{tikzpicture}\;.
\end{equation}
If the mapping is given, e.g., by the first case in Eq.~\eqref{eq:finegrain_extended_triangle_corner}, the upper corner of this move looks like
\begin{equation}
\begin{tikzpicture}
\atoms{void}{0/, 1/p=-135:1.5, 2/p=-90:1.5, 3/p=-45:1.5}
\fadecorner{0}{1}{2}
\fadecorner{0}{2}{3}
\atoms{circ}{{x/p={-112.5:0.9},lab={t=$\scriptstyle{6}$,p=60:0.3}}, {y/p={-67.5:0.9},lab={t=$\scriptstyle{6}$,p=120:0.3}}}
\draw (x)edge[ind=a_x]++(135:0.6) (y)edge[ind=b_0]++(45:0.6) (x)--(y);
\end{tikzpicture}
=
\begin{tikzpicture}
\atoms{void}{0/, 1/p=-135:1.5, 2/p=-45:1.5}
\fadecorner{0}{1}{2}
\atoms{circ}{{x/p={-90:0.6},lab={t=$\scriptstyle{6}$,p=90:0.3}}}
\draw (x)edge[ind=a_x]++(135:0.7) (x)edge[ind=b_0]++(45:0.7);
\end{tikzpicture}
\;.
\end{equation}
However, this move cannot be derived from the moves of the extended triangle liquid, as it changes the number of edges in the plaquette containing $a_x$ and $b_0$ by one. So in order to be derivable from the moves of the extended triangle liquid, either the $6$ on the right would have to be a $5$, or the $6$s on the left would have to be $7$s. Similar arguments apply for any other choice of the mapping.

\section{Moves of the vertex liquid}
\label{sec:vertex_moves}
In this appendix, we discuss the moves of the vertex liquid in more detail.
To get a better feeling for
those moves, it is instructive to go to the dual lattice, where $n$-valent vertices become $n$-gon faces, and triangles become 3-valent vertices. Ignoring all the decorations for a moment, the Pachner move in Eq.~\eqref{eq:vertex_pachner_move} becomes
\begin{equation}
\begin{tikzpicture}
\atoms{vertex}{0/p={-0.3,0}, 1/p={0.3,0}, 2/p={1,0.8}, 3/p={-1,0.8}, 4/p={-1,-0.8}, 5/p={1,-0.8}, a0/p={$(0,0.8)+(30:0.8)$}, a1/p={$(0,0.8)+(70:0.8)$}, a2/p={$(0,0.8)+(110:0.8)$}, a3/p={$(0,0.8)+(150:0.8)$}, a4/p={$(-1,0)+(180:0.5)$}, a5/p={$(0,-0.8)+(-90:0.5)$}}
\draw (0)--(1) (1)--(2) (1)--(5) (0)--(3) (0)--(4) (2)--(a0)--(a1)--(a2)--(a3)--(3)--(a4)--(4)--(a5)--(5)--(2);
\end{tikzpicture}
\quad\leftrightarrow\quad
\begin{tikzpicture}
\atoms{vertex}{0/p={0,-0.3}, 1/p={0,0.3}, 2/p={1,0.8}, 3/p={-1,0.8}, 4/p={-1,-0.8}, 5/p={1,-0.8}, a0/p={$(0,0.8)+(30:0.8)$}, a1/p={$(0,0.8)+(70:0.8)$}, a2/p={$(0,0.8)+(110:0.8)$}, a3/p={$(0,0.8)+(150:0.8)$}, a4/p={$(-1,0)+(180:0.5)$}, a5/p={$(0,-0.8)+(-90:0.5)$}}
\draw (0)--(1) (1)--(2) (1)--(3) (0)--(5) (0)--(4) (2)--(a0)--(a1)--(a2)--(a3)--(3)--(a4)--(4)--(a5)--(5)--(2);
\end{tikzpicture}\;.
\end{equation}
This is a $3$-valent re-cellulation where the number of internal edges adjacent to each individual boundary vertex (which is either $0$ or $1$) does not change, such that all vertices remain $3$-valent. Intuitively speaking, the move does not change the vicinity of the boundary. This is the key property which allows us to map to \emph{any} topological liquid from the vertex liquid in Section~\ref{sec:2d_universal_finegrain}. When we reconsider Eq.~\eqref{eq:vertex_not_move} in this picture, it becomes
\begin{equation}
\begin{tikzpicture}
\atoms{vertex}{0/p={0,-0.5}, 1/p={0,0.5}, 2/p={-0.6,-0.4}, 3/p={-0.6,0.4}, 4/p={0.6,0}}
\draw (0)--(1) (0)--(4)--(1)--(3)--(2)--(0);
\end{tikzpicture}
\quad\leftrightarrow\quad
\begin{tikzpicture}
\atoms{vertex}{0/p={0,-0.5}, 1/p={0,0.5}, 2/p={-0.6,-0.4}, 3/p={-0.6,0.4}, 4/p={0.6,0}}
\draw (3)to[bend right](4) (0)--(4)--(1)--(3)--(2)--(0);
\end{tikzpicture}\;
\end{equation}
and we immediately see that this recellulation does \emph{not} define a move, since there are vertices which have no adjacent internal edge on the left side but one adjacent internal edge on the right side, and vice versa.

The dual representation makes it easier to come up with more general recellulations fulfilling the conditions for the vertex-liquid moves. As another example, the recellulation
\begin{equation}
\begin{tikzpicture}
\atoms{vertex}{0/, 1/p={1.5,0}, 2/p={0,1.5}, 3/p={1.5,1.5}}
\draw[looseness=1.5] (0)--(1)--(3)--(2)--(0) (0)to[bend left=45](1) (2)to[bend right=45](3);
\end{tikzpicture}
\quad\leftrightarrow\quad
\begin{tikzpicture}
\atoms{vertex}{0/, 1/p={1.5,0}, 2/p={0,1.5}, 3/p={1.5,1.5}}
\draw[looseness=1.5] (0)--(1)--(3)--(2)--(0) (1)to[bend left=45](3) (0)to[bend right=45](2);
\end{tikzpicture}
\end{equation}
defines the following valid move of the vertex liquid,
\begin{equation}
\begin{tikzpicture}
\atoms{vertexatom}{0/, 1/p={0,0.6}, 2/p={0,-0.6}}
\draw (0)edge[mark={ar,s}](1) (0)edge[mark={ar,s}](2) (0)--++(180:0.5) (0)--++(0:0.5) (1)--++(90:0.5) (2)--++(-90:0.5);
\end{tikzpicture}
=
\begin{tikzpicture}
\atoms{vertexatom}{0/, 1/p={0.6,0}, 2/p={-0.6,0}}
\draw (0)edge[mark={ar,s}](1) (0)edge[mark={ar,s}](2) (0)--++(90:0.5) (0)--++(-90:0.5) (1)--++(0:0.5) (2)--++(180:0.5);
\end{tikzpicture}\;,
\end{equation}
and the recellulation
\begin{equation}
\begin{tikzpicture}
\atoms{vertex}{0/p={0.5,0}, 1/p={-0.5,-0.5}, 2/p={-0.5,0.5}}
\draw (0)--(1) (1)to[bend left=30](2) (2)--(0) (1)to[bend right=45](2);
\end{tikzpicture}
\quad\leftrightarrow\quad
\begin{tikzpicture}
\atoms{vertex}{0/p={0.5,0}, 1/p={-0.5,-0.7}, 2/p={-0.5,0.7}, 3/p={0,0}, 4/p={-0.3,0}}
\draw (0)--(1) (1)to[bend left=40,looseness=1.4](2) (2)--(0) (1)--(3)--(2) (3)--(4);
\draw[looseness=1.8] (4)to[out=135,in=90]++(-0.4,0)to[out=-90,in=-135](4);
\end{tikzpicture}\;.
\end{equation}
corresponds to the following move
\begin{equation}
\begin{tikzpicture}
\atoms{vertexatom}{0/p={-0.5,0}, 1/p={0.5,0}}
\draw (0)edge[mark={ar,s}](1) (0)edge[ind=a,mark={ar,s}]++(180:0.6) (1)edge[ind=b,mark={ar,s}]++(45:0.6) (1)edge[ind=c]++(-45:0.6);
\end{tikzpicture}
=
\begin{tikzpicture}
\atoms{vertexatom}{0/p={-0.6,0}, 1/p={0.6,0}, 2/p={0,0}}
\draw[looseness=1.5,mark={ar,s}] (0)to[out=-45,in=-90]++(0.9,0)edge[out=90,in=45](0);
\draw (0)edge[bend right=80,looseness=2,mark={ar,s}](1) (0)edge[bend left=80,looseness=2,mark={ar,s}](1) (0)edge[ind=a,mark={ar,s}]++(180:0.6) (1)edge[ind=b,mark={ar,s}]++(45:0.6) (1)edge[ind=c]++(-45:0.6) (0)edge[mark={ar,e}](2);
\end{tikzpicture}\;.
\end{equation}
As we can see it is allowed that the non-dual networks do not contain any triangle plaquettes, like on the left-hand side above. The fact that the boundary bonds separating $a$ and $b$ as well as $a$ and $c$ are the `two sides of the same bond' is not a problem.

\section{Operator ansatz for the vertex liquid}
\label{sec:vertex_operator_ansatz}
In this appendix, we take a closer look at the equations of the vertex liquid in $1+1$ dimensions. While there does not appear to be much room for simplifying the equations on a purely combinatorial/diagrammatic level, we can do so by using properties specific to array tensors. We will refer to the method of simplification as \emph{operator ansatz}, since instead of specifying all vertex tensors separately, we specify operators which map between those tensors. More precisely, the operators map between tensors whose stars differ only by local moves. We will use a \emph{support truncation argument} to show that the operators do not depend on all of the two stars they map between, but only on the local move.

%The main mechanism we use for further simplification is the following: Let $A$ be one of the vertex tensors reshaped into a matrix by dividing the indices into two groups. Assume there is a set $C_0$, $C_1$, $\ldots$, $C_{n-1}$ of networks (each blocked into a matrix) such that in any network, $A$ occurs together with one of the $C$s, i.e., in the form $AC_i$ for some $i$.

Let us start by introducing an abbreviated notation. Imagine adding three indices $a,b,c$ between two consecutive indices of a vertex tensor, connecting $a$ and $c$ by a bond, and connecting $b$ with a new $1$-index vertex tensor, e.g.,
\begin{equation}
\begin{tikzpicture}
\atoms{vertexatom}{0/}
\draw (0)--++(45:0.7) (0)edge[mark={ar,s}]++(135:0.7) (0)edge[mark={ar,s}]++(-30:0.7) (0)--++(-150:0.7);
\end{tikzpicture}
\quad\rightarrow\quad
\begin{tikzpicture}
\atoms{vertexatom}{0/, a/p={0,-0.6}}
\draw (0)--++(45:0.7) (0)edge[mark={ar,s}]++(135:0.7) (0)edge[mark={ar,s}]++(-30:0.7) (0)--++(-150:0.7) (0)edge[mark={ar,e}](a);
\draw[mark={ar,s},looseness=1.4] (0)to[out=-60,in=0]++(0,-1)to[out=180,in=-120](0);
\end{tikzpicture}
\coloneqq
\begin{tikzpicture}
\atoms{vertexatom,loopinsertion=-90:}{0/}
\draw (0)--++(45:0.7) (0)edge[mark={ar,s}]++(135:0.7) (0)edge[mark={ar,s}]++(-30:0.7) (0)--++(-150:0.7);
\end{tikzpicture}\;.
\end{equation}
We will refer to such an operation as a \emph{loop insertion} between the corresponding indices, and denote it by a little flag between the index lines. The direction of this flag indicates the direction of the bond between $a$ and $c$.

Consider the following two vertex tensors which we will refer to as $A_0$ and $\widetilde A_0$,
\begin{equation}
\label{eq:21operator_a0_tensors}
A_0:\quad
\begin{tikzpicture}
\atoms{vertexatom}{0/}
\draw (0)edge[mark=vor]++(180:0.6) (0)edge[mark=vor,ind=a]++(0:0.6);
\end{tikzpicture}
\quad\leftarrow\quad
\widetilde A_0:\quad
\begin{tikzpicture}
\atoms{vertexatom}{0/}
\draw (0)edge[mark=vor]++(180:0.6) (0)edge[mark=vor,ind=b]++(-30:0.6) (0)edge[mark=vor,ind=c]++(30:0.6);
\end{tikzpicture}\;.
\end{equation}

Let $X$ be a way to extend $A_0$ by one further vertex tensor which is connected to the index $a$. For each such extension we can find an extension $\widetilde X$ of $\widetilde A_0$ by one further vertex tensor which is connected to the both indices $b$ and $c$ as follows. $\widetilde X$ can be obtained from $X$ by replacing the index connected to $a$ by two indices connected to $b$ and $c$, with a loop insertion in between. The crucial property of $X$ and $\widetilde X$ is that $A_0$ and $\widetilde{A}_0$ together with their respective extension evaluate to the same tensor due to the moves of the vertex liquid. For example, consider the following two extensions,
\begin{equation}
\begin{tikzpicture}
\atoms{vertexatom}{0/, 1/p={1.5,0}}
\draw (0)edge[mark=vor]++(180:0.5) (0)edge[mark=vor,mark={ar,e}](1) (1)edge[mark={ar,s}]++(0:0.6) (1)edge[]++(45:0.6) (1)--++(-45:0.6) (1)edge[mark={ar,s}]++(-90:0.6);
\draw[green,dashed] (1.6,0)ellipse(0.7cm and 0.7cm);
\node[green] at ($(1.6,0)+1.3*(40:0.7cm and 0.7cm)$){$X$};
\draw[green,dashed] (-0.1,0)ellipse(0.7cm and 0.7cm);
\node[green] at ($(-0.1,0)+1.4*(140:0.7cm and 0.7cm)$){$A_0$};
\end{tikzpicture}
=
\begin{tikzpicture}
\atoms{vertexatom}{0/, {1/p={1.5,0}, loopinsertion=180:r}}
\draw (0)edge[mark=vor]++(180:0.5) (0)edge[mark=vor,mark={ar,e},bend left=40](1) (0)edge[mark=vor,mark={ar,e},bend right=40](1) (1)edge[mark={ar,s}]++(0:0.6) (1)edge[]++(45:0.6) (1)--++(-45:0.6) (1)edge[mark={ar,s}]++(-90:0.6);
\draw[green,dashed] (1.6,0)ellipse(0.7cm and 0.7cm);
\node[green] at ($(1.6,0)+1.3*(40:0.7cm and 0.7cm)$){$\widetilde X$};
\draw[green,dashed] (-0.1,0)ellipse(0.7cm and 0.7cm);
\node[green] at ($(-0.1,0)+1.4*(140:0.7cm and 0.7cm)$){$\widetilde A_0$};
\end{tikzpicture}\;.
\end{equation}
The equation can be derived from the moves of the vertex liquid as described in Section~\ref{sec:vertex_liquid}. We can interpret the tensors $A_0$, $\widetilde A_0$, $X$ and $\widetilde X$ above as linear operators from the left to the right. Then the above equation reads 
\begin{equation}
\label{eq:extension_equation}
A_0 X = \widetilde A_0 \widetilde X\;.
\end{equation}
Let $\mathcal X$ denote the set of all possible extensions of $A_0$. We can combine all the equations of the form Eq.~\eqref{eq:extension_equation} into a single equation,
\begin{equation}
\label{eq:support_equation_stack}
\begin{gathered}
A_0\mathbf X=\widetilde A_0\widetilde{\mathbf X}\;,\\
\mathbf X: \operatorname{dom}(A_0) \rightarrow \bigoplus_{X\in \mathcal X} \operatorname{codom}(X)\;,\quad \mathbf X=\bigoplus_{X\in \mathcal X} X\;,\\
\widetilde{\mathbf X}: \operatorname{dom}(A_0) \rightarrow \bigoplus_{X\in \mathcal X} \operatorname{codom}(\widetilde X)\;,\quad \widetilde{\mathbf X}=\bigoplus_{X\in \mathcal X} \widetilde X\;.\\
\end{gathered}
\end{equation}
Every occurrence of $A_0$ in a network always comes together with some extension $X\in \mathcal X$. Thus, we can without loss of generality assume that the image of the operator $A_0$ is inside the union of the supports of all possible extensions $X\in \mathcal X$. Otherwise, restricting $A_0$ to this union of supports constitutes an invertible domain wall. We will call this line of reasoning the \emph{support truncation argument}. Thus, the image of $A_0$ is inside the support of $\mathbf X$. Therefore, we can cancel $\mathbf X$ in Eq.~\eqref{eq:support_equation_stack} with some $\mathbf X^-$,
\begin{equation}
\label{eq:operator_a0relation}
A_0 = A_0\mathbf X\mathbf X^- = \widetilde A_0\widetilde{\mathbf X} \mathbf X^-= \widetilde A_0 Z\;.
\end{equation}
The 3-index tensor $Z\coloneqq \widetilde{\mathbf X} \mathbf X^-$ will be called \emph{2-1 operator}, and is denoted by
\begin{equation}
\begin{tikzpicture}
\atoms{21operator}{1/rot=-90}
\draw (1-cr)edge[mark={ar,s}]++(-170:0.4) (1-cl)edge[mark={ar,s},mark={vor,r}]++(170:0.4) (1-ct)--++(0:0.4);
\end{tikzpicture}\;.
\end{equation}
Eq.~\eqref{eq:operator_a0relation} then becomes
\begin{equation}
\begin{tikzpicture}
\atoms{vertexatom}{0/}
\draw (0)edge[mark=vor]++(180:0.6) (0)edge[mark=vor]++(0:0.6);
\end{tikzpicture}
=
\begin{tikzpicture}
\atoms{vertexatom}{0/}
\atoms{21operator}{{1/p={0.8,0},rot=-90}}
\draw (1-cr)edge[out=-170,in=-30,mark={ar,s}](0) (1-cl)edge[out=170,in=30,mark={ar,s},mark={vor,r}](0) (1-ct)--++(0:0.4);
\draw (0)edge[mark=vor]++(180:0.6);
\end{tikzpicture}\;.
\end{equation}

Now, consider two stars which differ by inserting/removing an edge,
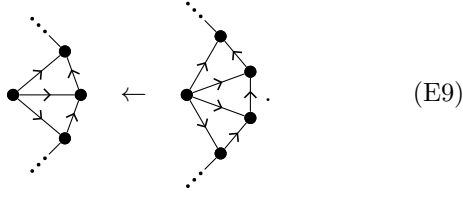
\begin{equation}
\label{eq:operator_link_21}
\begin{tikzpicture}
\atoms{vertex}{0/, 1/p=-40:0.9, 2/p=0:0.9, 3/p=40:0.9}
\draw (1)edge[mark={arr,-}](0) (2)edge[mark={arr,-}](0) (3)edge[mark={arr,-}](0) (1)edge[mark=arr](2) (2)edge[mark=arr](3) (1)edge[enddots]++(-135:0.3) (3)edge[enddots]++(135:0.3);
\end{tikzpicture}
\quad\leftarrow\quad
\begin{tikzpicture}
\atoms{vertex}{0/, 1/p=-60:0.9, 2/p=-20:0.9, 3/p=20:0.9, 4/p=60:0.9}
\draw (1)edge[mark={arr,-}](0) (2)edge[mark={arr,-}](0) (3)edge[mark={arr,-}](0) (4)edge[mark={arr,-}](0) (1)edge[mark=arr](2) (2)edge[mark=arr](3) (3)edge[mark=arr](4) (1)edge[enddots]++(-135:0.3) (4)edge[enddots]++(135:0.3);
\end{tikzpicture}\;.
\end{equation}
Associated to them is a pair of vertex tensors $A$ and $\widetilde A$,
\begin{equation}
\label{eq:21operator_general_tensors}
A:\quad
\begin{tikzpicture}
\atoms{vertexatom}{0/}
\draw (0)edge[mark=vor]++(-120:0.6) (0)edge[mark=vor]++(120:0.6) (0)edge[mark=vor,ind=a]++(0:0.6);
\node at (-0.5,0){$\ldots$};
\end{tikzpicture}
\quad\leftarrow\quad
\widetilde A:\quad
\begin{tikzpicture}
\atoms{vertexatom}{0/}
\draw (0)edge[mark=vor]++(-120:0.6) (0)edge[mark=vor]++(120:0.6) (0)edge[mark=vor,ind=b]++(-20:0.6) (0)edge[mark=vor,ind=c]++(20:0.6);
\node at (-0.5,0){$\ldots$};
\end{tikzpicture}\;,
\end{equation}
such that $\ldots$ stands for an arbitrary sequence of indices which is the same on both sides. Let $Y$ be an extension of $\widetilde A$ at all indices except $b$ and $c$. That is, $Y$ consists of the vertex tensors connected to those indices inside some network and the bonds between those vertex tensors. For every such extension, there is an extension $Y_0$ of $\widetilde A_0$ at all indices except for $b$ and $c$, such that the two extended networks are related by the vertex-liquid moves. E.g., if $\ldots$ 
%\je{[What is this?]} that's explained in the first sentence of this paragraph
in Eq.~\eqref{eq:21operator_general_tensors} consists of one further index, a possible extension $Y$ of $\widetilde{A}$ and the corresponding $Y_0$ of $\widetilde{A}_0$ are given by
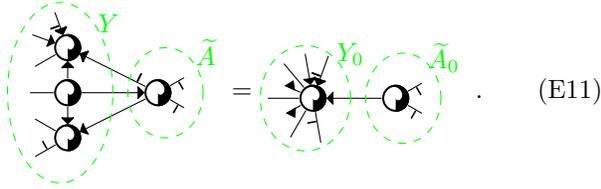
\begin{equation}
\label{eq:21operator_same_operator}
\begin{tikzpicture}
\atoms{vertexatom}{0/, 1/p={-1.2,0.6}, 2/p={-1.2,0}, 3/p={-1.2,-0.6}}
\draw (0)edge[mark=vor]++(30:0.4) (0)edge[mark=vor]++(-30:0.4) (0)edge[mark={ar,e},mark=vor](1) (0)edge[mark={ar,s}](2) (0)edge[mark={ar,e}](3) (1)edge[mark={ar,s}](2) (2)edge[mark={ar,e}](3) (1)edge[mark={ar,s},mark={vor,r}]++(110:0.5) (1)edge[mark={ar,s}]++(160:0.5) (1)--++(-150:0.5) (2)--++(180:0.5) (3)--++(150:0.5) (3)edge[mark={vor}]++(-150:0.5);
\draw[green,dashed] (-1.3,0)ellipse(0.7cm and 1.2cm);
\node[green] at ($(-1.3,0)+1.2*(40:0.7cm and 1.2cm)$){$Y$};
\draw[green,dashed] (0.1,0)ellipse(0.5cm and 0.6cm);
\node[green] at ($(0.1,0)+1.4*(40:0.5cm and 0.6cm)$){$\widetilde A$};
\end{tikzpicture}
=
\begin{tikzpicture}
\atoms{vertexatom}{0/, {2/p={-1.1,0},loopinsertion=155:r, loopinsertion=-150:}}
\draw (0)edge[mark=vor]++(30:0.4) (0)edge[mark=vor]++(-30:0.4) (0)edge[mark={ar,e}](2) (2)edge[mark={ar,s},mark={vor,r}]++(70:0.6) (2)edge[mark={ar,s}]++(100:0.6) (2)--++(130:0.6) (2)--++(180:0.6) (2)--++(-120:0.6) (2)edge[mark=vor]++(-80:0.6);
\draw[green,dashed] (-1.2,0)ellipse(0.6cm and 0.7cm);
\node[green] at ($(-1.2,0)+1.3*(40:0.6cm and 0.7cm)$){$Y_0$};
\draw[green,dashed] (0.1,0)ellipse(0.5cm and 0.6cm);
\node[green] at ($(0.1,0)+1.4*(40:0.5cm and 0.6cm)$){$\widetilde A_0$};
\end{tikzpicture}\;.
\end{equation}
Interpreting $Y$, $\widetilde A$, $Y_0$, and $\widetilde A_0$ as linear operators, this can be written as
\begin{equation}
Y\widetilde A = Y_0 \widetilde A_0\;.
\end{equation}
Now, $Y$ and $Y_0$ are at the same time extensions of $A$ and $A_0$ at all indices except for $a$, such that the analogous equation in Eq.~\eqref{eq:21operator_same_operator} holds with one instead of two open indices on the right. So we also get the equation
\begin{equation}
Y A = Y_0 A_0\;.
\end{equation}

Similar to Eq.~\eqref{eq:support_equation_stack}, we can stack all possible extensions $Y\in \mathcal Y$ into a single operator
\begin{equation}
\mathbf Y: \bigoplus_{Y\in \mathcal Y} \operatorname{codom}(Y) \rightarrow \operatorname{dom}(\widetilde A)\;,
\end{equation}
and the same for $Y_0$. We get two combined equations,
\begin{equation}
\mathbf Y A = \mathbf Y_0  A_0\;,
\end{equation}
and
\begin{equation}
\mathbf Y\widetilde A = \mathbf Y_0 \widetilde A_0\;.
\end{equation}
Applying the support truncation argument, the co-image of both $A$ and $\widetilde A$ has to be inside the co-support of $\mathbf Y$. Thus, there is an operator $\mathbf Y^-$ cancelling $\mathbf Y$, and we find
\begin{equation}
\begin{multlined}
\widetilde A Z = \mathbf Y^- \mathbf Y \widetilde A Z = \mathbf Y^- \mathbf Y_0 \widetilde A_0 Z\\
= \mathbf Y^- \mathbf Y_0 A_0 =\mathbf Y^- \mathbf Y A = A\;.
\end{multlined}
\end{equation}
So, we have in general,
\begin{equation}
\begin{tikzpicture}
\atoms{vertexatom}{0/}
\draw (0)edge[mark=vor]++(-120:0.6) (0)edge[mark=vor]++(120:0.6) (0)edge[mark=vor]++(0:0.6);
\node at (-0.5,0){$\ldots$};
\end{tikzpicture}
=
\begin{tikzpicture}
\atoms{vertexatom}{0/}
\atoms{21operator}{{1/p={0.8,0},rot=-90}}
\draw (1-cr)edge[out=-170,in=-30,mark={ar,s}](0) (1-cl)edge[out=170,in=30,mark={ar,s},mark={vor,r}](0) (1-ct)--++(0:0.4);
\draw (0)edge[mark=vor]++(-120:0.6) (0)edge[mark=vor]++(120:0.6);
\node at (-0.5,0){$\ldots$};
\end{tikzpicture}\;,
\end{equation}
representing the tensor $Z$ by a triangle shape. Note that $Z$ cannot be used to fuse any two indices of a vertex tensor to one, but can only be applied to index pairs such that the edge directions of the $4$ neighbouring triangles are as in Eq.~\eqref{eq:operator_link_21}. For other choices of edge directions we need a different operator $Z$.

To summarize, we considered pairs of vertex tensors whose stars differed by the local move in Eq.~\eqref{eq:operator_link_21}. We found that every such pair of tensors are related by applying the same operator to the indices around where the move happens. The analogous arguments can also be applied to all other local moves of stars.

For example, consider the opposite direction of the move in Eq.~\eqref{eq:operator_link_21}. To this end, we exchange what we call $A_0$ and $\widetilde{A}_0$ in Eq.~\eqref{eq:21operator_a0_tensors}, and the extensions $X$ consist of two tensors whereas $\widetilde{X}$ consists of one tensor with a loop insertion on the opposite side of where it is connected to $A_0$. E.g., a pair of extensions is given by
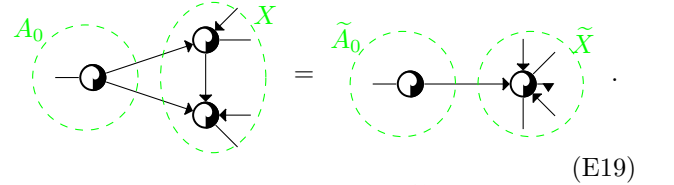
\begin{equation}
\begin{tikzpicture}
\atoms{vertexatom}{0/, 1/p={1.5,-0.5}, 2/p={1.5,0.5}}
\draw (0)edge[]++(180:0.5) (0)edge[mark={ar,e}](1) (0)edge[mark={ar,e}](2) (1)edge[mark={ar,s}](2) (1)edge[mark={ar,s}]++(0:0.6) (1)edge[]++(-45:0.6) (2)--++(0:0.6) (2)edge[mark={ar,s}]++(45:0.6);
\draw[green,dashed] (1.6,0)ellipse(0.7cm and 1cm);
\node[green] at ($(1.6,0)+1.3*(40:0.7cm and 1cm)$){$X$};
\draw[green,dashed] (-0.1,0)ellipse(0.7cm and 0.7cm);
\node[green] at ($(-0.1,0)+1.4*(140:0.7cm and 0.7cm)$){$A_0$};
\end{tikzpicture}
=
\begin{tikzpicture}
\atoms{vertexatom}{0/, {1/p={1.5,0}, loopinsertion=0:r}}
\draw (0)edge[]++(180:0.5) (0)edge[mark={ar,e}](1) (1)edge[mark={ar,s}]++(-45:0.6) (1)edge[]++(-90:0.6) (1)--++(45:0.6) (1)edge[mark={ar,s}]++(90:0.6);
\draw[green,dashed] (1.6,0)ellipse(0.7cm and 0.7cm);
\node[green] at ($(1.6,0)+1.3*(40:0.7cm and 0.7cm)$){$\widetilde X$};
\draw[green,dashed] (-0.1,0)ellipse(0.7cm and 0.7cm);
\node[green] at ($(-0.1,0)+1.4*(140:0.7cm and 0.7cm)$){$\widetilde A_0$};
\end{tikzpicture}\;.
\end{equation}
Again we can combine all the $X$ and $\widetilde{X}$ into a single operator each, and apply the support truncation argument to show that $A_0$ is obtained from $\widetilde{A}_0$ via some operator $Z$. We can then look at arbitrary $A$ and $\widetilde{A}$ like in Eq.~\eqref{eq:21operator_general_tensors} with $A$ and $\widetilde{A}$ exchanged. The equations Eq.~\eqref{eq:21operator_same_operator} and its version without tilde are the same. From them, it follows that $Z$ is the same for all pairs of $A$ and $\widetilde{A}$, and we get an equation
\begin{equation}
\begin{tikzpicture}
\atoms{vertexatom}{0/}
\draw (0)edge[mark=vor]++(45:0.6) (0)edge[mark=vor]++(-45:0.6);
\draw (0)edge[mark=vor]++(-120:0.6) (0)edge[mark=vor]++(120:0.6);
\node at (-0.5,0){$\ldots$};
\end{tikzpicture}
=
\begin{tikzpicture}
\atoms{vertexatom}{0/}
\atoms{12operator}{{1/p={0.8,0},rot=90}}
\draw (1-cl)--++(-30:0.5) (1-cr)edge[mark=vor]++(30:0.5) (1-ct)edge[mark={ar,s}](0);
\draw (0)edge[mark=vor]++(-120:0.6) (0)edge[mark=vor]++(120:0.6);
\node at (-0.5,0){$\ldots$};
\end{tikzpicture}\;,
\end{equation}
where the black triangle will be called the \emph{1-2 operator}.

As another example, consider two stars differing by reversing the direction of an outer edge,
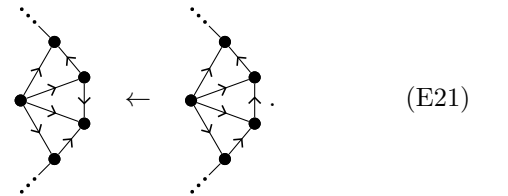
\begin{equation}
\label{eq:operator_link_edgereversal}
\begin{tikzpicture}
\atoms{vertex}{0/, 1/p=-60:0.9, 2/p=-20:0.9, 3/p=20:0.9, 4/p=60:0.9}
\draw (1)edge[mark={arr,-}](0) (2)edge[mark={arr,-}](0) (3)edge[mark={arr,-}](0) (4)edge[mark={arr,-}](0) (1)edge[mark=arr](2) (2)edge[mark={arr,-}](3) (3)edge[mark=arr](4) (1)edge[enddots]++(-135:0.3) (4)edge[enddots]++(135:0.3);
\end{tikzpicture}
\quad\leftarrow\quad
\begin{tikzpicture}
\atoms{vertex}{0/, 1/p=-60:0.9, 2/p=-20:0.9, 3/p=20:0.9, 4/p=60:0.9}
\draw (1)edge[mark={arr,-}](0) (2)edge[mark={arr,-}](0) (3)edge[mark={arr,-}](0) (4)edge[mark={arr,-}](0) (1)edge[mark=arr](2) (2)edge[mark=arr](3) (3)edge[mark=arr](4) (1)edge[enddots]++(-135:0.3) (4)edge[enddots]++(135:0.3);
\end{tikzpicture}\;.
\end{equation}
In this case we need to not only include the two neighbouring tensors in $X$ and $\widetilde{X}$, but also the other tensor connected to both,
\begin{equation}
\begin{tikzpicture}
\atoms{vertexatom}{0/, 1/p={1.2,-0.5}, 2/p={1.2,0.5}, 3/p={1.9,0}}
\draw (0)edge[]++(180:0.5) (0)edge[mark={ar,e}](1) (0)edge[mark={ar,e}](2) (1)edge[mark={ar,e}](3) (2)edge[mark={ar,e}](3) (1)edge[mark={ar,s}](2) (1)edge[mark={ar,s}]++(-135:0.6) (1)edge[]++(-45:0.6) (2)--++(135:0.6) (2)edge[mark={ar,s}]++(45:0.6) (3)edge[mark={ar,s},mark=vor]++(45:0.6) (3)edge[mark={ar,s}]++(-45:0.6);
\draw[green,dashed] (1.6,0)ellipse(0.9cm and 1cm);
\node[green] at ($(1.6,0)+1.3*(40:0.9cm and 1cm)$){$X$};
\draw[green,dashed] (-0.1,0)ellipse(0.7cm and 0.7cm);
\node[green] at ($(-0.1,0)+1.4*(140:0.7cm and 0.7cm)$){$A_0$};
\end{tikzpicture}
=
\begin{tikzpicture}
\atoms{vertexatom}{0/, 1/p={1.2,-0.5}, 2/p={1.2,0.5}, 3/p={1.9,0}}
\draw (0)edge[]++(180:0.5) (0)edge[mark={ar,e}](1) (0)edge[mark={ar,e}](2) (1)edge[mark={ar,e}](3) (2)edge[mark={ar,e}](3) (1)edge[mark={ar,e}](2) (1)edge[mark={ar,s}]++(-135:0.6) (1)edge[]++(-45:0.6) (2)--++(135:0.6) (2)edge[mark={ar,s}]++(45:0.6) (3)edge[mark={ar,s},mark=vor]++(45:0.6) (3)edge[mark={ar,s}]++(-45:0.6);
\draw[green,dashed] (1.6,0)ellipse(0.9cm and 1cm);
\node[green] at ($(1.6,0)+1.3*(40:0.9cm and 1cm)$){$\widetilde{X}$};
\draw[green,dashed] (-0.1,0)ellipse(0.7cm and 0.7cm);
\node[green] at ($(-0.1,0)+1.4*(140:0.7cm and 0.7cm)$){$\widetilde{A}_0$};
\end{tikzpicture}\;.
\end{equation}
In the end, we get an equation,
\begin{equation}
\begin{tikzpicture}
\atoms{vertexatom}{0/}
\draw (0)edge[]++(45:0.6) (0)edge[mark={vor,r},mark=vor]++(-45:0.6);
\draw (0)edge[mark=vor]++(-120:0.6) (0)edge[mark=vor]++(120:0.6);
\node at (-0.5,0){$\ldots$};
\end{tikzpicture}
=
\begin{tikzpicture}
\atoms{vertexatom}{0/}
\atoms{barflip}{{1/p={0.8,0},rot=90}}
\draw (1-bl)edge[mark={vor,r}]++(-30:0.5) (1-br)--++(30:0.5) (0)edge[bend left=30,mark={ar,e},mark=vor](1-tr) (0)edge[bend right=30,mark={ar,e},mark=vor](1-tl);
\draw (0)edge[mark=vor]++(-120:0.6) (0)edge[mark=vor]++(120:0.6);
\node at (-0.5,0){$\ldots$};
\end{tikzpicture}\;.
\end{equation}
We will call the tensor $Z$ represented by a square the \emph{outer edge flip operator}.

As a last example, consider changing the direction of an internal edge of the star,
\begin{equation}
\begin{tikzpicture}
\atoms{vertex}{0/, 1/p=-40:0.9, 2/p=0:0.9, 3/p=40:0.9}
\draw (1)edge[mark={arr,-}](0) (2)edge[mark={arr}](0) (3)edge[mark={arr,-}](0) (1)edge[mark={arr,-}](2) (2)edge[mark=arr](3) (1)edge[enddots]++(-135:0.3) (3)edge[enddots]++(135:0.3);
\end{tikzpicture}
\quad\leftarrow\quad
\begin{tikzpicture}
\atoms{vertex}{0/, 1/p=-40:0.9, 2/p=0:0.9, 3/p=40:0.9}
\draw (1)edge[mark={arr,-}](0) (2)edge[mark={arr,-}](0) (3)edge[mark={arr,-}](0) (1)edge[mark={arr,-}](2) (2)edge[mark=arr](3) (1)edge[enddots]++(-135:0.3) (3)edge[enddots]++(135:0.3);
\end{tikzpicture}\;.
\end{equation}
Examples for extensions are
\begin{equation}
\begin{tikzpicture}
\atoms{vertexatom}{0/, 1/p={1.5,-0.6}, 2/p={1.5,0}, 3/p={1.5,0.6}}
\draw (0)edge[]++(180:0.5) (0)edge[mark={ar,e}](1) (0)edge[mark={ar,e}](2) (0)edge[mark={ar,e}](3) (1)edge[mark={ar,s}](2) (2)edge[mark={ar,e}](3) (1)edge[mark={ar,s}]++(0:0.6) (1)edge[]++(-45:0.6) (3)edge[mark={ar,s}]++(0:0.6) (3)edge[]++(45:0.6)  (2)edge[mark={ar,s}]++(0:0.6);
\draw[green,dashed] (1.6,0)ellipse(0.7cm and 1.1cm);
\node[green] at ($(1.6,0)+1.3*(40:0.7cm and 1.1cm)$){$X$};
\draw[green,dashed] (-0.1,0)ellipse(0.7cm and 0.7cm);
\node[green] at ($(-0.1,0)+1.4*(140:0.7cm and 0.7cm)$){$A_0$};
\end{tikzpicture}
=
\begin{tikzpicture}
\atoms{vertexatom}{0/, 1/p={1.5,-0.6}, 2/p={1.5,0}, 3/p={1.5,0.6}}
\draw (0)edge[]++(180:0.5) (0)edge[mark={ar,e}](1) (0)edge[mark={ar,s}](2) (0)edge[mark={ar,e}](3) (1)edge[mark={ar,s}](2) (2)edge[mark={ar,e}](3) (1)edge[mark={ar,s}]++(0:0.6) (1)edge[]++(-45:0.6) (3)edge[mark={ar,s}]++(0:0.6) (3)edge[]++(45:0.6)  (2)edge[mark={ar,s}]++(0:0.6);
\draw[green,dashed] (1.6,0)ellipse(0.7cm and 1.1cm);
\node[green] at ($(1.6,0)+1.3*(40:0.7cm and 1.1cm)$){$\widetilde{X}$};
\draw[green,dashed] (-0.1,0)ellipse(0.7cm and 0.7cm);
\node[green] at ($(-0.1,0)+1.4*(140:0.7cm and 0.7cm)$){$\widetilde{A}_0$};
\end{tikzpicture}\;.
\end{equation}
We get another equation
\begin{equation}
\begin{tikzpicture}
\atoms{vertexatom}{0/}
\draw (0)edge[]++(45:0.6) (0)edge[mark=vor]++(-45:0.6) (0)edge[mark={ar,s}]++(0:0.6);
\draw (0)edge[mark=vor]++(-120:0.6) (0)edge[mark=vor]++(120:0.6);
\node at (-0.5,0){$\ldots$};
\end{tikzpicture}
=
\begin{tikzpicture}
\atoms{vertexatom}{0/}
\atoms{orientflip}{{1/p={0.8,0},rot=90}}
\draw (1-bl)edge[]++(-30:0.5) (1-br)edge[]++(30:0.5) (1-b)edge[mark={ar,s}]++(0:0.5) (0)edge[bend left=30,mark={ar,e},mark=vor](1-tr) (0)edge[bend right=30,mark={ar,e},mark=vor](1-tl) (0)edge[mark={ar,e},mark=vor](1-t);
\draw (0)edge[mark=vor]++(-120:0.6) (0)edge[mark=vor]++(120:0.6);
\node at (-0.5,0){$\ldots$};
\end{tikzpicture}\;.
\end{equation}
We will call the tensor $Z$ represented by a square the \emph{inner edge flip operator}.

Given the 1-2 operator together with some variants of edge flip operators, we can generate all vertex tensors from one very simple vertex tensor, say, a 1-index tensor. Even though the set of all those operators is more complicated than the set of tensors specifying common algebraic structures, such as Hopf algebras, it is still considerably easier than the set of all vertex tensors for a sufficient set of stars.

The operator ansatz does not only simplify the set of tensors, but also the set of axioms. Applying different sequences of operators can have the same effect on the stars, yielding coherence axioms for the operators. For example, we have
\begin{equation}
\begin{tikzpicture}
\atoms{vertexatom}{v/p={-0.8,0}}
\atoms{rot=90,12operator}{0/, 1/p={0.6,-0.6}}
\draw (v)edge[mark={ar,e}](0-ct) (0-cl)edge[mark={ar,e}](1-ct) (0-cr)edge[mark=vor]++(30:0.4) (1-cr)edge[mark=vor]++(30:0.4) (1-cl)--++(-30:0.4);
\draw (v)edge[mark=vor]++(-120:0.6) (v)edge[mark=vor]++(120:0.6);
\node at (-1.3,0){$\ldots$};
\end{tikzpicture}
=
\begin{tikzpicture}
\atoms{vertexatom}{0/}
\draw (0)edge[mark=vor]++(45:0.6) (0)edge[mark=vor]++(-45:0.6) (0)edge[mark=vor]++(0:0.6);
\draw (0)edge[mark=vor]++(-120:0.6) (0)edge[mark=vor]++(120:0.6);
\node at (-0.5,0){$\ldots$};
\end{tikzpicture}
=
\begin{tikzpicture}
\atoms{vertexatom}{v/p={-0.8,0}}
\atoms{rot=90,12operator}{0/, 1/p={0.6,0.6}}
\draw (v)edge[mark={ar,e}](0-ct) (0-cr)edge[mark={ar,e},mark=vor](1-ct) (0-cl)edge[]++(-30:0.4) (1-cr)edge[mark=vor]++(30:0.4) (1-cl)--++(-30:0.4);
\draw (v)edge[mark=vor]++(-120:0.6) (v)edge[mark=vor]++(120:0.6);
\node at (-1.3,0){$\ldots$};
\end{tikzpicture}
\;.
\end{equation}
This holds for any sequence of indices $\ldots$, that is, for any way to extend the index between the vertex tensor and the 1-2 operator by some network to the left. Thus, via the support truncation argument, we can deduce the coherence axiom
\begin{equation}
\label{eq:operator_associativity}
\begin{tikzpicture}
\atoms{rot=90,12operator}{0/, 1/p={0.6,-0.6}}
\draw (0-ct)edge[mark={ar,s}]++(180:0.4) (0-cl)edge[mark={ar,e}](1-ct) (0-cr)edge[mark=vor]++(30:0.4) (1-cr)edge[mark=vor]++(30:0.4) (1-cl)--++(-30:0.4);
\end{tikzpicture}
=
\begin{tikzpicture}
\atoms{rot=90,12operator}{0/, 1/p={0.6,0.6}}
\draw (0-ct)edge[mark={ar,s}]++(180:0.4) (0-cr)edge[mark={ar,e},mark=vor](1-ct) (0-cl)edge[]++(-30:0.4) (1-cr)edge[mark=vor]++(30:0.4) (1-cl)--++(-30:0.4);
\end{tikzpicture}
\;.
\end{equation}
Another example for such a coherence axiom is
\begin{equation}
\begin{tikzpicture}
\atoms{rot=90,12operator}{0/}
\atoms{rot=-90,21operator}{1/p={0.7,0}}
\draw (0-ct)edge[mark={ar,s}]++(180:0.4) (0-cl)edge[bend right=30,mark={ar,e}](1-cr) (0-cr)edge[bend left=30,mark={ar,e}](1-cl) (1-ct)--++(0:0.4);
\end{tikzpicture}
=
\begin{tikzpicture}
\draw (0,0)--++(1,0);
\end{tikzpicture}
\;.
\end{equation}
Furthermore there is a bunch of coherence axioms involving the edge flip operators. We cannot write down any of them here since we so far only defined two specific edge flip operators which are not compatible. In addition to the coherence axioms, we still need to impose the Pachner moves. Instead of having one 2-2 Pachner move for every set of $4$ stars, we can reduce to a much smaller set. If we ignore all the edge directions for a moment, then the Pachner moves can be implemented by only one single equation,
\begin{equation}
\begin{tikzpicture}
\atoms{12operator}{{0/p={0,-0.9},rot=180}, {1/p={0,0.9}}, {2/p={-1.1,0},rot=90}, {3/p={-0.6,-0.2},rot=90}, {4/p={1.1,0},rot=-90}, {5/p={0.6,-0.2},rot=-90}}
\draw (0-cr)--(3-cl) (0-cl)--(5-cr) (1-cl)--(2-cr) (1-cr)--(4-cl) (2-cl)--(3-ct) (4-cr)--(5-ct) (3-cr)--(5-cl) (0-ct)--++(-90:0.3) (1-ct)--++(90:0.3) (2-ct)--++(180:0.3) (4-ct)--++(0:0.3);
\end{tikzpicture}
=
\begin{tikzpicture}
\atoms{12operator}{{0/p={0,-1.1},rot=180}, {1/p={0,1.1}}, {2/p={-0.9,0},rot=90}, {3/p={0.2,0.6}}, {4/p={0.9,0},rot=-90}, {5/p={0.2,-0.6},rot=180}}
\draw (0-cr)--(2-cl) (0-cl)--(5-ct) (1-cl)--(2-cr) (1-cr)--(3-ct) (4-cl)--(3-cr) (4-cr)--(5-cl) (5-cr)--(3-cl) (0-ct)--++(-90:0.3) (1-ct)--++(90:0.3) (2-ct)--++(180:0.3) (4-ct)--++(0:0.3);
\end{tikzpicture}\;.
\end{equation}

The axiom in Eq.~\eqref{eq:operator_associativity} is familiar. If we interpret the 1-2 operator as a bi-linear operator from the right to the left, it defines an associative algebra. Furthermore, the cyclic symmetry of the $3$-index tensor yields
\begin{equation}
\begin{tikzpicture}
\atoms{vertexatom}{0/}
\atoms{indexgrow,vflip}{1/p={0,0.6}}
\draw (0)edge[mark={ar,s},mark=vor](1-ct) (0)edge[mark={ar,s},mark=vor,ind=c]++(-90:0.5) (1-cr)edge[ind=b]++(60:0.4) (1-cl)edge[ind=a]++(120:0.4);
\end{tikzpicture}
=
\begin{tikzpicture}
\atoms{vertexatom}{0/}
\draw (0)edge[mark={ar,s},mark=vor,ind=a]++(150:0.5) (0)edge[mark={ar,s},mark=vor,ind=c]++(-90:0.5) (0)edge[mark={ar,s},mark=vor,ind=b]++(30:0.5);
\end{tikzpicture}
=
\begin{tikzpicture}
\atoms{vertexatom}{0/}
\atoms{indexgrow,vflip}{1/p={0,0.6}}
\draw (0)edge[mark={ar,s},mark=vor](1-ct) (0)edge[mark={ar,s},mark=vor,ind=b]++(-90:0.5) (1-cr)edge[ind=a]++(60:0.4) (1-cl)edge[ind=c]++(120:0.4);
\end{tikzpicture}\;.
\end{equation}
If we interpret the $2$-index vertex tensor as a bilinear form, the equation above is just the equation fulfilled by a Frobenius form (apart from being non-degenerate). We further have
\begin{equation}
\begin{tikzpicture}
\atoms{vertexatom}{0/}
\atoms{indexgrow,vflip}{1/p={0,0.6}}
\draw (0)edge[mark={ar,s},mark=vor](1-ct) (1-cr)edge[ind=b]++(60:0.4) (1-cl)edge[ind=a]++(120:0.4);
\end{tikzpicture}
=
\begin{tikzpicture}
\atoms{vertexatom}{0/}
\draw (0)edge[mark={ar,s},mark=vor,ind=a]++(180:0.5) (0)edge[mark={ar,s},mark=vor,ind=b]++(0:0.5);
\end{tikzpicture}
\;,
\end{equation}
so the $1$-index vertex tensor can be interpreted as a linear form generating the bilinear form through the product. If the bilinear form is non-degenerate, the corresponding matrix has an inverse,
\begin{equation}
\begin{tikzpicture}
\atoms{vertexatom}{0/}
\atoms{invtwoform}{1/p={0.7,0}}
\draw (0)edge[mark={ar,s},mark=vor](1) (0)edge[mark={ar,s},mark=vor]++(180:0.5) (1-r)--++(0:0.3);
\end{tikzpicture}
=
\begin{tikzpicture}
\atoms{vertexatom}{0/}
\atoms{invtwoform}{1/p={-0.7,0}}
\draw (0)edge[mark={ar,s},mark=vor](1) (0)edge[mark={ar,s},mark=vor]++(0:0.5) (1-l)--++(180:0.3);
\end{tikzpicture}
=
\begin{tikzpicture}
\draw (0,0)--(1,0);
\end{tikzpicture}\;,
\end{equation}
which allows us to also construct a unit,
\begin{equation}
\begin{tikzpicture}
\atoms{frobunit}{0/}
\draw (0)--++(0.5,0);
\end{tikzpicture}
\coloneqq
\begin{tikzpicture}
\atoms{vertexatom}{0/}
\atoms{invtwoform}{1/p={0.7,0}}
\draw (0)edge[mark={ar,s},mark=vor](1) (1-r)--++(0:0.3);
\end{tikzpicture}
\;.
\end{equation}
We note that the Frobenius form (the $2$-index vertex tensor) as well as its inverse are symmetric. We indeed find that the above defined vector is a left unit,
\begin{equation}
\begin{tikzpicture}
\atoms{indexgrow,vflip}{0/}
\atoms{frobunit}{1/p={-0.5,0.5}}
\draw (0-cl)--(1) (0-cr)--++(60:0.4) (0-ct)--++(-90:0.3);
\end{tikzpicture}
=
\begin{tikzpicture}
\atoms{indexgrow}{{0/vflip}}
\atoms{invtwoform}{i0/p={$(0-ct)+(-90:1)$}, {i1/p={$(0-cl)+(135:0.4)$},rot=135}}
\atoms{vertexatom}{v0/p={$(0-ct)+(-90:0.5)$}, {v1/p={$(0-cl)+(135:1)$}}}
\draw (0-cl)--(i1) (v1)edge[mark={ar,s},mark=vor](i1) (v0)edge[mark={ar,s},mark=vor](0-ct) (v0)edge[mark={ar,s},mark=vor](i0) (0-cr)--++(45:0.4) (i0)--++(-90:0.4);
\end{tikzpicture}
=
\begin{tikzpicture}
\atoms{indexgrow}{{0/rot=-120,vflip}}
\atoms{invtwoform}{i0/p={$(0-cr)+(-90:0.4)$}, {i1/p={$(0-ct)+(135:1)$},rot=135}}
\atoms{vertexatom}{v0/p={$(0-ct)+(135:0.5)$}, {v1/p={$(0-ct)+(135:1.5)$}}}
\draw (v0)edge[mark={ar,s},mark=vor](0-ct) edge[mark={ar,s},mark=vor](i1) (v1)edge[mark={ar,s},mark=vor](i1) (0-cr)--(i0) (0-cl)--++(45:0.4) (i0)--++(-90:0.4);
\end{tikzpicture}
=
\begin{tikzpicture}
\atoms{indexgrow}{{0/rot=-120,vflip}}
\atoms{invtwoform}{i0/p={$(0-cr)+(-90:0.4)$}}
\atoms{vertexatom}{{v1/p={$(0-ct)+(135:0.5)$}}}
\draw (v1)edge[mark={ar,s},mark=vor](0-ct) (0-cr)--(i0) (0-cl)--++(45:0.4) (i0)--++(-90:0.4);
\end{tikzpicture}
=
\begin{tikzpicture}
\atoms{vertexatom}{0/}
\atoms{invtwoform}{1/p={0,-0.7}}
\draw (0)edge[mark={ar,s},mark=vor](1) (0)edge[mark={ar,s},mark=vor]++(90:0.5) (1-b)--++(-90:0.3);
\end{tikzpicture}
=
\begin{tikzpicture}
\draw (0,0)--++(0,-0.7);
\end{tikzpicture}
\;,
\end{equation}
and analogously a right unit. Thus, the 1-2 operator forms a symmetric Frobenius algebra.

\section{Universality mappings in other dimensions}
\subsection{Edge liquid and universality mapping in \texorpdfstring{$0+1$}{0+1} dimensions}
\label{sec:universal_1d}
In this appendix we discuss the rather trivial case of liquids in one spacetime dimension. We refer to the $0+1$-dimensional analogue of the triangle liquid as the $0+1$-dimensional \emph{edge liquid}. It associates one $2$-index tensor to every edge of a triangulation of a 1-manifold, that is, a decomposition of a collection of circles into edges. The 1-manifolds are oriented, and the edges carry directions, which we assume (without loss of generality) to be aligned with the orientation,
\begin{equation}
\begin{tikzpicture}
\atoms{vertex}{0/, 1/p={1,0}}
\draw (0)edge[or](1);
\end{tikzpicture}
\quad\rightarrow\quad
\begin{tikzpicture}
\atoms{1dedge}{0/}
\draw (0-l)--++(180:0.4) (0-r)--++(0:0.4);
\end{tikzpicture}\;.
\end{equation}
For Hermitian models, the tensor associated to an edge is a Hermitian matrix.

There is only one Pachner move, which replaces two edges with a single edge,
\begin{equation}
\begin{tikzpicture}
\atoms{vertex}{0/, 1/p={1,0}, 2/p={2,0}};
\draw (0)edge[or](1) (1)edge[or](2);
\end{tikzpicture}\quad\leftrightarrow\quad
\begin{tikzpicture}
\atoms{vertex}{0/, 1/p={1,0}};
\draw (0)edge[or](1);
\end{tikzpicture}
\;
\end{equation}
and is implemented by 
\begin{equation}
\label{eq:1d_edge_move}
\begin{tikzpicture}
\atoms{1dedge}{0/, 1/p={0.8,0}};
\draw (0-r)--(1-l) (0-l)edge[ind=a]++(180:0.4) (1-r)edge[ind=b]++(0:0.4);
\end{tikzpicture}=
\begin{tikzpicture}
\atoms{1dedge}{0/}
\draw (0-l)edge[ind=a]++(180:0.4) (0-r)edge[ind=b]++(0:0.4);
\end{tikzpicture}\;.
\end{equation}
Thus, on the level of models, the 2-index tensor is a (Hermitian) projector.

In $0+1$ dimensions, the analogues of the triangle and vertex liquid are actually equivalent, and both are universal. In the following we provide a universality mapping which shows that any liquid $\calb$ describing one-dimensional topological manifolds is equivalent to the edge liquid. We start by filling some interval with $\calb$-network such that the combinatorial distance of the two ends is $\lambda$. Then, we spilt the $\calb$-network into two halves,
\begin{equation}
\label{eq:1d_edge_cutting}
\begin{multlined}
\begin{tikzpicture}
\atoms{vertex}{0/, 1/p={1,0}}
\draw (0)edge[or](1);
\end{tikzpicture}
\quad\rightarrow\quad
\begin{tikzpicture}
\draw[1dmanifold] (0,0)--(2,0);
\draw (0,-0.2)--(0,0.2) (2,-0.2)--(2,0.2);
\end{tikzpicture}
\\
\rightarrow\quad
\begin{tikzpicture}
\draw[1dmanifold] (0,0)--(1,0) (1.5,0)--(2.5,0);
\draw (0,-0.2)--(0,0.2) (2.5,-0.2)--(2.5,0.2);
\draw[dotted] (1,-0.2)--(1,0.2) (1.5,-0.2)--(1.5,0.2);
\end{tikzpicture}\;.
\end{multlined}
\end{equation}
The $\calb$-network associated to the edge tensor is constructed by taking the left half  of the split network and gluing it to the right half by filling the gap between them,
\begin{equation}
\begin{tikzpicture}
\draw[1dmanifold] (0,0)--(1,0) (1.3,0)--(2.3,0);
\draw (1,-0.2)--(1,0.2) (1.3,-0.2)--(1.3,0.2);
\draw[dotted] (0,-0.2)--(0,0.2) (2.3,-0.2)--(2.3,0.2);
\end{tikzpicture}
\quad\rightarrow\quad
\begin{tikzpicture}
\draw[1dmanifold] (0,0)--(1,0) (1.3,0)--(2.3,0) (1,0)--(1.3,0);
\draw (1,-0.2)--(1,0.2) (1.3,-0.2)--(1.3,0.2);
\draw[dotted] (0,-0.2)--(0,0.2) (2.3,-0.2)--(2.3,0.2);
\end{tikzpicture}\;,
\end{equation}
\begin{equation}
\begin{tikzpicture}
\atoms{1dedge}{0/}
\draw (0-l)edge[ind=a]++(180:0.4) (0-r)edge[ind=b]++(0:0.4);
\end{tikzpicture}
\coloneqq
\begin{tikzpicture}
\draw[1dmanifold] (0,0)edge[startind=a](1,0) (1.3,0)edge[ind=b](2.3,0) (1,0)--(1.3,0);
\draw (1,-0.2)--(1,0.2) (1.3,-0.2)--(1.3,0.2);
\draw[dotted] (0,-0.2)--(0,0.2) (2.3,-0.2)--(2.3,0.2);
\end{tikzpicture}\;.
\end{equation}

If we apply the mapping to an edge-liquid network, we simply re-glue patches that were previously cut according to Eq.~\eqref{eq:1d_edge_cutting}. Thus, the mapping yields valid $\calb$-networks. Moreover, applying the mapping to the move in Eq.~\eqref{eq:1d_edge_move} yields
\begin{equation}
\begin{tikzpicture}
\draw[1dmanifold] (0,0)--(4,0);
\draw (0.9,-0.2)--(0.9,0.2) (1.1,-0.2)--(1.1,0.2) (2.9,-0.2)--(2.9,0.2) (3.1,-0.2)--(3.1,0.2);
\draw[dotted] (2,-0.2)--(2,0.2) (0,-0.2)--(0,0.2) (4,-0.2)--(4,0.2);
\draw[fill, fill opacity=0.3, red,rc] (0.8,-0.2)rectangle (3.2,0.2);
\end{tikzpicture}
=
\begin{tikzpicture}
\draw[1dmanifold] (0,0)--(2,0);
\draw (0.9,-0.2)--(0.9,0.2) (1.1,-0.2)--(1.1,0.2);
\draw[dotted] (2,-0.2)--(2,0.2) (0,-0.2)--(0,0.2);
\draw[fill, fill opacity=0.3, red,rc] (0.8,-0.2)rectangle (1.2,0.2);
\end{tikzpicture}
\;.
\end{equation}
We observe that the two $\calb$-networks only differ in the red shaded regions, which are separated from the open boundary by a combinatorial distance scaling linearly in the fine-graining scale $\lambda$. Thus, we if choose the fine-graining scale $\lambda$ large enough, the two sides can be transformed into each other using the $\calb$-moves.

\subsection{Universality mapping with boundary in \texorpdfstring{$2+1$}{2+1} dimensions}
\label{sec:universality_boundary_3d}
In this appendix, we describe in detail the universality mapping from the simplex liquid in $2+1$ dimensions to an arbitrary topological liquid with boundary $\calb$, outlined in Section~\ref{sec:universality_higherdim}. The generalization to arbitrary dimensions is relatively straight-forward. Similar to the construction in $1+1$ dimensions given in Section~\ref{sec:finegrain_boundary}, we need to generalize the boundary liquid $\calb$ slightly and require the existence of defects and topology-changing moves similar to the move in Eq.~\ref{eq:2d_puncture_healing_move}. Concomitantly, we also need to generalize the simplex liquid and consider a variant which includes \emph{vertex weights} as well as \emph{edge weights}. I.e., around each edge of the triangulation we need to include one edge weight associated to one of the contracted index pairs 'encircling' the edge, and around each vertex we need to include one vertex weight associated to one of the surrounding index pairs (cf. Ref.~\cite{liquid_intro}).

For the boundary liquid $\calb$ we require the following additional properties. We need to introduce a $0+0$-dimensional defect within the $1+1$-dimensional boundary, whose link (cf.~Appendix~\ref{sec:defects}) is a disk,
\begin{equation}
\label{eq:2d_boundary_point_defect}
\begin{tikzpicture}
\fill[manifoldbdfull] (0,0)--(2,0.5)--(2,2)--(0,1.5)--cycle;
\atoms{defectvertex}{x/p={1,1}}
\draw[\manifoldbdcol,manifold] (1,1)ellipse (0.4cm and 0.6cm);
\end{tikzpicture}
\quad\rightarrow\quad
\begin{tikzpicture}
\draw[manifoldboundary, manifold] (0,0)circle(0.4);
\end{tikzpicture}\;,
\end{equation}
a $0+1$-dimensional defect within the boundary, whose link is an interval,
\begin{equation}
\label{eq:boundary_defect_line}
\begin{tikzpicture}
\fill[manifoldbdfull] (0,0)--(2,0.5)--(2,2)--(0,1.5)--cycle;
\draw[defectline] (0,0.75)--(2,1.25);
\draw[midedge] (1,1.5)arc(90:270:0.5);
\end{tikzpicture}
\quad\rightarrow\quad
\begin{tikzpicture}
\atoms{boundaryvertex}{0/, 1/p={0,0.7}}
\draw[1dmanifold] (0)--(1);
\end{tikzpicture}\;,
\end{equation}
and a $0+0$-dimensional defect within the aforementioned $0+1$-dimensional defect, whose link is a disk with two defect points,
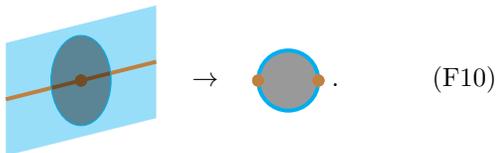
\begin{equation}
\label{eq:bounday_point_line_defect}
\begin{tikzpicture}
\fill[manifoldbdfull] (0,0)--(2,0.5)--(2,2)--(0,1.5)--cycle;
\draw[defectline] (0,0.75)--(2,1.25);
\atoms{defectvertex}{x/p={1,1}}
\draw[\manifoldbdcol,manifold] (1,1)ellipse (0.4cm and 0.6cm);
\end{tikzpicture}
\quad\rightarrow\quad
\begin{tikzpicture}
\draw[manifoldboundary, manifold] (0,0)circle(0.4);
\atoms{defectvertex}{0/p={-0.4,0}, 1/p={0.4,0}}
\end{tikzpicture}\;.
\end{equation}
We then require the following two topology-changing moves. The first move 
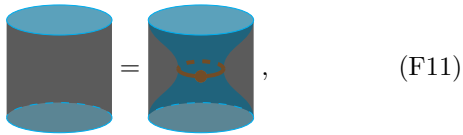
\begin{equation}
\label{eq:31_defect_surgery_move}
\begin{tikzpicture}
\fill[manifold] (180:0.7) arc (180:0:0.7cm and 0.2cm)--++(0,1.3)arc(0:180:0.7cm and 0.2cm) --cycle;
\fill[manifoldbdfull] (0,1.3) ellipse (0.7cm and 0.2cm);
\fill[manifoldbdfull] (0,0) ellipse (0.7cm and 0.2cm);
\fill[manifold] (180:0.7) arc (-180:0:0.7cm and 0.2cm)--++(0,1.3)arc(0:-180:0.7cm and 0.2cm) --cycle;
\draw[\manifoldbdcol,dashed] (0:0.7) arc (0:180:0.7cm and 0.2cm);
\draw[\manifoldbdcol] (0:0.7) arc (0:-180:0.7cm and 0.2cm);
\draw[\manifoldbdcol] (0,1.3) ellipse (0.7cm and 0.2cm);
\end{tikzpicture}
=
\begin{tikzpicture}
\fill[manifold] (180:0.7) arc (180:0:0.7cm and 0.2cm)--++(0,1.3)arc(0:180:0.7cm and 0.2cm) --cycle;
\fill[manifoldbdfull] (180:0.7) arc (180:0:0.7cm and 0.2cm)to[out=90,in=-90](0.3,0.65)to[out=90,in=-90](0.7,1.3)arc(0:180:0.7cm and 0.2cm)to[out=-90,in=90](-0.3,0.65)to[out=-90,in=90]cycle;
\fill[manifoldbdfull] (180:0.7) arc (-180:0:0.7cm and 0.2cm)to[out=90,in=-90](0.3,0.65)to[out=90,in=-90](0.7,1.3)arc(0:-180:0.7cm and 0.2cm)to[out=-90,in=90](-0.3,0.65)to[out=-90,in=90]cycle;
\draw[defectline,dashed] (0.3,0.65) arc (0:180:0.3cm and 0.1cm);
\draw[defectline] (0.3,0.65) arc (0:-180:0.3cm and 0.1cm);
\atoms{defectvertex}{0/p={0,0.55}}
\fill[manifold] (180:0.7) arc (-180:0:0.7cm and 0.2cm)--++(0,1.3)arc(0:-180:0.7cm and 0.2cm) --cycle;
\draw[\manifoldbdcol,dashed] (0:0.7) arc (0:180:0.7cm and 0.2cm);
\draw[\manifoldbdcol] (0:0.7) arc (0:-180:0.7cm and 0.2cm);
\draw[\manifoldbdcol] (0,1.3) ellipse (0.7cm and 0.2cm);
\end{tikzpicture}\;,
\end{equation}
equates a cylinder of $\calb$-network with physical boundaries on the top and the bottom and an open boundary on the side, with a solid torus whose boundary is split into a physical boundary on the inside and an open boundary on the outside.  The physical boundary on the inside is decorated with a loop of the defect line that hosts one point defect.

The second move
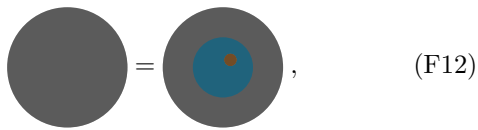
\begin{equation}
\label{eq:30_defect_surgery_move}
\begin{tikzpicture}
\fill[manifold] (0,0)circle(0.8);
\fill[manifold] (0,0)circle(0.8);
\end{tikzpicture}
=
\begin{tikzpicture}
\fill[manifold] (0,0)circle(0.8);
\fill[manifoldbdfull] (0,0)circle(0.4);
\fill[manifoldbdfull] (0,0)circle(0.4);
\atoms{defectvertex}{x/p={0.1,0.1}}
\fill[manifold] (0,0)circle(0.8);
\end{tikzpicture}\;,
\end{equation}
equates an open-boundary 3-ball with a 3-annulus ($S^2 \times [0,1]$), that has an open-boundary sphere on the outside and a physical-boundary sphere containing a single point defect on the inside. 

From a practical point of view, it is important to know when a given $\calb$-model with boundary can be extended to a model with additional defects such that the equations above hold. We will answer this question by relating the defects and moves to their lower dimensional counterparts obtained via compactification.

We first consider the move in Eq.~\eqref{eq:31_defect_surgery_move} which involves a boundary line defect containing a point defect. To find the latter two, we apply the idea of compactification mappings from Appendix~\ref{sec:compactification}. The line defect in Eq.~\eqref{eq:boundary_defect_line} corresponds to the boundary of the $1+1$-dimensional model given by compactifying the 2+1-dimensional model with the link in Eq.~\eqref{eq:boundary_defect_line}.
Using that same compactification, the point defect on the line defect becomes a point defect on the boundary of the $1+1$-dimensional model.
If we apply the compactification mapping to Eq.~\eqref{eq:2d_puncture_healing_move}, we precisely obtain the move in Eq.~\eqref{eq:31_defect_surgery_move}, which in the reverse direction can be imagined by squeezing the latter in the vertical direction. So the existence of line and point defects fulfilling Eq.~\eqref{eq:31_defect_surgery_move} is equivalent to the existence of a boundary with point defect of the compactified $1+1$-dimensional model fulfilling Eq.~\eqref{eq:2d_puncture_healing_move}. The latter are exactly the conditions for the $1+1$-dimensional universality mapping with boundary in Section~\ref{sec:finegrain_boundary}. As we argued in that section, those conditions can be seen to be satisfied for any physically relevant phase using another compactification.

We now consider the second move of the generalized liquid stated in Eq.~\eqref{eq:30_defect_surgery_move} and the point defect needed to define it. Using the compactification mapping for the link of the point defect in Eq.~\eqref{eq:2d_boundary_point_defect}, we establish an equivalence between point defects and boundaries of the resulting $0+1$-dimensional model. Those $0+0$-dimensional boundaries form a vector space, which from a physical point of view, corresponds to a ground state space of the non-compactified model on a disk. Applying the compactification mapping to Eq.~\eqref{eq:interval_healing_equation} yields Eq.~\eqref{eq:30_defect_surgery_move}, so the latter condition reduces to the former. Analogously to our argumentation in Section~\ref{sec:finegrain_boundary}, the former equation holds automatically for any type of robust topological order (including symmetry-breaking order if we do impose the symmetry) with non-degenerate boundary, as the ground space on a disk is one-dimensional.

Let us be more concrete and take a look at the situation for $2+1$-dimensional intrinsic bosonic topologically ordered phases whose anyons form a UMTC $\mathcal{M}$. In this case different boundaries are in one-to-one correspondence with different fusion categories $\mathcal{F}$ whose Drinfel'd centre is $\mathcal{M}$. The compactification with Eq.~\eqref{eq:boundary_defect_line} yields a 
non-robust symmetry-breaking-without-symmetry $1+1$-dimensional model, i.e., a GHZ-like superposition of trivial sectors, with one sector for each simple object of $\mathcal{F}$.
More precisely, each sector consists of an $\alpha^\chi$ Euler-characteristic classical invariant (cf.~Section~\ref{sec:classical_appendix}) with $\alpha$ being the \emph{quantum dimension} of the simple object. Thus, the irreducible $0+1$-dimensional boundaries of the compactified model and the boundary line defects of the $2+1$-dimensional model are in one-to-one correspondence with the simple objects of $\mathcal{F}$. In order to guarantee the existence of a boundary defect fulfilling Eq.~\eqref{eq:31_defect_surgery_move}, we need to take as boundary a direct sum in which every simple object occurs at least once. The canonical choice is a boundary which contains each simple object exactly once, such that the vector space of point defects on that line defect is spanned by the simple objects. Then there is a unique choice of the point defect fulfilling Eq.~\eqref{eq:31_defect_surgery_move}, namely the vector formed by the quantum dimensions $d_i$.

Having analysed the requirements that allow for additional boundary defects and topology-changing moves, we are now in the position to formulate the invertible mapping from the simplex liquid to $\calb$. 
We start by considering two tetrahedra glued together at one of their faces, and consider them as a patch of 3-manifold with open boundary. Then we remove a neighbourhood of all edges and vertices which are shared by both tetrahedra. The boundary created by the removal is a physical boundary. Next, we place a defect line running across the physical boundary near the centre of every edge shared by both tetrahedra. We then fill this manifold including all the defects with $\calb$-network such that all corners have a combinatorial distance of at least $\lambda$. Finally, we cut the network into two halves corresponding to the original tetrahedra,
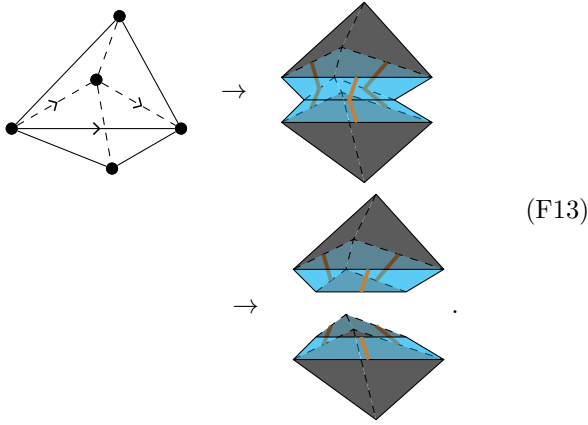
\begin{equation}
\label{eq:3d_universal_boundary_cut}
\begin{multlined}
\begin{tikzpicture}
\atoms{vertex}{0/, 1/p={-150:1.3}, 2/p={-30:1.3}, 3/p={70:0.9}, 4/p={-80:1.2}}
\draw[backedge] (1)edge[or](0) (0)edge[or](2) (3)--(0)--(4);
\draw (1)edge[or](2) (1)--(3)--(2) (1)--(4)--(2);
\end{tikzpicture}
\quad\rightarrow\quad
\begin{tikzpicture}
\atoms{void}{x0/p={-0.3,0.3}, x1/p={-0.7,0}, x2/p={0.5,0}, 3/p={0.1,1.3}, 4/p={0.1,-1.1}}
\atoms{void}{y03/p={-0.2,0.7}, y13/p={-1,0.3}, y23/p={1,0.3}, y04/p={-0.2,0.1}, y14/p={-1,-0.3}, y24/p={1,-0.3}}
\fill[manifold] (y03)--(3-c)--(y13)--cycle (y03)--(3-c)--(y23)--cycle (y04)--(4-c)--(y14)--cycle (y04)--(4-c)--(y24)--cycle;
\fill[manifoldbdfull] (x0)--(y03)--(y13)--(x1)--cycle (x0)--(y04)--(y14)--(x1)--cycle (x0)--(y03)--(y23)--(x2)--cycle (x0)--(y04)--(y24)--(x2)--cycle;
\draw[backedge] (x1)--(x0)--(x2) (y13)--(y03)--(y23) (y14)--(y04)--(y24) (3)--(y03)--(x0)--(y04)--(4);
\draw[defectline] ($(y03)!0.5!(y13)$)--($(x0)!0.5!(x1)$)--($(y04)!0.5!(y14)$) ($(y03)!0.5!(y23)$)--($(x0)!0.5!(x2)$)--($(y04)!0.5!(y24)$);
\fill[manifold] (y13)--(3-c)--(y23)--cycle (y14)--(4-c)--(y24)--cycle;
\fill[manifoldbdfull] (x2)--(y23)--(y13)--(x1)--cycle (x2)--(y24)--(y14)--(x1)--cycle;
\draw (3)--(y13)--(x1)--(y14)--(4) (3)--(y23)--(x2)--(y24)--(4) (y13)--(y23) (y14)--(y24) (x1)--(x2);
\draw[defectline] ($(y23)!0.5!(y13)$)--($(x2)!0.5!(x1)$)--($(y24)!0.5!(y14)$);
\end{tikzpicture}\\
\quad\rightarrow\quad
\begin{tikzpicture}
\atoms{void}{x0a/p={-0.3,0.3}, x1a/p={-0.7,0}, x2a/p={0.5,0}, 3/p={0.1,1.3}}
\atoms{void}{x0b/p={-0.3,-0.3}, x1b/p={-0.7,-0.6}, x2b/p={0.5,-0.6}, 4/p={0.1,-1.7}}
\atoms{void}{y03/p={-0.2,0.7}, y13/p={-1,0.3}, y23/p={1,0.3}}
\atoms{void}{y04/p={-0.2,-0.5}, y14/p={-1,-0.9}, y24/p={1,-0.9}}
\fill[manifold] (y03)--(3-c)--(y13)--cycle (y03)--(3-c)--(y23)--cycle (y04)--(4-c)--(y14)--cycle (y04)--(4-c)--(y24)--cycle;
\fill[manifoldbdfull] (x0a)--(y03)--(y13)--(x1a)--cycle (x0b)--(y04)--(y14)--(x1b)--cycle (x0a)--(y03)--(y23)--(x2a)--cycle (x0b)--(y04)--(y24)--(x2b)--cycle;
\fill[manifold] (x1a)--(x0a)--(x2a)--cycle;
\draw[backedge] (x1a)--(x0a)--(x2a) (x1b)--(x0b)--(x2b) (y13)--(y03)--(y23) (y14)--(y04)--(y24) (3)--(y03)--(x0a) (x0b)--(y04)--(4);
\draw[defectline] ($(y03)!0.5!(y13)$)--($(x0a)!0.5!(x1a)$) ($(x0b)!0.5!(x1b)$)--($(y04)!0.5!(y14)$) ($(y03)!0.5!(y23)$)--($(x0a)!0.5!(x2a)$) ($(x0b)!0.5!(x2b)$)--($(y04)!0.5!(y24)$);
\fill[manifold] (y13)--(3-c)--(y23)--cycle (y14)--(4-c)--(y24)--cycle;
\fill[manifoldbdfull] (x2a)--(y23)--(y13)--(x1a)--cycle (x2b)--(y24)--(y14)--(x1b)--cycle;
\draw (3)--(y13)--(x1a) (x1b)--(y14)--(4) (3)--(y23)--(x2a) (x2b)--(y24)--(4) (y13)--(y23) (y14)--(y24) (x1a)--(x2a) (x1b)--(x2b);
\draw[defectline] ($(y23)!0.5!(y13)$)--($(x2a)!0.5!(x1a)$) ($(x2b)!0.5!(x1b)$)--($(y24)!0.5!(y14)$);
\fill[manifold] (x1b)--(x0b)--(x2b)--cycle;
\end{tikzpicture}
\;.
\end{multlined}
\end{equation}
Note, that the triangle shared by both tetrahedra carries a branching structure, which is used to make an unambiguous choice of the cut.

Next, we take four of the tetrahedra whose bottom edges are truncated, bring their top vertices near each other, and glue them by filling all the spaces between them. We obtain a tetrahedron where all edges are truncated,
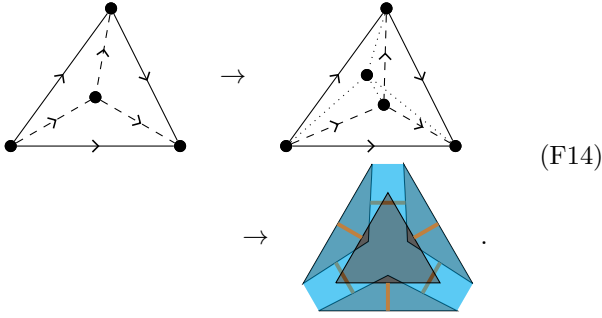
\begin{equation}
\label{eq:truncated_tetrahedron}
\begin{multlined}
\begin{tikzpicture}
\atoms{vertex}{0/, 1/p={-150:1.3}, 2/p={-30:1.3}, 3/p={80:1.2}}
\draw[backedge] (1)edge[or](0) (0)edge[or](2) (3)edge[ior](0);
\draw (1)edge[or](2) (1)edge[or](3) (3)edge[or](2);
\end{tikzpicture}
\quad\rightarrow\quad
\begin{tikzpicture}
\atoms{vertex}{0/p=-30:0.2, 1/p={-150:1.3}, 2/p={-30:1.3}, 3/p={80:1.2}, 4/p={100:0.3}}
\draw[backedge] (1)edge[or](0) (0)edge[or](2) (3)edge[ior](0);
\draw (1)edge[or](2) (1)edge[or](3) (3)edge[or](2);
\draw[midedge] (0)--(4) (1)--(4) (2)--(4) (3)--(4);
\end{tikzpicture}\\
\rightarrow\quad
\begin{tikzpicture}
%\atoms{vertex}{0/, 1/p={-150:1.5}, 2/p={-30:1.5}, 3/p={90:1.5}}
\atoms{void}{a0/p=-150:0.8, a1/p=-30:0.8, a2/p=90:0.8, b0/p=-140:1.2, b1/p=-40:1.2, b2/p=-90:0.3, c0/p=-20:1.2, c1/p=80:1.2, c2/p=30:0.3, d0/p=100:1.2, d1/p=-160:1.2, d2/p=150:0.3}
\draw[manifold] (b0)--(b1)--(b2)--(b0) (c0)--(c1)--(c2)--(c0) (d0)--(d1)--(d2)--(d0);
\fill[manifoldbdfull] (b0)--(b2)--(b1)--(c0)--(c2)--(c1)--(d0)--(d2)--(d1)--cycle;
\draw[defectline] ($(b1)!0.5!(b2)$)--($(c0)!0.5!(c2)$) ($(c1)!0.5!(c2)$)--($(d0)!0.5!(d2)$) ($(d1)!0.5!(d2)$)--($(b0)!0.5!(b2)$);
\fill[even odd rule, manifoldbdfull] (a0)--(a1)--(a2)--cycle (b0)--(b1)--(c0)--(c1)--(d0)--(d1)--cycle;
\draw[manifold] (a0)--(a1)--(a2)--(a0);
\draw[defectline] ($(a0)!0.5!(a1)$)--($(b0)!0.5!(b1)$) ($(a1)!0.5!(a2)$)--($(c0)!0.5!(c1)$) ($(a0)!0.5!(a2)$)--($(d0)!0.5!(d1)$);
%\draw (1)--(2)--(3)--(1);
%\draw[backedge] (0)--(1) (0)--(2) (0)--(3);
\end{tikzpicture}\;.
\end{multlined}
\end{equation}

In order to make an unambiguous choice of filling, we can use the branching structure of the reference tetrahedron on the left. The resulting $\calb$-network is what we associate to the tetrahedron tensor,
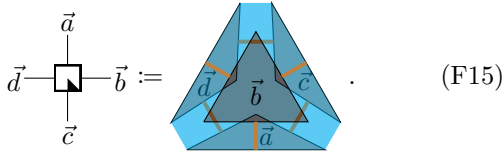
\begin{equation}
\begin{tikzpicture}
\atoms{tetrahedron}{0/}
\draw (0-b)edge[ind=\vec c]++(-90:0.4) (0-r)edge[ind=\vec b]++(0:0.4) (0-t)edge[ind=\vec a]++(90:0.4) (0-l)edge[ind=\vec d]++(180:0.4);
\end{tikzpicture}
\coloneqq
\begin{tikzpicture}
\atoms{void}{a0/p=-150:0.8, a1/p=-30:0.8, a2/p=90:0.8, b0/p=-140:1.2, b1/p=-40:1.2, b2/p=-90:0.3, c0/p=-20:1.2, c1/p=80:1.2, c2/p=30:0.3, d0/p=100:1.2, d1/p=-160:1.2, d2/p=150:0.3}
\draw[manifold] (b0)--(b1)--(b2)--(b0) (c0)--(c1)--(c2)--(c0) (d0)--(d1)--(d2)--(d0);
\fill[manifoldbdfull] (b0)--(b2)--(b1)--(c0)--(c2)--(c1)--(d0)--(d2)--(d1)--cycle;
\node at (13:0.65){$\vec c$};
\node at (170:0.7){$\vec d$};
\node at (-77:0.6){$\vec a$};
\draw[defectline] ($(b1)!0.5!(b2)$)--($(c0)!0.5!(c2)$) ($(c1)!0.5!(c2)$)--($(d0)!0.5!(d2)$) ($(d1)!0.5!(d2)$)--($(b0)!0.5!(b2)$);
\fill[even odd rule, manifoldbdfull] (a0)--(a1)--(a2)--cycle (b0)--(b1)--(c0)--(c1)--(d0)--(d1)--cycle;
\draw[manifold] (a0)--(a1)--(a2)--(a0);
\draw[defectline] ($(a0)!0.5!(a1)$)--($(b0)!0.5!(b1)$) ($(a1)!0.5!(a2)$)--($(c0)!0.5!(c1)$) ($(a0)!0.5!(a2)$)--($(d0)!0.5!(d1)$);
\node at (0,0){$\vec b$};
\end{tikzpicture}\;.
\end{equation}
$\vec a$ stands for the composite of the open indices located on the corresponding open-boundary triangle, whose ordering can be chosen unambiguously using the branching structure of the corresponding face of the tetrahedron. The same holds for $\vec b$, $\vec c$, and $\vec d$.

The edge weight, which is inserted at some of the bonds corresponding to the triangles of the triangulation, is mapped to a triangle prism which fills the gap between two tetrahedra after performing the cut in Eq.~\eqref{eq:3d_universal_boundary_cut}. This triangle prism has an open boundary on the bottom and top, and a physical boundary on the sides. Across each side there is a boundary defect line, and a boundary line point defect is placed on the side which corresponds to the edge of the simplex triangulation to which the edge weight is associated. I.e., there are three types of triangle prism, corresponding to the $01$, $12$ and $02$ edge weights as defined in Ref.~\cite{liquid_intro}. As an example, the $02$ edge weight mapping is depicted as follows,
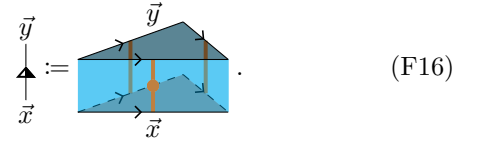
\begin{equation}
\label{eq:mapping_edge_weight}
\begin{tikzpicture}
\atoms{weight02}{0/}
\draw (0-mb)edge[ind=\vec x]++(-90:0.3) (0-ct)edge[ind=\vec y]++(90:0.3);
\end{tikzpicture}
\coloneqq
\begin{tikzpicture}
\atoms{void}{0/, 1/p={2,0}, 2/p={1.4,0.5}, 0x/p={0,0.7}, 1x/p={2,0.7}, 2x/p={1.4,1.2}}
\fill[manifoldbdfull] (0)--(2)--(1)--(1x)--(2x)--(0x)--cycle;
\draw[defectline] ($(0)!0.5!(2)$)--($(0x)!0.5!(2x)$) ($(2)!0.5!(1)$)--($(2x)!0.5!(1x)$);
\fill[manifold] (0)--(1)--(2)--cycle (2x)--(1x)--(0x)--cycle;
\draw[dashed] (0)edge[mark={arr,p=0.4}](2) (2)edge[mark={arr,p=0.4}](1);
\draw[] (0x)edge[mark={arr,p=0.4}](2x) (2x)edge[mark={arr,p=0.4}](1x);
\fill[manifoldbdfull] (0)--(1)--(1x)--(0x)--cycle;
\atoms{void}{m/p={$(0)!0.5!(1)$}, mx/p={$(0x)!0.5!(1x)$}}
\draw[defectline] (m)--(mx);
\draw[] (0)edge[mark={arr,p=0.4}](1) (0x)edge[mark={arr,p=0.4}](1x);
\atoms{defectvertex}{0/p={$(m)!0.5!(mx)$}}
\node at (1,1.3){$\vec y$};
\node at (1,-0.2){$\vec x$};
\end{tikzpicture}\;.
\end{equation}

The vertex weight is mapped to the same prism, just that instead of the point defect on the line defect, we add a point defect on the boundary
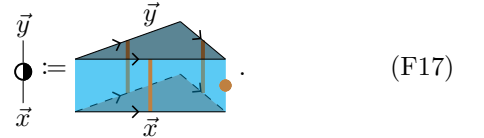
\begin{equation}
\label{eq:mapping_vertex_weight}
\begin{tikzpicture}
\atoms{circ,small,rhalf}{0/}
\draw (0-b)edge[ind=\vec x]++(-90:0.3) (0-t)edge[ind=\vec y]++(90:0.3);
\end{tikzpicture}
\coloneqq
\begin{tikzpicture}
\atoms{void}{0/, 1/p={2,0}, 2/p={1.4,0.5}, 0x/p={0,0.7}, 1x/p={2,0.7}, 2x/p={1.4,1.2}}
\fill[manifoldbdfull] (0)--(2)--(1)--(1x)--(2x)--(0x)--cycle;
\draw[defectline] ($(0)!0.5!(2)$)--($(0x)!0.5!(2x)$) ($(2)!0.5!(1)$)--($(2x)!0.5!(1x)$);
\fill[manifold] (0)--(1)--(2)--cycle (2x)--(1x)--(0x)--cycle;
\draw[dashed] (0)edge[mark={arr,p=0.4}](2) (2)edge[mark={arr,p=0.4}](1);
\draw[] (0x)edge[mark={arr,p=0.4}](2x) (2x)edge[mark={arr,p=0.4}](1x);
\fill[manifoldbdfull] (0)--(1)--(1x)--(0x)--cycle;
\atoms{void}{m/p={$(0)!0.5!(1)$}, mx/p={$(0x)!0.5!(1x)$}}
\draw[defectline] (m)--(mx);
\draw[] (0)edge[mark={arr,p=0.4}](1) (0x)edge[mark={arr,p=0.4}](1x);
\atoms{defectvertex}{0/p={$(1)!0.5!(1x)$}}
\node at (1,1.3){$\vec y$};
\node at (1,-0.2){$\vec x$};
\end{tikzpicture}\;.
\end{equation}
Similar to the edge weights, we can define vertex weights corresponding to the other two corners of the triangle by placing the point defect there.

If we apply the mapping to a triangulation of a 3-manifold, we have to replace every tetrahedron by the truncated-tetrahedron 3-manifold in Eq.~\eqref{eq:truncated_tetrahedron}. This results in a valid $\calb$-network since we just re-glue parts in the same way we cut them in Eq.~\eqref{eq:3d_universal_boundary_cut}. However, this $\calb$-network does not represent the original 3-manifold, but one where the neighbourhood of all the vertices and edges of the triangulation has been removed, with the 3-manifold terminating at a physical boundary. Along each edge of the triangulation, there is a tube of vacuum surrounded by a physical boundary, and around this tube wraps a line like defect with one point defect on it. We can use the move in Eq.~\eqref{eq:31_defect_surgery_move} to fill the tube of vacuum with a cylinder of 3-manifold. Then we are left with a $3$-ball of vacuum at every vertex, with one boundary point-defect on the physical boundary. We can use the move in Eq.~\eqref{eq:30_defect_surgery_move} to fill this ball of vacuum with a ball of 3-manifold. We thus see that the mapping yields a valid $\calb$-network representing the same $3$-manifold.

Also, it is easy to see that the mapped Pachner moves yield equations between $\calb$-networks which differ only at places with a certain distance from the open boundary. By choosing a larger and larger fine-graining scale $\lambda$ for the mapping, we are eventually guaranteed that the two networks are related by $\calb$-moves. In addition, we need the move in Eq.~\eqref{eq:31_defect_surgery_move} in the derivation of the mapped 2-3 Pachner move, and $4$ times Eq.~\eqref{eq:31_defect_surgery_move} plus one time Eq.~\eqref{eq:30_defect_surgery_move} in the derivation of a mapped 1-4 Pachner move.

Note that for any robust topological order (including intrinsic, symmetry-breaking, SPT/SET, fermionic models), the space of point-like defects on the boundary is equal to the (symmetry-preserving, even-parity) ground state space on a disk and its dimension is equal to $1$. Hence, the vertex weight is just a number. For intrinsic topological order given by a UMTC $\mathcal{M}$, this number is the inverse square root of the total quantum dimension of $\mathcal{M}$ (which is the inverse total quantum dimension of the fusion category $\mathcal{F}$ describing the boundary). Furthermore, for intrinsic topological order, consider the right hand side of Eq.~\eqref{eq:mapping_edge_weight} or Eq.~\eqref{eq:mapping_vertex_weight} without the point defect as an operator from the bottom to the top. This operator is a projector, and its support can be identified with the fusion space of the three boundary line defects in $\mathcal{F}$. This way we directly obtain the Turaev-Viro-Barrett-Westbury state sum, which is a slightly refined formulation of simplex liquid models.

\subsection{Universality mapping for the vertex liquid in \texorpdfstring{$2+1$}{2+1} dimensions}
\label{sec:higher_dimensions_vertex_finegrain}
In this appendix, we describe in detail the universality mapping from the simplex liquid in $2+1$ dimensions to an arbitrary topological liquid $\calb$, outlined in Section~\ref{sec:universality_higherdim}. In the end, we will also sketch the mapping in arbitrary spacetime dimensions $n$. The mapping for $n=3$ proceeds in four steps.

In step one, we fill the branching-structure tetrahedron with a $\calb$-network with fine-graining scale $\lambda$. We then cut the tetrahedron into $4$ `kite' volumes, each of which is the convex hull of one of the corner vertices and the centres of each adjacent edge, triangle, and of the tetrahedron itself. Geometrically, a kite is the same as a deformed cube, where one cube vertex is a corner of the tetrahedron and the cube vertex on the opposite side is the centre of the tetrahedron. The cut can be chosen unambiguously using the branching structure or the original tetrahedron. The following picture shows the edges of the cut in red, as well as what it looks like removing one of the kites,
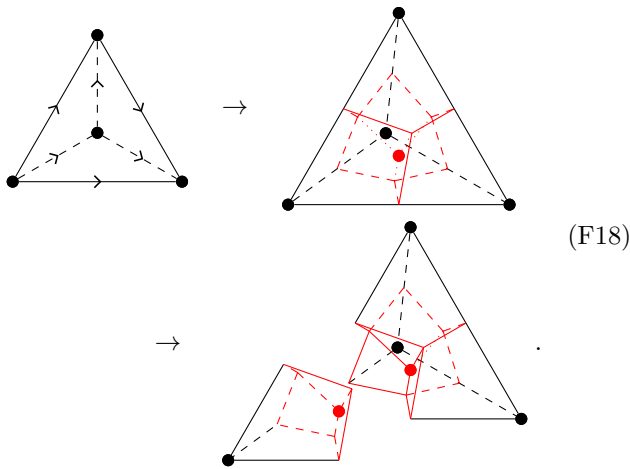
\begin{equation}
\label{eq:tetrahedrin_kite}
\begin{multlined}
\begin{tikzpicture}
\atoms{vertex}{0/, 1/p=-150:1.3, 2/p=-30:1.3, 3/p=90:1.3}
\draw[backedge] (1)edge[or](0) (0)edge[or](2) (3)edge[ior](0);
\draw (1)edge[or](2) (1)edge[or](3) (3)edge[or](2);
\end{tikzpicture}
\quad\rightarrow\quad
\begin{tikzpicture}
\atoms{vertex}{0/p={150:0.2}, 1/p=-150:1.7, 2/p=-30:1.7, 3/p=90:1.7}
\atoms{void}{01/p={$(0)!0.5!(1)$}, 02/p={$(0)!0.5!(2)$}, 03/p={$(0)!0.5!(3)$}, 12/p={$(1)!0.5!(2)$}, 13/p={$(1)!0.5!(3)$}, 23/p={$(2)!0.5!(3)$}, 012/p={intersection of 1--02 and 0--12}, 013/p={intersection of 1--03 and 0--13}, 023/p={intersection of 2--03 and 0--23}, 123/p={30:0.2}, c/p={-90:0.2}}
\draw[backedge] (1)--(0) (0)--(2) (3)--(0);
\draw (1)--(2) (1)--(3) (3)--(2);
\draw[red] (123)--(12) (123)--(13) (123)--(23);
\draw[red,backedge] (012)--(01) (012)--(02) (012)--(12) (013)--(01) (013)--(03) (013)--(13) (023)--(02) (023)--(03) (023)--(23);
\draw[red,midedge] (012)--(c) (013)--(c) (023)--(c) (123)--(c);
\atoms{vertex,bdastyle=red,decstyle=red}{x/p=c}
\end{tikzpicture}\\
\rightarrow\quad
\begin{tikzpicture}
\tikzset{s/.style={shift={(-150:0.8)}}}
\atoms{vertex}{0/p={150:0.2}, 2/p=-30:1.7, 3/p=90:1.7}
\atoms{void}{1/p=-150:1.7}
\atoms{void}{01/p={$(0)!0.5!(1)$}, 02/p={$(0)!0.5!(2)$}, 03/p={$(0)!0.5!(3)$}, 12/p={$(1)!0.5!(2)$}, 13/p={$(1)!0.5!(3)$}, 23/p={$(2)!0.5!(3)$}, 012/p={intersection of 1--02 and 0--12}, 013/p={intersection of 1--03 and 0--13}, 023/p={intersection of 2--03 and 0--23}, 123/p={30:0.2}, c/p={-90:0.2}}
\draw[backedge] (01)--(0) (0)--(2) (3)--(0);
\draw (12)--(2) (13)--(3) (3)--(2);
\draw[red] (123)--(12) (123)--(13) (123)--(23);
\draw[red,backedge] (012)--(02) (013)--(03) (023)--(02) (023)--(03) (023)--(23);
\draw[red] (012)--(01) (012)--(12) (013)--(01) (013)--(13);
\draw[red] (012)--(c) (013)--(c) (123)--(c);
\draw[red,midedge] (023)--(c);
\atoms{vertex,bdastyle=red,decstyle=red}{x/p=c}
\begin{scope}[transform canvas={shift={(-150:1.1)}}]
\draw[backedge] (1)--(01);
\draw (1)--(12) (1)--(13);
\draw[red,backedge] (012)--(c) (013)--(c) (123)--(c);
\draw[red] (123)--(12) (123)--(13);
\draw[red,backedge] (012)--(01) (012)--(12) (013)--(01) (013)--(13);
\atoms{vertex}{x/p={1}}
\atoms{vertex,bdastyle=red,decstyle=red}{x/p=c}
\end{scope}
\useasboundingbox (-150:3)--(2)--(3);
\end{tikzpicture}\;.
\end{multlined}
\end{equation}

In step two, for each star of a triangle, consider the three adjacent kites of each of the two adjacent tetrahedra, separated by a gap where the triangle is. We then fill the gap and choose a way to cut the result into three pieces, in the same way as we cut the triangle into $2$-dimensional kites. The filling can be chosen unambiguously using the orientation of the 3-manifold, and the cut can be chosen using th branching structure on the triangle.

In step three, for each star of an edge, consider the two adjacent kites of each adjacent tetrahedron, with gaps between all of the tetrahedra. Fill the gaps along the triangles between the kites of different tetrahedra using the choices from step two, such that a gap along the edge itself remains. Then, fill this gap with $\calb$-network, and cut it into two pieces along the perpendicular plane going through the centre of the edge. Note that in order to fill the gap unambiguously we need the dual branching of the edge which removes any symmetries of its star, and the branching structure of the edge is needed to unambiguously define the cut.

Finally, in step four, for each star of a vertex, consider the adjacent kite of each adjacent tetrahedron, with gaps between the different tetrahedra. Fill the gaps between the tetrahedra along the triangles and the edges using the choices from step two and step three, such that only a gap at the vertex itself remains. Then, fill the remaining gap with $\calb$-network. Note that the choice of filling does not matter since it is distant from the boundary of the surrounding kites, so we do not need a dual branching of the vertices. The so-obtained volume can be identified with the volume dual to the vertex in the dual cellulation. E.g., if the link is an octahedron, then the surrounding kites yield a volume which looks like a cube,
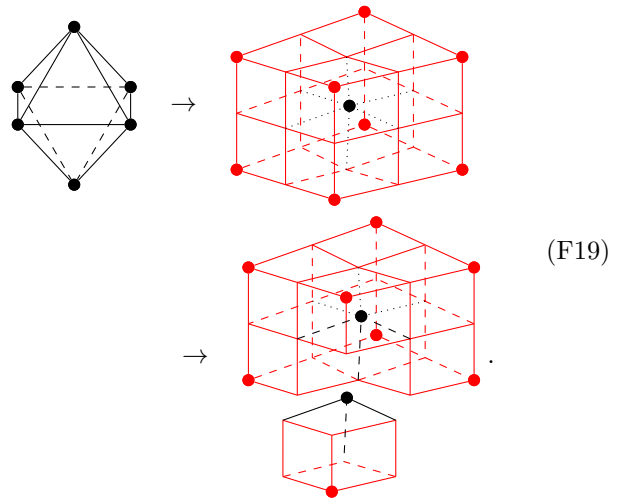
\begin{equation}
\begin{multlined}
\begin{tikzpicture}
\atoms{vertex}{0/, 1/p={1.5,0}, 2/p={0,0.5}, 3/p={1.5,0.5}, 4/p={0.75,-0.8}, 5/p={0.75,1.3}}
\draw (2)--(0)--(1)--(3) (0)--(4) (1)--(4) (0)--(5) (1)--(5) (2)--(5) (3)--(5);
\draw[backedge] (2)--(3) (2)--(4) (3)--(4);
\end{tikzpicture}
\quad\rightarrow\quad
\begin{tikzpicture}
\atoms{vertex,bdastyle=red,decstyle=red}{0/, 1/p={3,0}, 2/p={1.7,0.6}, 3/p={1.3,-0.4}, 0x/p={0,1.5}, 1x/p={3,1.5}, 2x/p={1.7,2.1}, 3x/p={1.3,1.1}}
\atoms{void}{02/p={$(0)!0.5!(2)$}, 03/p={$(0)!0.5!(3)$}, 00x/p={$(0)!0.5!(0x)$}, 12/p={$(1)!0.5!(2)$}, 22x/p={$(2)!0.5!(2x)$}, 13/p={$(1)!0.5!(3)$}, 33x/p={$(3)!0.5!(3x)$}, 02x/p={$(0x)!0.5!(2x)$}, 12x/p={$(1x)!0.5!(2x)$}, 13x/p={$(1x)!0.5!(3x)$}, 03x/p={$(0x)!0.5!(3x)$}, 11x/p={$(1)!0.5!(1x)$}}
\atoms{void}{b/p={intersection of 0--1 and 2--3}, t/p={intersection of 0x--1x and 2x--3x}, l/p={intersection of 0--2x and 0x--2}, r/p={intersection of 3--1x and 1--3x}, f/p={intersection of 0--3x and 3--0x}, ba/p={intersection of 2--1x and 1--2x}}
\draw[backedge, red] (0)--(2)--(1) (2)--(2x);
\draw[red] (0)--(3)--(1) (0x)--(3x)--(1x)--(2x)--(0x) (0)--(0x) (1)--(1x) (3)--(3x);
\atoms{vertex}{c/p={$(3x)!0.5!(2)$}}
\draw[red] (00x)--(f)--(33x)--(r)--(11x) (03)--(f)--(03x)--(t)--(12x) (13)--(r)--(13x)--(t)--(02x);
\draw[red,backedge] (03)--(b)--(12)--(ba)--(12x) (13)--(b)--(02)--(l)--(02x) (11x)--(ba)--(22x)--(l)--(00x);
\draw[midedge] (b)--(c) (t)--(c) (l)--(c) (r)--(c) (f)--(c) (ba)--(c);
\end{tikzpicture}
\\
\rightarrow\quad
\begin{tikzpicture}
\atoms{vertex,bdastyle=red,decstyle=red}{0/, 1/p={3,0}, 2/p={1.7,0.6}, 0x/p={0,1.5}, 1x/p={3,1.5}, 2x/p={1.7,2.1}, 3x/p={1.3,1.1}}
\atoms{void}{3/p={1.3,-0.4}}
\atoms{void}{02/p={$(0)!0.5!(2)$}, 03/p={$(0)!0.5!(3)$}, 00x/p={$(0)!0.5!(0x)$}, 12/p={$(1)!0.5!(2)$}, 22x/p={$(2)!0.5!(2x)$}, 13/p={$(1)!0.5!(3)$}, 33x/p={$(3)!0.5!(3x)$}, 02x/p={$(0x)!0.5!(2x)$}, 12x/p={$(1x)!0.5!(2x)$}, 13x/p={$(1x)!0.5!(3x)$}, 03x/p={$(0x)!0.5!(3x)$}, 11x/p={$(1)!0.5!(1x)$}}
\atoms{void}{b/p={intersection of 0--1 and 2--3}, t/p={intersection of 0x--1x and 2x--3x}, l/p={intersection of 0--2x and 0x--2}, r/p={intersection of 3--1x and 1--3x}, f/p={intersection of 0--3x and 3--0x}, ba/p={intersection of 2--1x and 1--2x}}
\draw[backedge,red] (0)--(2)--(1) (2)--(2x);
\draw[red] (0)--(03) (13)--(1) (0x)--(3x)--(1x)--(2x)--(0x) (0)--(0x) (1)--(1x) (33x)--(3x);
\atoms{vertex}{c/p={$(3x)!0.5!(2)$}}
\draw[red] (00x)--(f)--(33x)--(r)--(11x) (03)--(f)--(03x)--(t)--(12x) (13)--(r)--(13x)--(t)--(02x) (03)--(b) (13)--(b);
\draw[red,backedge] (b)--(12)--(ba)--(12x) (b)--(02)--(l)--(02x) (11x)--(ba)--(22x)--(l)--(00x);
\draw[backedge] (r)--(c) (f)--(c) (b)--(c);
\draw[midedge] (t)--(c) (l)--(c) (ba)--(c);
\begin{scope}[transform canvas={shift={(-100:1.1)}}]
\draw[red] (3)--(03) (3)--(13) (3)--(33x);
\draw[red] (03)--(f)--(33x)--(r)--(13);
\draw (f)--(c-c)--(r);
\draw[red,backedge] (03)--(b) (13)--(b);
\draw[backedge] (c-c)--(b);
\atoms{vertex}{z/p={c}}
\atoms{vertex,bdastyle=red,decstyle=red}{w/p=3}
\end{scope}
\useasboundingbox ([shift={(-100:1.2)}]3)--(0x)--(2x)--(1x);
\end{tikzpicture}
\;.
\end{multlined}
\end{equation}

The obtained volume is what we associate to the vertex tensor with the same star. Each index of the vertex tensor corresponds to an internal edge of the star. Each such index is mapped to the composite of all the indices on the face of the volume which is dual to this internal edge. In the above example, the open indices at each face of the cube form a composite index. Due to the dual branching of the internal edge dual to that face, we can choose an unambiguous ordering of indices when forming their composite.

We need to show that the $\calb$-network obtained by applying the mapping to a triangulation is valid. To construct this $\calb$-network we replace each vertex by the $\calb$-network filling its dual $3$-cell, and then glue all these $3$-cells together. In doing so, we simply re-glue patches of $\calb$-network in the same way they have been cut before in the construction. More precisely, at each face of the dual cellulation we re-glue the cut made in step three of the construction. At every edge of the dual cellulation we re-glue the cut from step two, and at every vertex of the dual cellulation we re-glue the cut from step one. We thus see that the resulting $\calb$-network is valid everywhere.

Next we need to check whether the mapped moves of the vertex liquid can be derived from the $\calb$-moves. Recall that the moves of the vertex liquid are dual to re-cellulations, where both sides agree at the boundary as well as at the immediate vicinity of the boundary, as shown in Eq.~\eqref{eq:vertex_3d_move}. Now decompose all the volumes of the re-cellulation into kites. We will find that all the kites which have contact with the boundary of the re-cellulation remain the same. The two cellulations only differ at kites in the interior, which are separated from the open boundary by a certain distance. This distance can be made arbitrarily large by increasing the fine-graining scale $\lambda$. Thus, no matter what the $\calb$-moves are, they can be used eventually for deriving the mapped vertex liquid moves if we pick a large enough $\lambda$.

At this point, the generalization to $n>3$ is straight-forward: First consider the intersection of an $n$-dimensional triangulation with its dual cellulation, yielding a decomposition of all $x$-simplices and all dual $x$-cells into kites, which are deformed cubes. Then, the construction proceeds in $n+1$ steps $0,\ldots,n$. In the $i$th step, for every star of an $n-i$-simplex, we consider the $n-i+1$ adjacent kites of each adjacent $n$-simplex (for $i=0$ there are none), with gaps in between the kites at different tetrahedra. We fill the gaps at the $y$-simplices ($n-i<y<n$) between the kites at different $n$-simplices using the result of step $n-y$, such that only a gap along the $n-i$-simplex remains. We fill this gap with valid $\calb$-network, and choose a way to cut the resulting network into $n-i$ parts such that the $n-i$-simplex itself is cut into $n-i$-dimensional kites. When filling, we take the combinatorial distance between different corners to be at least $\lambda$.

\section{All-scalar vertex-liquid models and characteristic classes}
\label{sec:classical_appendix}
In Section~\ref{sec:unphysical}, we have announced that there are all-scalar vertex-liquid models which cannot be expressed as all-scalar triangle-liquid models. In this section, we will describe those models in more detail. We will also give an introduction to characteristic classes and cohomology operations in simplicial cohomology, tools which allow us to construct a wealth of such vertex-liquid models.
%Specifically, in Section~\ref{sec:simplex_euler}, we argue that all-scalar simplex-liquid models are all equivalent to models determined by the Euler characteristic. Then we give a simple example of an all-scalar vertex-liquid model in $1+1$ dimensions which is not equivalent to an Euler characteristic model in Section~\ref{sec:simple_11d}. We then turn to the general case introducing the basic vocabulary of simplicial cohomology, introducing local combinatorial formulas for generating characteristic classes and cohomology operations, and showing how to combine them into all-scalar liquid models. Finally, we will discuss which of the obtained liquid models are in the same exact phases, and which only yield the same invariants when evaluated on a manifold.

At this point we would like to stress once more that all-scalar liquid models are physically trivial in many different ways. The tensor networks of such models are just products of scalars, so there are no degrees of freedom, or equivalently, the overall Hilbert space is 1-dimensional. Furthermore, those models become even mathematically trivial if we resort to projective tensors. Yet another way in how all the models presented in this appendix become mathematically trivial is by restricting to regular, translation-invariant lattices. There, the models either directly associate the number $1$ to every vertex, or possess an invertible domain wall to vacuum. This is in contrast to the notion of a non-trivial ``phase'' in condensed matter theory, which is determined by the model on a regular lattice. Last, many of the presented models are again trivial if we define phases via continuous gapped paths (where ``gapped'' for all-scalar models the same as non-zero on all ball-like patches). This is because many of them come in continuous families containing the trivial model. Even though all-scalar models are themselves physically trivial, they can model \emph{gravitational anomalies} of other non-trivial models, and they demonstrate that there is no general scheme which tells us that simplex-liquid models are always sufficient.

%All-scalar liquid models are equivalent to what is often referred to as \emph{classical invariants} in the study of continuum manifolds. Roughly speaking, classical invariants are computed by integrating a quantity over the spacetime manifold which only depends on the local presentation of that manifold (e.g., on the metric when expressed in a specific coordinate system). If we discretize the integral to a sum on a triangulation, which we then exponentiate, we obtain an all-scalar liquid model. In contrast, for \emph{quantum invariants}, the local quantity also depends on a field configuration, and we sum an exponentiated action over all field configurations. In this section we look at two particularly simple examples of classical invariants in $1+1$ dimensions, and argue that one of them needs to be formulated as a vertex-liquid model. A more systematic approach to obtain all-scalar vertex-liquid models from so-called \emph{characteristic classes} can be found in Appendix~\ref{sec:classical_appendix}.

\subsection{All-scalar simplex-liquid models and the Euler characteristic}
\label{sec:simplex_euler}
In this section we point out that every all-scalar simplex-liquid model is equivalent to an exponentiated Euler characteristic. The most general variant of this liquid does not only associate tensors (here, scalars) to $n$-simplices, but also to $i$-simplices for all $0\leq i \leq n$. E.g., the generalized triangle liquid in $n=2$ includes vertex weights and edge weights as described in Section~\ref{sec:finegrain_boundary}.

The \emph{Euler characteristic} $\chi$ for a triangulation $T$ of an $n$-manifold is given by
\begin{equation}
\label{eq:euler_characteristic}
\begin{multlined}
\chi(T) = |S^0(T)|-|S^1(T)|+|S^2(T)|-\ldots \\= \sum_{0\leq i\leq n}\sum_{s\in S^i(T)} (-1)^i\;,
\end{multlined}
\end{equation}
where $S^i(T)$ denotes the set of $i$-simplices of $T$. Instead of $\chi$, we can equivalently compute $\alpha^\chi$ for some $\alpha\neq 0$, given by
\begin{equation}
\label{eq:euler_characteristic_product}
\begin{multlined}
\alpha^{\chi(T)} = \alpha^{\sum_{0\leq i\leq n}\sum_{s\in S^i(T)} (-1)^i}\\=  \prod_{0\leq i\leq n}\prod_{s\in S^i(T)} \alpha^{(-1)^i}\;,
\end{multlined}
\end{equation}
This is the same as the evaluation of an all-scalar model of a simplex liquid associating the scalar $\alpha^{(-1)^i}$ to the $i$-simplex. The Pachner topological moves are easily observed to hold. E.g., in $n=2$ spacetime dimensions, the 1-3 Pachner move
\begin{equation}
\begin{tikzpicture}
\atoms{vertex}{0/, 1/p={-150:0.6}, 2/p={-30:0.6}, 3/p={90:0.6}}
\draw (1)--(2)--(3)--(1) (0)--(1) (0)--(2) (0)--(3);
\end{tikzpicture}
\quad\rightarrow\quad
\begin{tikzpicture}
\atoms{vertex}{1/p={-150:0.6}, 2/p={-30:0.6}, 3/p={90:0.6}}
\draw (1)--(2)--(3)--(1);
\end{tikzpicture}
\end{equation}
holds due to
\begin{equation}
\alpha (\alpha^{-1})^3 \alpha^3 = \alpha\;.
\end{equation}
The two sides of the equation are given by the product of all scalars at the vertices, edges and faces shown in the two pictures above, except for the boundary edges and vertices which are the same for both sides. For $n=2$, the model is in a non-trivial exact phase since its evaluation is not equal to $1$ on some 2-manifolds, such as $\alpha^{-2}$ on a sphere.

Let us now show that in fact, Euler-characteristic models capture \emph{all} simplex-liquid models up to equivalence, and restrict ourselves to the case $n=2$ and unoriented triangulations for simplicity. An arbitrary all-scalar triangle-liquid model is then given by a triangle tensor $a$, edge weight $b$, and vertex weight $c$, i.e., a triple $(a, b, c)$ of scalars. The 1-3 Pachner move yields the only non-trivial equation,
\begin{equation}
\label{eq:scalar_triangle_constraint}
a^3b^3c = a\;.
\end{equation}
On the other hand, we can insert the identity $1=xx^{-1}$ for some non-zero scalar $x$ at every pair of adjacent edge and triangle, and associate every factor $x$ to the triangle and every factor $x^{-1}$ to the edge. Since each triangle has $3$ edges and each edge has $2$ adjacent triangles, this defines an invertible domain wall between the models
\begin{equation}
\label{eq:scalar_triangle_equivalence}
(a,b,c)\sim (ax^3, bx^{-2}, c)\;.
\end{equation}
Now, using Eq.~\eqref{eq:scalar_triangle_constraint}, we see that every model is of the form $(a, b, a^{-2}b^{-3})$. Applying Eq.~\eqref{eq:scalar_triangle_equivalence} with  $x=(ab)^{-1}$, we get
\begin{equation}
(a, b, a^{-2}b^{-3}) \sim (a^{-2}b^{-3}, a^2b^3, a^{-2}b^{-3})\;.
\end{equation}
We notice that the right-hand side is the $\alpha^\chi$ model with $\alpha=a^{-2}c^{-3}$. Thus, every all-scalar (unoriented, or real and hermitian) triangle-liquid model is equivalent to the $\alpha^\chi$ model for some $\alpha$. The situation is similar for the oriented case or in higher dimensions.

\subsection{An all-scalar vertex-liquid model in \texorpdfstring{$1+1$}{1+1} dimensions}
\label{sec:all_scalar_vertex_11d}

In this section, we give a first example for a non-trivial unoriented-vertex-liquid model in $1+1$ dimensions. By non-trivial we mean that it is not in the same exact phase as any Euler-characteristic model, and hence, any all-scalar triangle-liquid model. Also, it does not have any all-scalar topological boundary. We will refer to the model as the $(-1)^{\omega_1^2}$-model, as it is based on a discretized version of the square of the first Stiefel-Whitney characteristic class $\omega_1$, as we will see in the following sections. The unoriented version of the vertex liquid comes with additional dual directions for all edges of the two-dimensional unoriented branching-structure triangulation. The model associates a number $\pm 1$ to each star of a vertex as follows. We temporarily choose an orientation of the star and mark all internal edges which are between a clock-wise and a counter-clockwise triangle. Then we consider the number $\#r$ of marked edges whose dual direction points clockwise, and the number $\#l$ of marked edges pointing counter-clockwise. The number associated to the vertex is then given by
\begin{equation}
(-1)^{(\#l-\#r)/2}\;.
\end{equation}
The result is always $\pm 1$ since the total number of marked edges $\#l+\#r$ is always even and thus $\#l-\#r$ is divisible by $2$. Furthermore, it is easy to see that the result is independent of the temporary choice of orientation.

The topological invariance can be seen as follows. When applying the Pachner move consisting of only clockwise (or only counter-clockwise) triangles shown in Eq.~\eqref{eq:22pachner_move}, the edge in the interior of the move is not marked, and the markings of the boundary edges do not change, so the numbers $\#l$ and $\#r$ at the involved vertices are not altered. When changing the direction of an edge (if allowed), this changes the orientation of the two adjacent triangles, and thus the marking of the four outer edges of those two triangles. Every edge is a clockwise dually oriented edge in the star of one of its vertices, and a counter-clockwise edge of the other, so those contributions cancel. Finally, changing the dual direction of an edge yields a factor of $-1$ at both of its adjacent vertices if it is marked, and no change otherwise.

The evaluation of the $(-1)^{\omega_1^2}$ model on unoriented 2-manifolds is the same as the evaluation of $(-1)^\chi$, namely $-1$ on manifolds of odd non-orientable genus, and $1$ on manifolds with even non-orientable genus. However, the two models are not in the same exact phase. Roughly speaking, this can be seen by considering the evaluation of the two models on an open disk. For $(-1)^{\omega_2}$, we get $-1$ since the Euler characteristic of the disk is $1$ and thus odd, but for $(-1)^{\omega_1^2}$ we get $1$ (which as we will understand in the following sections, corresponds to the fact that the disk is orientable and so $\omega_1$ has a trivialization). Knowing that the $(-1)^{\omega_1^2}$ model is inequivalent to the $(-1)^\chi$ model, we know that it is inequivalent to \emph{any} triangle-liquid model.

Of course, the evaluation of a liquid model on the open disk (or any other manifold with open boundary) is ambiguous since there can be tensors (in our case scalars) right on the open boundary for which we have to specify whether they belong to the disk or not. Let us provide a more detailed explanation of why there is no invertible domain wall between the $(-1)^{\omega_1^2}$ and $(-1)^{\omega_2}$ models. First, we observe that both models are trivial, i.e., associate $1$ to every vertex, on a flat, regular, translation-invariant triangulation obtained by dividing all plaquettes of a square lattice into two triangles. Now take such an $n\times m$ square lattice yielding a triangulation of a cylinder/annulus with periodic boundary conditions in $m$, and close off one of the $m$-edge boundary circles with a triangulation of a disk. Padded with a depth-$n$ margin where both models are trivial, the evaluation on the disk is now well-defined, and disagrees for both models as we mentioned earlier. More precisely, assume there was an invertible domain wall between the two models, and place this domain wall around the non-contractible loop on the cylinder translation-invariant in $m$. Since the models themselves are trivial on the regular triangulations,
the domain wall itself is a 1-dimensional liquid model given by a vector space, whose evaluation on a circle yields the dimension of that vector space, which is a positive integer $x$. Per definition, we can contract the invertible domain wall over the disk transforming the $(-1)^{\omega_1^2}$ model into the $(-1)^{\omega_2}$ model, yielding
\begin{equation}
x = x\cdot (-1)^{\omega_1^2}[\text{disk}] = (-1)^{\omega_2}[\text{disk}] = -1\;,
\end{equation}
which contradicts $x$ being a positive integer.

Note that the arguments in Section~\ref{sec:finegrain_boundary} imply that if there was an all-scalar boundary for the $(-1)^{\omega_1^2}$ model, then the latter would be equivalent to a triangle-liquid model. Since we showed that this is not the case, there also cannot be any all-scalar boundary. We would like to point out that, however, one can define non-scalar boundaries for the model. As we will understand in the following sections, those boundaries correspond to summing over trivializations of the 1-cycle $\omega_1$ inside the boundary. The arguments in Section~\ref{sec:finegrain_boundary} then further imply that there are non-scalar triangle-liquid models equivalent to the $(-1)^{\omega_1^2}$ model.

\subsection{Simplicial homology}
Simplicial (co-)homology is a powerful tool to construct concrete liquid models related to finite groups, including non-trivial all-scalar liquid models. In this section, we will introduce the basic notions of simplicial (co-)homology.
In general, the central entities of ordinary manifold (co-)homology are \emph{$i$-(co-)cycles} (or, (co-)cycles of \emph{degree} $i$), which intuitively can be thought of embedded sets of points, closed loops, closed membranes, or generally embedded closed $i$-manifolds ($n-i$-manifolds). Embeddings are equivalent if they can be continuously deformed into each other. Moreover, the embedded closed manifolds are coloured by elements of an Abelian \emph{coefficient group} $A$, those coefficients are added at places where the embedding overlaps, and places coloured by $0$ may be removed from the embedding. In order to define cycles, we first need to define \emph{chains}. On an $n$-dimensional branching-structure triangulation $T$ an $A$-valued \emph{$i$-chain} for $0\leq i\leq n$ is a map
\begin{equation}
c: S^i(T)\rightarrow A\;,
\end{equation}
where $S^i$ is the set of $i$-simplices of $T$. The \emph{boundary} of an $i$-chain is the $i-1$-chain
\begin{equation}
(\delta c)(s) = \sum_{t \in S^{i+}(s)} \epsilon(t|s) c(t)\;,
\end{equation}
for all $s\in S^{i-1}(T)$, where $S^{i+}(s)$ is the set of $i$-simplices containing $s$, and $\epsilon(t|s)$ is $\pm 1$ depending on the orientation of $s$ relative to $t$. If $A=\zz_2$ (as will often be the case in this work), we can think about $c$ in terms of the subset of $i$-simplices $s$ with $c(s)=1$ (with $\zz_2$ written additively), which form an embedded $i$-manifold with boundary. Furthermore, the orientation $\epsilon$ has no effect and the boundary of $c$ is literally its boundary as an embedded manifold.

We can shift the above notions to the Poincar\'e dual cellulations such that what used to be the $i$-simplices are now $n-i$-simplices. An \emph{$i$-cochain} is the same as an $i$-chain, however, we define its boundary (which is then called \emph{coboundary}) as the $i+1$-cochain
\begin{equation}
(d c)(s) = \sum_{t \in S^{i-}(s)} \epsilon(s|t) c(t)\;,
\end{equation}
for all $s\in S^{i+1}(T)$, where $S^{i-}(s)$ is the set of $i$-simplices contained in $s$. In the $\zz_2$-valued case, $c$ forms an embedded $n-i$-manifold with boundary given by the $n-i$-cells dual to the $i$-simplices of $c$, and the coboundary is literally this boundary.

The (co-)boundary maps on the sets of (co-)chains of different degree define \emph{chain complexes}, meaning for any (co-)chain $c$, ${d}^2c=0$ ($\delta^2c=0$). $c$ is called \emph{(co-)cycle} if $\delta c=0$ (${d}c=0$) and \emph{(co-)boundary} if $c=\delta x$ ($c={d}x$) for some chain $x$. We will then call $x$ a \emph{trivialization} of $c$. The sets of (co-)chains, (co-)cycles, and (co-)boundaries form groups under simplex-wise $A$-addition. The group of $i$-(co-)cycles modulo the group of $i$-(co-)boundaries is known as the $i$th \emph{(co-)homology group}, and the corresponding equivalence class of a (co-)cycle is known as its \emph{(co-)homology class}.

\subsection{Characteristic classes, cohomology operations, and classifying spaces}
In this section we will give a rough overview of the ideas of characteristic classes, cohomology operations, and classifying spaces in conventional algebraic topology. The discussion here will be on a rather informal and abstract level since we will attempt to define everything explicitly and combinatorially in terms of simplicial homology later anyways. It is still insightful to match the predictions from algebraic topology with the simplicial formulas.

A \emph{fiber bundle} with \emph{fiber} $i$-manifold $F$ and \emph{base} $n$-manifold $B$ is a $in$-manifold which locally looks like $B\times F$, but where the copies of $F$ at different points of $B$ are not identified trivially but via an element of a \emph{structure group} $G$ of homeomorphisms over $F$. The bundle we are interested in is the \emph{tangent bundle} of an $n$-manifold, which is a fiber bundle with fiber $\mathbb{R}^n$ and structure group $O(n)$ (or $SO(n)$ in the oriented case). Roughly speaking, characteristic classes are prescriptions to compute a non-trivial (co-)cycle locally from a presentation of fibre bundles with a fixed structure group, in our case the tangent bundle of the spacetime $n$-manifold. Now, there exists a so-called \emph{classifying space} $BG$ of the structure group $G$ (in our case $BO(n)$ or $BSO(n)$), which is some infinite-dimensional topological space, and a \emph{universal bundle} over that classifying space, with the following property. Isomorphism classes of bundles with base $B$ are in one-to-one correspondence with homotopy classes of \emph{classifying maps} (maps up to continuous deformations) from $B$ to $BG$, and are obtained as a \emph{pullback} of the universal bundle of $BG$ via the classifying map. Characteristic classes are then just cohomology classes of $BG$ which give rise to cohomology classes of $B$, also by pullback via the same classifying map.

Now, all-scalar vertex-liquid models are locally computable $\mathbb{C}/\{0\}$-valued $0$-cycles of a presentation of a $n$-manifold and its tangent bundle, namely a triangulation. This suggests that there is a correspondence between all-scalar liquid models and characteristic $0$-cycles/$n$-cocyles, i.e., degree-$n$ cohomology classes of $BO(n)$ or $BSO(n)$. Indeed, in Ref.~\cite{Levitt1978}, the authors explicitly construct simplicial combinatorial representations for $\mathbb{R}^i$-bundles, the classifying space, and classifying map, and show that for every degree-$i$ characteristic class there exist local formulas which determine the value of a representing $n-i$-cycle on a $n-i$-simplex depending only on its star. So the degree-$n$ characteristic classes precisely yield vertex-liquid models. Note that those local formulas also depend on an ordering of the vertices within each star which is consistent on overlapping stars, an additional decoration which is similar to our dual branching introduced in Section~\ref{sec:vertex_liquid_higher}.

In order to obtain explicit local formulas for (degree-$n$) characteristic classes, we follow a 2-step approach. We first give formulas for a set of generating characteristic classes. Then we obtain the remaining characteristic classes via so-called \emph{cohomology operations}. Roughly, those are prescriptions which locally compute a $j$-(co-)cycle from one (or more) cocycles of other degree $i$. Such a prescription has to be compatible with cohomology, in the sense that if we add a coboundary to one of the input cocycles, the output cocycles differ by a coboundary as well. In contrast to characteristic classes, cohomology operations are agnostic to the tangent bundle, and even fact that our spacetime is a $n$-manifold (triangulation) and not an arbitrary simplicial complex. Now, $A$-valued $i$-cocycles can themselves be described as pullbacks of a universal $i$-cocycle from a classifying space $B^i(A)$ via a classifying map from the spacetime manifold to $B^i(A)$. Then cohomology operations correspond to $j$-cocycles on $B^i(A)$, which are pulled back to a $j$-cocycle on the spacetime manifold via the classifying map. For discrete $A$, $B^i(A)$ can be explicitly constructed as a \emph{simplicial Abelian group} via the \emph{Dold-Kahn correspondence}. As a result, cohomology operations possess simple local combinatorial formulas in simplicial homology, where the output $j$-cochain of a cohomology operation on a $j$-simplex can be obtained from the input cochains on its sub-simplices.

In the following sections, we will give explicit realizations of local formulas predicted by the above considerations. In the end, we will obtain local formulas for $\zz_2$-valued or $\zz$-valued $0$-cycles $c$, which become vertex-liquid models after exponentiation. More concretely, such a vertex-liquid model is then given by assigning to a vertex $v$ the scalar $(-1)^{c(v)}$ for $\zz_2$-valued classes, or $\alpha^{c(v)}$ for $\zz$-valued classes.

Note that a $\alpha^c$ model for a $\zz$-valued characteristic class can be contiuously deformed to the trivial model at $\alpha=1$, and is thus in a trivial phase according to the continuous-path definition. Here it is reasonable to demand that the models on the continuous path should be liquid models as well, otherwise any all-scalar model based on complex numbers would be trivial. If we restrict to models based on real numbers (implementing a time-reversal symmetry), then the models for $\alpha>0$ are still trivial. However, the models for $\alpha<0$ are non-trivial as any path to the trivial model would have to go through $0$.

\subsection{Simplicial formulas for generating characteristic classes}
In this section, we discuss simplicial formulas for three $\zz_2$-valued or $\zz$-valued generating characteristic classes of the tangent bundle. First, the \emph{Euler class} $e$ is a $\zz$-valued degree-$n$ characteristic class in $n$ dimensions. A local formula which computes a dual simplicial $0$-cycle representing the Euler class on a vertex $v$ is given by
\begin{equation}
\label{eq:euler_class}
e(v)=\sum_{0\leq x\leq n} (-1)^x \sum_{t\in S^{x+}(v)} \delta_{v|t, 0}\;,
\end{equation}
where $v|t$ is the number $0\leq v|t\leq x$ of the vertex $v$ in the simplex $t$. So, $e(v)$ is the number of $x$-simplices which contain $v$ as their $0$-vertex (including $v$ itself), weighted by $(-1)^x$. Note that the summation of $e$ over all vertices is the same as the Euler characteristic in Eq.~\eqref{eq:euler_characteristic} where we moved the number $(-1)^x$ to the $0$-vertex of each $x$-simplex.

Next, the $i$th \emph{Stiefel-Whitney class} $\omega_i$ is a $\zz_2$-valued degree-$i$ characteristic class in any spacetime dimension $n$. There is a local formula which computes a dual simplicial $(n-i)$-cycle representing $\omega_i$. Following Ref.~\cite{Goldstein1976}, its value on an $n-i$-simplex $s$ is given by
\begin{equation}
\omega_i(s) = \sum_{n-i\leq x\leq n} \sum_{t\in S^{x+}(s)} E(s|t)\;.
\end{equation}
Here, $s|t$ contains the information about which of the subsimplices of $t$ coincides with $s$, and is specified by the ordered sequence of numbers $0\leq \{(s|t)_l\}_{0\leq l\leq n-i}\leq x$ corresponding to the vertices of $s$, when the vertices of $t$ are numbered according to the branching structure ordering. $E(s|t)$ is $1$ if $(s|t)_{2l}=(s|t)_{2l-1}+1$ for all $1\leq l< (n-i)/2$, and $(s|t)_0=0$, and $(s|t)_{n-i}=x$ for $n-i$ even, and $0$ otherwise. E.g., for $i=n$, $E(s|t)=\delta_{s|t, 0}$, so the formula coincides with the $\mod 2$ reduction of the Euler class in Eq.~\eqref{eq:euler_class}. For $i=1$, $E(s|t)=1$ if $s=t$, and $E(s|t)$ equals the orientation of $s$ relative to $t$ if $t$ a $n$-simplex, that is, $E(s|t)=i \mod 2$ if $s$ equals $t$ with the $i$th vertex missing. Thus, $\omega_1$ precisely consists of the $n-1$-simplices where the orientation of the $n$-simplices would change if we choose an orientation locally.

The $i$th \emph{Pontryagin class} $P_i$ is a $\zz$-valued degree-$4i$ characteristic class in any spacetime dimension $n$. Unfortunately, local simplicial formulas for the Pontryagin classes comparable to the ones above in simplicity do not exist to date. Nonetheless, it has been shown in Ref.~\cite{Gaifullin2004} that there exist formulas for computing a $\mathbb{Q}$-valued $n-4i$-cycle $P_i$ representing the $i$th Pontryagin class locally from a triangulation. The value of $P_i$ on an $n-4i$-simplex depends only on its link and the orientation on the latter. The formula depends neither on a branching structure nor any other decoration of the triangulation. Ref.~\cite{Gaifullin2018} presents a more or less explicit formula computing the $\mathbb{Q}$-value of $P_1$ from the link of a $n-4$-simplex. It is conceivable that a formula for a $\zz$-valued instead of $\mathbb{Q}$-valued $n-4i$-cycle is possible when it is allowed to depend on a branching structure or other decorations. Note that unlike the two characteristic classes above, the Pontryagin classes are defined on \emph{oriented} manifolds, and reversing the orientation corresponds to a $\zz$-inversion of the Pontryagin class.

\subsection{Simplicial formulas for cohomology operations}
In this section, we will discuss how to implement cohomology operations with concrete simplicial formulas. The simplest cohomology operation is given by going from $A$-valued to $B$-valued (co-)cycles by applying a simplex-wise group homomorphism $A\rightarrow B$. This includes taking the simplex-wise sum of two $A$-valued $i$-(co-)cycles with the same $i$. Note that for $0$-cycles corresponding to all-scalar liquid models, taking the sum is the same as stacking the models on top of each other.

The most important cohomology operation for our purposes takes two input (co-)cycles and is called the \emph{cup product}. Intuitively, the cup product of two (co-)cycles is simply the (co-)cycle formed by the intersections of the corresponding networks of submanifolds. Accordingly, the cup product of an $x$-(co-)cycle $X$ and and a $y$-(co-)cycle $Y$ is a $x+y-n$-cycle ($x+y$-cocycle) $X\cup Y$. If we want to define the intersection of $X$ and $Y$ on a triangulation, we run into an ambiguity if $X$ and $Y$ coincide on some simplex. This ambiguity can be removed by infinitesimally shifting either $X$ or $Y$ in a direction which is determined by some additional decoration, such as a branching structure. Indeed, Ref.~\cite{Steenrod1947} gives a very simple formula for the cup product of cocycles which can be interpreted as shifting $X$ towards the $0$-vertex of every simplex,
\begin{equation}
\label{eq:cup_product_cocycle}
(X \cup Y)(s) = x(S^{x-}_{(0,\ldots,x)}(s)) \cdot y(S^{y-}_{(x,\ldots x+y)}(s))\;,
\end{equation}
where $s$ is a $x+y$-simplex. Note that in order to define the cup product we need $A$ to not only carry the structure of a group but of a ring, and the above product is the multiplication of that ring. Unfortunately, we cannot directly apply this formula to the characteristic classes from the previous section, since those are given as cycles and not cocycles. To define the cup product on cycles we need an analogue of a branching structure on the cellulation dual to the triangulation, which is precisely given by the notion of a dual branching introduced in Section~\ref{sec:vertex_liquid_higher}.

We start by giving a recipe to turn an $i$-cycle into an $n-i$-cocycle inside a $0$-dual-branched triangulation (cf.~Section~\ref{sec:vertex_liquid_higher}). The construction depends on a couple of choices which yield different, but equivalent formulas. For every representative of a star $L$ of a $y$-simplex with $y\leq i$ and every $n-i+y+1$-coboundary $B$ on the interior simplices of $L$, we need to choose a trivialization, i.e., a $n-i+y$-cochain $I(L,B)$ on the interior simplices of $L$, such that
\begin{equation}
dI(L,B)=B\;.
\end{equation}
For $y=i$, $B$ would be an $n+1$-cocycle and is thus not defined. In this case, we choose $I(L, \{\})$ to be a $n$-cocycle consisting of a single $n$-simplex (it would also suffice for it to be in the homology class of $1$, e.g., consisting of an odd number of $n$-simplices for $\zz_2$-valued cycles). Note that such a choice of trivialization is a choice on the level of star representatives and not a choice for the stars of concrete simplices in a triangulation, and does therefore not correspond to adding a new decoration to the triangulations.

With the choice of $I$, given a $i$-cycle $c$, we can obtain a shifted $n-i$-cocycle $c^*$ as follows. For every $y$-simplex $Y$ with $y\leq i$, we construct a $n-i+y$-cochain $\Phi(Y)$ on the interior simplices of the star $\star(Y)$ in the following way. For $y=i$, we set
\begin{equation}
\Phi(Y)= c(Y)\cdot  I(\star(Y), \{\})\;.
\end{equation}
Then, we iteratively set
\begin{equation}
\Phi(Y) = I(\star(Y), \sum_{X\in S^{(n-i+y+1)+}(Y)} \Phi(X))\;,
\end{equation}
where we think of $\Phi(X)$ as embedded in $\star(Y)$. Then, we use
\begin{equation}
c^* = \sum_{v\in S_0(T)} \Phi(v)\;.
\end{equation}

Using the formula for the Poincar\'e dual above, we can define the \emph{linking number} $a \linking b$ between an $i$-boundary $a$ and a $n-i-1$-boundary $b$ inside a $0$-dual-branched triangulation of an $n$-sphere. It is obtained by constructing the $i+1$-coboundary $b^*$, choosing a trivialization $x$ of $b^*$, i.e., an $i$-coboundary $x$ such that $d x=b^*$, and then simply calculating the overlap of $a$ and $x$ on the $i$-simplices,
\begin{equation}
a\linking b = \sum_{s\in S_i} a(s)\cdot x(s)\;.
\end{equation}

Using the linking number, we can construct the cup product of an $a$-cycle $A$ and a $b$-cycle $B$ as the following $a+b-n$-cycle $A\cup B$. The link of an $a+b-n$-simplex $s$ in a $1$-dual-branched triangulation is a $a+b-1$-dimensional triangulation which is $0$-dual-branched. Within the link, $A$ becomes a $n-b-1$-boundary $\widetilde A$, and $B$ a $n-a-1$-boundary $\widetilde B$ by identifying each $n-b-1$-simplex of the link with the corresponding internal $a$-simplex of the star. We then set
\begin{equation}
(A\cup B)(s) = \widetilde{A}\linking \widetilde{B}\;.
\end{equation}

Let us give a small example to illustrate the construction of the cup product for cycles. The simplest example is the cup product of two $1$-cycles $A$ and $B$ in two dimensions. The star of an edge consists of the two adjacent triangles. If the branching structure of the two triangles is reflection-symmetric around the edge, there are two different possible identifications with the standard representative if there is no orientation of the $2$-manifold. An identification can be indicated by specifying a favourite adjacent triangle of the edge, i.e., giving it a dual direction. So a 1-dual-branched triangulation is one where all edges have dual directions. The link of a vertex $v$ is an $l$-gon, and $A$ and $B$ turn into $0$-boundaries $\widetilde A$ and $\widetilde B$ on that $l$-gon. The vertices of the link have favourite edges, due to the dual directions of the corresponding internal edges of the star. We can turn $\widetilde B$ into a $1$-boundary $\widetilde B^*$ by shifting it from each vertex to its favourite edge. Now imagine moving $\widetilde A$-coloured vertices and annihilating pairs of them till there are none left, and collecting a $\zz_2$-valued summand $1$ for every time moving an $\widetilde A$-vertex past a $\widetilde B^*$-edge. The resulting value is the linking number of $\widetilde A$ and $\widetilde B$ which is the value of $A\cup B$ on $v$. Using this procedure for both $A$ and $B$ equal to the characteristic class cycle $\omega_1$ yields exactly the $(-1)^{\omega_1^2} = (-1)^{\omega_1\cup \omega_1}$ vertex-liquid model from Section~\ref{sec:all_scalar_vertex_11d}.

Another cohomology operation is the so-called \emph{Bockstein homomorphism}, which can be defined for any short exact sequence
\begin{equation}
B\overset{f}{\rightarrow} C\overset{g}{\rightarrow} A\;.
\end{equation}
$A$, $B$ and $C$ are Abelian groups, $f$ is an injective and $g$ a surjective group homomorphism, and $g\circ f$ is the trivial homomorphism sending everything to the identity of $A$. The Bockstein homomorphism $\beta$ has a simple combinatorial formula in terms of triangulations, and maps an $i$-cocycle $c$ to an $i+1$-cocycle
\begin{equation}
\beta(c) = f^{-1}(d(g^{-1}(c)))\;,
\end{equation}
where both $g^{-1}$ and $f^{-1}$ are applied simplex-wise. Different choices for the right inverse $g^{-1}$ of $g$ yield different, but equivalent local formulas. The formula for cycles instead of cocycles is completely analogous. Of particular importance is the Bockstein homomorphism for the short exact sequence
\begin{equation}
\label{eq:bockstein_zz2}
\zz\overset{\cdot 2}{\rightarrow} \zz \overset{\mod 2}{\rightarrow} \zz_2\;.
\end{equation}
It can be used to map a $\zz_2$-valued $i$-cocycle to a $\zz$-valued $i+1$-cocycle.

Another famous example for a cohomology operation is the so-called $k$th \emph{Steenrod square} $\operatorname{Sq}^k(x)$, which turns a $\zz_2$-valued $i$-cocycle $x$ into an $i+k$-cocycle for $1\leq k\leq i$. An explicit combinatorial formula for cocycles can be found in terms of a \emph{higher-order cup product} in Ref.~\cite{Steenrod1947} and Ref.~\cite{Gaiotto2015},
\begin{equation}
\operatorname{Sq}^k(x)=x\cup_{i-k}x\;.
\end{equation}
By shifting cycles to cocycles and back using the dual branching and branching, we can obtain a formula transforming a $k$-cycle into a $k-i$-cycle.

\subsection{Local equivalence between combinatorial formulas}
\label{sec:characteristic_equivalence}
Characteristic classes correspond to cohomology classes of the classifying spaces $BO(n)$ or $BSO(n)$. Roughly, concrete simplicial formulas for characteristic classes correspond to specific cocycles of the classifying space representing those classes, which are pulled back via the classifying map. Different cocycles $c_1$ and $c_2$ representing the same cohomology class of the classifying space are related by a boundary,
\begin{equation}
c_1-c_2 = du\;.
\end{equation}
Pulling back this relation, we get a local formula for a cochain $u$ which trivializes the difference between $c_1$ and $c_2$ on any triangulation. In general, we will call a local formula for a (co-)cycle like $c_1-c_2$ which is trivialized by another local formula for a (co-)chain \emph{locally trivial}, and we will call local formulas such as $c_1$ and $c_2$ \emph{locally equivalent}. If two $0$-cycle/$n$-cocycle local formulas are locally equivalent, we can use the trivializing $1$-chain/$n-1$-chain local formula to construct an invertible domain wall between the corresponding all-scalar liquid models. So local equivalence of local formulas can be seen as a generalization of phase equivalence of all-scalar liquid models.

The $\zz_2$-valued or $\zz$-valued cohomology of the classifying spaces of $BO(n)$ and $BSO(n)$ are known and can be found in the literature. Composing local formulas of characteristic classes and cohomology operations is the same as applying those cohomology operations to the corresponding cocycles on the classifying space. As announced, all cohomology classes of $BO(n)$ or $BSO(n)$ can be obtained by applying cohomology operations to the Euler class, Stiefel-Whitney classes, and Pontryagin classes. Not all combinations of generating characteristic classes and cohomology operations are different, but some might result in the same cohomology classes of the classifying space, such that their local formulas are locally equivalent. Analogously, different combinations of cohomology operations alone might represent the same cohomology class of $B^i(A)$, and consequently, their local formulas can be trivialized by some (co-)chain local formula taking the same inputs.

Let us give some examples of combinations of cohomology operations resulting in the same cohomology class of (cartesian products of different) $B^i(A)$, and explicitly look at the local equivalence of the corresponding local formulas. First of all, according to its interpretation as intersections, the cup product should be associative,
\begin{equation}
(X\cup Y)\cup Z=X\cup (Y\cup Z)\;.
\end{equation}
This local equivalence is in fact an equality for cocycles if we use the cup product formula in Eq.~\eqref{eq:cup_product_cocycle}. Another equality for this cup product formula is given by its bilinearity,
\begin{equation}
X\cup (Y+Z)=X\cup Y+X\cup Z\;.
\end{equation}
Another expected local equivalence is the graded-commutativity of the cup product,
\begin{equation}
X\cup Y=(-1)^{ij}Y\cup X\;,
\end{equation}
where $X$ is an $i$-cocycle and $Y$ a $j$-cocycle. 
This local equivalence is \emph{not} an equality of cocycles since the cup product formula in Eq.~\eqref{eq:cup_product_cocycle} is not symmetric in $A$ and $B$. However, using the formula in theorem 5.1 of Ref.~\cite{Steenrod1947} we see that a trialization of is given in terms of the higher-order cup product,
\begin{equation}
X\cup Y-(-1)^{ij}Y\cup X = d((-1)^{i+j+1} X\cup_1 Y)\;.
\end{equation}
All in all, different orderings, placing brackets, or expansions of cup products and sums yield locally equivalent formulas. We will consequently drop the brackets and write $AB$ instead of $A\cup B$.

The $\zz_2$-valued cohomology classes of $BO(n)$ are given by polynomials in Stiefel-Whitney classes, where the addition is simplex-wise $\zz_2$ addition and multiplication is the cup product. For $BSO(n)$ we get the same except that $\omega_1$ is trivial. Accordingly, all other cohomology operations applied to Stiefel-Whitney classes can be expressed again as a polynomial of Stiefel-Whitney classes. E.g., for the Steenrod squares of Stiefel-Whitney classes, we have
\begin{equation}
\label{eq:wu_formula}
Sq^i(\omega_j) = \sum_{t=0}^i \binom{j-i-1+t}{t} \omega_{i-t} \omega_{j+t}\;,
\end{equation}
for all $i$ and $j$ which is known under the name \emph{Wu formula} (cf.~page 197 in Ref.~\cite{May1999}).

The $\zz$-valued cohomology of $BO(n)$ or $BSO(n)$ is more complicated to describe \cite{Brown1982}. It consists of polynomials generated by the Euler class, Pontryagin classes, but also of the Bockstein homomorphisms under Eq.~\eqref{eq:bockstein_zz2} of products of different even-degree Stiefel-Whitney classes $\omega_{2i}$, subject to several equivalences. The $\mod 2$ reduction (which is a cohomology operation corresponding to a group homomorphism) of $\zz$-valued characteristic classes are $\zz_2$-valued characteristic classes, and thus again equivalent to polynomials of Stiefel-Whitney classes. The $\mod 2$ quotient of a Bockstein homomorphism is known to be the same as the Steenrod square $Sq^1$, which applied to a product of Stiefel-Whitney classes can be obtained from Eq.~\eqref{eq:wu_formula}. For the Pontryagin classes, the relation is
\begin{equation}
\label{eq:pontryagin_mod2}
P_i \mod 2 = \omega_{2i}^2\;,
\end{equation}
cf.~page 181 in Ref.~\cite{Milnor1974}.

\subsection{Local versus global equivalence}
\label{sec:local_versus_global}
If two liquid models are in the same exact phase, then they yield the same numbers when evaluated on different manifolds. However, the converse it not necessarily true -- there can be different exact phases which still produce the same global invariants. The analogue of this for local formulas of characteristic classes is the following. The cycle that a locally trivial formula associates to different manifolds are always boundaries, but the converse is not true. There can be characteristic classes/local formulas which are not locally trivial, but always yield boundaries when applied to the tangent bundle of a manifold (they will, however, always be non-trivial on some bundle which is not a tangent bundle). In this section we will give some examples for this.

The so-called \emph{Wu relation} states that the $i$th Steenrod square of a $\zz_2$-valued $n-i$-cocycle $X$ in $n$ dimensions is equivalent to its cup product with a degree-$i$ characteristic \emph{Wu class} $\nu_i$ of the tangent bundle,
\begin{equation}
\label{eq:wu_relation}
\operatorname{Sq}^i(X)=\nu_i\cup X\;.
\end{equation}
$\nu_i$ can be expressed as a polynomial in Stiefel-Whitney classes, and for $i=1,2,3$ we have $\nu_1=\omega_1$, $\nu_2=\omega_1^2+\omega_2$, and $\nu_3=\omega_1\omega_2$. The equation only holds as an equation between the resulting homology classes for arbitrary manifolds and $X$. Both sides can be interpret as degree-$i$ cohomology classes in $B^{n-i}(\zz_2)\times BO(n)$, pulled back via the cartesian product of the classifying map of $X$ and the classifying map of the tangent bundle. Those cohomology classes are different since the left-hand side is trivial on the $BO(n)$ part in contrast to the right-hand side, and accordingly the equation only holds for the tangent bundle but not arbitrary vector bundles. Thus, two sides of the equation will not be be locally equivalent as local formulas. So two vertex-liquid models related by the application of Wu relation above need to evaluate to the same numbers on manifolds, but can be in distinct exact phases.

Part of the definition of the Steenrod square is $\operatorname{Sq}^i(X)=0$ if $X$ is a $j$-cocycle with $j<i$, where $0$ denotes the trivial chain. Thus, we must have $\nu_j=0$ on $n$-manifolds for $j>n/2$. This yields equations for the cohomology classes resulting from polynomials of Stiefel-Whitney classes applied to the tangent bundle, such as $\nu_2=\omega_1^2+\omega_2=0$ on 2-manifolds and 3-manifolds, or $\omega_1\omega_2=0$ on 3-, 4- and 5-manifolds. This implies, e.g., that the liquid models $(-1)^{\omega_1^2}$ and $(-1)^{\omega_2}$ evaluate to the same numbers on any 2-manifold. However, we have argued in Section~\ref{sec:all_scalar_vertex_11d} that the latter two models are in different exact phases. This can now be explained by the fact that different polynomials of Stiefel-Whitney classes are always different cohomology classes of $BO(n)$, and thus local formulas for $\omega_1^2$ and $\omega_2$ cannot be locally equivalent.

Another relation between characteristic classes which is not a local equivalence follows from the \emph{Hirzebruch signature theorem} (cf.~page 86 in Ref.~\cite{Hirzebruch1966}),
\begin{equation}
\sigma(X)=\int_X P_1/3\;,
\end{equation}
for all $4$-manifolds $X$, where $\sigma$ is the \emph{signature} of $X$. That is, it is the signature of the matrix whose rows and columns correspond to the generators of the degree-$2$ $\zz$-valued cohomology classes and whose entries are the integrals over cup products between the latter. Since the signature is an integer, we conclude that
\begin{equation}
\label{eq:p1mod3}
P_1 \mod 3 = 0\;.
\end{equation}
as an equation between the resulting cohomology classes of the tangent bundle of a $4$-manifold. However, this $P_1\mod 3$ is not trivial on the classifying space and also not trivial on arbitrary vector bundles. Hence, we cannot expect that there is a local formula for a trivialization of $P_1\mod 3$ on triangulations, and $e^{\frac23\pi i P_1}$ is an all-scalar vertex-liquid model which evaluates to $1$ on every $4$-manifold but which is not in a trivial exact phase.

%Interestingly, when we restrict to real numbers, then also the invertible boundary of the odd-dimensional $\alpha^\chi$ model with negative $\alpha$ fails, since, e.g., there is no real number $\alpha^{1/2}$.

%In summary, we have seen that there are three different notions of equivalence between expressions involving characteristic classes, cup products and Steenrod squares. Either, the local formulas are directly equal, or they are unequal but there is a local formula computing a trivialization of the difference of the resulting cycles, or are they in the same homology class but there is no locally computable trivialization.

\section{More discussion of the CYWW model}
\label{sec:CYWW_detail}
In this appendix, we look at the CYWW model in more detail. To start with, we will support the claim that the modular CYWW bulk is in a trivial topological phase. A good first indication is that the model is known to be \emph{invertible}. Even though there are examples of invertible but non-trivial phases with fermions (such as the Kitaev chain), with symmetries (such as SPTs), or with chirality (such as the intrinsic bosonic $E_8$ invertible phase), all invertible non-chiral intrinsic bosonic topological phases in $2+1$ and $1+1$ dimensions turn out to be trivial, and we expect that the same is true in $3+1$ dimensions. Note that invertibility implies that the ground state is non-degenerate on any manifold.

Second, the modular CYWW model does not possess any defects of co-dimension 2, 3, and 4. Along the lines of Section~\ref{sec:defects}, membrane defects (co-dimension 2) are in one-to-one correspondence with the boundaries of the $3$-dimensional model obtained by compactifying one of the $4$ dimensions into a circle. Roughly, if we look at the ground state space of the compactified model on $M\times S_1$ for a 2-manifold $M$, we can see that we can create pairs of small anyon loops around $S_1$ and move them around. As in a UMTC all anyons braid non-trivially with some other anyons, this forces the anyon network on $M$ to be trivial and leads to a trivial ground state space. Line defects (co-dimension 3) are in one-to-one correspondence with the boundaries of the $2$-dimensional model obtained by compactifying two of the $4$ dimensions into a sphere, corresponding to the ground states of the CYWW model on $S_1\times S_2$. Again, intuitively, we can create an anyon loop and pull it around $S_2$, which will force any loop around $S_1$ to be trivial. Point defects (co-dimension 4), potentially singular, are the same as ground states on different manifolds. As the ground states form a 1-dimensional vector space on any manifold, the point defects are trivial.

Third, the evaluation of the CYWW on a closed $4$-manifold $X$ is given by
\begin{equation}
Z(X)=e^{ic\pi \sigma(X)/4}\;,
\end{equation}
and is an exponential of the \emph{signature} $\sigma$ \cite{Walker2011}, defined in Appendix~\ref{sec:local_versus_global}, whose basis is determined by the chiral central charge $c$ of the input UMTC. Using the Hirzebruch signature theorem (cf.~Appendix~\ref{sec:classical_appendix}), we can express this in terms of the first Pontryagin class $P_1$,
\begin{equation}
Z(X) = (e^{2\pi i\frac{c}{24}})^{\int_X P_1}\;.
\end{equation}
So the evaluation of the modular CYWW model is precisely the same as that of the all-scalar anomaly liquid model to which we hope to find an invertible domain wall in the non-projective case, as depicted in Eq.~\eqref{eq:cyww_compactification_anomaly}. As we stressed Appendix~\ref{sec:local_versus_global}, the fact that the modular CYWW evaluates to the same numbers as the all-scalar anomaly model does not necessarily imply that the two are in the same exact phase, but it can be seen as a good indication. In any case, this provides a very neat interpretation of the modular CYWW as a physically non-trivial way of writing the chiral anomaly of the chiral model given by its standard boundary.

Furthermore, let us now argue why the compactified $2+1$-dimensional model obtained from the modular CYWW model indeed represents the chiral phase given by the input UMTC. We sketch how to show that the anyon content of the $2+1$-dimensional model above is described by precisely this UMTC. The anyons of the (compactified) $2+1$-dimensional model are in one-to-one correspondence with the boundaries of the $1+1$-dimensional model obtained by another compactification with a circle. The two compactifications can be combined into a single compactification of the CYWW with an annulus with one invertible and one cone boundary. Using the topology-changing moves corresponding to invertibility, the invertible-boundary puncture can be filled, such that we are left with the compactification with a cone-boundary disk,
\begin{equation}
\begin{tikzpicture}
\draw[blue, line width=2] (0,0)circle(0.6);
\draw[red, line width=2] (0,0)circle(0.7);
\draw[line width=2] (0,0)circle(0.65);
\end{tikzpicture}
=
\begin{tikzpicture}
\fill[even odd rule, opacity=0.3] (0,0)circle(0.8) (0,0)circle(0.4);
\draw[blue, line width=2] (0,0)circle(0.8);
\draw[red, line width=2] (0,0)circle(0.4);
\end{tikzpicture}
=
\begin{tikzpicture}
\fill[opacity=0.3] (0,0)circle(0.7);
\draw[blue, line width=2] (0,0)circle(0.7);
\end{tikzpicture}
\;.
\end{equation}

Thus, the anyons of the $2+1$-dimensional model are in one-to-one correspondence to the boundaries of the $1+1$-dimensional model obtained from the disk compactification above. The latter is a GHZ-type model, whose boundaries are in one-to-one correspondence with the ground states on $S_1$. Those ground states are the same as the ground states of the CYWW model on the solid torus $S_1\times B_2$. The ground states of the latter can be thought of as superpositions of networks of lines decorated with simple objects in the solid torus, modulo local relations given by the $F$- and $R$-symbols of the UMTC. Since $B_2$ is simply connected, a set of independent anyon networks spanning the ground state space of the CYWW model on the solid torus is given by a simple object along $S_1\times x$ with some fixed $x\in B_2$ for any simple object. So the anyons of the $2+1$-dimensional model are in one-to-one correspondence with the simple objects of the UMTC. Showing that also the fusion and braiding coincides would require a little more work.

\section{The three-fermion phase}
\label{sec:three_fermion}
%A particularily simple chiral modular CYWW model is the one for the \emph{3-fermion UMTC}, which has a $\zz_2$-valued flavor as its chiral central charge is $4\mod 8$. Concretely, it can be written as a $2$-cocycle gauge theory with an action depending on two $\zz_2$-valued simplicial $2$-cocycles. Furthermore, the all-scalar anomaly liquid for the conjectured 3-fermion liquid model takes the form $(-1)^{\omega_2^2}$, for which we can concretely write down the vertex scalars, in contrast to the general anomaly determined by the first Pontryagin class. We therefore expect the 3-fermion phase to be the easiest candidate for finding an invertible domain wall between the CYWW and anomaly model, or directly a vertex-liquid model.

In this appendix, we describe attempts to discover a liquid model for the possibly simplest chiral phase, the 3-fermion phase.  The latter is given by a UMTC with $4$ anyons, the trivial one and three fermions. The fusion rules are the same as for the $e$ and $m$ anyons in the toric code, but the braiding is different. The three-fermion phase is a promising candidate for posessing a reasonably simple fixed-point model for a number of reasons.

First, it is one of the simplest chiral UMTCs. There are only a few UMTCs with less anyons, including the semion UMTC, or the Fibonacci UMTC. Among them, only the semion UMTC and the $\zz_3$ UMTC have Abelian fusion rules \cite{Rowell2007}, i.e., unique fusion outcome.

Second, the $F$- and $R$-symbols of the three-fermion UMTC can be chosen purely real. Consequently, the three-fermion CYWW model, which is directly built from the $F$- and $R$ symbols, is a liquid model with real tensors on unoriented triangulations. Physically, reality of the tensors corresponds to imposing a time-reversal symmetry for bosonic models. Accordingly, it has been conjectured that the three-fermion CYWW is a non-trivial SPT phase protected by time-reversal symmetry \cite{Burnell2013}.

Third, in accordance with reality, the chiral central charge of the three-fermion UMTC is $4\mod 8$. If we pick the representative with $c=12$, the chiral anomaly also becomes a purely real all-scalar liuqid model,
\begin{equation}
\label{eq:3fermion_anomaly}
e^{2\pi i \frac{12}{24} P_1}=(-1)^{P_1} = (-1)^{\omega_2^2}\;,
\end{equation}
where we used Eq.~\eqref{eq:pontryagin_mod2} from Appendix~\ref{sec:characteristic_equivalence}, and the equations are to be understood as local equivalences of all-scalar liquid models. Note that in contrast to the general $P_1$ anomaly, a simple combinatorial formula is known for its mod $2$ reduction $\omega_2^2$.

Fourth, the $F$- an $R$-symbols, and hence the tensors of the three-fermion CYWW model, are \emph{stabilizer tensors}, that is, they are tensors which are unique ground states of some stabilizer code. Stabilizer tensors are a subset of array tensors for which tensor networks can be efficiently stored and evaluated even for large networks with many indices. It is thus tempting to think that, in our search for an invertible boundary of the three-fermion CYWW model, we can restrict to models consisting of stabilizer tensors as well.

Fifth, the three-fermion CYWW model can be formulated in terms of simplicial cohomology (as in fact all Abelian-MTC CYWW models, for general cohomology operations). In general, the CYWW state-sum for an UMTC involves a sum over different anyon labels at every triangle of a $4$-dimensional cellulation. At every $4$-simplex, we take the $3$-dimensional triangulation corresponding to its boundary and thread the corresponding anyon world lines perpendicular through each triangle. Then, we connect the anyon world lines coming from the four adjacent triangles at every tetrahedron with fusions. Evaluating the resulting anyon-network in a $3$-sphere yields the $15$j-symbol associated to the $4$-simplex \cite{Williamson2016}. For the three-fermion UMTC, the anyon labels on the faces form the group $\zz_2\times\zz_2$. The fusion rules imply that the two $\zz_2$ factors of the anyons labelling the faces must be two $\zz_2$ $2$-cocycles (cf.~Appendix~\ref{sec:classical_appendix}) $A$ and $B$. The evaluation of the anyon network corresponding to the $15$j-symbol on the $4$-simplex is easy since the $F$-symbols are trivial. It suffices to apply the $R$-matrix once to the anyon world lines perpendicular to the triangles $012$ and $234$ of the branching-structure $4$-simplex. This results in a weight
\begin{equation}
(-1)^{A(012)A(234)+B(012)B(234)+A(012)B(234)}\;,
\end{equation}
at every $4$-simplex \cite{Chen2021}, which is the same as
\begin{equation}
\label{eq:three_fermion_cup}
(-1)^{A\cup A+B\cup B+A\cup B}
\end{equation}
evaluated on the $4$-simplex when we use the cup-product formula from Ref.~\cite{Steenrod1947}.

Sixth, and most importantly, there is the aforementioned QCA disentangling the three-fermion CYWW model found in Ref.~\cite{Haah2018}. An extension of the QCA to arbitrary cubulations can be found in Ref.~\cite{Fidkowski2019}. Moreover, a fully explicit formula for the QCA on a regular cubic lattice can be found in the appendix of Ref.~\cite{Shirley2022}, and the interpretation of the resulting Pauli terms have a rather straight-forward interpretation in terms of higher-order cup products \cite{Chen2021}. This allows us to express them on arbitrary cellulations.

The ultimate goal would be to find an invertible boundary of the three-fermion CYWW model, or more precisely, an invertible domain wall to the all-scalar anomaly liquid model in Eq.~\eqref{eq:3fermion_anomaly}. Note that in principle, the models $e^{2\pi i \frac16 P_1}$ and $e^{2\pi i \frac56 P_1}$ are possible anomalies as well which are evaluate to the same number but are in distinct exact phases. In fact, for the three-fermion CYWW, we can show quite simply that it evaluates to the same number as the suggested anomaly models. Indeed, following Ref.~\cite{Chen2021}, we can use the Wu relation (cf. Eq.~\eqref{eq:wu_relation}) to obtain
\begin{equation}
\begin{multlined}
\sum_{A,B} (-1)^{\int A\cup A+B\cup B+A\cup B}\\
= \sum_{A,B} (-1)^{\int (\omega_1^2+\omega_2+B)\cup A + B\cup B)} \\= \sum_B (-1)^{\int B\cup B} \sum_A (-1)^{\int (\omega_1^2+\omega_2+B)\cup A}
\\= \sum_B (-1)^{\int B\cup B} \delta_{0, \omega_1^2+\omega_2+B}\\= (-1)^{\int (\omega_1^2+\omega_2)^2} =
(-1)^{\int \omega_1^4+\omega_2^2} = (-1)^{\int \omega_2^2}\;.
\end{multlined}
\end{equation}
Note that the first equality is only on the level of evaluations and does not say that the two liquid models are in the same exact phase, since the Wu relation does not correspond to a local equivalence as we stressed in Appendix~\ref{sec:local_versus_global}. All other equality signs do in fact correspond to phase equivalences. The last equation holds for combinatorial oriented 4-manifolds, where $\omega_1$ has a trivializing $4$-cycle by definition such that any term containing $\omega_1$ disappears.

Due to the triviality of the $4$th Wu class on $4$-manifolds, we have
\begin{equation}
\nu_4 = \omega_1^4+\omega_2^2+\omega_1\omega_3+\omega_4=0
\end{equation}
as equation between cohomology classes. Thus,
\begin{equation}
\omega_2^2 = \omega_4 = \chi \mod 2
\end{equation}
on oriented manifolds, where the first equality is between cohomology classes whereas the second equality is a local equivalence. Thus, judging from the evaluation only, the $(-1)^\chi$ all-scalar model would be another candidate for being in the same phase as the three-fermion CYWW as well. However, this model cannot represent any non-trivial anomaly, since it does have an all-scalar topological boundary as the Euler characteristic is defined on manifolds with boundary. If $(-1)^\chi$ was the anomaly of a $2+1$-dimensional liquid model, we could compactify with an interval of $(-1)^\chi$ with the $2+1$-dimensional model on the one and the all-scalar boundary on the other side. This would yield the $2+1$-dimensional model without anomaly. Contrary, attempts to construct all-scalar boundaries seem to consistently fail for the $(-1)^{\omega_2^2}$ model, though we cannot give any proof for this to date.

As we have expressed the 3-fermion CYWW model as a sum over 2-cocycles weighted by a cup product formula, one might hope that the hypothetical invertible boundary can also be expressed in this language. However, this cannot be exactly the case, since the cup product has a formula where its value on an $x$-simplex only depends on the values of sub-simplices, and consequently can be written as a simplex-liquid model. This would thus yield a simplex-liquid model for the three-fermion phase, which is impossible as argued in the previous section. However, it is possible for the similar toric-code CYWW model. The partition function of the latter can also be written as a sum over two 2-cocycles, this time only weighted by a single cup product,
\begin{equation}
\label{eq:toric_code_cup}
\sum_{A,B} (-1)^{\int A\cup B}
\end{equation}
(Co-)cycles and the corresponding boundary maps can be extended to manifolds with boundary in two distinct ways. For \emph{closed} cocycles, the boundary map at the boundary is the same as in the bulk and given by summing over values on adjacent simplices. So intuitively, closed cocycles are still given by patterns of closed loops, membranes, etc., inside a manifold with boundary. On the other hand, for \emph{free} cocycles, the boundary map is trivial at the boundary. Thus, free cocycles are patterns of loops, membranes, etc. which are closed in the inside of the manifold, but are allowed to end freely at the boundary.

The cup product of two closed cocycles inside a manifold with boundary can be defined as before. It yields a closed cocycle, and is still compatible with the closed boundary map, i.e., the resulting cohomology class only depends on the cohomology class of the two input cocycles. The same is true for the cup product of a closed and a free cocycle. In contrast, the cup product of two free cocycles is not compatible with the boundary map, as the following example of a green and red free 1-cocycle inside a 2-manifold with blue boundary shows,
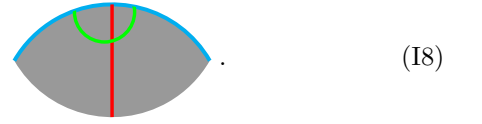
\begin{equation}
\begin{tikzpicture}
\fill[manifold] (30:1.5)arc(30:150:1.5)arc(-150:-30:1.5);
\draw[manifoldboundary] (30:1.5)arc(30:150:1.5);
\clip (30:1.5)arc(30:150:1.5)arc(-150:-30:1.5);
\draw[line width=1.3,red] (0,-1)--(0,2);
\draw[line width=1.3,green] (-0.1,1.4)circle(0.4);
\end{tikzpicture}\;.
\end{equation}
The green cocycle is a boundary as a free cocycle and thus in the trivial cohomology class, yet it has an odd intersection number with the red cocycle.

The standard CYWW boundary of both the three-fermion CYWW model in Eq.~\eqref{eq:three_fermion_cup} and the toric-code CYWW model in Eq.~\eqref{eq:toric_code_cup} corresponds to extending both $A$ and $B$ as closed 2-cocycles. For the toric-code CYWW model, the invertible boundary corresponds to extending $A$ as closed 2-cocycle, but $B$ as free 2-cocycle (or vice versa). In principle, the same type of boundary would also work for the three-fermion CYWW model. However, the term $B\cup B$ in the action is not consistent with cohomology for a free $B$. The cup product $B\cup B$ can be obtained by shifting one copy of $B$ from a 2-cocycle to a 2-cycle using the branching structure, and then taking the intersection of $B$ with the shifted $B$. Let $B_\partial$ be the $1$-cocycle we get from restricting $B$ to the boundary. For $B$ being a 2-boundary, $B\cup B$ is given by the linking number of $B_\partial$ with its shifted copy. So we would have to find a path integral defined on $B_\partial$-coloured $3$-triangulations which evaluates to the linking number of $B_\partial$ with its shifted copy if $B_\partial$ is a boundary.

Given a trivialization $X$ of $B_\partial$, the self-linking number can be obtained by $X\cup B_\partial$, and by summing over all possible $X$ we would obtain a path integral evaluating to the self-linking number. However, summing over trivializations of $B_\partial$ inside the boundary is equivalent to having $B$ closed instead of free. So the resulting boundary would be the standard boundary which is not invertible. The latter would be guaranteed by using an all-scalar path integral evaluating to the self-linking of $B_\partial$. However, this cannot exist since the self-linking of a single large loop $B_\partial$ can be changed by applying a Dehn twist along another loop linked with the $B_\partial$-loop which can be arbitrarily far away from $B_\partial$. Thus, the self-linking cannot be determined only by the triangulation in a constant-size environment of $B_\partial$. So in order to construct an invertible boundary we would need something in between summing over trivializations of $B_\partial$ and an all-scalar path integral depending on $B_\partial$. Moreover, the path integral cannot be a state sum of a simplex-liquid type. Attempting to find such a path integral will be left for future work.

\end{document}